\documentclass[12pt,tightenlines,superscriptaddress]{revtex4PATCH}

\usepackage[page,title]{appendix}
\usepackage{url}
\usepackage{graphicx}
\usepackage{amssymb}
\usepackage{bm}
\usepackage{mathrsfs}
\usepackage{amsmath}
\usepackage{amsfonts}
\usepackage{color}
\usepackage{xspace}
\usepackage{tikz}
\usepackage{slashed}
\usepackage{cancel}
\usepackage{hyperref}
\usepackage[utf8]{inputenc}											
\usepackage{slashed}

\usetikzlibrary{arrows}
\usetikzlibrary{shapes}
\usetikzlibrary{trees}
\usetikzlibrary{matrix}
\usetikzlibrary{arrows} 			
\renewcommand{\draft}[1]{{\color{blue}#1}}

\usepackage{graphicx,epsfig,psfrag,bm,amssymb}
\usepackage{dcolumn}
\usepackage{bm}
\usepackage{color}
\usepackage{mathrsfs,amsfonts,hepunits,color}
\usepackage[utf8]{inputenc}
\usepackage{epsfig,latexsym,cancel,amssymb,amsmath,mathrsfs}
\usepackage{graphicx}
\usepackage{epstopdf}
\usepackage{mciteplus}
\usepackage{latexsym}
\usepackage{amsthm}
\usepackage{amsmath}
\usepackage{amssymb}
\usepackage{hepunits}
\usepackage{hyperref}
\usepackage{bbm}
\usepackage{bm}
\usepackage{xfrac}
\usepackage{colordvi}
\usepackage{comment}
\usepackage{dcolumn}
\usepackage{times,latexsym,graphicx,wrapfig}
\usepackage{epsfig,lineno,bm}
\usepackage{slashed}
\usepackage{booktabs}
\usepackage{relsize}

\newcommand{\be}{\begin{eqnarray}}
\newcommand{\ee}{\end{eqnarray}}

\newcommand{\benum}{\begin{enumerate}}
\newcommand{\eenum}{\end{enumerate}}
\newcommand{\bi}{\begin{itemize}}
\newcommand{\ei}{\end{itemize}}
\newcommand{\geant}{\textsc{Geant4}\xspace}

\newcommand{\ecal}{ECal\xspace}
\newcommand{\hcal}{HCal\xspace}
\newcommand{\pn}{photo-nuclear\xspace}
\newcommand{\en}{electro-nuclear\xspace}
\newcommand{\aprime}{$A^{\prime}$\xspace}
\def\babar{\mbox{\slshape B\kern-0.1em{\smaller A}\kern-0.1emB\kern-0.1em{\smaller A\kern-0.2em R}}}

\newcommand{\ednote}[1]{} 
\newcommand{\people}[1]{} 
\newcommand{\morepeople}[1]{} 

\colorlet{RED}{red}
\colorlet{BLUE}{blue}
\colorlet{ORANGE}{orange}

\begin{document}


%
%
\title{Light Dark Matter eXperiment (LDMX)}

\date{\today}
\hspace*{0pt}\hfill 
{\small FERMILAB-PUB-18-324-A, SLAC-PUB-17303}

\bigskip

\author{Torsten~Åkesson}
\affiliation{Lund University, Department of Physics, Box 118, 221 00 Lund, Sweden}

\author{Asher Berlin}
\affiliation{SLAC National Accelerator Laboratory, Menlo Park, CA 94025, USA}

\author{Nikita Blinov}
\affiliation{SLAC National Accelerator Laboratory, Menlo Park, CA 94025, USA}

\author{Owen~Colegrove}
\affiliation{University of California at Santa Barbara, Santa Barbara, CA 93106, USA}

\author{Giulia~Collura}
\affiliation{University of California at Santa Barbara, Santa Barbara, CA 93106, USA}

\author{Valentina~Dutta}
\affiliation{University of California at Santa Barbara, Santa Barbara, CA 93106, USA}

\author{Bertrand~Echenard}
\affiliation{California Institute of Technology, Pasadena, CA 91125, USA}

\author{Joshua~Hiltbrand}
\affiliation{University of Minnesota, Minneapolis, MN 55455, USA}

\author{David~G.~Hitlin}
\affiliation{California Institute of Technology, Pasadena, CA 91125, USA}

\author{Joseph~Incandela}
\affiliation{University of California at Santa Barbara, Santa Barbara, CA 93106, USA}

\author{John~Jaros}
\affiliation{SLAC National Accelerator Laboratory, Menlo Park, CA 94025, USA}

\author{Robert~Johnson}
\affiliation{Santa Cruz Institute for Particle Physics, University of California at Santa Cruz, Santa Cruz, CA 95064, USA}

\author{Gordan~Krnjaic}
\affiliation{Fermi National Accelerator Laboratory, Batavia, IL 60510, USA}

\author{Jeremiah~Mans}
\affiliation{University of Minnesota, Minneapolis, MN 55455, USA}

\author{Takashi~Maruyama}
\affiliation{SLAC National Accelerator Laboratory, Menlo Park, CA 94025, USA}

\author{Jeremy~McCormick}
\affiliation{SLAC National Accelerator Laboratory, Menlo Park, CA 94025, USA}

\author{Omar~Moreno}
\affiliation{SLAC National Accelerator Laboratory, Menlo Park, CA 94025, USA}

\author{Timothy~Nelson}
\affiliation{SLAC National Accelerator Laboratory, Menlo Park, CA 94025, USA}

\author{Gavin~Niendorf}
\affiliation{University of California at Santa Barbara, Santa Barbara, CA 93106, USA}

\author{Reese~Petersen}
\affiliation{University of Minnesota, Minneapolis, MN 55455, USA}

\author{Ruth~Pöttgen}
\affiliation{Lund University, Department of Physics, Box 118, 221 00 Lund, Sweden}

\author{Philip~Schuster}
\affiliation{SLAC National Accelerator Laboratory, Menlo Park, CA 94025, USA}

\author{Natalia~Toro}
\affiliation{SLAC National Accelerator Laboratory, Menlo Park, CA 94025, USA}

\author{Nhan~Tran}
\affiliation{Fermi National Accelerator Laboratory, Batavia, IL 60510, USA}

\author{Andrew~Whitbeck}
\affiliation{Fermi National Accelerator Laboratory, Batavia, IL 60510, USA}


\begin{abstract}
We present an initial design study for LDMX, the Light Dark Matter Experiment, a small-scale accelerator experiment having broad sensitivity to both direct dark matter and mediator particle production in the sub-GeV mass region.  LDMX employs  missing momentum and energy techniques in multi-GeV electro-nuclear fixed-target collisions to explore couplings to electrons in uncharted regions that extend down to and below levels that are  motivated by direct thermal freeze-out mechanisms. LDMX would also be sensitive to a wide range of visibly and invisibly decaying dark sector particles, thereby addressing many of the science drivers highlighted in the 2017 US Cosmic Visions New Ideas in Dark Matter Community Report. LDMX would achieve the required sensitivity by leveraging existing and developing detector technologies from the CMS, HPS and Mu2e experiments. In this paper, we present our initial design concept, detailed GEANT-based studies of detector performance, signal and background processes, and a preliminary analysis approach. We demonstrate how a first phase of LDMX could expand sensitivity to a variety of light dark matter, mediator, and millicharge particles by several orders of magnitude in coupling over the broad sub-GeV mass range.

\end{abstract}

\maketitle
\newpage
\tableofcontents
\newpage


\clearpage
\section{Introduction}
Discovering the particle nature of dark matter (DM) is perhaps the most pressing challenge facing elementary particle physics today. Among the simplest possibilities is one in which dark matter arose as a thermal relic from the hot early Universe, which only requires small non-gravitational interactions between dark and familiar matter, and is robustly viable over the MeV to TeV mass range. Testing the hypothesis that the dark matter abundance arises from weak boson-mediated interactions has been the primary focus of direct and indirect detection experiments to date, which are most sensitive to dark matter particles with masses ranging from a few GeV to a TeV. However, the lower mass range of MeV to GeV, where the most stable forms of ordinary matter are found, has remained stubbornly difficult to explore with existing experiments. 

In the last several years, powerful ideas to probe ``light dark matter'' (LDM) in the sub-GeV mass range have emerged from efforts to test the intriguing possibility that dark matter is part of a dark sector that is neutral under all Standard Model (SM) forces (see \cite{Alexander:2016aln,Battaglieri:2017aum} for recent reviews). As with Weakly Interacting Massive Particles (WIMPs), an attractive sensitivity milestone is motivated by the requirement that thermal freeze-out reactions give rise to an appropriate abundance of dark matter. This casts a spotlight on dark matter interactions with electrons that is only a few orders of magnitude beyond existing accelerator-based sensitivity. To reach this exciting goal, our aim is to use an electron beam to produce dark matter in fixed-target collisions, making use of missing energy and momentum to identify and measure dark matter reactions \cite{Izaguirre:2014bca,Izaguirre:2015yja}. The NA64 experiment at CERN has already carried out a first physics run for a fixed-target electron beam experiment using missing energy as the identifying signature~\cite{Banerjee:2016tad,Banerjee:2017hhz}. That experiment promises to reach a sub-GeV dark matter sensitivity surpassing all existing constraints by 2020 \cite{Alexander:2016aln,Gninenko:2016kpg}, but will nonetheless fall short of the required sensitivity primarily due to luminosity limitations. At masses above $\mathcal{O}(500)$ MeV, Belle II's future missing mass measurements might provide the required sensitivity, but will certainly fall short at lower masses due to luminosity and background limitations.

The ``Light Dark Matter eXperiment'' (LDMX) described in this note is designed to meet the following science goals: 
\begin{itemize}
\item Provide a high-luminosity measurement of missing momentum in multi-GeV electron fixed-target collisions, through {\bf both} direct dark matter and mediator particle production. This measurement would provide broad sensitivity to dark matter interactions over the entire sub-GeV mass range, extending below the $\mathcal{O}(\keV)$ warm dark matter mass bound \cite{Baur:2015jsy}, and would circumvent limitations inherent to non-relativistic probes of dark matter. LDMX would aim to extend sensitivity by three orders of magnitude beyond the expected reach of NA64 in the near future. Combined with Belle II's future missing mass measurements, this would largely provide the sensitivity needed to test most scenarios of dark matter freeze-out via annihilation into Standard Model final states below a few GeV, among many others. This goal was highlighted in \cite{Battaglieri:2017aum}; it will provide LDMX with excellent discovery potential, and is the primary science driver for the experiment. 
\item Using missing momentum measurements, explore broad and important new territory for secluded dark matter models, millicharge particles, invisibly decaying dark photons, axions, and dark higgs particles. By extension, explore significant new territory for SIMP \cite{Hochberg:2014kqa,Hochberg:2015vrg,Berlin:2018tvf}, ELDER \cite{Kuflik:2015isi,Kuflik:2017iqs}, 
asymmetric \cite{Petraki:2013wwa,Zurek:2013wia}, and freeze-in \cite{Dodelson:1993je,Hall:2009bx,Chu:2011be} dark matter scenarios.  
\item Using LDMX as a short baseline beam dump, provide sensitivity to displaced visibly decaying dark photons, axions, inelastic dark matter, dark higgs, and other long-lived dark sector particles.  Variations of LDMX with a {\it muon} beam can also explore
dark sectors whose particles couple preferentially to the second
generation \cite{Kahn:2018cqs}.
\end{itemize}
As a multi-purpose experiment, LDMX will be able to address an especially broad range of the dark sector science highlighted in the US Cosmic Vision New Ideas in Dark Matter Community Report \cite{Battaglieri:2017aum}, with special emphasis on the simplest thermal sub-GeV dark matter scenarios. We believe that LDMX, along with an appropriate set of complementary experiments, would therefore provide the foundation for a successful light dark matter program in the US or abroad. 

\begin{figure}
\includegraphics[width=16cm]{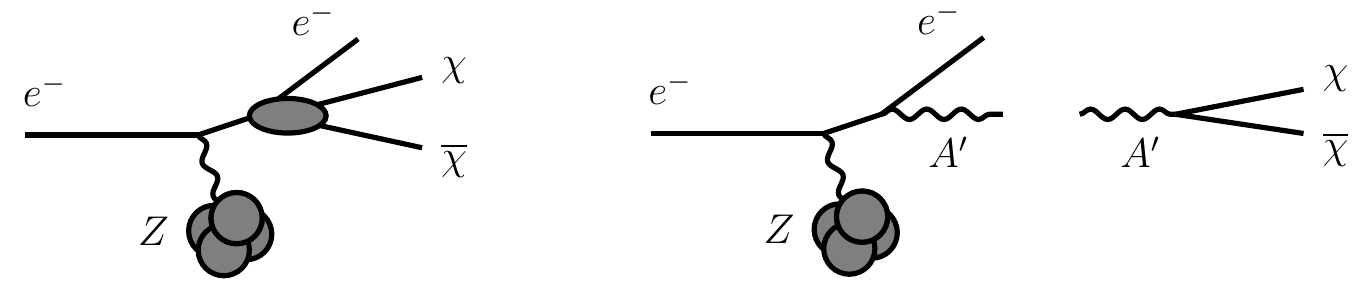}
\caption{\label{fig:BothSignalReactions}
Left panel: Feynman diagram for direct dark matter particle-antiparticle production. Right panel:
Feynman diagram for radiation of a mediator particle off a beam electron, followed by its decay into dark matter particles.
Measuring both of these (and similar) reactions is the primary science goal of LDMX, and will provide broad and powerful sensitivity to light dark matter and many other types of dark sector physics.}
\end{figure}

The design considerations for LDMX are as follows. An electron beam incident on a thin target can produce dark matter particles through a ``dark bremsstrahlung'' process, in which most of the incident electron's energy is typically carried away by the invisible dark matter. This can occur either through direct dark matter production (left panel of Fig.~\ref{fig:BothSignalReactions}), or through production of mediator particles that decay to dark matter (right panel of Fig.~\ref{fig:BothSignalReactions}).  
To search for either process, LDMX  reconstructs the kinematics of each beam electron both up- and down-stream of the target using low-mass tracking detectors. The up-stream tracker tags the incoming beam electrons while the down-stream tracker selects the low-energy, moderate transverse-momentum recoils of the beam electrons. Calorimetry is then used to veto events with an energetic forward photon or any additional forward-recoiling charged particles or neutral hadrons.  Because each electron passes through the detector, the experiment must contend with high event rates in the tracker and electromagnetic calorimeter.  Therefore, LDMX requires low-mass tracking that provides high-purity tagging for incoming electrons and clean, efficient reconstruction of recoils in a high-rate environment.  The calorimetry for LDMX must simultaneously be fast enough to support this high rate of background events, most of which are ``straightforward'' to reject based on their high electromagnetic energy deposition, and sensitive enough to reject rare but subtle processes where a hard bremsstrahlung photon undergoes a photo-nuclear reaction in the target or in the calorimeter itself. These simultaneous requirements call for a high-speed, high-granularity calorimeter with minimum-ionizing particle (MIP) sensitivity to identify photo-nuclear products, used in conjunction with a hadron calorimeter that experiences much lower event rates.  As described in this paper, LDMX plans to meet these technical challenges by leveraging technology under development for the HL-LHC and Mu2e, as well as  experience from the Heavy Photon Search (HPS) experiment.

To achieve sufficient statistics, LDMX proposes to use a low-current ($\sim$pA) but high bunch-repetition ($\sim 40$ MHz) electron beam with multi-GeV energy.  A beam with 10$^8$ electrons/second on target and energy in the 4 to 16 GeV range can  explore most of the sub-GeV dark matter parameter space, while remaining below threshold for production of neutrinos,  which are an irreducible background. Three options for such a beam are currently under consideration -- a proposed 4-8 GeV beam transfer line at SLAC~\cite{Raubenheimer:2018mwt}, the 11 GeV CEBAF beam at Jefferson Laboratory~\cite{CEBAF}, or a proposed 3.5-16 GeV beam derived from slow SPS extraction at CERN~\cite{CERNeBeamFacility}. For historical reasons, our design study has focused on a 4 GeV incoming electron beam. This is conservative, as beams with higher energy, in the 8-16 GeV range, alleviate most of the potential backgrounds that LDMX will need to contend with. 

Our general strategy is to run the experiment in two phases. In Phase I, we target a modest total required luminosity of 0.8 pb$^{-1}$ corresponding to $4 \times 10^{14}$ tagged electrons on target.  During this phase, we would produce strong physics results with only minor deviations from established detector technologies. The studies and results in this paper primarily focus on Phase I performance at a beam energy of 4~\GeV. Our results are strongly encouraging, and imply that our approach has a healthy margin of safety for contending with potential backgrounds. These results also suggest that higher (by factor of $\sim 25$) luminosity running at $8-16 \ \GeV$ energies will be possible. We therefore discuss requirements and implications for a Phase II of the experiment with up to $\sim 10^{16}$ tagged electrons on target at a beam energy of 8 \GeV or higher.

We begin with a discussion of the dark matter science goals for LDMX (Section II); a summary of the signal and background processes (Section III); a description of the detector design concept and ongoing developments (Section IV); details of the physics simulation and basic detector performance (Section V); studies of the background rejection strategies (Section VI) and the corresponding signal efficiency (Section VII); a discussion of signal sensitivity in phase I (Section VIII); simple methods to extend sensitivity beyond phase I (Section IX); and then a brief discussion of the broader physics reach of LDMX (Section X).  

While this paper will focus on missing momentum and energy signatures exclusively, a companion phenomenology paper~\cite{LDMXSciencePaper} explores applications of LDMX to a broader variety of dark matter and other dark sector scenarios, including displaced decay signatures. Investigation of the detailed experimental aspects of displaced decay signatures, among others, is left to future work. We also note that with proper design of the trigger and data acquisition systems, LDMX can provide greatly improved data on final states in electron-nuclear scattering in the multi-GeV region, which is of much interest to the neutrino scattering community. This will also be discussed in more detail in future publications.


\clearpage
\section{Science Goals}
\label{sec:goals}
The primary science driver for LDMX is to search for dark matter particles with mass ranging from $\GeV$ to well below the $\keV$-scale --- a region that is simultaneously well motivated by the thermal freeze-out hypothesis (among others) for the origin of dark matter and experimentally open and accessible territory.  In simple models, a thermal origin for dark matter implies a \emph{minimum} interaction strength between dark and ordinary matter (as does any thermal contact between dark and ordinary matter in the early Universe). For DM masses below several hundred MeV, LDMX aims to provide the sensitivity needed to explore most of the scalar and Majorana dark matter coupling range compatible with thermal freeze-out into Standard Model final states, and to cover a significant part of the fermion DM parameter space in a first phase. LDMX will then probe the remaining fermion parameter space in its second phase. Exploring light dark matter to this level of sensitivity is a science priority highlighted in recent community reports \cite{Alexander:2016aln,Battaglieri:2017aum}, and resonates with the dark matter priorities emphasized in the P5 report \cite{ParticlePhysicsProjectPrioritizationPanel(P5):2014pwa}. This section provides a self-contained discussion of thermal dark matter, realizations of sub-GeV thermal dark matter in particular, and the precise sensitivity targets that this general hypothesis motivates. A more comprehensive discussion of light dark matter may be found in \cite{Izaguirre:2015yja,Alexander:2016aln}.

Of course, missing momentum measurements have powerful applications beyond thermal dark matter, and it's worth emphasizing that this signature provides sensitivity to sub-GeV dark matter, mediator particles, and other sub-GeV physics {\it in general}. For example, such measurements can provide tremendous new sensitivity to secluded, SIMP \cite{Hochberg:2014kqa,Hochberg:2015vrg,Berlin:2018tvf}, ELDER \cite{Kuflik:2015isi,Kuflik:2017iqs}, asymmetric \cite{Petraki:2013wwa,Zurek:2013wia}, and freeze-in  dark matter scenarios \cite{Dodelson:1993je,Hall:2009bx,Chu:2011be}, as well as millicharge particles \cite{Davidson:2000hf}, invisibly decaying dark photons, axions, and dark Higgs particles \cite{Alexander:2016aln}. These applications, as well as the use of displaced decay signatures in LDMX are described in \cite{LDMXSciencePaper}. 

\subsection{Thermal Dark Matter:  A Target of Opportunity}
\label{sec:goalsTestingThermalDM}


Perhaps the simplest possibility for the origin of dark matter is that its abundance arises from non-gravitational interactions with Standard Model particles in the hot early Universe. This compelling scenario is largely model independent and only requires the DM-SM interaction rate exceed the Hubble expansion rate at some point in the early Universe. 
If such a condition is satisfied, dark matter automatically reaches thermal equilibrium with visible matter and a residual DM abundance is guaranteed. 
This mechanism for generating the DM abundance is attractive and important for many reasons. 

\begin{figure}[t!]
\center
\includegraphics[width=15cm]{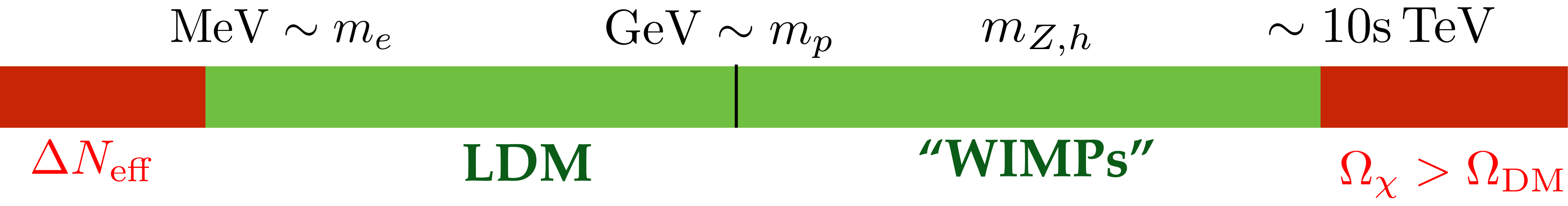}
\caption{The allowed mass range over which DM can thermalize with the SM in the early universe and yield the observed relic abundance via 
annihilation. For DM masses below $\lesssim $ MeV, freeze out occurs after neutrinos decouple from the SM, so the DM annihilation heats SM photons relative to neutrinos (or vice versa), thereby changing $N_{\rm eff}$, the effective number of relativistic species. Independently of this issue, for masses below $\sim 10$ keV, DM is too hot to accommodate the observed matter power spectrum  \cite{Viel:2013apy} and for masses above $\gtrsim 10$ TeV, a perturbative annihilation rate cannot achieve the correct relic abundance in simple models \cite{Griest:1989wd}. }
\label{fig:schematic-top}
\end{figure}

First, this mechanism is simple and generically realized because equilibrium is hard to avoid even for tiny couplings between dark and visible matter. Therefore, most {\it discoverable} models of DM with non-gravitational interactions fall into this category. Second, this mechanism implies a minimum annihilation rate $\langle \sigma v \rangle \sim 10^{-26}$ cm$^3$s$^{-1}$ in order to avoid producing an overabundance of dark matter at ``freeze-out''.  This minimum annihilation rate defines a minimum cross section which must be experimentally probed to rule out dark matter of thermal origin. Finally,  the allowed thermal DM mass range is well defined -- see Fig. \ref{fig:schematic-top}. The most natural mass range is comparable to that of Standard Model matter, so that DM annihilation into SM final states is generically possible and well-motivated. 

\subsection{Light Thermal Dark Matter (LDM)}


If DM is realized in the upper half of the thermal mass window $\sim$ GeV - 10 TeV, it can be a WIMP charged under the electroweak force. This has been the traditional focus of dark matter direct and indirect detection experiments, driven in part by the well known connection between WIMPs and supersymmetry (SUSY), whose DM candidates realize this paradigm. However, powerful null results from direct and indirect detection experiments have largely ruled out the simplest WIMP scenarios by several orders of magnitude, and the remaining parameter space is largely cornered by upcoming experiments. 


The lower half of the thermal mass window, $\sim$ MeV - GeV, has remained stubbornly difficult to test with traditional experiments designed to probe WIMPs and is not well explored. 
This is unfortunate because the sub-GeV mass range is well-motivated by ``hidden sector'' (or ``dark sector'') scenarios in which dark matter is simply a particle with its own forces and interactions, neutral under the Standard Model, but with sufficient coupling to visible matter that thermal equilibrium is achieved in the early Universe. This mass range is also independently motivated by asymmetric dark matter scenarios, in which dark matter carries a net particle number in analogy with the baryon asymmetry observed in visible matter. The particle physics community has highlighted these scenarios as among the most important to test in the P5 report \cite{ParticlePhysicsProjectPrioritizationPanel(P5):2014pwa}, the recent Dark Sectors 2016 community report \cite{Alexander:2016aln}, and the US Cosmic Visions New Ideas in Dark Matter community report \cite{Battaglieri:2017aum}. For these reasons, and to help focus our design efforts, the primary science driver for LDMX is the exploration of dark matter interactions with electrons to a level of sensitivity needed to decisively test most predictive thermal dark matter scenarios over nearly the entire sub-GeV mass range. At the same time, many other dark matter and dark sector scenarios will also be explored, as discussed in \cite{LDMXSciencePaper}. 



\subsection{A Benchmark Scenario for LDM}

\begin{figure}[t!]
\center
\includegraphics[width=15cm]{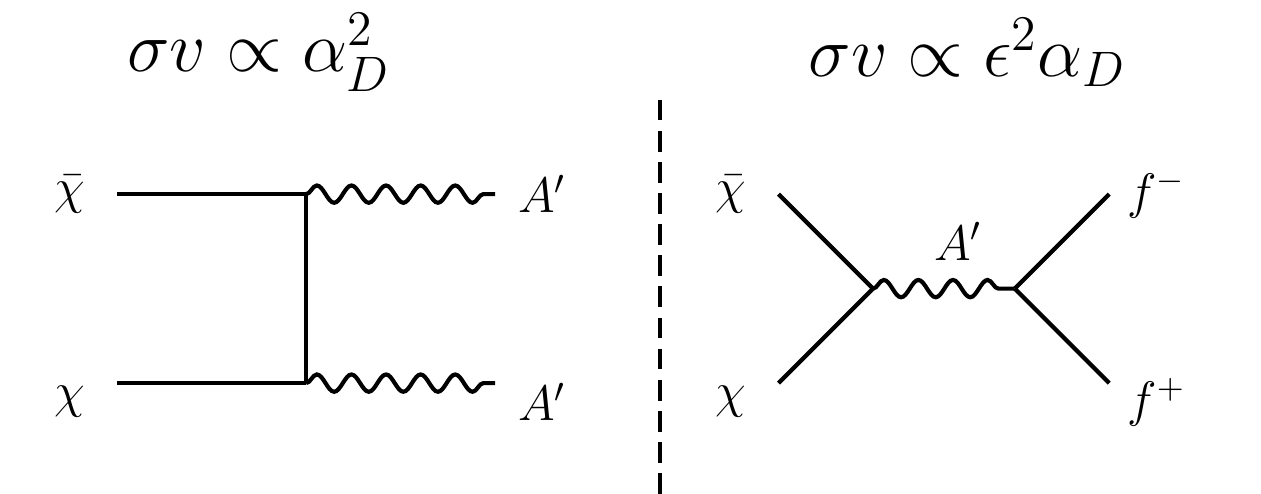}
\caption{ Feynman diagrams for LDM with secluded annihilation (left) with $m_\chi > m_{A^\prime}$ and direct annihilation (right) with $m_\chi < m_{A^\prime}$. In the secluded
regime, the dark photon decays visibly to kinematically accessible SM final states and motivates experimental searches for hidden forces (see \cite{Alexander:2016aln}), but 
the DM annihilation cross section is independent of the $A^\prime$ coupling to visible matter. 
  In the direct annihilation regime, the cross section for achieving the correct relic density depends on the parameter $\epsilon$ which couples the ${A^\prime}$ to charged SM particles, so
there is a minimum value of this coupling for each choice of $\chi$ mass that realizes a thermal history in the early universe. These minimum values define experimental targets for discovery or falsification (see Fig. \ref{fig:mainplot}).}
\label{fig:diagrams}
\end{figure}

In the MeV-GeV mass range, viable models of LDM have the following properties: 
\begin{itemize}
\item {\bf Light Forces:} There have to be comparably light force carriers to mediate an efficient annihilation rate 
for thermal freeze-out (this follows from a simple generalization of the Lee-Weinberg bound \cite{Lee:1977ua,Boehm:2003hm}). 
\item {\bf Neutrality:} Both the DM and the mediator must be singlets under the full SM gauge group; otherwise they 
 would have been produced and detected at LEP or at hadron colliders \cite{ALEPH:2010aa}. 
\end{itemize}
These properties single out the hidden sector scenario highlighted in \cite{ParticlePhysicsProjectPrioritizationPanel(P5):2014pwa, Alexander:2016aln, Battaglieri:2017aum}, which is the focus of considerable experimental activity. 
Thus, for the remainder of this note, we will use one of the simplest and most representative hidden sector models in the literature -- a dark matter particle charged under a $U(1)$ gauge field (i.e. ``dark QED''). Sensitivity to a variety of other new physics models, mediator particles, and dark matter is described in a companion paper \cite{LDMXSciencePaper}. 

We define the LDM particle to be $\chi$ and the $U(1)$ gauge boson $A^\prime$ (popularly called a ``dark photon" mediator).  
 The general Lagrangian for this family of models contains
 \be
 \label{eq:master-lagrangian}
\mathscr{L} \supset -\frac{1}{4} {F^\prime}^{\mu\nu}F^\prime_{\mu\nu} +  \frac{m^2_{A^\prime}}{2}A^\prime_\mu {A^\prime}^\mu -   A^\prime_\mu    (\epsilon e J_{\rm EM}^\mu +
 g_D J_{D }^\mu),
 \ee
 where $\epsilon$ is the kinetic mixing parameter, $m_{A^\prime}$ is the dark photon mass, and $J^\mu_{\rm EM} \equiv \sum_f Q_f \bar f \gamma^\mu f$ is the SM electromagnetic current where $f$ is an SM fermion with charge $Q_f$, $g_D \equiv \sqrt{4\pi\alpha_D}$ is the $U(1)_D$ coupling constant, and $J_D$ is the dark matter current. Although each possible choice for $\chi$ 
  has a different form for $J_D$, the relic density has the same dependence on the four model parameters $\{\epsilon, g_D, m_\chi, m_{A^\prime} \}$ and can be captured in full generality 
  with this setup. 
%

This framework permits two qualitatively distinct annihilation scenarios depending on the $A^\prime$ and $\chi$ masses.  

\begin{figure}[t!]
\center
\includegraphics[width=6.5cm]{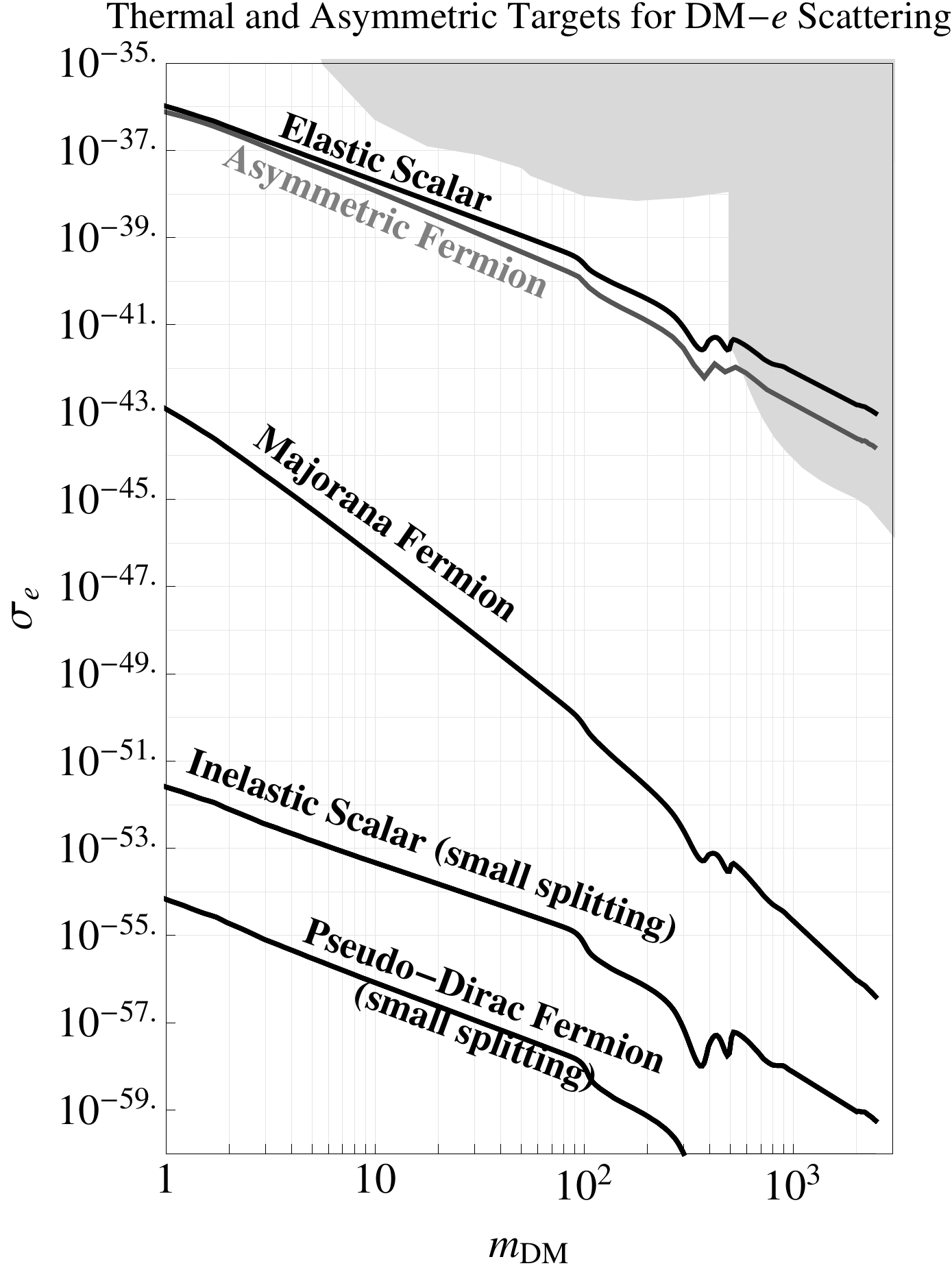}
\includegraphics[width=9.5cm]{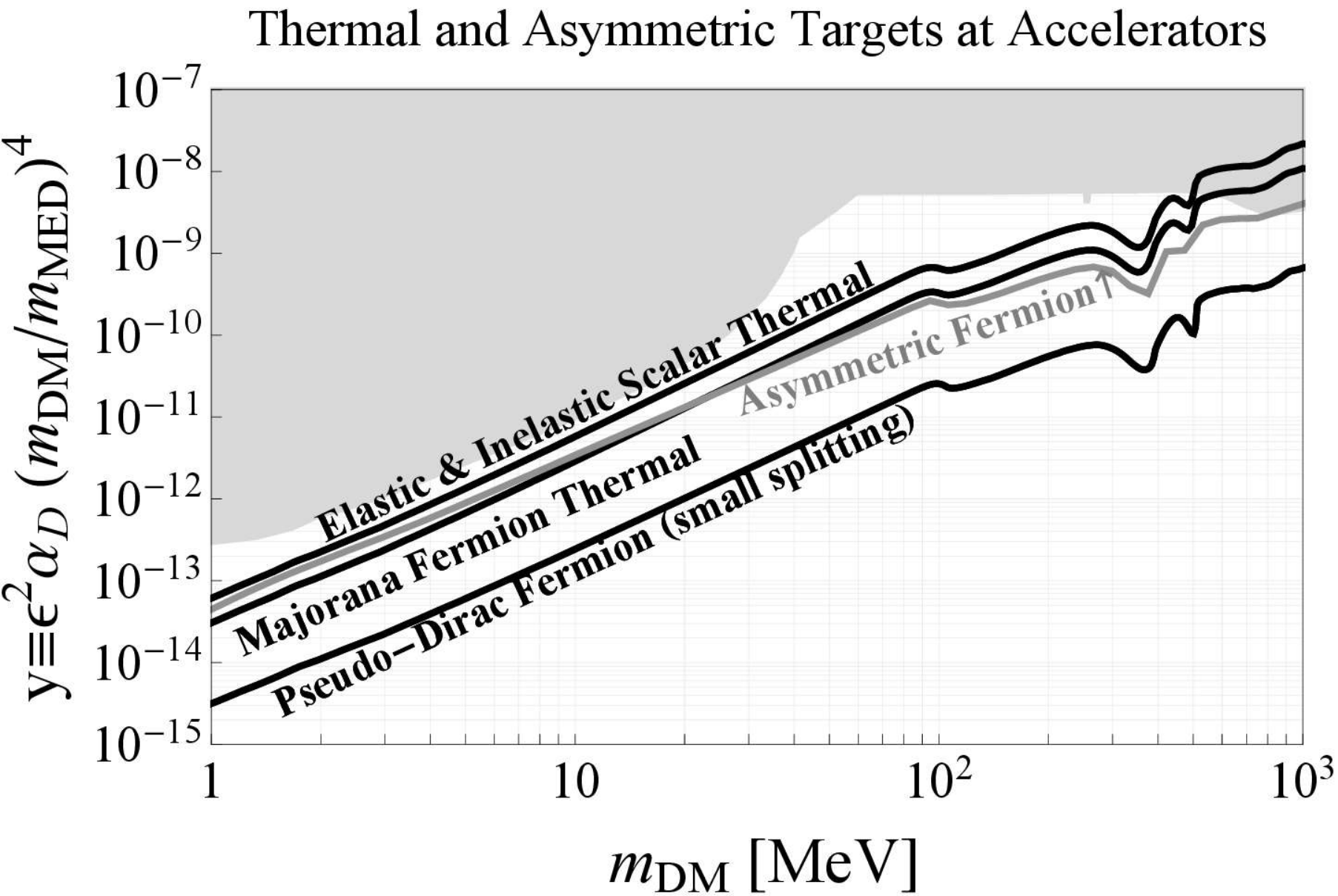}
\caption{ {\bf Left:} Thermal targets for representative DM candidates plotted in terms of the electron-recoil direct detection cross section $\sigma_e$ vs. mass $m_{DM}$. The appropriate thermal freeze-out curve for each scenario differs by many orders of magnitude in the $\sigma_e$ plane due to DM velocity suppression factors, loop-level factors, or spin suppression, any of which are significant for non-relativistic scattering. This is entirely analogous to the variation in scattering rates predicted by canonical WIMP models, where the spread in cross-sections is similarly due to spin, velocity, and loop-level factors. {\bf Right:}  By contrast, the dimensionless couplings (captured by $y$) motivated by thermal freeze-out do not differ by more than a couple orders of magnitude from one another, as shown in the $y$ vs. $m_\chi$ plane. Probing couplings at this magnitude is readily achievable using accelerator techniques, which involve DM production and/or detection, as well as mediator production, all in a relativistic setting. Both plots above are taken from \cite{Battaglieri:2017aum} and also show a target for asymmetric fermion dark matter, a commonly discussed variation on the thermal-origin framework.  This scenario and the resulting target are discussed further in \cite{Battaglieri:2017aum}. 
}
\label{fig:collapsing}
\end{figure}

\begin{itemize}
\item {\bf Secluded Annihilation:} For $m_{A^\prime}  < m_{\chi}$, DM annihilates predominantly into $A^\prime$ pairs \cite{Pospelov:2007mp} as depicted on the left panel of Fig. \ref{fig:diagrams}.  This annihilation rate is independent of the SM-$A^\prime$ coupling $\epsilon$. The simplest version of this scenario is robustly constrained by CMB data \cite{madhavacheril:2013cna}, which rules out DM masses below $\mathcal{O}(10)$ GeV for simple secluded annihilation models. More complex secluded models remain viable for low DM masses; these are potentially discoverable by LDMX, and are discussed in \cite{LDMXSciencePaper}, but they do not provide a sharp parameter space target.  
\item  {\bf Direct Annihilation:} For $m_{A^\prime} >  m_{\chi}$, annihilation 
 proceeds via  $\chi \chi \to {A^\prime}^* \to ff$ to SM fermions $f$ through a virtual mediator as depicted on the right panel of Fig. \ref{fig:diagrams}. This scenario is quite predictive, because the SM-$A^\prime$ coupling $\epsilon$ must be large enough, and the $A^\prime$ mass small enough, in order to achieve the thermal relic cross-section.  Unlike the secluded regime, no robust constraint on this scenario can be extracted from CMB data.  Therefore, the observed DM abundance implies a \emph{minimum} DM production rate at accelerators.  
 The detailed phenomenology depends on the ratio $m_{A^\prime} /m_\chi$.  If $m_\chi < m_{A^\prime} < 2 m_\chi$,
the mediator decay to DM is kinematically forbidden so that visible dark photon searches and DM searches like LDMX are both relevant.  If instead $m_\chi < m_{A^\prime}/2$, the mediator decays primarily to DM, making searches for DM production even more sensitive and absolutely essential.  This case represents most of the parameter space for direct annihilation, and will be the focus of the remaining discussion.  
\end{itemize}
 
Since the Feynman diagram that governs direct annihilation (right panel of Figure \ref{fig:diagrams}) can be rotated to yield a scattering process off SM particles, the  direct detection cross section is uniquely predicted by the annihilation rate in the early universe for each choice of DM mass. Thus, direct annihilation models define thermal targets in the $\sigma_e$ vs. $m_{\chi}$ plane shown in Fig. \ref{fig:collapsing} (left). Since non-relativistic direct detection cross sections can often be loop or velocity suppressed in many models, these targets vary by {\it dozens} of orders of magnitude in some cases. However, these vast differences in the direct detection plane mask the underlying similarity of these models in relativistic contexts where both the scattering and annihilation cross sections differ only by order-one amounts. 

To comprehensively study all direct annihilation models on an equal footing, we follow conventions in the literature (see \cite{Alexander:2016aln}), and introduce the dimensionless interaction strength $y$ as
\be
\sigma v (\chi \chi \to {A^\prime}^*  \to f f) \propto  \epsilon^2 \alpha_D \frac{m_\chi^2}{m_{A^\prime}^4} =  \frac{y}{m_\chi^2}~~~~,~~~~ y \equiv 
\epsilon^2 \alpha_D  \left(     \frac{m_\chi}{m_{A^\prime}  }    \right)^4~~.
\label{eq:ydef}
\ee
This is a convenient variable for quantifying sensitivity because for each choice of $m_\chi$ there is a unique value of $y$ compatible with thermal freeze-out independently of the individual  values of $\alpha_D, \epsilon$ and $m_\chi/m_{A^\prime}$. On the right panel of Fig.\ref{fig:collapsing} we show thermal targets for various direct annihilation models plotted in the  $y$ vs $m_{\chi}$ plane. Although these are the same models shown on the left 
panel of this figure, this parametrization reveals the underlying similarity of these targets and their relative proximity to existing accelerator bounds (shaded regions). Reaching experimental sensitivity to these benchmarks for masses between MeV -- GeV would provide nearly decisive coverage of this class of models \footnote{Near $m_{A^\prime} \approx 2m_\chi$, the $\chi \chi \to \bar f f$ annihilation process is near resonance, so the relic target for this region of parameter space becomes sharply sensitive to the DM/mediator mass ratio, so the relic targets presented in Fig.~\ref{fig:mainplot} move down in the $y$ parameter space (see Supplemental Material in \cite{Izaguirre:2015yja} and \cite{Feng:2017drg} for more details). However, away from the resonant region where $m_{A^\prime} \gtrsim 2m_\chi$, the relic targets in Fig.~\ref{fig:mainplot} are robustly insensitive to the mass and coupling ratios that define the $y$ variable}. 
\begin{figure}[t!]
\center
\includegraphics[width=8.1cm]{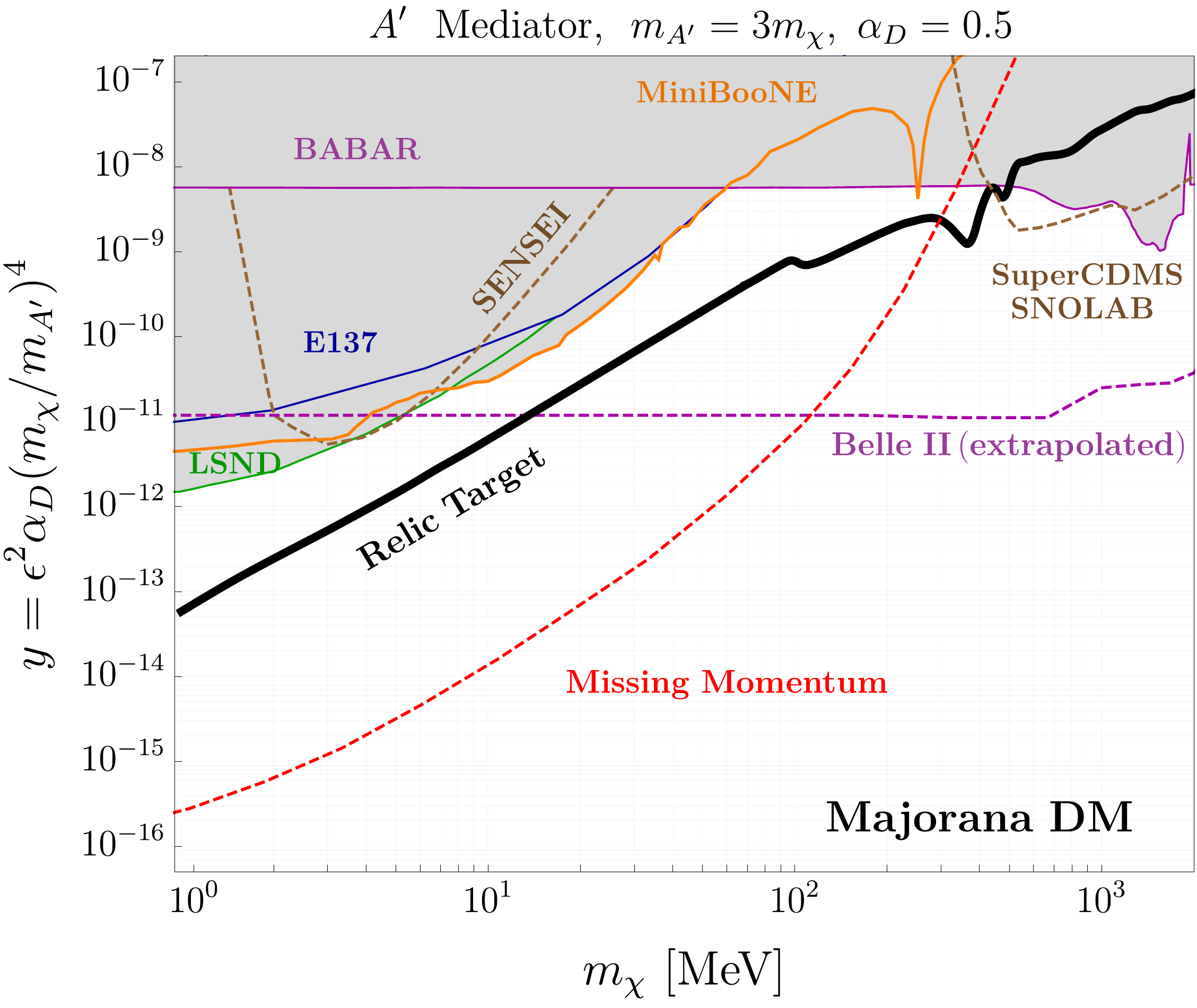}
\includegraphics[width=8.1cm]{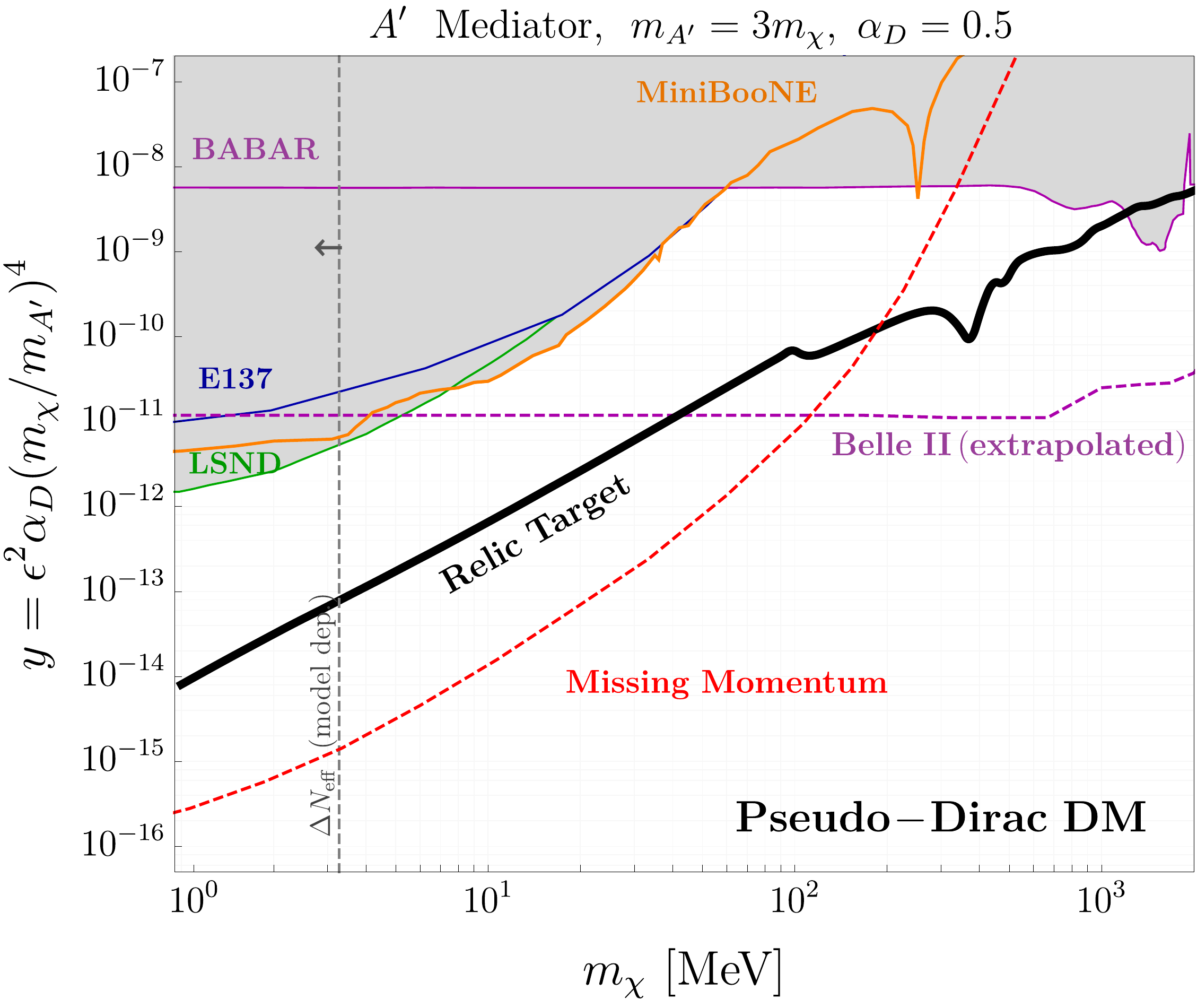}
\medskip
\includegraphics[width=8.1cm]{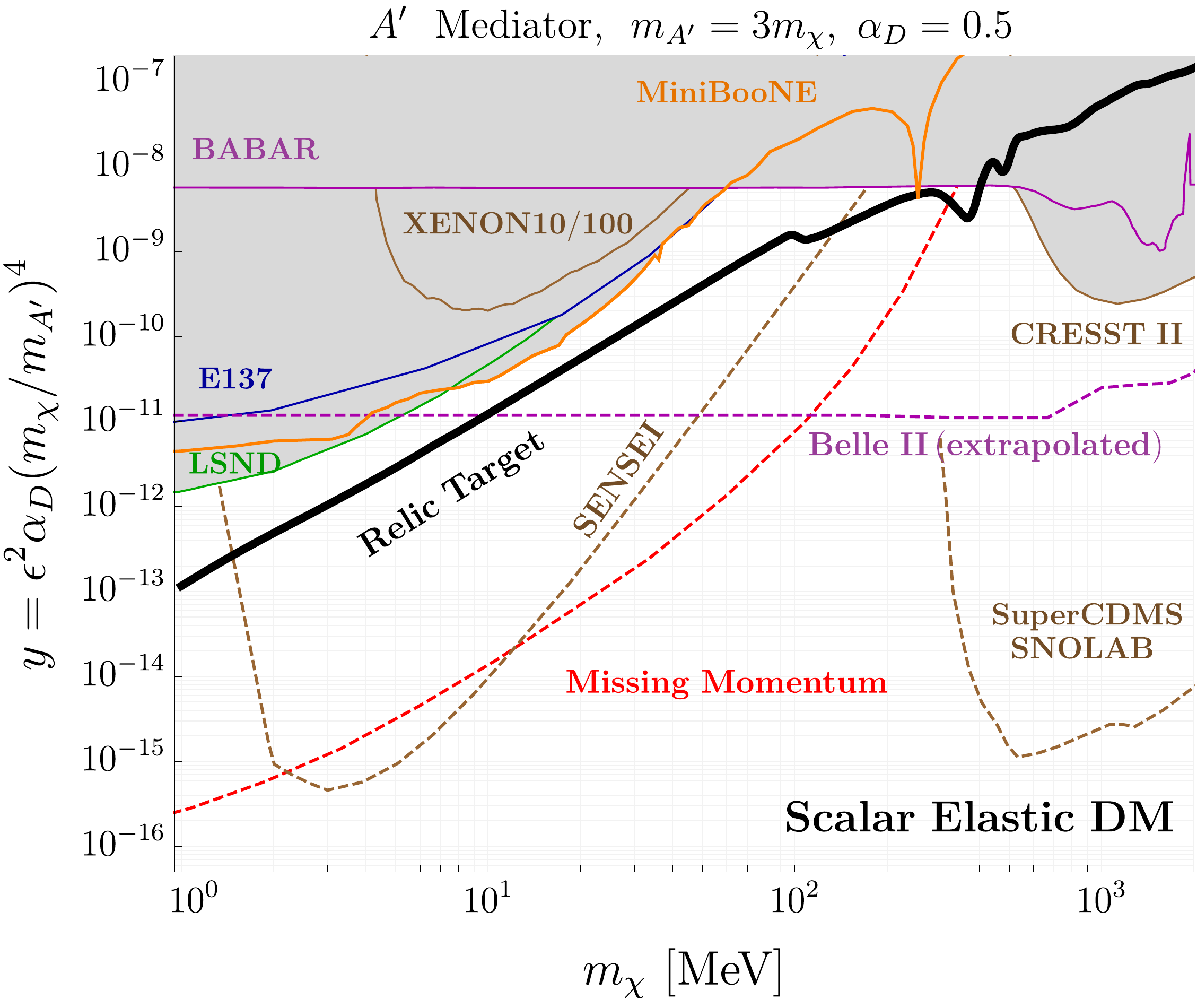}
\includegraphics[width=8.1cm]{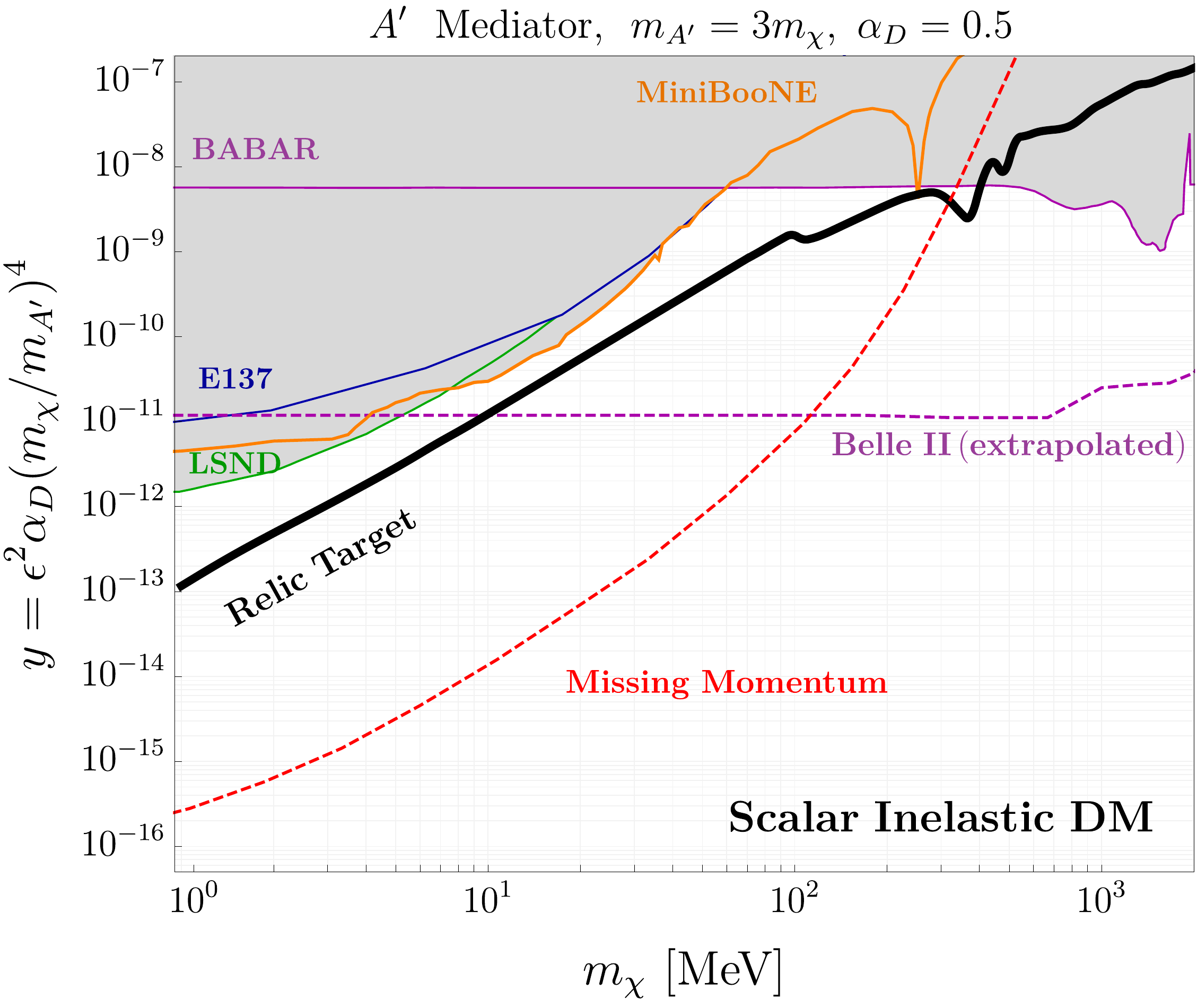}
\caption{ The parameter space for LDM and future experimental projections in the $y$ vs. $m_\chi$ plane plotted against
 the thermal relic targets for representative scalar and fermion DM candidates coupled to a dark photon $A^\prime$ -- see text for a discussion. The red dashed curve represents the ultimate reach of an LDMX-style missing momentum experiment.}
\label{fig:mainplot}
\end{figure}

The viable choices for $\chi$ (which vary according to spin and type of mass) whose relic density arises from direct annihilation can be simply enumerated 
\begin{itemize}

\item{\bf Majorana Dark Matter:} In this scenario, $\chi$ is a Majorana fermion. It therefore couples through an axial-vector current
\be
J_D^\mu = \frac{1}{2}\overline \chi \gamma^\mu \gamma^5 \chi,
\ee
which introduces velocity dependence in nonrelativistic scattering 
processes $\sigma_{\rm scat} \propto v^2$ and suppresses 
signals which involve DM interactions at low velocities.  
 The thermal target and parameter space for this 
model are presented in the upper left panel of Fig.~\ref{fig:mainplot}.

 \item{\bf ``Pseudo'' Dirac Dark Matter:} The DM $\chi$ can be a Dirac fermion. If the mass term for $\chi$ is $U(1)_D$ preserving, then the model is already constrained by CMB data, unless there is a particle-antiparticle asymmetry. If the mass terms for $\chi$ include $U(1)_D$ breaking (by analogy to the $SU(2)_W$ breaking mass terms of particles in the Standard Model), then $\chi$ splits into two Majorana fermions (in the mass basis) which couple off-diagonally to the $A^\prime$ through
 \be
J_{D}^\mu =  i \overline \chi_1 \gamma^\mu \chi_2 ~~      ~\rm (mass~basis).
\ee
as well as a small diagonal axial current coupling to the $A'$.  The thermal target and parameter space for this 
model are presented in the upper right panel of Fig.~\ref{fig:mainplot}.

\item{\bf Scalar Elastic: } In this scenario, $\chi$ is a complex scalar particle with $U(1)_D$ preserving mass terms, and 
current
\be
J^\mu_D \equiv i (\chi^* \partial^\mu \chi - \chi \partial^\mu \chi^*).
\ee
The annihilation cross section for this model is $p$-wave suppressed, so  $\sigma v(\chi \chi^* \to f \bar f) \propto v^2$ and therefore requires a slightly larger coupling to achieve freeze out relative to other scenarios. 
This model also yields elastic signatures at direct detection experiments, so 
it can be probed with multiple complementary techniques. The thermal target and parameter space for this 
model are presented in the lower left panel of Fig.~\ref{fig:mainplot}.

\item{\bf Scalar Inelastic Dark Matter: } In this scenario, $\chi$ is a complex scalar particle with $U(1)_D$ breaking mass terms (by analogy to the $SU(2)_W$ breaking mass terms of particles in the Standard Model). 
Therefore, $\chi$ couples to $A^\prime$ inelastically and must transition to a slightly heavier state in order to scatter through the current 
\be
J_D^\mu = i( \chi^*_1 \partial^\mu \chi_2 - \chi^*_2 \partial^\mu \chi_1)~, 
\ee
which typically suppresses direct detection signals even for small 
mass differences between $\chi_{1,2}$. Thus, this model can only be probed robustly at accelerators. 
The thermal target and parameter space for this model are presented in the lower right panel of Fig.~\ref{fig:mainplot}.

\end{itemize}

For each of these models, there is a slightly different $y$ vs. $m_\chi$ thermal target  in  Fig.~\ref{fig:mainplot}, but 
they all share the same parametric dependence on the product $\alpha_D \epsilon^2 (m_\chi/m_{A^\prime})^4$. Furthermore, 
all of the models on this list are safe from  bounds on late-time energy injection around $T \sim$ eV from Planck measurements
of the CMB temperature anisotropies \cite{ade:2015xua}. 

\begin{figure}[t!]
\center
\includegraphics[width=8.1cm]{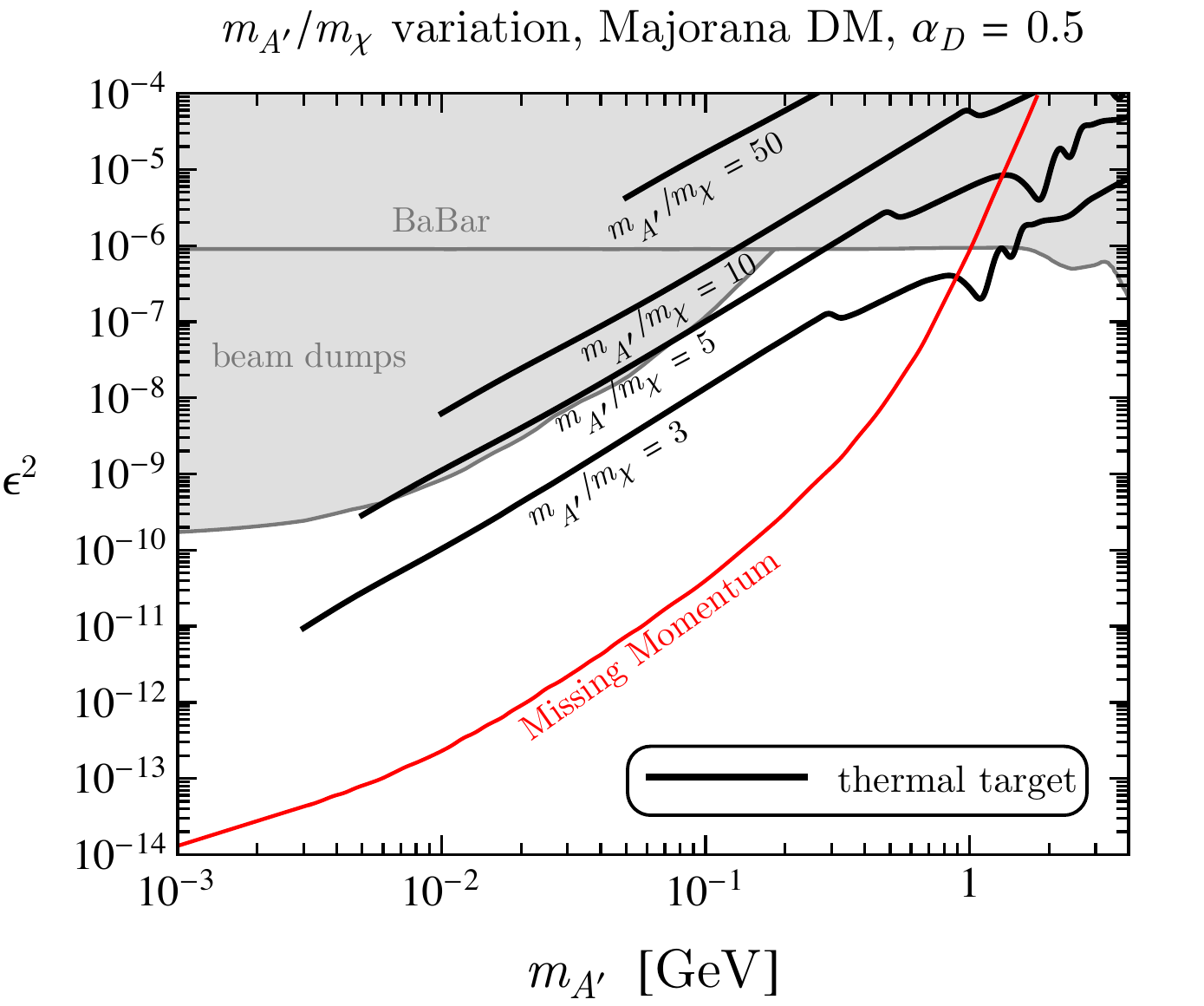}
\includegraphics[width=8.1cm]{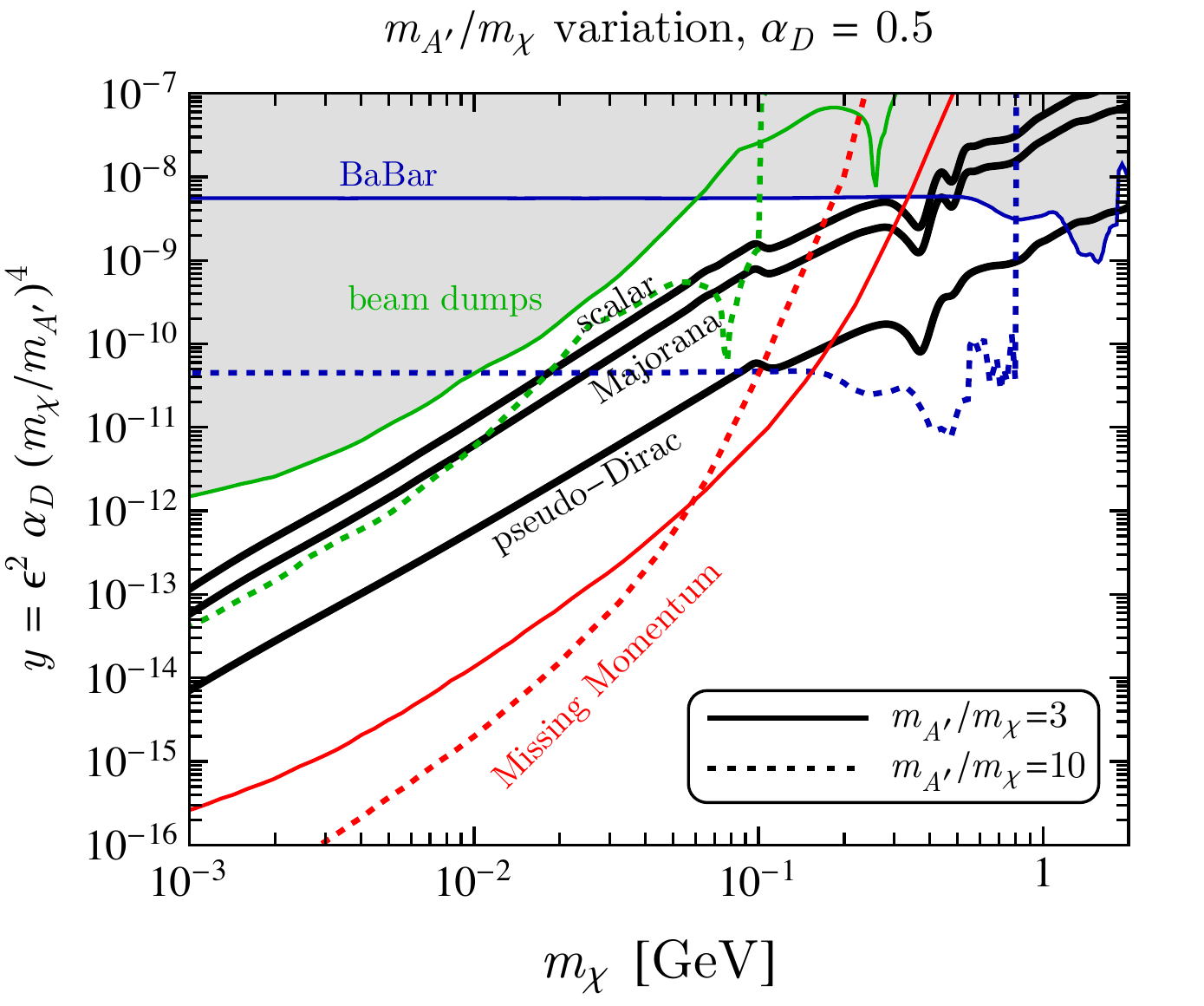}
\includegraphics[width=8.1cm]{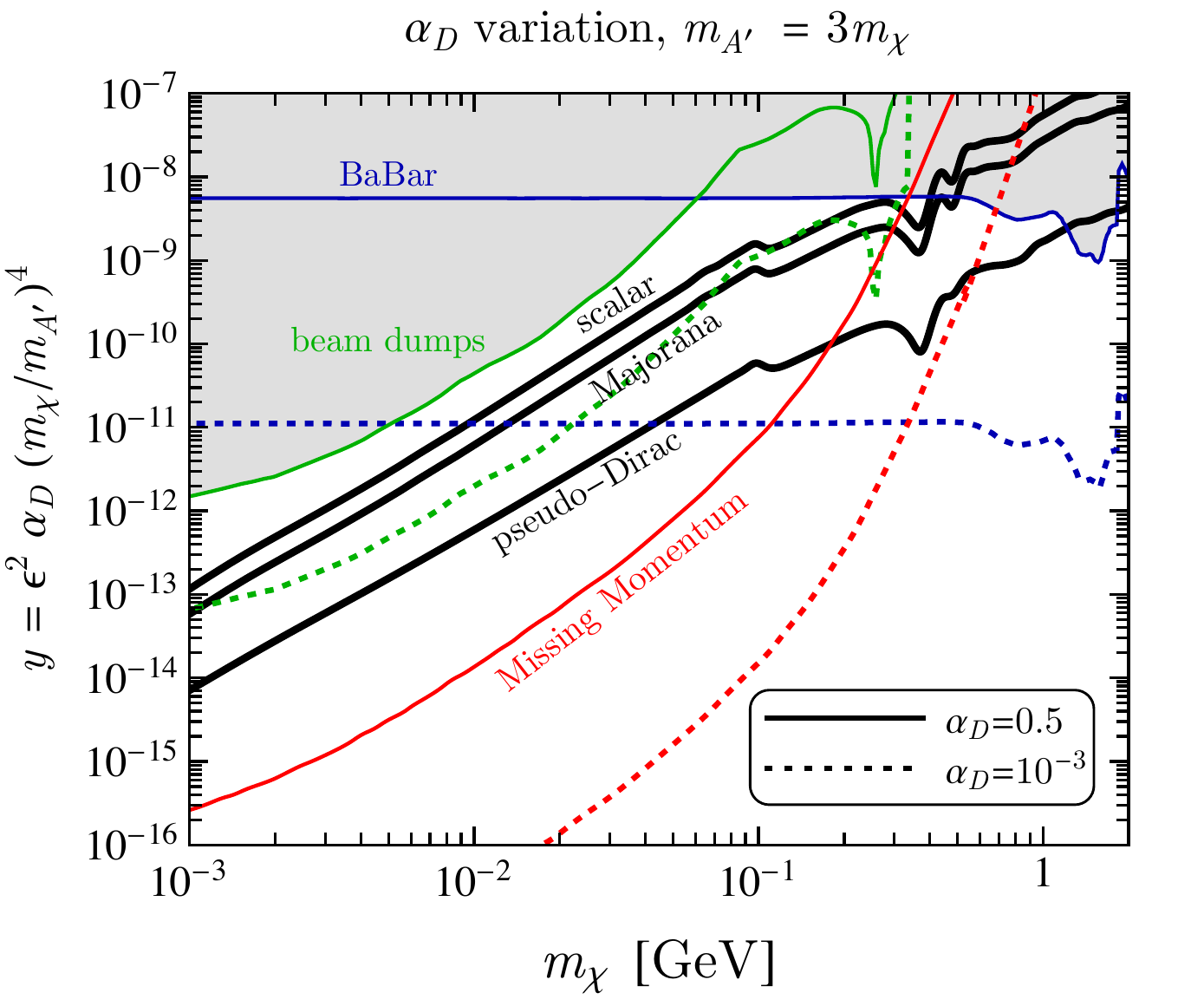}
\caption{The panels above demonstrate that the $\alpha_D = 0.5$ and $m_{A^\prime} = 3 m_\chi$ benchmarks shown in Fig. \ref{fig:mainplot} are conservative choices; deviations away from these vales largely {\it improve} the coverage of missing momentum style techniques -- except near the $m_{A^\prime} \approx 2m_\chi$ resonance (see Fig. \ref{fig:NearResonance}) {\bf Top Left:} $\alpha_D$ is fixed as the $m_{A^\prime}/m_\chi$ ratio varies relative to the benchmarks in Fig.~\ref{fig:mainplot}. The sensitivity of accelerators is shown in the $\epsilon^2$ vs. $m_{A'}$ plane. Note that the thermal freeze-out curves move to larger values of $\epsilon^2$ as $m_{A^\prime}/m_\chi$ is increased, while the accelerator sensitivity does not change. {\bf Top Right:} $\alpha_D$ is fixed as the $m_{A^\prime}/m_\chi$ ratio varies relative to the benchmarks in Fig.~\ref{fig:mainplot}, but now shown in the $y$ vs. $m_{\chi}$ plane. Thermal freeze-out curves do not vary, but the accelerator sensitivity shifts to lower values of $y$ and lower $m_{\chi}$ as $m_{A^\prime}/m_\chi$ is increased. {\bf Bottom:} parameter space in the $y$ vs. $m_\chi$ plane where the solid curves are identical to those shown in Fig. \ref{fig:mainplot} (with $\alpha_D=0.5$), but the dotted curves show how the constraints and projections vary for the choice $\alpha_D = 10^{-3}$. For fixed values of $y$, a smaller $\alpha_D$ requires a larger $\epsilon^2$ (i.e. larger mediator coupling), which makes that parameter point {\it easier} to constrain. Hence, accelerator sensitivity generally improves in the $y$ vs. $m_{\chi}$ plane for smaller $\alpha_D$. Note that the thermal freeze-out curves in this plane are identical for both values of $\alpha_D$ shown here because the thermal abundance scales with $y$. 
}
\label{fig:mainplot-variations}
\end{figure}
 
All existing constraints, collected in the $y$ vs $m_\chi$ parameter space, are depicted in Fig. \ref{fig:mainplot}  alongside projections for various experimental techniques (see \cite{Battaglieri:2017aum} for a discussion). 
As explained above, the thermal targets for each of the LDM candidates (indicated by the solid black line in each panel) are largely invariant in this parameter space regardless of the values of $\epsilon, \alpha_D$, and $m_\chi/m_{A^\prime}$ separately. Direct detection constraints are also  naturally expressed as functions of $y$ and $m_{\chi}$, but accelerator-based constraints are not --- for example, collider production bounds depend on $m_{A^\prime}$ and $\epsilon$.  However, within the predictive framework of direct DM annihilation, both $\alpha_D$ and $m_{\chi}/m_{A^\prime}$ are bounded from above, accelerator-based constraints typically become \emph{stronger} as either $\alpha_D$ or $m_{\chi}/m_{A^\prime}$ is decreased (specifically, lowering $\alpha_D$ expands the constrained regions downward in Figure \ref{fig:mainplot} while lowering $m_{\chi}/m_{A^\prime}$ moves them downward and to the left).   Therefore, we plot conservative versions of these constraints with $\alpha_D = 0.5$ and $m_{\chi}/m_{A^\prime}$ = 1/3 near their upper limits to reveal the significant gaps in sensitivity to the DM parameter space consistent with direct annihilation freeze-out \footnote{Though keep in mind that the freeze-out targets rapidly decreases on the $y$ vs. $m_{\chi}$ plot in the fine tuned region of parameter space near $2m_{\chi}=m_{A^\prime}$.}. Figure~\ref{fig:mainplot-variations} illustrates how accelerator sensitivity and the thermal freeze-out curves vary with changes to $\alpha_D$ or $m_{\chi}/m_{A^\prime}$, and justify $\alpha_D = 0.5$ and $m_{\chi}/m_{A^\prime}$ = 1/3 as a reasonably conservative choice outside of the fine-tuned resonance region. For more details about this parameter space see \cite{Izaguirre:2015yja}.

So long as $m_{A^\prime} > 2m_\chi$, the dominant DM production process in LDMX is direct production of the mediator 
$A^{\prime}$ followed by its decay to $\chi$ particles. This occurs in most of the parameter space, and therefore we have focused on this case in our design studies. In the region $m_{\chi} < m_{A^\prime} < 2m_\chi$, DM is instead produced 
in LDMX through a virtual $A^{\prime}$ with comparable sensitivity (see \cite{Izaguirre:2015yja} and Sec.~\ref{sec:breadth}). For $A^{\prime}$ masses 
that are significantly larger than a few GeV, DM production is dominated by virtual $A^{\prime}$ mediated reactions with contact operator behavior. 

\begin{figure}[t!]
\center
\includegraphics[width=11.cm]{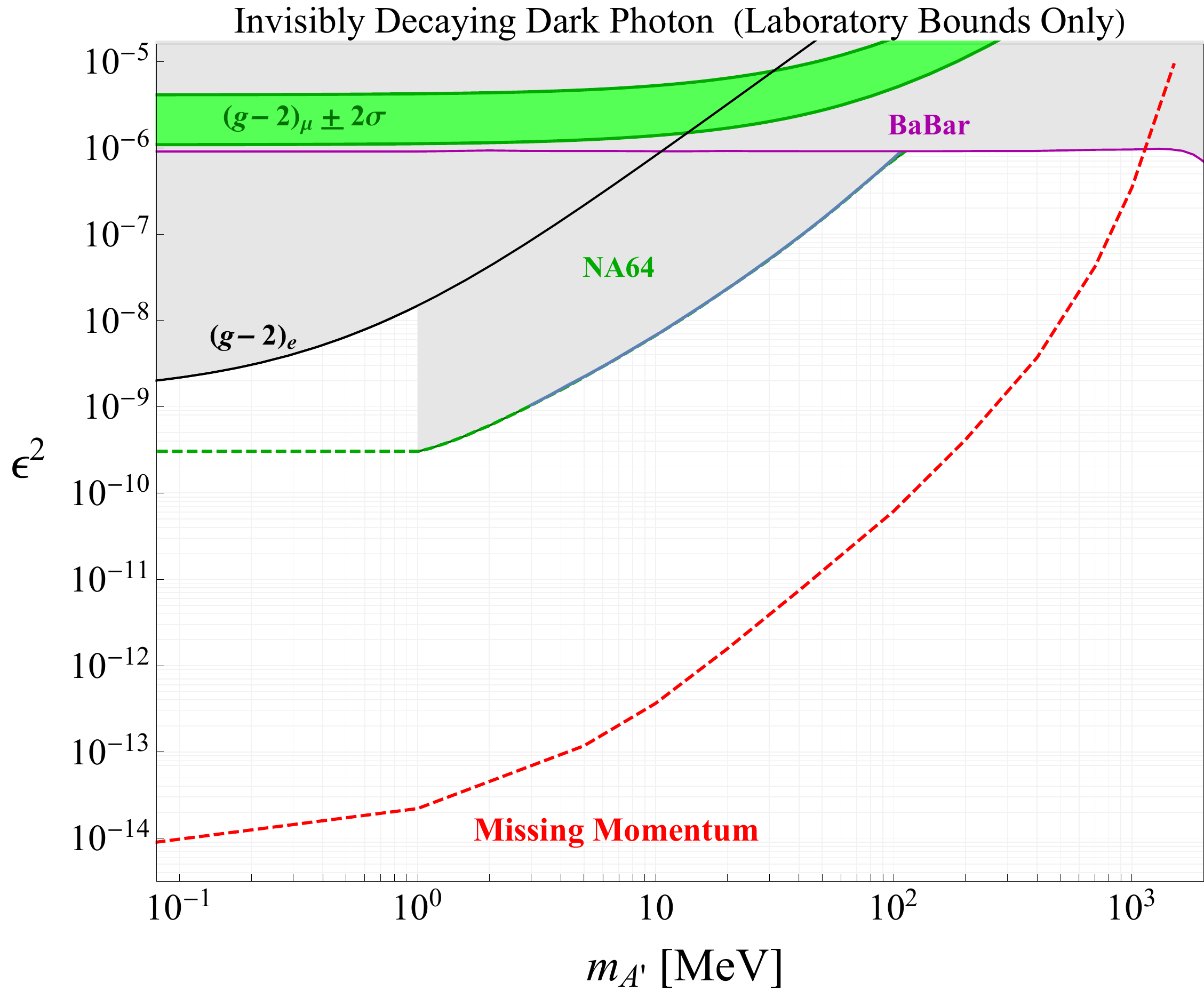}
\caption{ The parameter space for an invisibly decaying dark photon $A^\prime$ in the $\epsilon^2$ vs $m_{A^\prime}$ plane. Also shown 
are various experimental constraints from \cite{Battaglieri:2017aum} and the updated NA64 constraint from \cite{Banerjee:2017hhz}. The 
red curve represents the ultimate reach of an LDMX-style missing momentum experiment.}
\label{fig:mediatorplot}
\end{figure}

Independent of the details of $\chi$, LDMX is sensitive to invisible decay channels of mediators that couple to electrons, such as an $A^{\prime}$. Fig.~\ref{fig:mediatorplot} shows the available parameter space for an invisibly decaying $A^{\prime}$ in the $\epsilon^2$ vs. $m_{A^\prime}$ plane alongside projections for future experiments; this figure makes
no assumptions about dark matter as long as the $A^\prime$ decays invisibly with unit branching fraction.   

Although our emphasis in this section has been on various models of DM with direct annihilation through a dark photon mediator $A^\prime$, the missing momentum technique can probe multiple other mediator scenarios with equally powerful sensitivity to the corresponding theoretical targets. For example, both dark and visible matter could be directly charged under a new $U(1)$  group which gauges an anomaly-free combination of SM quantum numbers (e.g. baryon minus lepton number). Such new forces can also mediate DM direct annihilation to SM particles with thermal targets analogous to those presented in Fig.~\ref{fig:mainplot}. 
Some of these are discussed in Sec.~\ref{sec:breadth}.
Furthermore, missing momentum techniques can also probe strongly interacting dark sectors \cite{Berlin:2018tvf}, millicharged particles, minimal dark photons, minimal $U(1)$ gauge bosons, axion like particles, and light new leptophilic scalars particles \cite{LDMXSciencePaper}.

\subsection{Searching for Light Dark Matter Production}\label{ssec:strategies}

As discussed above, the primary science goal for LDMX is to search for the process depicted in Fig.~\ref{fig:BothSignalReactions}, wherein a  DM particle-antiparticle pair is radiated off a beam electron as it scatters off a target nucleus.  The DM production is mediated by a kinetically mixed dark photon -- see Eq.~\eqref{eq:master-lagrangian}.  
%
%
Depending on the dark photon and DM masses, the leading contribution to this process may be \emph{decay} of a dark photon into a DM particle-antiparticle pair ($m_{A^\prime}>2 m_\chi$) or pair-production through a virtual dark photon.  In either case, a constraint on the DM particle production rate can be used to infer a robust bound on the interaction strength $y$, which can in turn be directly compared to the targets from thermal freeze-out shown in Figure \ref{fig:mainplot}.  This note will focus, for concreteness, on the former case, though the final-state kinematics is extremely similar for either on- or off-shell dark photons and the same search strategies apply to both.

LDMX will search for this process using the \emph{energy-angle kinematics} of the recoiling electron, or ``missing momentum'' approach. As discussed in detail in Section III, this kinematics is distinctive, with the recoiling electron typically carrying a small fraction of the beam energy (the rest is carried by the DM and hence invisible) and receiving an appreciable transverse kick from the DM production process. Thus, the experimental signature for the signal comprises three basic features: (i) a reconstructed recoiling electron with energy substantially less than beam energy but also (ii) detectable, with measurable transverse momentum, and (iii) an absence of any other activity in the final state.  

This search strategy has distinct advantages over other approaches that have been used to detect DM production: 
\begin{itemize}
\item \textbf{Missing mass} (as in \babar, PADME, MMAPS$^*$, and VEPP-3$^*$ --  ($^*$ indicates proposed experiments) relying on full reconstruction of \emph{all} recoiling particles, is only practical in $e^+e^-$ collisions, and requires a much lower luminosity, greatly reducing production yield and hence sensitivity.
\item \textbf{DM re-scattering} in a detector downstream of the production point (as in LSND~\cite{deniverville:2011it}, E137~\cite{batell:2014mga}, MiniBoone~\cite{deniverville:2016rqh,Aguilar-Arevalo:2017mqx,Aguilar-Arevalo:2018wea}, and BDX$^*$~\cite{izaguirre:2013uxa,battaglieri:2016ggd}) can use very intense beams of either protons or electrons, but the low probability of DM scattering weakens sensitivity relative to what is possible in a kinematic search -- whereas the kinematic signals of DM production scale as the square of the weak SM-DM coupling, re-scattering signals scale as the fourth power. As a result, even the most aggressive proposals with intense beams fall short of the anticipated LDMX reach.
\item \textbf{Missing energy} (as in NA64), reconstructing only the energy (not the angle or 3-momentum), is closely related to the missing-momentum approach but with fewer kinematic handles to reject SM backgrounds and measure veto inefficiencies \emph{in situ}.  In addition, missing energy experiments lack the ability to distinguish final-state electrons from one or more photons, introducing irreducible neutrino backgrounds to high-rate missing energy experiments \cite{izaguirre:2013uxa}).
\end{itemize}

\begin{figure}
  \centering
  \includegraphics[width=4in]{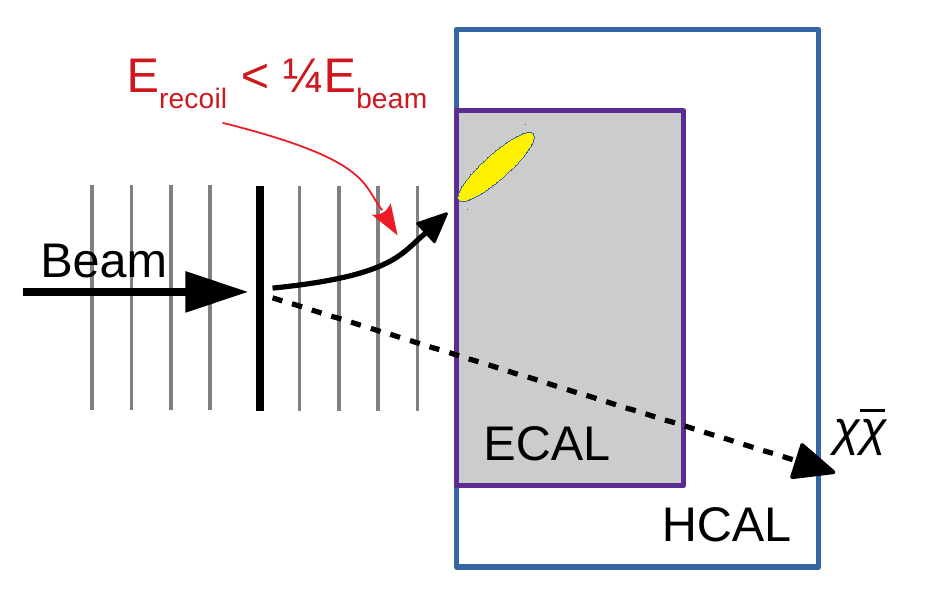}
  \caption{Conceptual drawing of the LDMX experiment, showing the electron beam passing through a tagging tracker, impacting on a thin tungsten target, the recoil tracker, the electromagnetic calorimeter, and hadron calorimeter.}\label{fig:ldmx_cartoon}
  \end{figure}

However, reaching the full potential of this technique places
demanding constraints on the experiment and beamline supporting it.  A
high repetition rate of electrons is required ($\sim50$M $\rm e^-$/sec on
target for Phase I, and as much as $\sim1$G $\rm e^-$/sec on target for Phase II), and so also a fast detector that can individually resolve the
energies and angles of electrons incident on the detector, while
simultaneously rejecting a variety of potential background processes varying in
rate over many orders of magnitude.  A conceptual cartoon diagram of the
proposed experimental design is shown in Fig.~\ref{fig:ldmx_cartoon},
showing the alignment of the beam, the thin target, a tracker for the
recoil electrons, and the required electromagnetic and hadron
calorimeters to confirm the missing momentum signature.
This cartoon will be helpful to the reader for understanding the discussion of 
signal and potential background reactions in Section III. 
The remainder of this note, from Section IV onward, describes the design in greater detail. 

\clearpage
\section{Signal and Background Processes}
\subsection{Dark Matter Signal Production}
\label{sec:proc_sig}
This section briefly summarizes cross-sections and kinematic features of the DM production process in the case of an on-shell  dark photon. More detail can be found in~\cite{Izaguirre:2014bca,Bjorken:2009mm}.
The DM production cross-section $\sigma_{DM}$ scales as the square of the kinetic mixing parameter $\epsilon$, therefore, the quantity $\sigma_{DM} /\epsilon^2$ and the kinematic distributions of the recoiling electron depend only on particle masses. Throughout this section, we assume the common benchmark value $m_{A'} = 3 m_{\chi}$. 

\begin{figure}[t]
\includegraphics[width=0.5\textwidth]{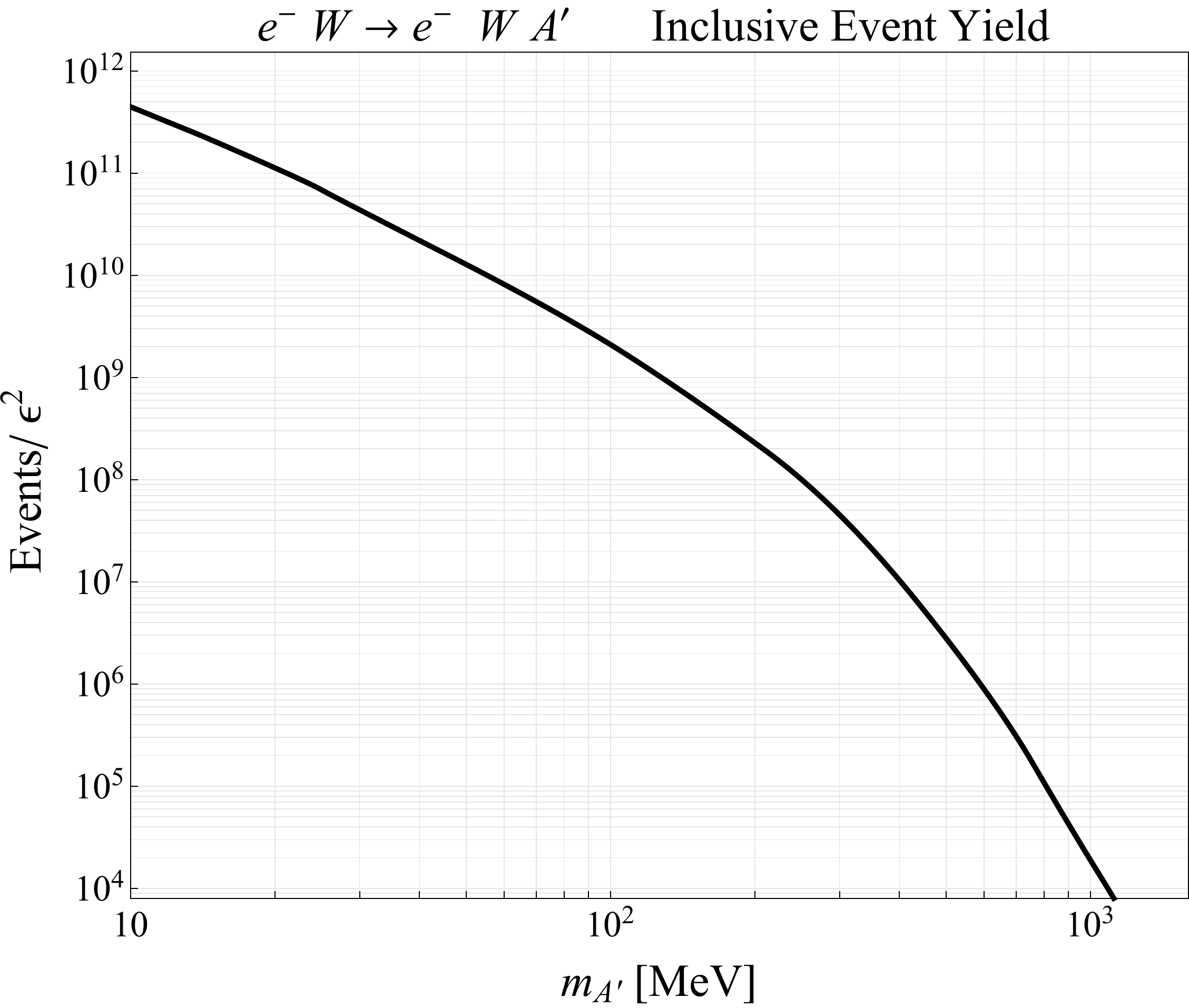}
\caption{\label{fig:LDMyield}
Inclusive cross-section normalized to $\epsilon =1$ for dark matter pair production off a tungsten target. The figure is adapted from \cite{Izaguirre:2014bca}, but utilizes 
signal samples for a 4 GeV electron beam and normalized to the Phase I LDMX luminosity of 4 $\times 10^{14}$ EOT and target thickness $0.1\,X_0$.
}
\end{figure}

The normalized DM production cross-section on a tungsten target, as a function of DM mass, is shown in Fig.~\ref{fig:LDMyield}.  For low masses, the production cross-section scales as $1/m_{A^\prime}^2$; for higher masses, it is further suppressed by the steeply falling form-factor for scattering off the nucleus with high enough momentum exchange to produce an on-shell mediator. 

\begin{figure}[!htb]
\includegraphics[width=0.45\textwidth]{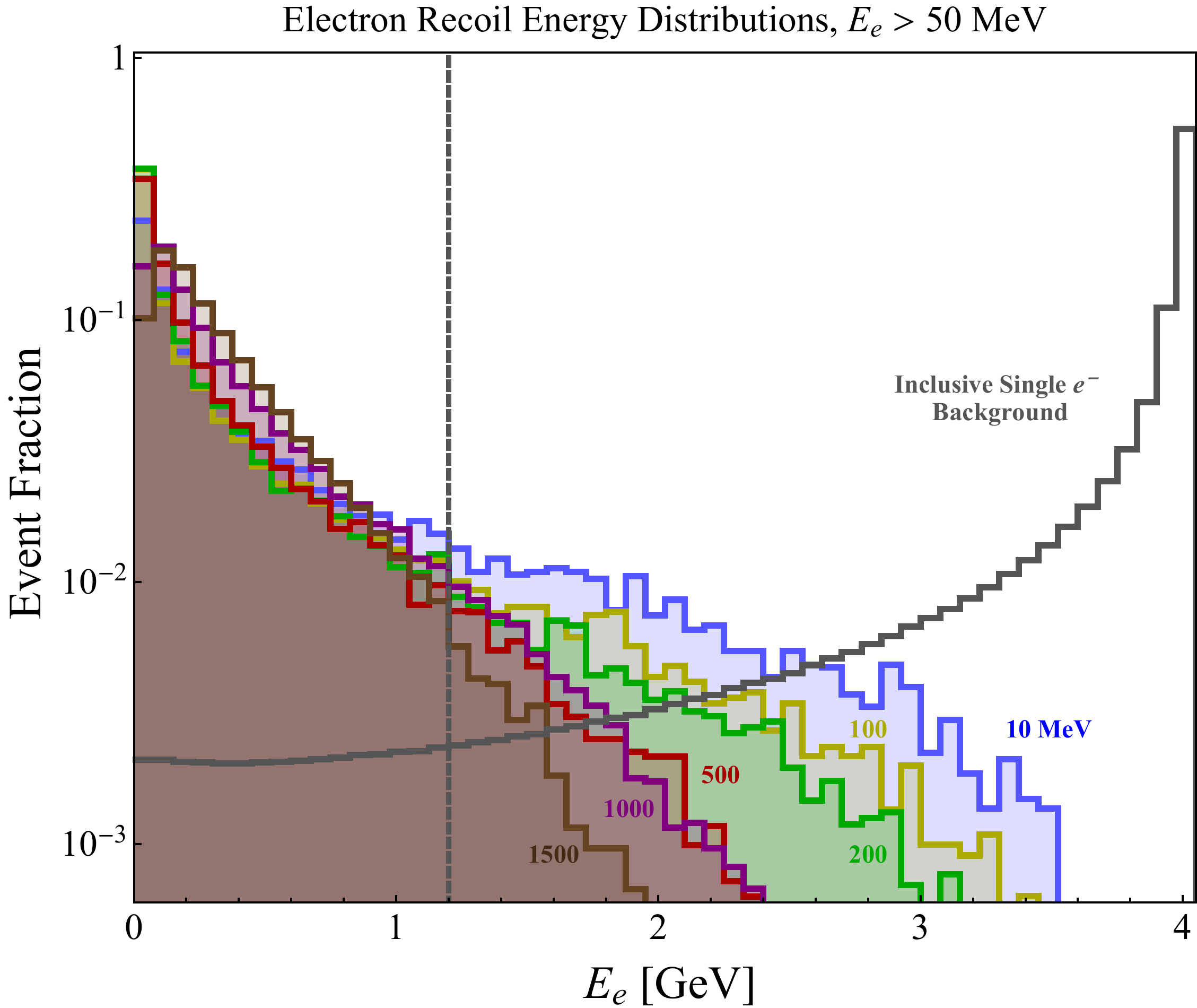}
\includegraphics[width=0.45\textwidth]{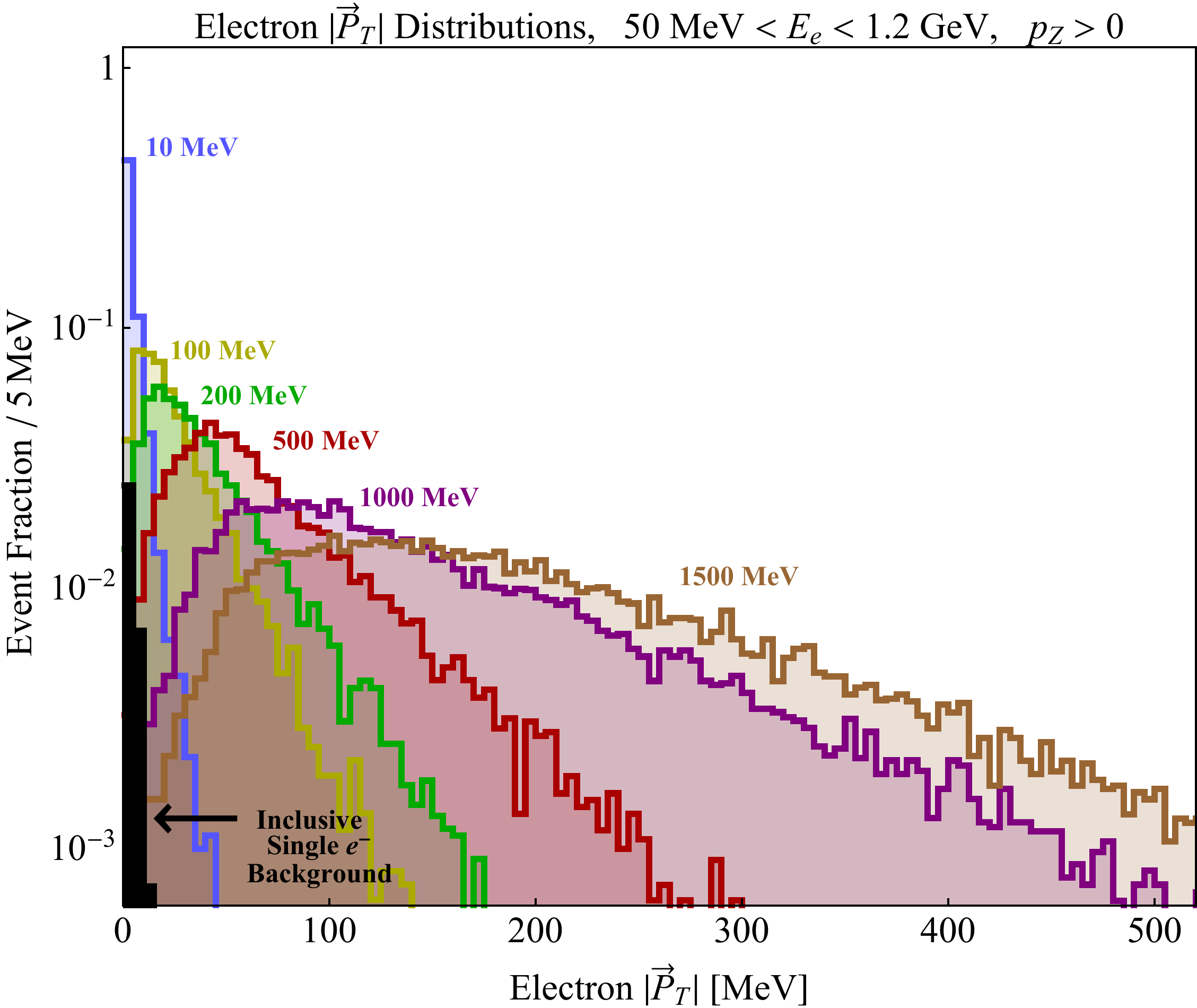}\\
\includegraphics[width=0.45\textwidth]{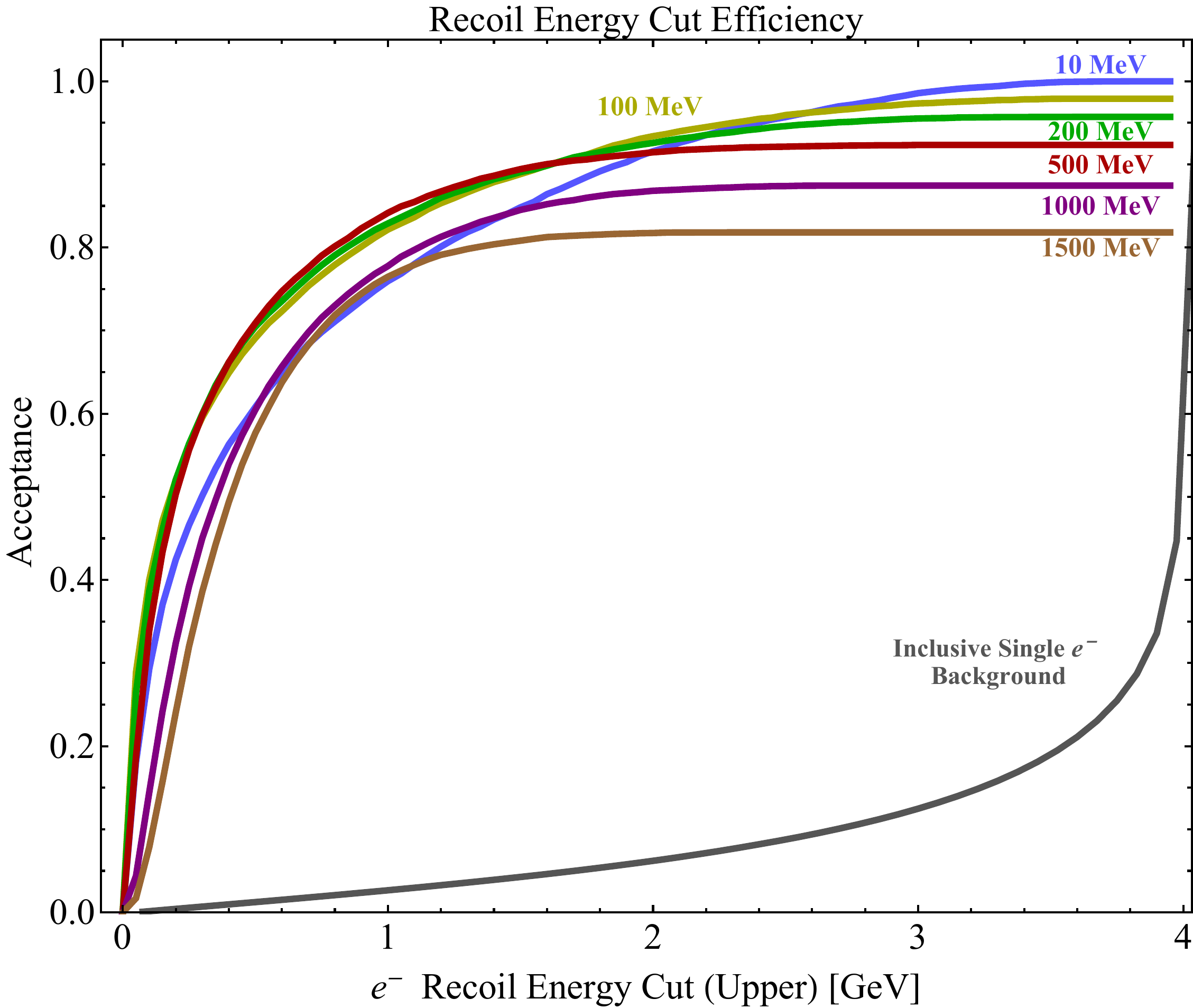}
\includegraphics[width=0.45\textwidth]{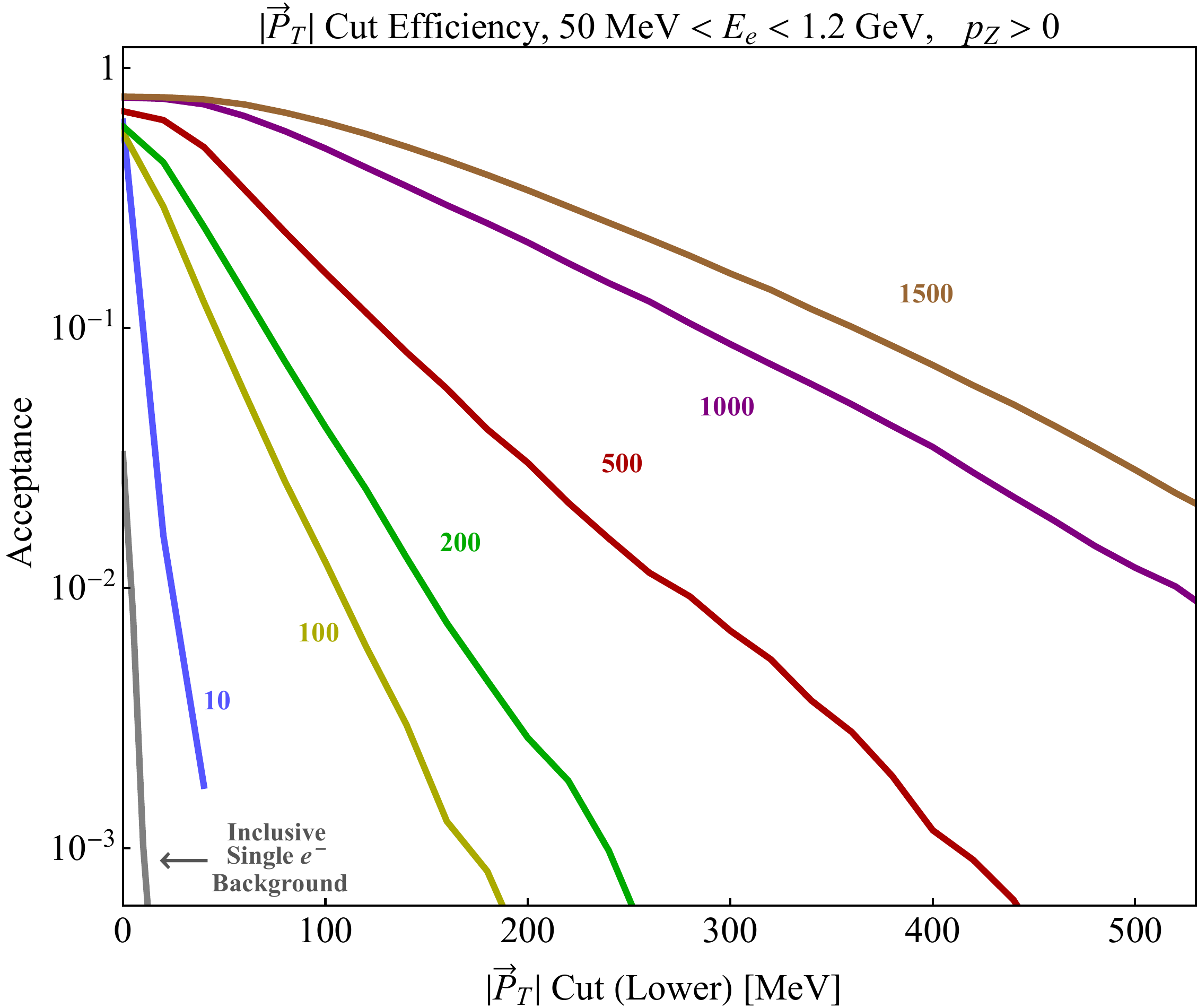}
\caption{\label{fig:LDMspectra}  
Top: Electron energy (left) and $p_T$ (right) spectra for DM pair radiation process, at various dark matter masses.  Bottom Left: Selection efficiency for energy cut $E_e < E_{cut}$,  as a function of $E_{cut}$, on inclusive signal events, The nominal cut is $E_{cut} = 0.3 E_{beam} $. 
Bottom Right: Selection efficiency for $p_T$ cut $p_{T,e}>p_{T,cut}$, as a function of $p_{T,cut}$, on events with $50 \ \MeV<E_e<E_{cut}$.  In all panels, the numbers next to each curve indicate $A'$ mass. Also included in each plot is the corresponding inclusive single electron background distribution.}
\end{figure}

The kinematics of the electron in DM production also depend on the $A'$ mass.  In general, as long as the $A'$ or $\chi\bar\chi$ pair that is produced is heavy relative to the electron mass, the differential cross-section for DM production is peaked in the phase space where the DM carries away the majority of the beam energy and the electron carries relatively little.  This is, of course, the exact opposite of the structure for the kinematics of the vast majority of bremsstrahlung events (with the tiny remainder constituting the main LDMX backgrounds, as discussed in the next section), and the primary kinematic handle for a missing momentum search.  This qualitative behavior is accentuated for larger $\chi$ and $A'$ masses, and less dramatic for lower masses comparable to $m_e$, as shown in the top left panel of Figure \ref{fig:LDMspectra}.  The selection efficiency of a requirement  $E_e < E_{cut}$ is shown on the bottom left panel as a function of $E_{cut}$ (the fraction of events accepted saturates below 100\% for some mass points because events with backward-going recoil electrons are not included for this plot).  A nominal value $E_{cut} = 0.3 E_{beam} = 1.2 $ GeV (for 4 GeV beam energy) is assumed as a signal selection throughout this note.  While this cut has not yet been optimized, it does keep 75-90\% of signal events over the full range of masses considered.  Reducing it by a factor of 2 would only slightly degrade signal efficiency.  In the remainder of this section we also impose a cut $E_e > 50 \ \MeV$ motivated by the degradation of tracking acceptance at lower energies. This too could be relaxed in future studies, but is included now to be conservative. 

The top right panel of Figure \ref{fig:LDMspectra} shows the distribution of the electron transverse momentum ($p_T$) for events that pass this $E_e <1.2$ GeV cut for a range of DM and $A'$ masses.  These are to be contrasted with the sharply falling $p_T$ distribution from bremsstrahlung, which (even after accounting for multiple scattering in a $10\%\,X_0$ target) falls off as $1/p_T^3$ for $p_T \gtrsim 4$ MeV.  As we will see below, bremsstrahlung-originated events dominate the background, so this kinematic difference is a powerful one. 

The optimal strategy for using angular/transverse momentum information in a DM search will depend on the ultimate rejection power of the experiment's visible-particle veto, and varies depending on the DM mass of interest. Our approach to designing LDMX is to require that {\bf no} kinematic information be needed to mitigate potential backgrounds that descend from bremsstrahlung photons. Instead, our goal is to fully measure the kinematics of the recoiling electron in as unbiased a manner as possible. In this way, any candidate signal events can be cross-checked against the expected features of dark matter production and its mass measured, thereby strengthening our discriminating power in the case of a discovery. 
If, however, a substantial number of events pass the energy selection and overall detector veto, either due to unexpected background reactions or poorer than planned performance, then the recoil electron $p_T$ measurement can be used either to \emph{reject} these events or to measure the bremsstrahlung component. Requiring a \emph{minimum} recoil electron $p_T$ of 20--50 MeV offers substantial kinematic rejection of bremsstrahlung background, but substantially degrades the experiment's acceptance  for  $A'$ masses below $\sim 100$ MeV (i.e. DM masses below $\sim 30 \ \MeV$ in our plots).   At lower masses, a 2D selection using both energy and angle information, or a multi-component signal+background fit to the 2D $p_T$-energy distribution may be the more effective use of this measurement.

\begin{figure}[htbp]
\includegraphics[height=0.4\textwidth]{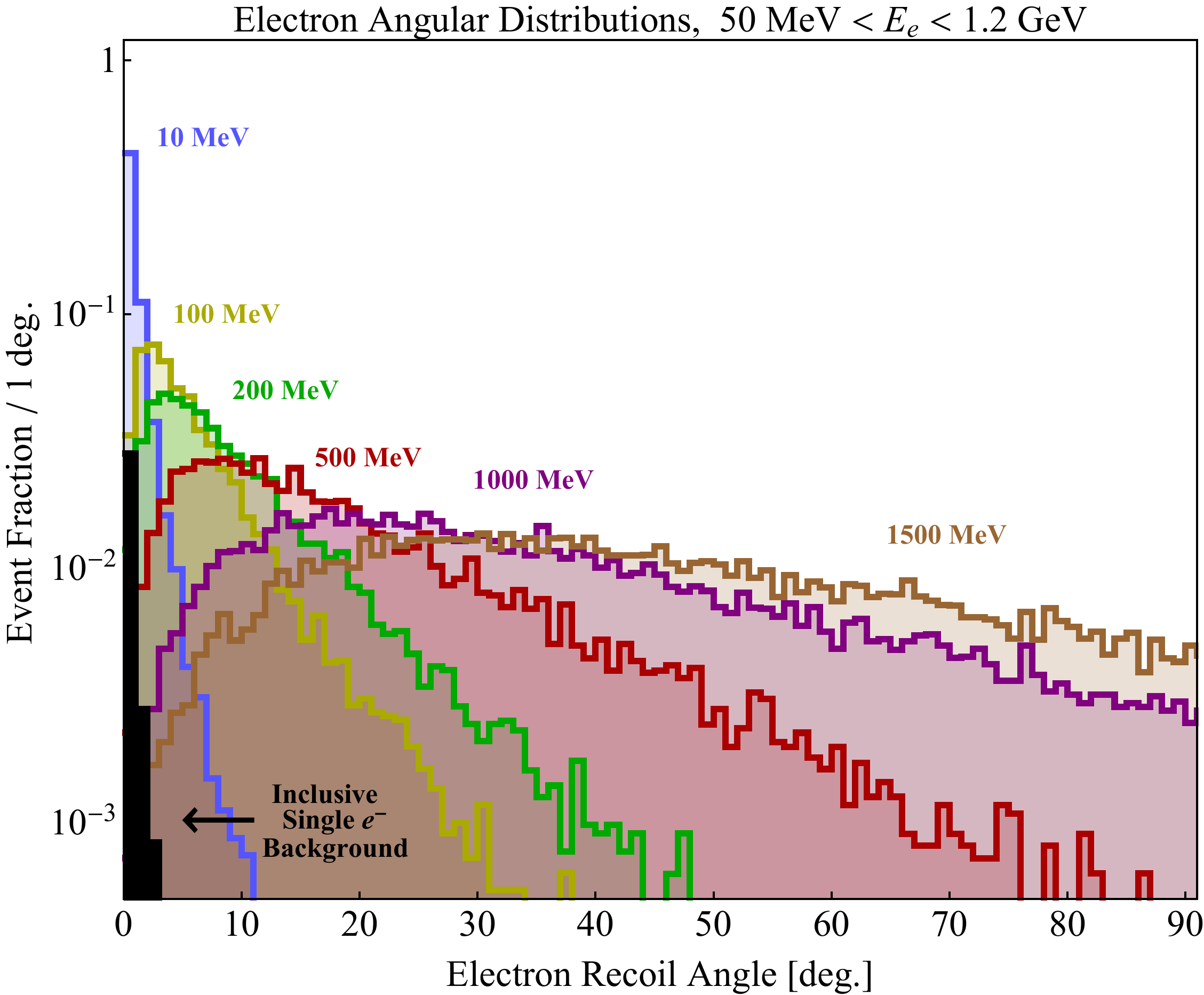}
\hspace{0.4in}
\includegraphics[height=0.4\textwidth]{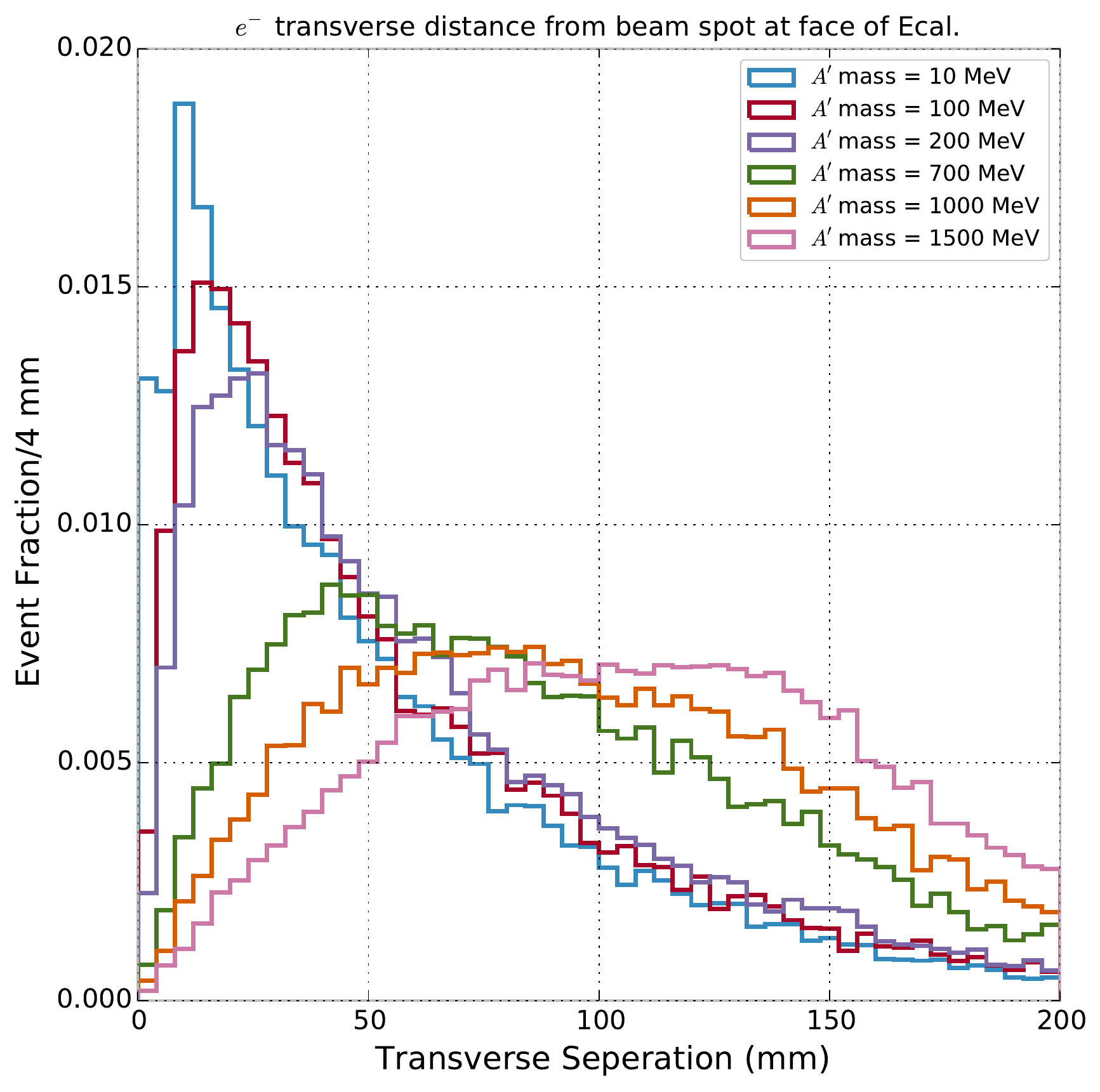}
\caption{\label{fig:angularSpectrum}
Left: Angular distribution of signal events passing the energy cut $E_e < 1.2$ GeV.  The numbers next to each curve indicate $A'$ mass.
Right: Transverse separation of the recoiling electron from the beam spot at the face of the ECAL  for the detector geometry described in Section~\ref{sec:detector}.
}
\end{figure}

Another very important kinematic feature is the angular/transverse momentum separation between the recoiling electron and the direction of the beam.  The larger this separation, the better the spatial separation between the recoil electron's shower and other final-state particles that one would like to veto.  Figure \ref{fig:angularSpectrum} (left) shows the angle of the recoiling electron at the target for events passing the energy cut, while Figure \ref{fig:angularSpectrum} (right) shows, for the same events, the transverse separation of the recoiling electron from the beam spot at the face of the ECAL for the detector geometry described in Section~\ref{sec:detector}.

\subsection{Background Processes \morepeople{Philip, David}}
\label{sec:proc_bkg}
This section summarizes and briefly describes the important background processes for a dark photon signal.
This is an introduction to the main challenges presented by each background, and later in Sec.~\ref{sec:bkgrej}, a detailed study of the rejection of the backgrounds will be presented.

\begin{figure}[tbh]
\includegraphics[width=0.90\textwidth]{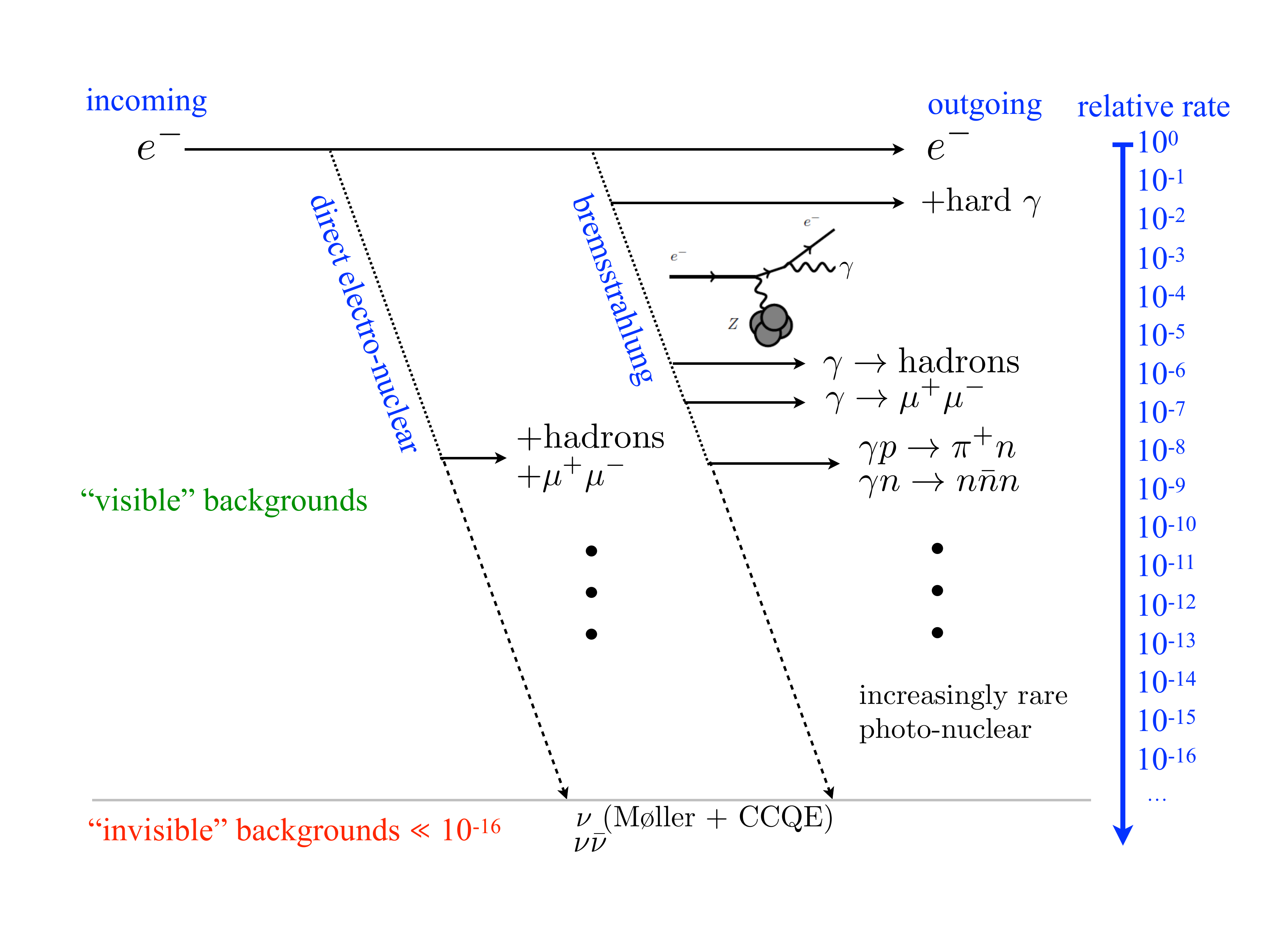}
\caption{\label{fig:BackgroundsChart} 
\ednote{Tim: The Keynote file is in the figures directory if more changes are desired.} 
Flow of important (veto design driving) potential background processes and their raw rates relative to the number of beam electrons incident on the target.}
\end{figure}

As an orientation, the highest rate process is the situation when the electron does not interact in the target and showers in the calorimetry. If the electron does undergo bremsstrahlung then the next set of processes are related to the potential fate of the photon. In rarer instances, the photon can interact without undergoing a typical electromagnetic shower, and therefore presents unique detector challenges. A similar process can happen for electrons undergoing a nuclear interaction. Finally, we discuss the rates of neutrino backgrounds. These processes and their relative rates are summarized in Fig.~\ref{fig:BackgroundsChart}.

\ednote{Give an explanation of the approximate rates, add more feynman diagrams and schematics?}

\subsubsection{Incident low-energy particles/beam impurities\people{Tim, Omar}} 

At the level of of $10^{16}$ incident electrons, a large number of electrons with low energies - consistent with signal recoils - are certain to intersect the target even for the most stringent requirement on the purity of the incoming beam. Such electrons, reconstructed correctly by the detectors downstream of the target, cannot be distinguished from signal.

For this reason, LDMX employs a tagging tracker to measure the trajectories of incoming electrons to ensure that each recoil corresponds to a clean 4~GeV beam electron entering the apparatus on the expected trajectory to veto any apparent signal recoils originating from beam impurities.

\subsubsection{Electrons that do not interact in the target \people{Mans}} 

These electrons experience some straggling in the trackers and target, but do not lose appreciable energy.  These events feature a hard track through both trackers and typically include a high-energy ($\approx 4$ GeV) shower in the \ecal.  Occasionally, such events may have lower energy in the \ecal due to electro-nuclear or photo-nuclear interactions occurring during the shower development in \ecal. 

\subsubsection{Hard bremsstrahlung \people{all}}

\ednote{Philip and Omar should go through and switch all numbers to relative rate.}

These occur in the target (or, less frequently, in a tracking layer), such that the electron track through the recoil tracker falls below $E_{cut} = 1.2$ GeV, with a relative rate of $3\times 10^{-2}$ per incident electron. These events have two showers in the \ecal, with combined shower energy $\approx 4$ GeV, separated by $1-2$ cm or more depending on the electron energy. In rejecting these backgrounds, it becomes important to resolve the electron from the photon and measure the photon energy with high resolution.  In this context, a photon that is very poorly measured can give rise to a fake dark photon signal.

The discussion above assumes that the bremsstrahlung photon does not interact until it hits the \ecal.  Of course, the photon may convert in the target (or, less frequently, in a tracking layer).  This background topology occurs at a relative rate of $1.5\times 10^{-3}$ per incident electron. It is generally easier to detect than the non-interacting bremsstrahlung: in addition to the energy deposition in the \ecal, the $e^+e^-$ pair from the conversion produces  one or two additional tracks in the recoil tracker, which can be vetoed.   The same considerations apply to the background of trident reactions $e^-\rightarrow e^- e^+e^-$ in the target ($\sim 1.5\times 10^{-4}$ per incident electron) where at least one of the outgoing $e^-$ has energy below $E_{cut}=1.2$ GeV.

A more challenging variation of hard bremsstrahlung is when, rather than producing a typical electromagnetic shower, the bremsstrahlung photon undergoes a rare process such as a photo-nuclear reaction or conversion to $\mu^+\mu^-$ (or $\pi^+\pi^-$).  These processes typically deposit less detectable energy in the \ecal, leading to a significant undermeasurement of the photon energy.  We now discuss these photon processes which represent a challenging class of rarer backgrounds.  These additional photon interactions can occur in the target or the calorimeter, with each case presenting unique challenges.


\subparagraph*{Hard bremsstrahlung + \pn reaction in the target or \ecal} These are events where the hard bremsstrahlung photon does not undergo conversion but instead undergoes a \pn reaction in the target area or one of the first layers of the \ecal (typically in a tungsten layer).  The \pn cross-section is roughly a thousand times smaller than the conversion cross-section, so these events will occur with overall relative rate of $1.7\times 10^{-5}$ per incident electron.

The \pn processes initiated by $2.8-4$ GeV photons can result in a wide range of final states.  When pions are produced in the interaction and escape the nucleus, they typically give rise to either ``tracks'' ($\pi^\pm$) or substantial energy deposition ($\pi^0$) in the \ecal.   In many cases, a large number of low-energy protons and neutrons are liberated from a heavy nucleus; some of the protons may deposit energy in the \ecal, and some of the neutrons can be detected by the \hcal.  Two rare but important classes of these events are characterized by only two to three ${\cal O}$(GeV), moderate-angle neutrons escaping from the nucleus, or a single forward $\sim 3$ GeV neutron. These events must be rejected by the \hcal. 

The same \pn reactions can also occur in the target.  These events have
50$\times$ lower rate (relative to reactions in the \ecal) because of the thin target material, and tracking provides
additional handles to reject them, but they are also more difficult to detect
calorimetrically because the effective angular coverage of the \hcal is reduced.

\begin{figure}[tbh]
\includegraphics[height=0.2\textheight]{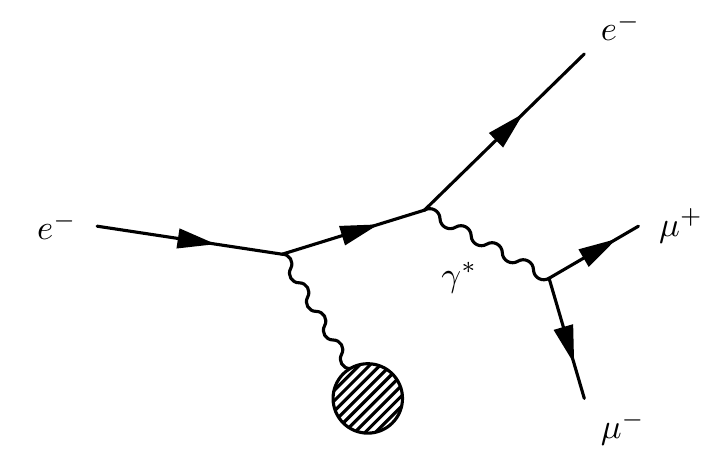} \hspace{0.5cm}
\includegraphics[height=0.2\textheight]{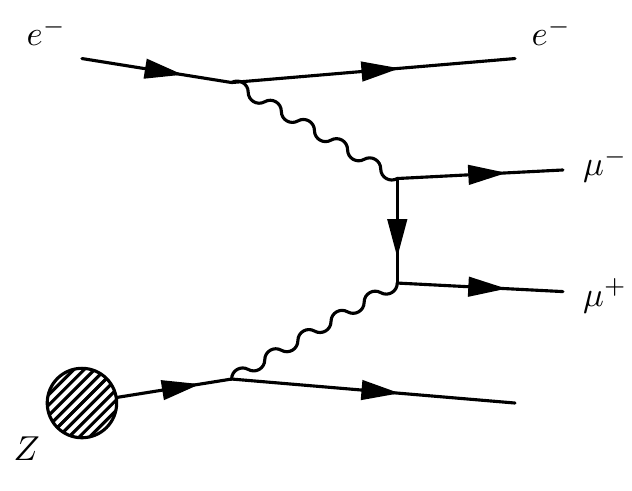}
\caption{\label{fig:muonfeyn}
Reducible potential background reactions, photoconversion (left) and Bethe-Heitler (right), with pair-produced muons in the final state.}
\end{figure}

\subparagraph*{Photon conversion to muons} 
Analogous to the usual conversion reaction, photons can instead convert to a pair of muons with a rate suppressed by $m_e^2/m_\mu^2$.  Such events occur at a relative rate of $9.1\times 10^{-7}$ per incident electron.   Muon events leave a qualitatively different detector signature than other background processes, namely one or two ``tracks'' passing through the calorimeter.  
 Such processes can occur in the target, tracking system, or the photoconversion process can occur in the calorimeters.  To ensure that wide-angle muon pair production is well modeled, we have included both coherent conversion $\gamma Z \rightarrow \mu^+\mu^-Z$  off a high-$Z$ nucleus and the rarer incoherent conversion off a single nucleon. In addition, muon pairs can be produced in the absence of a preceding hard bremsstrahlung by trident reactions, illustrated in Fig.~\ref{fig:muonfeyn}.

\subsubsection{Electro-nuclear interactions in target \people{Omar}}
\begin{figure}[tbh]
\includegraphics[width=0.80\textwidth]{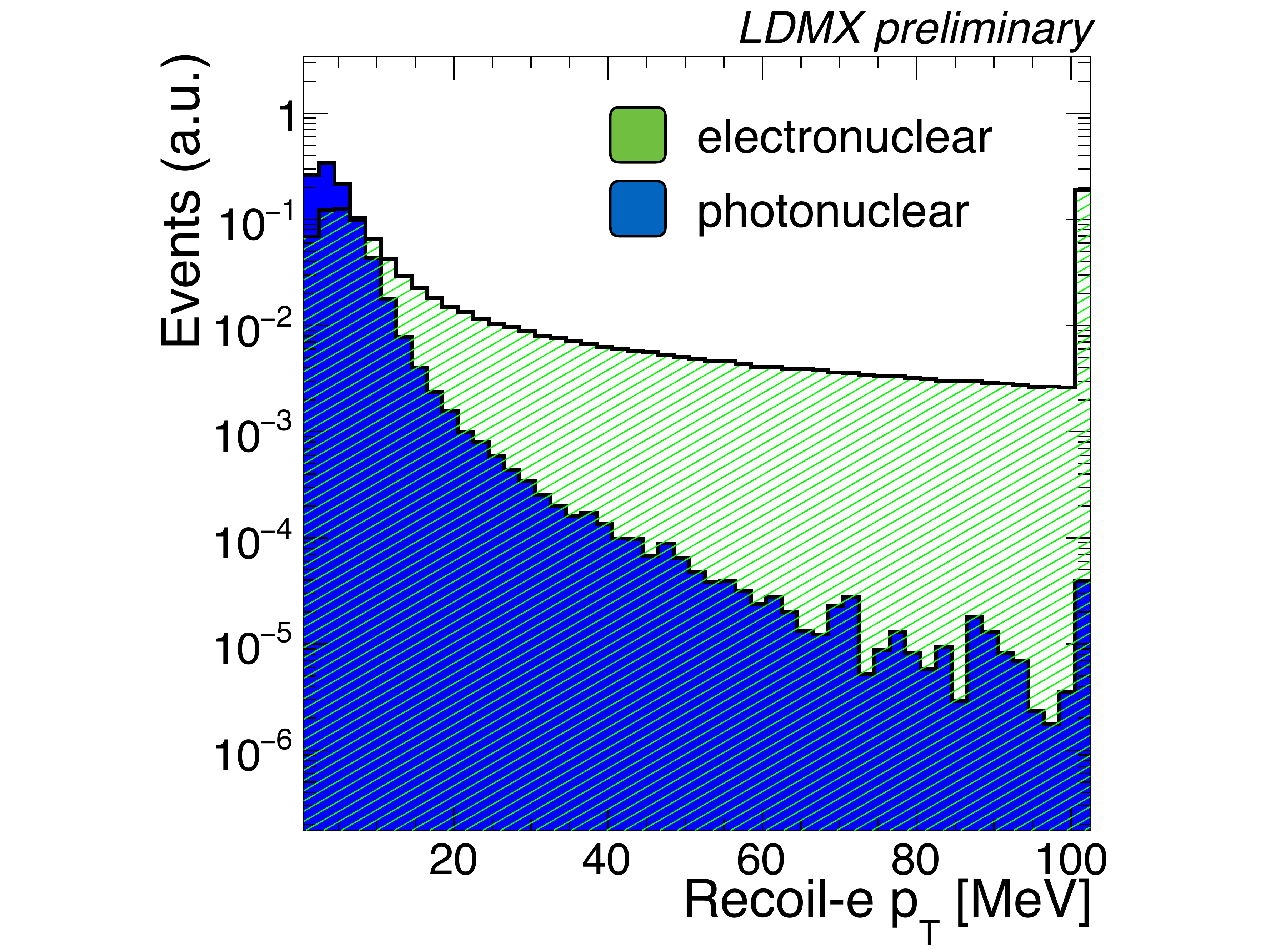}
\caption{\label{fig:recoil_pt_shapes}Distributions of recoil transverse momentum for photo-nuclear events originating in the ECAL and electro-nuclear events originating in the target. The last bin shows the integrated yield for $p_{\text{T}}>100$~MeV. } 
\end{figure}

Electro-nuclear interactions present a similar composition to \pn 
reactions but occur at a lower rate and have a broader $p_T$ distribution. Figure~\ref{fig:recoil_pt_shapes} shows the expected distribution of the recoil electron transverse momentum distribution for photo-nuclear and electro-nuclear events. Because electro-nuclear reactions in the \ecal have the additional handle of 
the tracking system which can reject full energy electrons, the primary 
concern becomes electro-nuclear reactions that occur in the target area.  
However, because they are kinematically similar to the \pn background originating inside the target,
the same rejection strategy is used for both interactions.

\subsubsection{Neutrino backgrounds}
The backgrounds discussed above are all ``instrumental'' in that they rely on mismeasurement of energy escaping in visible particles (e.g. hadrons or muons) to fake a missing energy signal.  Neutrino production, by contrast, transfers energy from the incident electron to particles that are (essentially) undetectable at LDMX.  The rates for these processes are extremely small.  

The dominant neutrino-production process, charged-current electron-nuclear scattering, has an inclusive cross-section of $\sim 8$ fb/GeV/nucleon leading to $\mathcal{O}(5)$ such events expected in Phase I of LDMX.  This is not a background for LDMX because there is no recoiling electron in the final state.  However, rarer processes and/or multiple interactions can mimic the  LDMX signal. Although the rates are small, it is worth elaborating on these background yields to ensure that they can be neglected and/or measured.  In particular, the latter two kinds of processes are reducible but may need to be considered in scaling up LDMX by 1-2 orders of magnitude in luminosity. 
\ednote{NT: THERE'S STILL SOME CONFUSION IN MY MIND ABOUT THE OVERALL RATE, ELECTRON VS. NEUTRINO INCIDENT...}

Neutrino backgrounds can be divided into three classes depending on the origin of the apparent recoil electron (from the neutrino-production process, from a second interaction in the target, or as a fake):

\subparagraph{Neutrino trident processes} $e N\rightarrow  e\,\nu\,\bar\nu\,N$ are an \emph{irreducible physics background} --- the final state has both a low-energy recoil electron (with broad $p_T$ distribution) and significant missing energy carried by invisible particles.  Indeed, neutrino trident events are a direct counterpart of our signal process, mediated by electroweak bosons rather than a new force carrier.  However, their rate in LDMX or even the most ambitious future upgrades is negligible.  Parametrically, this reaction is suppressed relative to hard bremsstrahlung by $\frac{\alpha_2}{\pi} \frac{m_e^2 M_{\max}^2}{M_Z^4} \sim 10^{-18}$ where $M_{\max}^2 \sim E_{beam}/R_{nuc}$ is the pair mass scale at which the nuclear form factor strongly suppresses the trident process.  For LDMX at 4 GeV, a neutrino trident cross-section of 20 ab was found using MadGraph/MadEvent, leading to a yield of $10^{-5}$ such events originating from the target in Phase I.  While this background is clearly irrelevant for the currently proposed experiment, it does set an in-principle floor to the sensitivity of the missing momentum search program. 

\subparagraph{Multiple interactions in target + charged-current scattering}  Supposing that the charged-current scattering produces no visible final-state particles, a background to the LDMX signal could still arise from the combination of two processes occurring within the target --- for example, production of a Moller electron at $\mathcal{O}(100)$ MeV followed by the charged-current scattering of the incident electron, leading to a ``low-energy $e$ + high-energy $\nu$'' final state.  Another similar final state arises from (1) soft bremsstrahlung by the incident electron, (2) asymmetric conversion of the resulting photon, with low-energy positron absorbed in the target and high-energy electron escaping, and (3) charged-current scattering of the hard incident electron.  These processes are both expected to lead to $\mathcal{O}(0.01)$ background events at LDMX.

\subparagraph{Inelastic charged-current scattering (fake electron).} The charged-current scattering cross-section for multi-GeV electrons (or neutrinos) is dominated by resonant excitation of a nucleon and deep inelastic scattering, which can lead to charged pions in the final state.  There is some region of phase space where these pions, if they fake an electron, would mimic the LDMX signal.  It is difficult to estimate this rate quantitatively. However, since it represents only a fraction of the final-state phase space for inclusive CC scattering, the yield should be no more than $\mathcal{O}(1)$ event and likely smaller.  Furthermore, since electrons and pions can be distinguished calorimetrically, this background is expected to be negligible.

\subsection{Experimental Strategy}
\label{sec:proc_strategy}
This section summarizes the high level design considerations for LDMX. We have two goals -- to measure the distinguishing properties of dark matter production (see Section~\ref{sec:proc_sig}), and to reject potential backgrounds for this process (see section~\ref{sec:proc_bkg}). 

\begin{figure}[h]
  \begin{minipage}{0.47\linewidth}
    \begin{center}
      \includegraphics[width=\linewidth]{sections/detector/figures/detector_cartoon_signal.pdf}
      \\
      (a)
    \end{center}
  \end{minipage}\hfill
  \begin{minipage}{0.51\linewidth}
    \begin{center}
      \includegraphics[width=\linewidth]{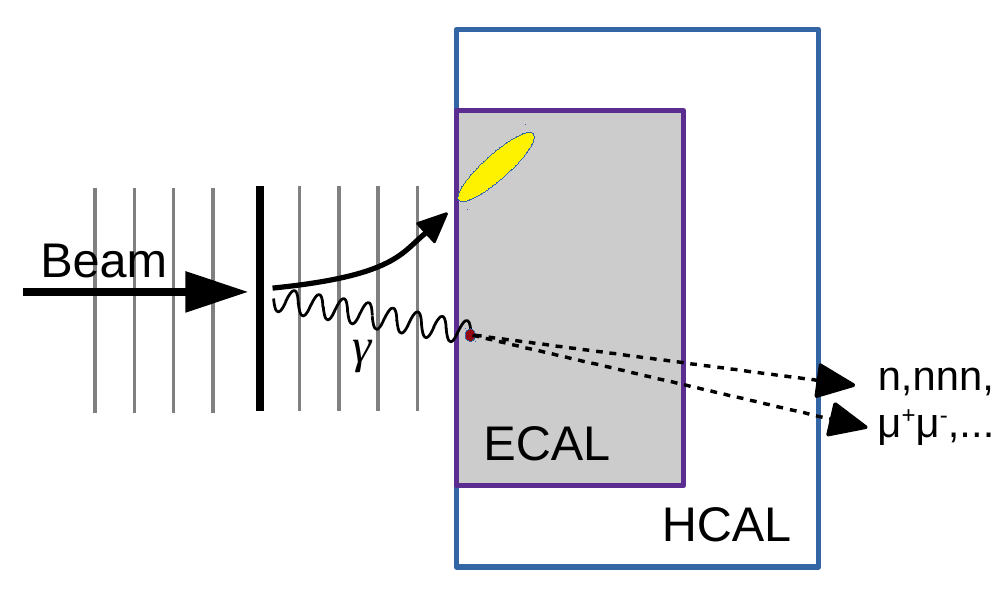}
      \\
      (b)
    \end{center}
  \end{minipage}

\begin{flushleft}
\caption{\label{fig:strategy}Conceptual schematic of a signal process (a) and dominant background (b) processes.}
\end{flushleft}
\end{figure}

To measure the signal process, we need to precisely measure the momentum of both the incoming and recoiling electron. The relative transverse momentum provides a measure of the mass of the produced dark matter, and the overall relative momentum provides a measure (to a high degree) of its direction. Thus, we want to use a thin target and low material tracking (minimize the impact of multiple scattering) in a magnetic field to measure the signal process. This is illustrated by the left cartoon in Fig.~\ref{fig:strategy}.

Having measured the momentum change across the target, our main goal is then to capture the fate of bremsstrahlung photons and non-interacting electrons. The fate of such photons is illustrated in the background flow chart shown in Fig.~\ref{fig:BackgroundsChart}. A typical candidate background reaction is illustrated by the right cartoon in Fig.~\ref{fig:strategy}.

Both non-interacting electrons, and electrons that undergo appreciable bremsstrahlung in the target give rise to EM showers (totalling the beam energy) in the ECAL system, but these are easily rejected with good shower energy reconstruction resolution. Therefore, these events can be readily rejected at trigger level, thereby limiting the event rate of the experiment to a manageable $\mathcal{O}(5)$ kHz or lower rate. 

At the level of roughly $\sim 2\times 10^{-5}$ per incident electron, photo-nuclear (PN) processes occur, and at roughly $\sim 10^{-6}$ muon conversion processes occur.  
The ECAL will continue to reject the majority of PN events due to energy deposition from $\pi^0$ production, but good MIP sensitivity is required 
to reject the small subset of events that contain no energetic $\pi^0$, but rather energetic $\pi^{\pm}$, $p$, or photon conversions to muon pairs.
In the system described in this paper, this can in principle provide another factor of $\sim 10^3$ rejection, depending on the type of final state. 

In practice, the most difficult potential background for the ECAL to reject is one in which the hard bremsstrahlung photon converts all of its energy into a single (or few body) neutral hadronic final state. This occurs at a relative rate of $\sim 10^{-3}$ per incident hard photo-nuclear reaction (on W), but these usually have a hard charged pion or proton in the final state. Thus, the region of phase space where the MIP is soft and invisible poses the largest threat of producing a background, and this is expected at the $\sim 4\times 10^{-4}$ per hard photo-nuclear interaction (on W). Per incident 4 GeV electron on Tungsten absorber, this corresponds to $\sim 10^{-8}$ in relative rate. For a benchmark of $1\times 10^{14}$ electrons on target, we would face up to $\sim 10^6$ events with a single hard forward neutron and very little else in the ECAL (other than the recoil electron). 
This drives the performance requirement of the hadronic veto -- we require better than $10^{-6}$ neutron rejection inefficiency in the few GeV energy range. In practice, an HCAL veto meeting this requirement is also sufficiently sensitive to muons to veto the remainder of the photon conversions to muon pairs (and by extension, pion pairs). Moreover, this level of inefficiency provides a great deal of redundancy against potential failures of the ECAL veto with respect to photo-nuclear, electro-nuclear, or MIP conversion events. 

\ednote{Philip: Do we want some kind of high level summary table of the performance requirements? Would that actually be useful?}

\ednote{Philip: Omar -- can you check some of the numbers above? Consult Natalia as well, as she is checking the GEANT rate for forward going neutrons.}


\clearpage
\section{Detector Concept}
\label{sec:detector}

As described in Sec.~\ref{sec:proc_sig}, the signature of DM production used by LDMX involves (i) substantial energy loss by the electron (e.g. recoil with $\lesssim 30\%$ of incident energy), (ii) a potentially large transverse momentum kick, and (iii) the absence of any additional visible final-state particles that could carry away energy lost by the electron. In the first phase of LDMX we will search for this signature in a sample of $4 \times 10^{14}$ incident electrons, delivered onto a $10\%$ radiation-length (0.1~$X_0$) target with a 46 MHz bunch spacing and average charge of one $e^-$~per bunch.  The target thickness has not been precisely optimized, but represents a reasonable compromise between event rate and transverse momentum resolution over a wide range of dark matter mass.  Performing the measurement of each kinematic component (i and ii above) on every incident electron requires a tracker and electromagnetic calorimetry (\ecal) downstream of the target, with a sensitive area that extends over the nominal beam axis.  Placing the tracker in a weak magnetic field allows us to use some of the most striking of these events in which a soft, wide-angle recoil electron does not penetrate into the \ecal.  In addition,  stray low-energy particles from beam halo must be rejected with very high efficiency, as these could mimic the DM signal.  This motivates another tracker with stronger B-field upstream of the target that precisely measures the momentum and trajectory of the incoming beam electrons. 
While the energy and angle resolution requirements for LDMX are modest, the experiment poses two main challenges. First, every incident electron passes through the trackers and showers in the \ecal; the detectors placed directly in the beam line must therefore contend with high radiation doses and a large event rate. In order to mitigate this issue, the beamspot for the experiment must be quite large - of order 10-100 cm$^2$. Second, the requirement of an event veto for any additional visible final-state particles to take advantage of (iii), that is robust enough to handle a variety of rare backgrounds is not only an important physics performance driver for the tracking and \ecal, but also calls for the addition of a dedicated hadron veto system surrounding the sides and back of the \ecal. 

\begin{figure}[htp]
    \centering
    \includegraphics[width=\textwidth]{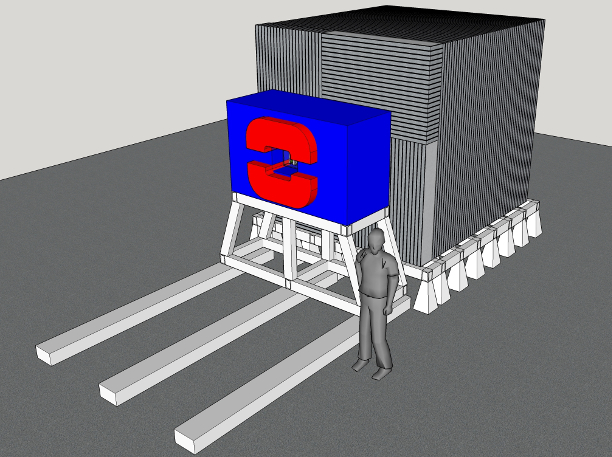}
    \caption{\small{An overview of the LDMX detector showing the full detector apparatus with a person for scale.}}
    \label{fig:LDMX_overview}
\end{figure}

\begin{figure}[htp]
    \centering
    \includegraphics[width=\textwidth]{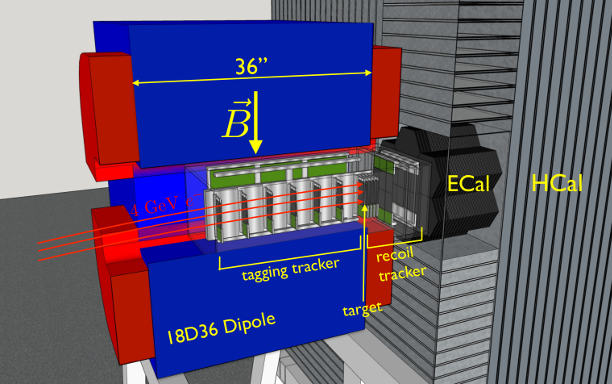}
    \caption{\small{A cutaway overview of the LDMX detector showing, from left to right, the trackers and target inside the spectrometer dipole, the forward \ecal, and the \hcal. Although not yet studied, the side \hcal may be extended forward (transparent region surrounding the \ecal) to provide better coverage for wide-angle bremsstrahlung and neutral hadrons originating from photo-nuclear reactions in the target and the front of the \ecal.}}
    \label{fig:LDMX_cutaway}
\end{figure}

These four detector systems -- the ``tagging tracker'' upstream of the target, ``recoil tracker'' downstream, a radiation hard forward \ecal, and the hadronic veto system, form the majority of the LDMX experimental concept.  To keep the detector compact and the field in the \ecal minimal, we place the tagging tracker inside the bore of a dipole magnet and the recoil tracker in its fringe field. The layout for LDMX is illustrated in Figures  \ref{fig:LDMX_overview} and \ref{fig:LDMX_cutaway}. 

The tracking proposed for LDMX borrows heavily from the Silicon Vertex Tracker (SVT) of the HPS experiment \cite{HPS_proposal_2014}, a fixed target search for visibly decaying dark photons using the CEBAF beam at the Thomas Jefferson National Accelerator Facility (JLab).  The HPS SVT was designed to provide high-purity, high-precision tracking for low-momentum ($\lesssim$~3~GeV) electrons at high occupancies (up to 4 Mhz/mm$^2$) with the nearly-continuous (2~ns bunch spacing) CEBAF beam.  The SVT meets these requirements with low-mass construction (0.7\%~$X_0$ per 3d~hit) and excellent hit-time resolution ($\sim2$~ns). Given very similar requirements, the solutions developed for HPS are also optimal for LDMX.

Similarly, a natural solution to the granularity, timing, and radiation hardness required of the \ecal is to exploit the silicon calorimetry designed for the CMS Phase-II high granularity calorimeter (HGC) upgrade in the forward region of the detector \cite{Contardo:2020886}.  The Tungsten-Silicon sampling calorimeter is an ideal technology with very good energy resolution for electromagnetic showers, potential for MIP tracking, and the ability to withstand high radiation levels from non-interacting electrons. In addition, having been designed for 25 ns bunch spacing with up to 200 proton-proton collisions per bunch crossing in the High-Luminosity phase of the CERN LHC, the HGC has rate capabilities that are ideal for the LDMX \ecal. The primary physics trigger will also require a positive signal in a scintillator pad overlaying the target that is coincident with \emph{low} (or no) energy deposition in the \ecal in order to avoid triggering on empty buckets.   Finally, the hadronic veto is designed to be a Steel-Plastic Scintillator sampling calorimeter (\hcal) with high sampling fraction for extremely high neutron detection efficiency and good angular coverage for large-angle background processes. Like the \ecal, the \hcal could take advantage of technology being developed for other experiments, such as high speed SiPM detector readout designed for the current CMS hadronic calorimeter or the Mu2e experiment. 

In the following sub-sections, we describe the LDMX beamline and detector sub-systems in more detail. 


\subsection{Beamline and Spectrometer Magnet \people{Omar,Tim} \morepeople{Robert,David}}
The LDMX beamline consists of a large-diameter beampipe terminating in a thin vacuum window immediately upstream of an analyzing magnet
inside of which the tagging and recoil trackers are installed. 
The analyzing magnet is a common 
18D36 dipole magnet with a 14-inch vertical gap and operated at a central field
 of 1.5~T.
The magnet is rotated by approximately 100 mrad about the vertical axis with
respect to the upstream beamline so that as the incoming 4~GeV beam is deflected by the field, it follows the desired trajectory to the target. In particular, the incoming beam arrives at normal incidence to, and centered on, the target, which is laterally centered in the magnet bore at $z=40$~cm relative to the center of the magnet. Although the magnet gap differs, this arrangement is very similar to that employed by the HPS experiment at JLab.

A number of 18D36 magnets, not currently in use, are available at SLAC, along with the steel required to adjust the magnet gap as may be required to suit our purposes. 
These include a magnet that is already assembled with the 14-inch gap planned for LDMX.  It was tested to 1.0~T in 1978, at which point 199~kW of power was dissipated.  
Based on the current capacity of the other similar magnets with smaller gaps, it is expected that this magnet can be operated at 1.5~T, resulting in a power dissipation of approximately 450~kW and requiring an approximately 55~gpm flow of cooling water. If this magnet proves to be suitable for LDMX, it will be split, cleaned up, and reassembled before testing and carefully mapping the field in the tracking volumes. Although the final location of LDMX has not been determined, the downstream end of the SLAC ESA beamline would permit construction of a large \hcal and could accommodate LDMX operation along with other experiments.



\subsection{Overview of Tracking and Target Systems \people{Omar,Tim,Mans} \morepeople{Robert,David}} 
\label{section:target}
\begin{figure}[htp]
    \centering
    \includegraphics[width=\textwidth]{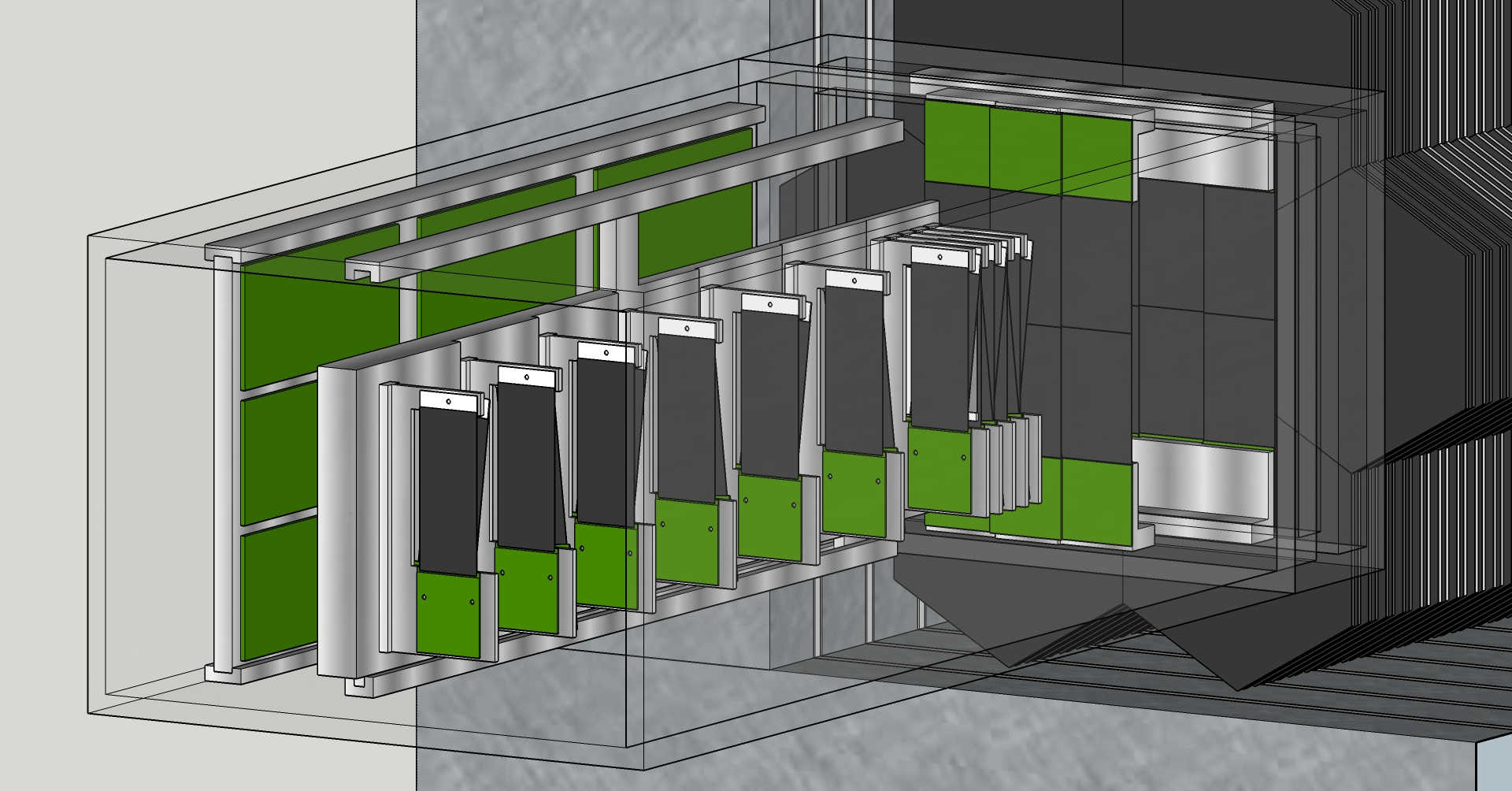}
    \caption{\small{An overview of the tracking systems and target inside the LDMX magnet.}}
    \label{fig:tracking_overview}
\end{figure}

Although the tagging and recoil trackers function as two distinct systems, they use common technologies and share the same support structures and data acquisition hardware.  In particular, the first four layers of the recoil tracker are identical to the layers of the tagging tracker and share a common support and cooling structure, as shown in Figure~\ref{fig:tracking_overview}.
The key element of this upstream support structure is a vertically-oriented aluminum plate onto which the stereo modules are mounted.  
To provide cooling, a copper tube through which coolant flows is pressed into a machined groove in the plate.  
Starting from the upstream end of the magnet, the plate slides into precision kinematic mounts in a support box that is aligned and locked in place within the magnet bore.  
Another similar plate slides into the support box on the positron side and hosts the Front End Boards (FEBs) that distribute power and control signals from the DAQ and also digitize raw data from the modules for transfer to the external DAQ.  
The last two layers of the recoil tracker, being much larger, are supported on another structure: a cooled support ring onto which the single-sided, axial-only modules are mounted. 
This support ring is installed from the downstream end of the magnet, engaging precision kinematic mounts in the support box for precise alignment to the upstream stereo modules. 
The cooling lines for all three cooled structures---the upstream and downstream tracker supports and the FEB support---are routed to a cooling manifold at the upstream end of the magnet which, in turn, connects to a cooling feedthrough with dielectric breaks on the outside of an environmental enclosure which shields the detector from light and RF and maintains an environment of dry gas. 

Overall, this design is similar to that of the HPS tracker, although with some important simplifications.  
First, because the radiation dose in the LDMX tracking system is modest, cooling is only  needed to remove heat from the readout electronics. It is not necessary to keep the silicon itself cold.  
Therefore, cooling water that is close to room temperature can be used and there are no significant issues of differential thermal expansion to be concerned with.  
Second, the LDMX detector is in no danger from the nominal LCLS-2 beam, so it does not need to be remotely movable, in contrast to HPS.  
Finally and most significantly, the LDMX detector does not need to operate inside the beam vacuum as is the case for HPS. This greatly simplifies many elements of the design, the material selection, and the construction techniques. 

\begin{figure}[htp]
    \centering
    \includegraphics[width=\textwidth]{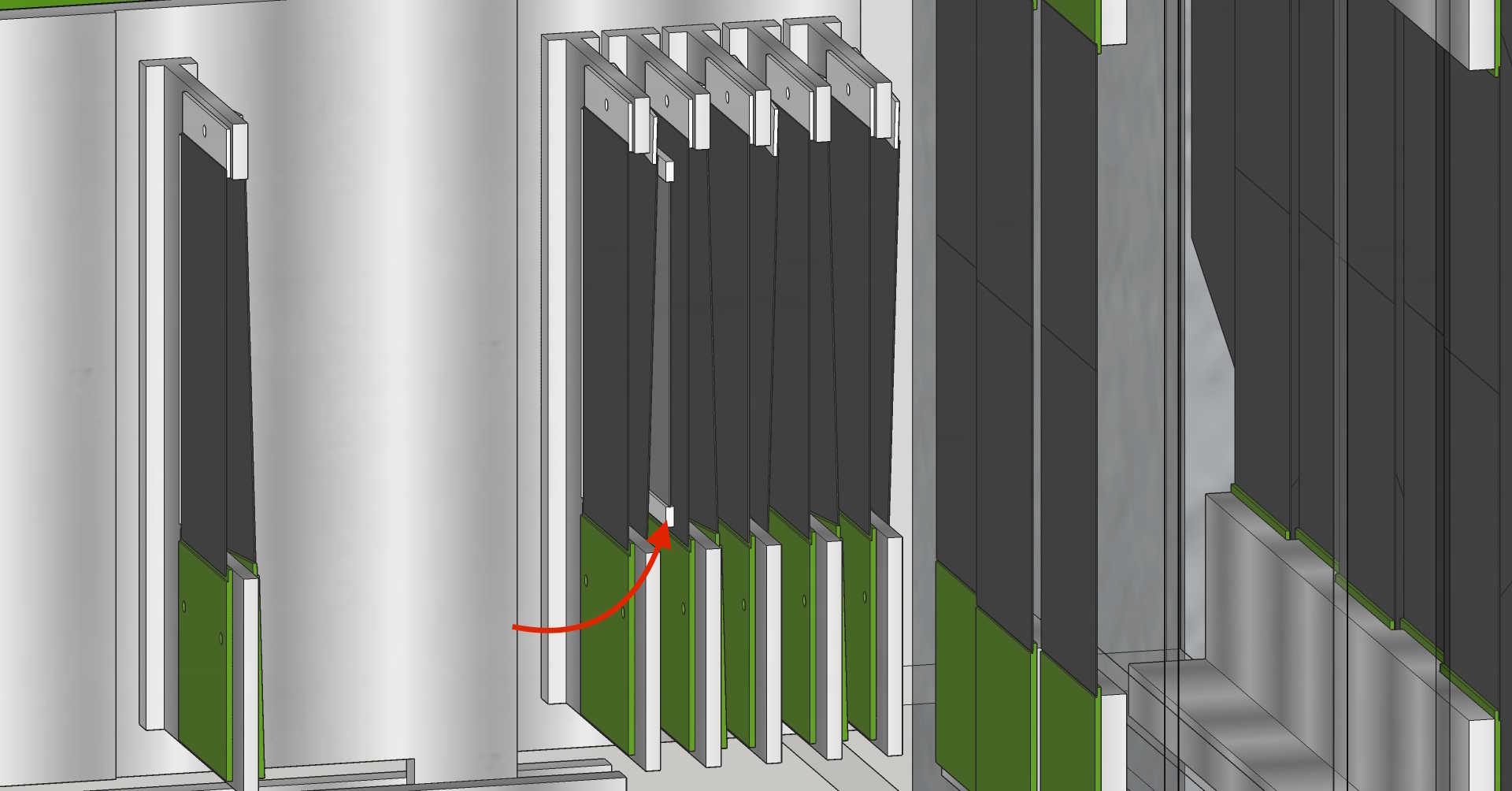}
    \caption{\small{The LDMX target, indicated by the red arrow, installed between the last tagging tracker layer and first recoil tracker layer.}}
    \label{fig:target}
\end{figure}

It is the target interposed between the last layer of the tagging tracker and the first layer of the recoil tracker that determines the different functional requirements of
the tagging and recoil tracking systems. 
The target is a 350 micron tungsten sheet, comprising 10\% of a radiation
length (0.1~$X_0$). 
This choice of thickness provides a good balance between
signal rate and transverse momentum transfer due to multiple
scattering, which limits the potential utility of transverse momentum as a
signal discriminator. 
The tungsten sheet is glued to a stack of two 2~mm planes of PVT scintillator that enables a fast count of the incoming electrons in each bunch as required to select the appropriate threshold employed by the \ecal trigger as discussed in the trigger description.
The scintillator-tungsten sandwich is mounted in an
aluminum frame that is inserted into the upstream tracker support
plate from the positron side to simplify the process of replacing or
swapping the target, as shown in Figure~\ref{fig:target}. 
With the scintillator mounted on the downstream
face of the target, it can also be used to help identify events in which a bremsstrahlung photon has undergone a photo-nuclear reaction in the target.  The structure and readout of the scintillator are discussed below as part of the trigger system.

\subsubsection{Tagging Tracker \people{Omar,Tim}}
The tagging tracker is designed to unambiguously identify incoming beam electrons that have the correct energy, as well as precisely measure their momentum, direction, and impact position at the target.  
The key elements of the design are determined by this goal.  
First, the long, narrow layout of the tagging tracker accepts only beam electrons that have roughly the momentum and trajectory expected for the incoming beam. This eliminates off-energy beam electrons.  
Second, the layers have low mass and are spaced far apart in a significant magnetic field to ensure a precise measurement of momentum.  
Finally, a large number of layers ensures the redundancy that is required for high-purity pattern recognition. 

The layout and resolution of the tagging tracker are summarized in Table~\ref{table:tagger_layout}. 
\begin{table}[!th]
  \caption{The layout and resolution of the tagging tracker.\label{table:tagger_layout}}
  \begin{tabular}{l|ccccccc} \hline
    Layer & 1 & 2 & 3 & 4 & 5 & 6 & 7 \\ \hline
    $z$-position, relative to target (mm) & -607.5 & -507.5 & -407.5 & -307.5 & -207.5 & -107.5 & -7.5 \\
    Stereo Angle (mrad) & -100 & 100 & -100 & 100 & -100 & 100 & -100 \\
    Bend plane (horizontal) resolution ($\mu$m) & $\sim$6 & $\sim$6 & $\sim$6 & $\sim$6 & $\sim$6 & $\sim$6 & $\sim$6 \\
    Non-bend (vertical) resolution ($\mu$m) & $\sim$60 & $\sim$60 & $\sim$60 & $\sim$60 & $\sim$60 & $\sim$60 & $\sim$60   \\
    \hline
    \end{tabular}
\end{table}
It consists of six double-sided, silicon microstrip modules arranged at 10~cm intervals upstream of the target, with the first module centered at $z=-7.5$~mm relative to the target.
The modules are positioned laterally within the magnet bore such that they are centered along the path of incoming 4~GeV beam electrons. Each module places a pair of 4~cm $\times$ 10~cm sensors back to back: one sensor with vertically oriented strips for the best momentum resolution, and the other at $\pm 100$~mrad stereo angle to improve pattern recognition and provide three-dimensional tracking. 
The sensors are standard $p^+$-in-$n$ type silicon microstrip sensors identical to those used for HPS.  
These sensors have $30$ ($60$) $\mu$m sensor (readout) pitch to provide excellent spatial resolution at high S/N ratios. They are operable to at least 350~V bias for radiation tolerance.

The sensors are read out with CMS APV25 ASICs operated in multi-peak mode, which provides reconstruction of hit times with a resolution of approximately 2~ns. 
At very high occupancies, six-sample readout can be used to distinguish hits that overlap in time down to 50~ns, which limits the readout rate, and therefore the trigger rate, to approximately 50~kHz.  
However, at the low hit occupancies anticipated in LDMX, three-sample readout may suffice, enabling a maximum trigger rate approaching 100 kHz.  
These sensors are mounted on standard FR4 hybrid circuit boards, which provide the power conditioning and I$^2$C control for the APV25, as well as a thermal path to the cooled support structure.

\begin{figure}[htp]
    \centering
    \includegraphics[width=\textwidth]{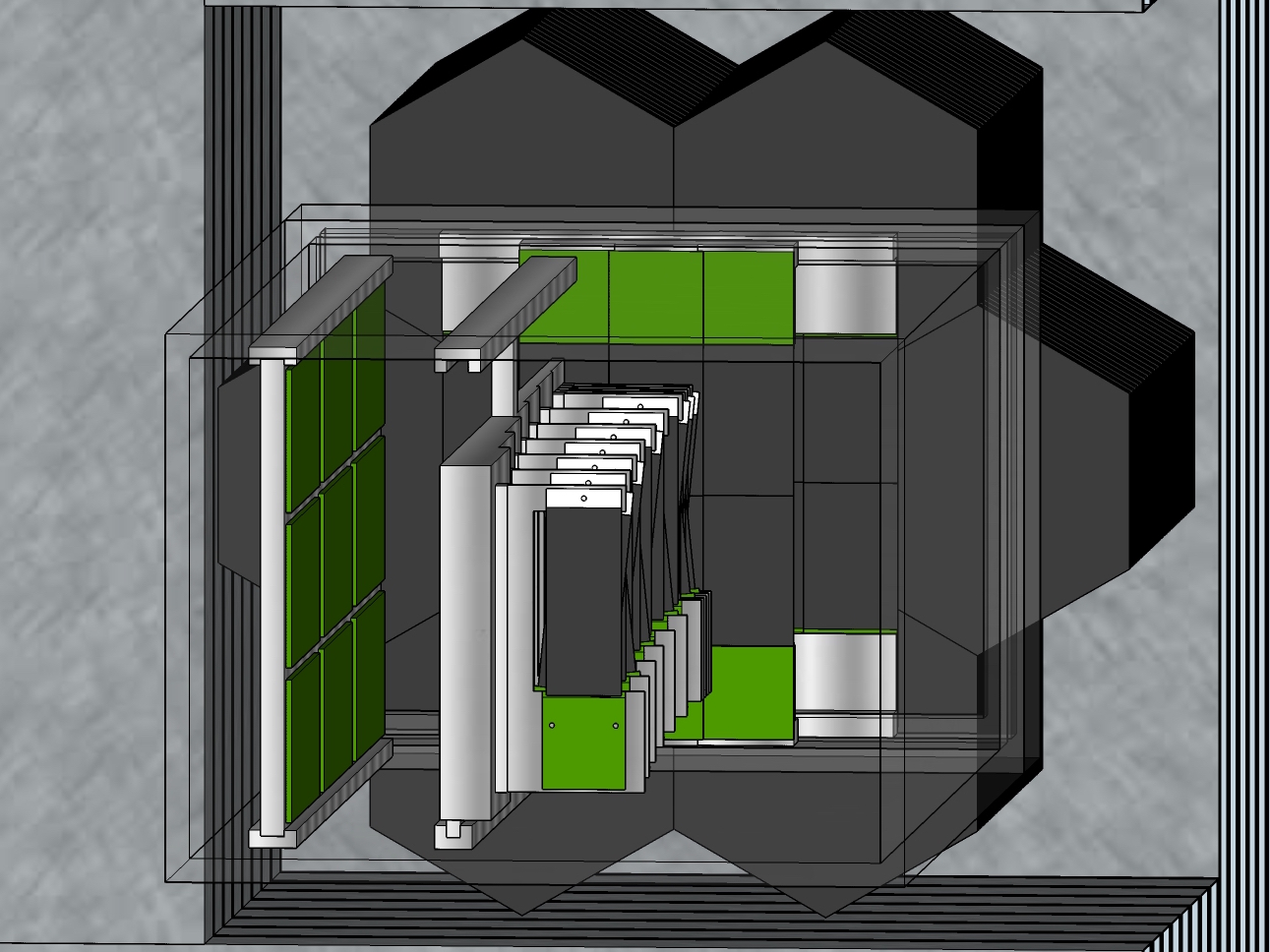}
    \caption{\small{A beam's eye view of the tagging and recoil trackers.}}
    \label{fig:tracker_beamseye}
\end{figure}

The sensors and hybrids are assembled into half-modules, which are the smallest non-reconfigurable units of the tagger tracker.  
Each half module consists of a single sensor that's glued to the hybrid at one end with conductive epoxy to provide bias voltage and support.  
A thin ceramic plate is glued to the other end of the sensor for support.  
A pair of half modules is attached to either side of an aluminum module support with screws to form a module.  
Finally these double-sided modules are attached to the upstream support plate described in Section~\ref{section:target} to position them along the beamline and provide cooling for the hybrids as shown in Figure~\ref{fig:tracker_beamseye}. 
This support arrangement places only silicon in the tracking volume and no material on the side of the tagger towards which electrons bend in order to avoid secondaries that would be generated by degraded beam electrons hitting material close to the tracking planes.  


\subsubsection{Recoil Tracker \people{Omar,Tim}}
The recoil tracker is designed to identify low-momentum (50~MeV to 1.2~GeV) recoil electrons and precisely determine their momentum, direction, and impact position at the target.  
In addition, it must work together with the calorimeters to correctly distinguish low-momentum signal recoils from scattered beam electrons and multi-particle backgrounds. 
The key elements of the design are determined by this goal.  
First, the recoil tracker is placed at the end of the magnet in the beginning of the fringe field to optimize tracking for particles up to two orders of magnitude softer than the beam-energy electrons measured by the tagging tracker.  
Second, the recoil tracker is wide for good acceptance in angle and momentum and is longitudinally compact to minimize the distance from the target to the calorimeters to maintain good angular coverage. 
Finally, the recoil tracker provides 3-d tracking near the target for measurement of both direction and impact parameter with good resolution, but emphasizes low mass density over the longest possible lever arm further downstream to deliver the best possible momentum resolution.  
This design delivers good momentum resolution for both multiple-scattering limited, low-momentum tracks and beam energy electrons that travel along a nearly straight path in the fringe field.

The layout and resolution of the recoil tracker are summarized in Table~\ref{table:recoil_layout}. 
It consists of four stereo layers located immediately downstream of the target and two axial layers at larger intervals in front of the \ecal.  
\begin{table}[!th]
  \caption{The layout and resolution of the recoil tracker.\label{table:recoil_layout}}
  \begin{tabular}{l|cccccc} \hline
    Layer & 1 & 2 & 3 & 4 & 5 & 6  \\ \hline
    $z$-position, relative to target (mm) & +7.5 & +22.5 & +37.5 & +52.5 & +90 & +180 \\
    Stereo Angle (mrad) & 100 & -100 & 100 & -100 & - & -  \\
    Bend plane (horizontal) resolution ($\mu$m) & $\approx$6 & $\approx$6 & $\approx$6 & $\approx$6 & $\approx$6 & $\approx$6 \\
    Non-bend (vertical) resolution ($\mu$m) & $\approx$60 & $\approx$60 & $\approx$60 & $\approx$60 & - & - \\
    \hline
    \end{tabular}
\end{table}
The stereo layers are double-sided modules of silicon microstrips arranged at~15 mm intervals downstream of the target, with the first module centered at $z=+7.5$~mm relative to the target.  
These modules are laterally centered on the target and the center of the magnet bore and are identical to the modules of the tagger tracker that are mounted upstream on the same support plate.  

\begin{figure}[htp]
    \centering
    \includegraphics[width=\textwidth]{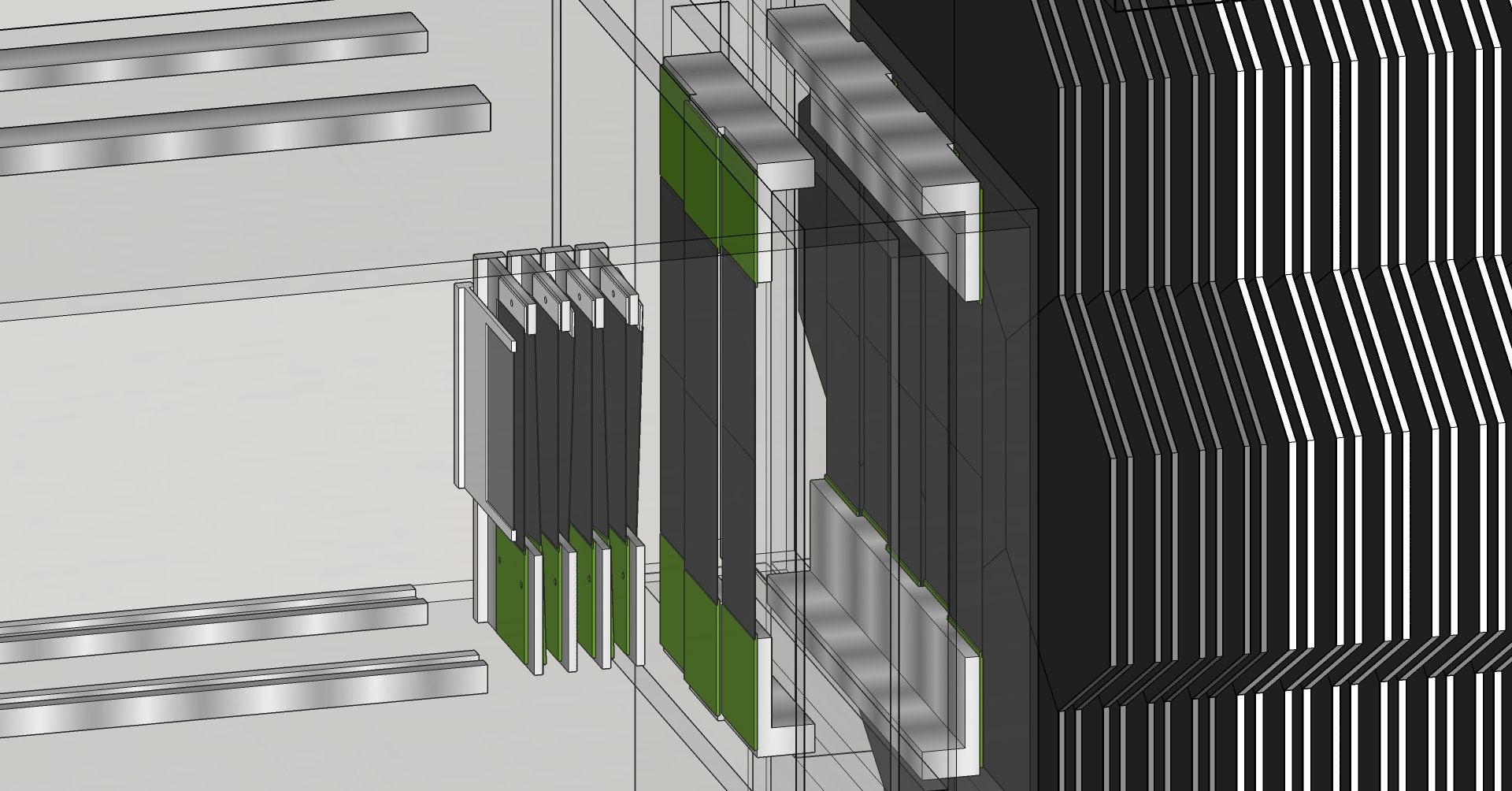}
    \caption{\small{The target and recoil tracker.}}
    \label{fig:recoil_tracker}
\end{figure}

The thinner axial-only layers, at $z=+90$~mm and $z=+180$~mm, are mounted on a separate support structure at the downstream end of the magnet bore as described in Section~\ref{section:target}.
They have a somewhat different module design as shown in Figure~\ref{fig:recoil_tracker}.
Each consists of a pair of sensors, glued end-to-end, with an APV25-based FR4 hybrid circuit board at each end of this structure to read out the two sensors. 
The sensors are standard p$^+$-~in~-~n silicon microstrip sensors, but are somewhat shorter and wider than those used in the stereo modules and therefore require six APV25 chips to read out each sensor instead of five. 
The modules are supported at both ends by screw attachment of the hybrids to castellated 
support blocks attached to the cooled support structure. 

\subsection{Forward Electromagnetic Calorimeter \people{Joe} \morepeople{Ruth,Mans}}
\label{sec:detector.ecal}

A high granularity, Si-W sampling calorimeter will be used for the \ecal of LDMX.  Its components and design draws on the High Granularity Calorimeter (HGC) for the forward calorimeter upgrade of the CMS experiment for the HL-LHC \cite{Contardo:2020886}. In particular, the hexagonal sensors, front-end readout electronics and front-end trigger architecture will be the same. This is facilitated by the fact that the sampling time for the CMS HGC is comparable to that planned for LDMX. 

An \ecal that is based on silicon pad sensors is well-suited to the LDMX experiment whose principal task for Phase~I, with $4 \times 10^{14}$ electrons on target (EOT), will be the identification of photons and electrons with very high efficiency and very good energy resolution.  The \ecal will also need to play a role in the identification of hadrons. However, a hadronic veto calorimeter, described below, will be used to identify hadrons from photo-nuclear and electro-nuclear reactions that pass through the \ecal without interactions that produce any identifiable signals. The high granularity of the \ecal provides the ability to track charged hadrons that cross multiple silicon layers and to identify isolated charged hadrons that range out in a single layer, depositing charge compatible with what is expected for several minimum ionizing particles (MIP).  

For LDMX Phase~II, we envision operation at higher energy and higher average number of electrons per 20~ns sampling time. While Phase I is the main topic of this note, we mention Phase II in the context of the \ecal because high granularity is needed even more in that higher rate environment and because a silicon based calorimeter has the radiation tolerance needed to sustain the higher fluences. 

With a reasonably good transverse spread of beam electrons, the granularity of the apparatus can be used to help identify overlapping showers. The central stack of hexagons would use the highest granularity available. The highest granularity being considered for CMS modules is $\sim$432 pads of area 0.52~cm$^2$ on a sensor made from an 8" wafer. The sensor would be read out by 6 HGCROC front end ASICs on a printed circuit board (PCB) that completely covers the sensor. The HGCROC front end readout chip comprises 78 channels (72 reading out standard cells, 2 reading out calibration cells, and 4 channels not connected to any sensor cells for common-mode noise estimation) and is designed in a radiation-hard 130~nm CMOS technology.  It may be possible to double the pad count to 864, with pads having area 0.26~cm$^2$, if it is determined that this would bring a significant improvement in performance for operating with multiple electrons per 20~ns sample time.  Note that independent of the sensor layout, LDMX would use 500 or 700~$\mu$m thick silicon as compared to a maximum of 320~$\mu$m thick silicon used by CMS in order to get better energy resolution and enhance the potential to detect hadrons. For sensors with the same granularity as CMS sensors, this could be done without the need for new masks to be designed and fabricated. 

As noted above, silicon is chosen to tolerate the fairly substantial radiation doses expected to be encountered in LDMX Phase II, and even in the central stack of the \ecal in Phase I.  Simulations of the radiation environment performed with FLUKA \cite{Battistoni:2007zzb} indicate that, while electromagnetic interactions are the dominant source of irradiation within the LDMX environment, a substantial hadronic component is also present.  In all operational scenarios under consideration, a diffuse beam would be used to spread the electrons over a cross-sectional area at the target of at least 16~cm$^2$.  

\begin{figure}[h]
\centering
\includegraphics[width=10 cm]{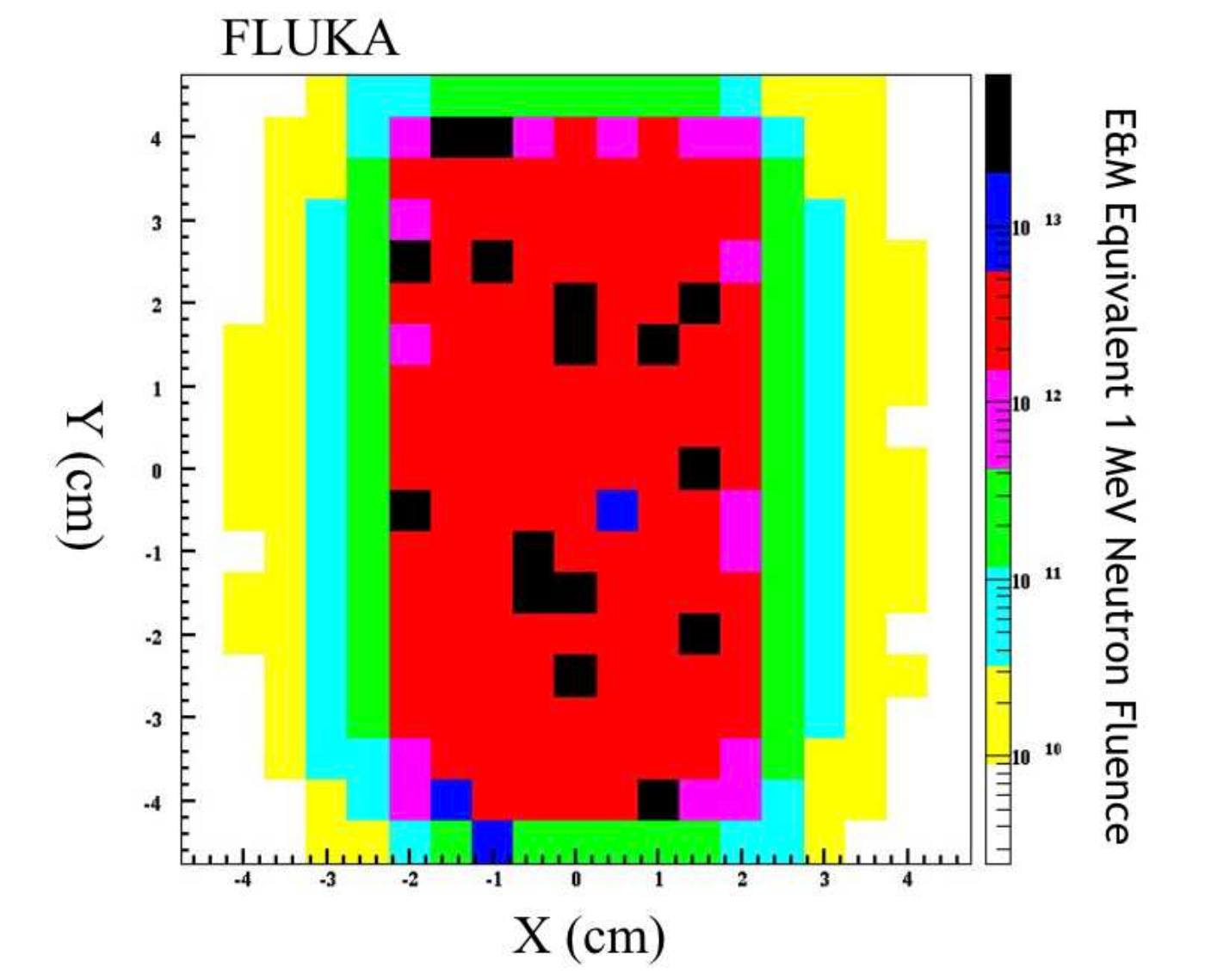}
\begin{flushleft}
\caption{\label{fig:fluence}Effective 1 MeV neutron equivalent fluence at shower max after $10^{14}$ electrons on target with a beamspot area of 32~cm$^{2}$.   The maximum effective fluence for this condition is found to be of order $3 \times 10^{13} ~n/$cm$^2$.  Scaling to a total fluence of $4 \times 10^{14}$ electrons with a beamspot of 16~cm$^{2}$, the expected fluence would be $2.4\times 10^{14}~n/$cm$^2$.}
\end{flushleft}
\end{figure}

The effective 1~MeV neutron-equivalent fluence at shower max for a 4~GeV electron beam spread to an area of 32~cm$^2$ is shown in Fig.~\ref{fig:fluence}. For $10^{14}$ EOT, the fluence is seen to peak at around $3\times 10^{13}~n/$cm$^2$. This can be linearly projected to $2.4~-~6\times 10^{14}~n/$cm$^2$  for $4~-~10\times 10^{14}$ EOT with a 16~cm$^2$ beamspot. This is certainly high enough to necessitate changes in the biasing of the silicon and require the silicon to be operated colder than room temperature. It is, however, substantially lower than the worst case expectation for the CMS HGC. The CMS experiment has performed radiation studies for 300, 200 and 120~$\mu$m sensors exposed to maximum fluences of roughly $9\times 10^{14}$, $4\times 10^{15}$ and $1.5\times 10^{16}$ 1~MeV equivalent neutrons per cm$^2$, respectively. The full depletion voltage of the sensor, $V_{d}$, is proportional to the square of the thickness, which can lead to a reduced operational lifetime for thicker sensors in a high radiation environment.  In the 120~$\mu$m case, the noise is much higher for irradiated sensors, reducing the signal-to-noise ratio to about 2. For the fluences expected in LDMX, the CMS studies suggest that a 432 channel sensor with thickness of 300~$\mu$m would have a signal-to-noise ratio of about 11 for $5\times 10^{14}~n/$cm$^2$. The radiation environment of the LDMX experiment may, however, be low enough to allow the use of 500~$\mu$m thick sensors, which would boost the signal by 67\%.  Moreover, with 500~$\mu$m thick sensors there would be 10\% improvement in the energy resolution of 4 GeV electrons over that obtained with 300~$\mu$m sensors. A drawback of thicker sensors is that they increase the space between absorbers, thereby increasing the Moli\`ere radius. In any case, the minimum signal-to-noise ratio expected at the end of operation of LDMX, even for 300~$\mu$m sensors, would be adequate to identify MIPs. 

\begin{figure}[t!]
\center
\includegraphics[width=12cm]{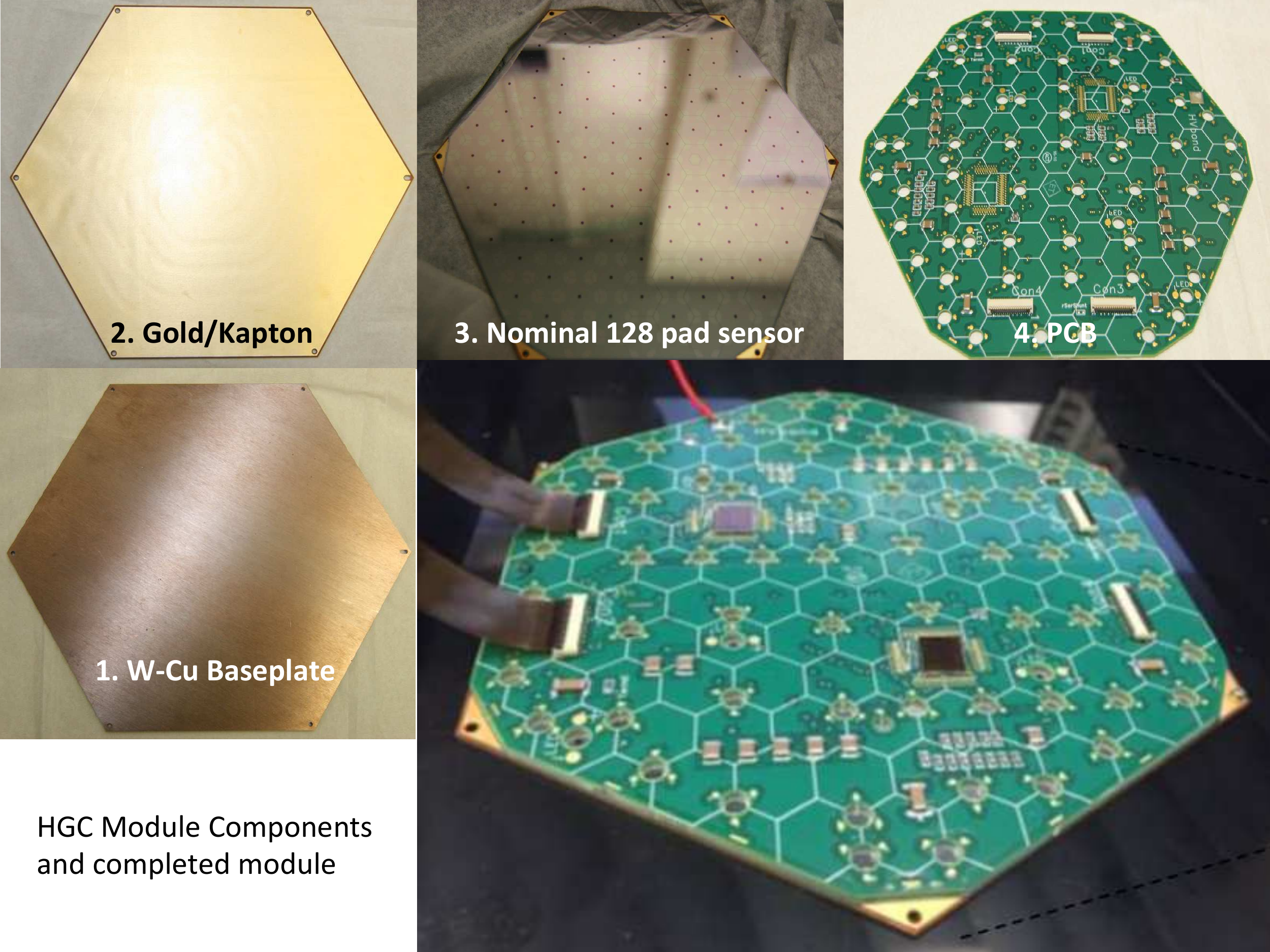}
\caption{Prototype CMS HGC 6 inch, 128 channel module with (1) a W/Cu (75\%/25\%) baseplate, to which (2) a thin Gold-on-Kapton sheet is glued for insulation and to enable biasing of (3) a Si sensor glued directly on top of it. Finally, (4) a PCB is glued on the sensor. Connections to each set of 3 adjacent pads on the sensor are made by wirebonds through 4 mm diameter holes. }
\label{fig:HGCModule}
\end{figure}

Module components and the first completed CMS prototype HGC, a nominal 6-inch, 128 channel module, are shown in Fig.~\ref{fig:HGCModule}. The current design of the \ecal used in the studies presented in this note makes use of 32 Silicon layers, each comprised of a central module surrounded by a ring of 6 modules.  The hexagonal sensors are to be produced on `8-inch' wafers, yielding a span of 170 mm between parallel sides.
The basic \ecal readout unit is a hexagonal module containing a single hexagonal sensor. The module has a metal baseplate that serves as electromagnetic absorber, and so has been chosen to be predominantly tungsten. It must also transport heat from the module electronics and sensor to a cooling/support plane. In order to enhance thermal transport without introducing an unacceptably large mis-match in thermal expansion coefficients between the silicon and the baseplate, CMS has chosen an alloy of 25\% Cu and 75\% W. In LDMX we have considered a pure W baseplate for the module, which however could be affixed to a Cu cooling/support plane similar to the CMS design. 

Note that in the module shown in Fig.~\ref{fig:HGCModule}, the corners of the sensor and PCB have been removed to give access to mounting holes on the baseplate that are used to position and attach the modules precisely on the cooling plate. In CMS the cooling surface is a large 5 mm--thick planar Cu sheet with embedded CO$_2$ cooling lines, where the Cu contributes to the absorber material budget. For LDMX we are planning to use a simpler, liquid cooling system, since the dark currents and thermal load are lower in the less severe radiation environment. We have begun to study \ecal performance with a very thin and lightweight cooling layer, made up of small diameter, thin-walled stainless steel cooling lines embedded in carbon fiber. A lightweight cooling plane could be useful in the upstream layers of the \ecal where we find that it is beneficial for the rejection of some relatively rare types of photo-nuclear interactions to start with a thin absorber, increasing the thickness to a nominal value near 1~$X_0$ over some 5 layers or so.

We first simulated \ecal and carried out performance studies for a system with 42 layers of silicon for which the first 10 layers have absorbers of 0.8$X_0$ and the remaining 32 layers having absorbers of 1.0~$X_0$.  We chose this geometry to allow us to understand shower containment and energy resolution very well, and we have found that at least for a 4 GeV electron beam, we can manage with fewer sensing layers. We have therefore begun studies and estimated costs for a 32 layer, 40~$X_0$ device. This would have 224 installed silicon modules. The active region of the \ecal would then be roughly 30 cm in depth and about 51 cm wide.

For the electronics, we plan to use the CMS HGCROC front-end chip and PCB design. The CMS development schedule is such that the final components will be produced in substantial quantities in time for the construction of the LDMX detector. The HGCROC, for instance, is expected to be ready for mass-production by the end of 2019. The PCB should be finalized even earlier, based on prototype, pin-compatible front-end chips.
For the DAQ and trigger, each plane of modules will be connected to a motherboard housing trigger and readout/control FPGAs as discussed in Section~\ref{sec:ecaldaq}.


The cooling planes would be stacked in an enclosed and thermally insulated box through which dry nitrogen gas or dry air would flow. The support planes with modules attached would be slotted into a box-like frame designed to allow for some level of reconfiguration of absorber and sensing layers. A simple approach with some of these same characteristics has been used for the CMS test-beam studies of the first HGC modules at FNAL and CERN. The box and support structure could be relatively simple and constructed from hard aluminum or stainless steel. Feed-throughs would be needed for the data and trigger fibers, the power and bias cables, and the cooling and gas lines. Fortunately, the density of services is relatively low in each plane. With appropriate care and attention in assembly and installation, we believe it will be possible to achieve a very good thermal and gas enclosure.

\subsection{Hadronic Veto System \people{David,Bertrand,Nhan,Andrew} \morepeople{Torsten}}
\label{sec:detector.hcal}
As the forward electromagnetic calorimeter contains most, but not all, of the energy of electromagnetic showers, and does not efficiently detect neutral hadrons or muons, a high efficiency hadronic veto calorimeter (\hcal) system is an important component of LDMX.
The main functions of the hadronic calorimeter are to detect neutral hadrons, mainly neutrons produced in photo-nuclear reactions in the \ecal or the target with very high efficiency, to be sensitive to electromagnetic showers that escape the \ecal, and to detect minimum-ionizing particles (MIPs) such as muons.  
A scintillator-based sampling calorimeter with a large number of nuclear interaction lengths of steel absorber meets these requirements.

Based on our studies of the backgrounds from hadronic processes, the hadronic veto system must identify neutral hadrons in the energy range from approximately $100~\MeV$ to several $\GeV$ with high efficiency. The most problematic events typically contain either a single high energy neutral hadron, or multiple lower energy neutral hadrons. 
The required efficiency for lower energy neutrons can be achieved with absorber plate sampling thicknesses in the range of 10\% to 30\% of a strong interaction length ($\lambda_A$). The sampling fluctuation contribution to the energy resolution of the \hcal varies inversely as the square root of the absorber thickness. In order to reduce the probability of a single high energy forward-going neutron to escape without interacting to the required negligible level, a total \hcal depth of approximately 16~$\lambda_A$ of the primary steel absorber is required. 
We also surround the \ecal with a Side \hcal in order to intercept neutral hadrons produced at large polar angles. 


As the \hcal rates are low, transverse segmentation is not a very demanding requirement for a veto, although some complementary measurements, such as studying hadronic systems in electro-nuclear scattering reactions, require reasonable angular resolution of the hadronic showers. A design with minimal cracks and/or dead spaces is important. We have, accordingly, compared both tile and bar realizations. A bar geometry requires many fewer channels than a tile geometry, and at the shower multiplicities we will face, does not suffer from a significant occupancy problem. We have therefore adopted a bar geometry, in an ($x,\ y$) configuration, with ambiguity resolution accomplished by means of timing correlations for signals arriving at the two ends of each bar. 
The \hcal dimensions and absorber segmentation are being optimized with detailed GEANT4 simulations of both single particle response (see Sec.~\ref{sec:hcalperf}) and various background processes (see Sec.~\ref{sec:bkgpn}). We have explored several geometries with different steel absorber thicknesses, ranging from 2~mm to 100~mm. Our initial background studies are based on the geometry depicted in Fig.~\ref{fig:LDMX_overview}. The Main (Side) \hcal size was conservatively set to 3~m~$\times$~3~m, enabling us to explore smaller transverse dimensions as well. The Main \hcal is 13~$\lambda_A$ in depth with absorber layers 50~mm thick (0.3~$\lambda_A$), which corresponds to a depth of roughly 3~m. The Side \hcal design is similar to that of the Main \hcal, but with finer sampling ($\sim 0.07~\lambda_A$, and is $\sim~3\lambda_A$ in depth). 

The final \hcal design will be optimized as we improve our understanding of background processes. For example, the transverse size will likely be less than 3~m, we may vary the sampling fraction with depth, with thinner absorber in the front of the \hcal for improved energy resolution, and the total \hcal depth can be reduced or extended in a modular way based on studies such as those of the photo-nuclear modeling presented in Sec.~\ref{sec:bkgmod}.



\subparagraph{Enabling \hcal technologies} 

\begin{figure}[thp]
\begin{minipage}{0.45\textwidth}\smallskip
    \includegraphics[width=7cm]{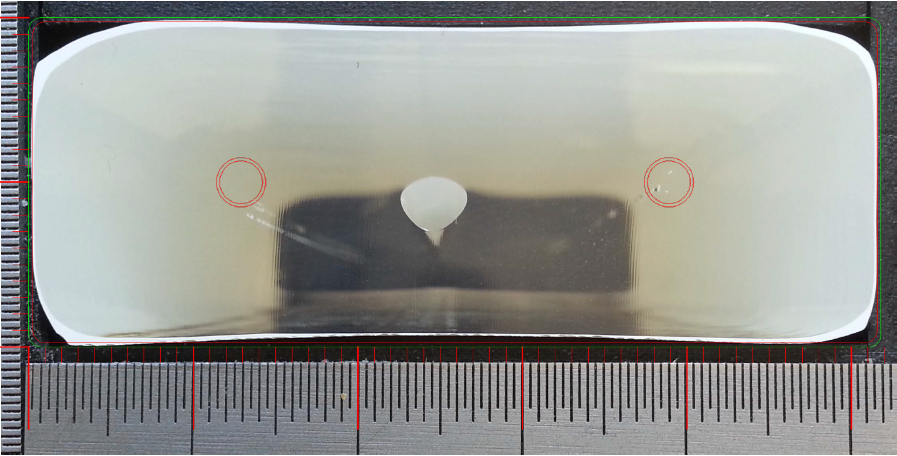}
    \caption{Photograph of the end of a 20~mm~$\times$~50~mm extruded polystyrene bar, coextruded with a TiO$_2$ diffuse reflecting layer and containing a single hole for a wavelength-shifting fiber.
We expect to use a 15~mm~$\times$~50~mm bar for the \hcal.    }
 \label{fig:bar}
\end{minipage}
\hskip 0.5cm
\begin{minipage}{0.45\textwidth}
    \includegraphics[width=7.6cm]
    {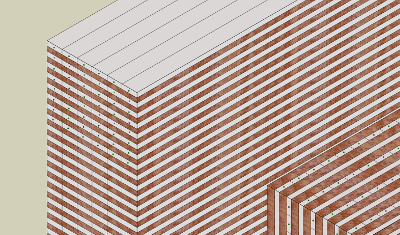}
    \bigskip\medskip
    \vskip-8mm
    \caption{Detail of the front corner of the \hcal, showing the 15~mm $\times$ 50~mm bars, each containing a single wavelength-shifting fiber.
    }
 \label{fig:corner}
\end{minipage}
\end{figure}

\begin{figure}[hbtp]
\begin{center}
    \includegraphics[width=0.7
    \textwidth]{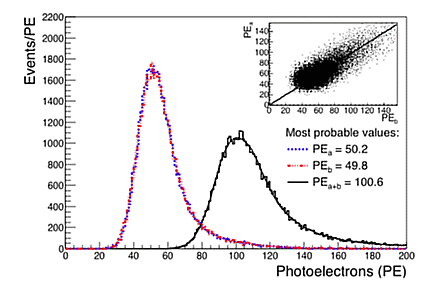}
    \caption{Measured photoelectron yield for 120 GeV protons on a 2 cm thick extruded polystyrene bar, read out with a 1.4 mm Kuraray scintillating fiber by a 2 mm $\times$ 2 mm Hamamatsu SiPM. The beam was 1 m from the end of the bar.}
 \label{fig:hcalMIP}
 \end{center}
\end{figure}

The \hcal scintillator design is based on that of the Mu2e Cosmic Ray Veto (CRV) system. The scintillator is in the form of doped polystyrene bars, of the type developed for the Mu2e CRV. Extruded polystyrene bars have been produced at Fermilab for several decades. The Mu2e bars are 20~mm thick $\times$ 50~mm wide, co-extruded with an integrated TiO$_2$ reflector. The extrusion also includes through holes into which a wavelength-shifting fiber is inserted (see Fig.~\ref{fig:bar} and Fig.~\ref{fig:corner}). The Mu2e system uses two 1.4~mm fibers per bar. The scintillator response to minimum ionizing particles (MIPs) has been measured for 120 GeV protons on a 20 mm thick extruded polystyrene bar, read out with a 1.4 mm Kuraray scintillating fiber and a 2 mm$\times$ 2 mm Hamamatsu SiPM~\cite{Artikov:2017lsc}. On average, each MIP at normal incidence deposited 3.9 MeV in a bar, corresponding to a response of 50 photo-electrons/fiber at each bar end (Fig.~\ref{fig:hcalMIP}). Based on these measurements, we expect that a 15~mm~$\times$~50~mm bar with a single fiber will yield 75 photoelectrons per fiber per bar (both ends), which should provide an adequate signal for LDMX. In the Side \hcal, due to space constraints, the readout will be single-ended. We have already produced prototype bars with a single hole, and we will assume a single fiber per bar in the following discussion. The transverse spatial resolution measured by Mu2e is 15~mm, while the resolution along a 2~m bar, determined by the relative timing of the signals at either end of the bar, has been measured to be 150~mm \cite{Artikov:2017lsc}. Slightly longer bars should offer similar performance, adequate for ambiguity resolution in the crossed $x,y$ configuration. We plan to attach the scintillator bars to the steel absorber, and to support this structure by an external frame. In this manner, there is ample access to the sides, top and bottom of the \hcal for signal, SiPM bias, and monitoring cabling.

Two existing readout electronics systems providing adequate performance have been identified. The merits of each are currently being explored; we briefly describe them below. The existence of multiple options demonstrates that an \hcal readout system is technically feasible.
Given the basic dimensions of the \hcal, the readout system must be able to service signals from approximately 5,000 wavelength-shifting fibers.

\begin{figure}[t!]
\begin{center}
    \includegraphics[width=10cm]{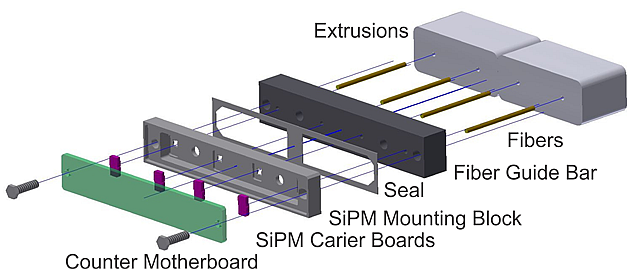}
    \caption{Exploded view of the Mu2e CRV Counter Mother Board (CMB), SiPM holder and fiber guide structure that is attached by two screws to the end of each 100~mm wide scintillator di-bar. The CMB provides a light-tight seal, holds the fiber ends in registration against the four 2~mm~$\times$~2~mm SiPMs, and has an HDMI receptacle to bring four channels to the Front End Boards (FEBs). The CMB also has a temperature monitor and two calibration LEDs. For an LDMX with one fiber per bar, the CMB would be modified to service four bars.}
 \label{fig:cmb}
 \end{center}
\end{figure}

The first system is that already developed for the Mu2e CRV. This could either be adopted as a more-or-less ``turn-key'' system or with modifications (such as a single fiber per bar) to reduce the number of readout channels. 
The basic unit is made of four bars glued together to form a 200~mm ``quad-bar'', containing four individual fibers, which are read out by SiPMs at each end of the bar, mounted on a modified CRV Counter Mother Board (CMB) (see below). Fig.~\ref{fig:corner} shows a corner of the \hcal, detailing the quad-bar and fiber structure. Fig.~\ref{fig:cmb} shows a CRV Counter Mother Board (CMB) mounted on each end of two bars, to readout the four fibers with 2~mm~$\times$~2~mm Hamamatsu SiPMs. For LDMX with a single fiber per bar, the CMB would be expanded in length to encompass four bars. 
 The CMB provides bias to the SiPMs, has a temperature monitor, and two flasher LEDs for independent monitoring and calibration of each 50 mm bar. In the event of a SiPM failure, the Counter Mother Boards, which hold the four SiPMs for each quad-bar, can be replaced by removing two mounting screws. The four SiPM signals are transmitted to a Front End Board (FEB) on four shielded twisted pairs via an HDMI-2 cable. 


A Front End Board (FEB) services 16 CMBs, digitizing a total of 64 SiPM signals. 
The Readout Controller (ROC) chassis, which receives the signals from 24 FEBs, also provides the 48 volt bias to the SiPMs and the power to the FEBs, all over a CAT 6 cable. 
The readout of the bars in the Side \hcal is similar to that described for the Main \hcal. As the energy of wide angle hadronic showers is lower, the sampling is reduced to 12.5~mm steel and the scintillator bar thickness is also reduced to 15~mm. This necessitates a design with a reduced thickness of the CMB. The rest of the FEB to ROC readout chain is unchanged.

The second readout system is based on the CMS hadronic calorimeter system.
In contrast to the Mu2e system, the CMS readout electronics system is a fiber plant scheme where fibers are taken from the scintillator to a centralized SiPM location called a readout box (RBX).
The RBX is described in more detail below. 
Signals from wavelength-shifting fibers in the scintillating bars are transported to the RBX via clear fiber cables.
The CMS system can optically gang up to 6 clear fibers onto a single large area SiPM thus reducing the channel count and effective segmentation.  
The light transmission efficiency of the wavelength-shifting fiber-to-clear-fiber combination is approximately 75\%.

\begin{figure}[t!]
\begin{center}
    \includegraphics[width=14cm]{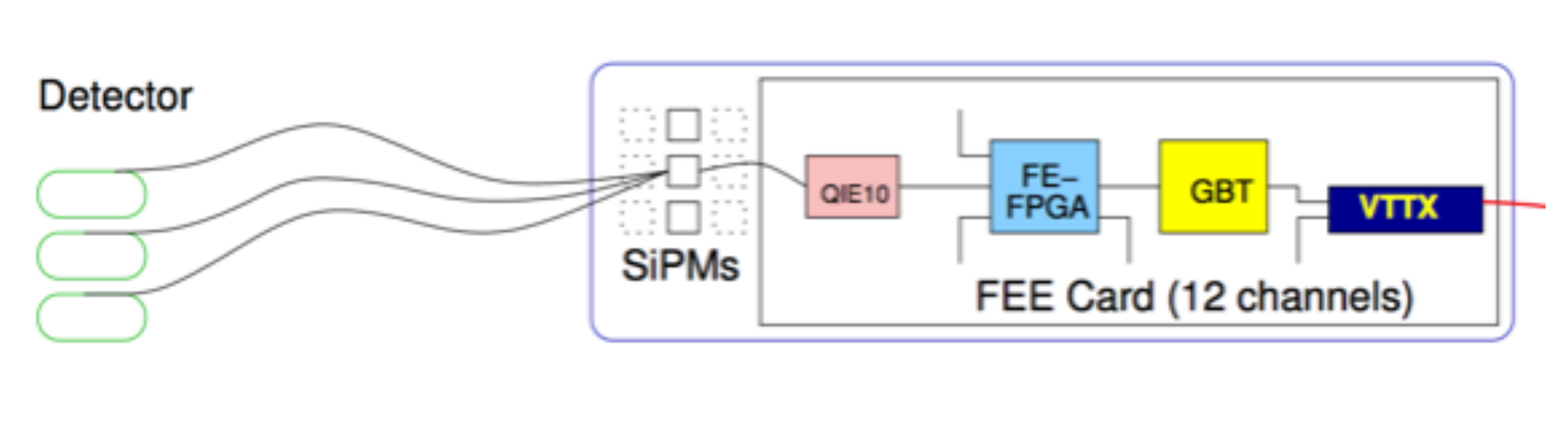}
    \caption{The CMS \hcal electronics readout chain}
 \label{fig:cms-scheme}
 \end{center}
\end{figure}

The clear fibers transport the signal from the scintillating fibers to a readout module (RM). 
An RM consists of an optical decoder unit (ODU) which organizes the clear fibers for ganging and is installed directly onto the SiPM mounting unit.
There are 64 SiPMs in a single RM. The SiPM signals are then sent to the QIE board which includes a QIE11 digitizer ASIC that digitizes the SiPM signal, which is then sent to the backend electronics via the CERN VTTX transceiver. Four RMs are contained in one RBX.
A schematic of the front-end electronics readout chain is given in Fig.~\ref{fig:cms-scheme}.
The current CMS \hcal system is designed for triggering at a 40~MHz rate.

\subsection{Trigger System \people{Mans} \morepeople{Bertrand,Nhan}}
The LDMX trigger system is designed to reduce the typical beam
particle arrival rate of 46~MHz to a rate of 5~kHz for storage and
analysis.  The selection is performed by a combination of dedicated
hardware and software running on general-purpose computers.

The first stage trigger is implemented in hardware and allows the
selection of both candidate events for dark matter production and
important samples for calibration and detector performance monitoring.
The overall trigger management is provided by a trigger manager board,
which receives inputs from the various triggering subsystems including
the \ecal and \hcal.  The latency
requirements of the trigger calculation are set by the tracker
readout ASIC to 2~$\mu$s.

The primary physics trigger is based on the \ecal and is designed to select events with energy deposition significantly lower
than the full beam energy.  The \ecal HGCROC calculates
the total energy in $2\times 2$ fundamental cells for every 46~MHz bucket.
The energy information is transferred by digital data link to the
periphery of the calorimeter, where sums are made over larger
regions and transferred by optical link to the trigger electronics.
The total energy is then used to select the events.  The details of
the electronics for the \ecal trigger are discussed along with the DAQ
for the \ecal below.

The use of a missing-energy calorimeter trigger requires information on the expected energy, which varies with the number of incoming beam electrons in each 
accelerator bucket.  The trigger scintillator system is designed to provide this information
to the trigger calculation.  It is important that this system makes an accurate count of the number of electrons.  
If the actual number of incoming electrons is systematically higher than the number reported by the trigger scintillators, the missing energy will appear to be small (or likely negative), resulting in all events being vetoed. Conversely, if the actual number of incoming electrons is lower than the number reported by the trigger scintillators, 
the missing energy will appear to be large, resulting in most events being triggered. 

\begin{figure}[!ht]
  \begin{center}
    \includegraphics[width=0.7\linewidth]{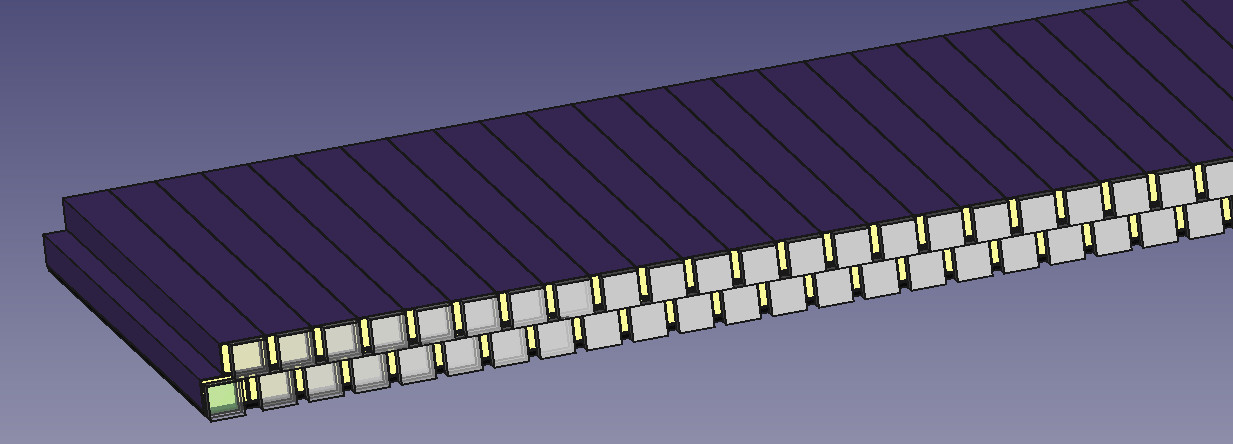}
  \end{center}
  \caption{Drawing of the concept for the target and trigger
    scintillator system, showing the relationship of the SiPMs to the
    scintillator strips. Each scintillator strip is wrapped in ESR foil.}
  \label{fig:target_trigger}
\end{figure}

The trigger scintillator system is constructed as two planes of thirty
$4~\mathrm{cm} \times 3~\mathrm{mm} \times 2~\mathrm{mm}$ strips of
scintillator each wrapped with 3M ESR film, which provides a very high
reflectivity.  Each strip is connected to a SiPM photodetector.  The structure is shown in
Fig.~\ref{fig:target_trigger}.  The strips are used to count the
number of incoming electrons.  The natural fluctuations in light
production and collection in the scintillator imply that it is
difficult to identify the number of electrons passing through a given
strip.  Instead, the logic is simply based on the number of strips
above threshold. The strips in a given row have a gap between them set
by the ESR film and construction tolerances to $\approx
200\,\mu\mathrm{m}$.  The second layer of strips is offset from the
first by 1.5 mm so that the gaps in the two layers do not align.

The count of incoming electrons can be lower than the actual number either due to multiple 
electrons hitting a single strip or by an electron slipping through a dead area.  The probability of 
such an issue occurring in a single layer with twenty-five illuminated 3~mm-width strips increases with the number of incoming electrons, such that 
it increases from a 7\% inefficiency for one initial electron to 40\% inefficiency for four initial electrons. 
The trigger counts the number of incoming electrons using algorithms which combine the information 
from the two different layers, to improve the efficiency from a single layer.  
Two different algorithms have been tested.  One algorithm simply takes the larger 
of the number of strips hit in either layer as the number of incoming electrons.  
The other algorithm attempts to identify patterns which are consistent with multiple
electrons per strip by comparing the data in the two layers.   The simpler algorithm improves the inefficiency for four electrons to 20\%, while the more-complex algorithm provides an additional enhancement to 15\%.

The trigger scintillator is read out using an array of $2\times 2$~mm SiPMs
connected to QIE11 electronics developed for the CMS \hcal upgrade.
The SiPMs are coupled directly to the edge of the scintillator strip.
The SiPMs and QIE11 readout cards will be located in the environmental enclosure for the tracker 
along with the tracker front end electronics.  The SiPMs will be
cooled in common with the tracker elements to achieve very low thermal
noise levels in the trigger scintillator readout.  The readout
electronics will continuously digitize the SiPM signal, providing an
integrated charge as well as time-of-arrival measurement for the pulse
with an LSB of 500~ps.  Both amplitude and timing information can be
provided to the trigger, allowing correction of the calorimeter
amplitude for time walk effects already at trigger level.


\subsection{DAQ \people{Mans,Tim} \morepeople{Bertrand,Nhan}}
The off-detector trigger and readout electronics are assumed to share
a single ATCA crate, using the Reconfigurable Cluster Element (RCE)
generic computational building block developed by SLAC.  The RCE is
based on the Xilinx Zynq 7 System-On-Chip technology, running a dual
core 1~GHz ARM processor with 1~GB of DDR3 memory tightly integrated
with on-chip programmable logic (FPGA).  The RCE blocks are integrated
into Data Processing Modules (DPM), each of which hosts two RCE
blocks.  The DPMs receive data from rear transition modules (RTM),
which are customized to adapt to the specific portion of the
experiment that is being read out.  Up to four DPMs can be mounted on
a single Cluster-On-Board (COB) ATCA blade, handling up to 96 optical
data links.  The modular ATCA design permits LDMX to re-use
architecture and functionality from other DAQ systems whose components
are similar to those used by LDMX.

\begin{figure}[htp]
    \centering
    \includegraphics[width=0.8\textwidth]{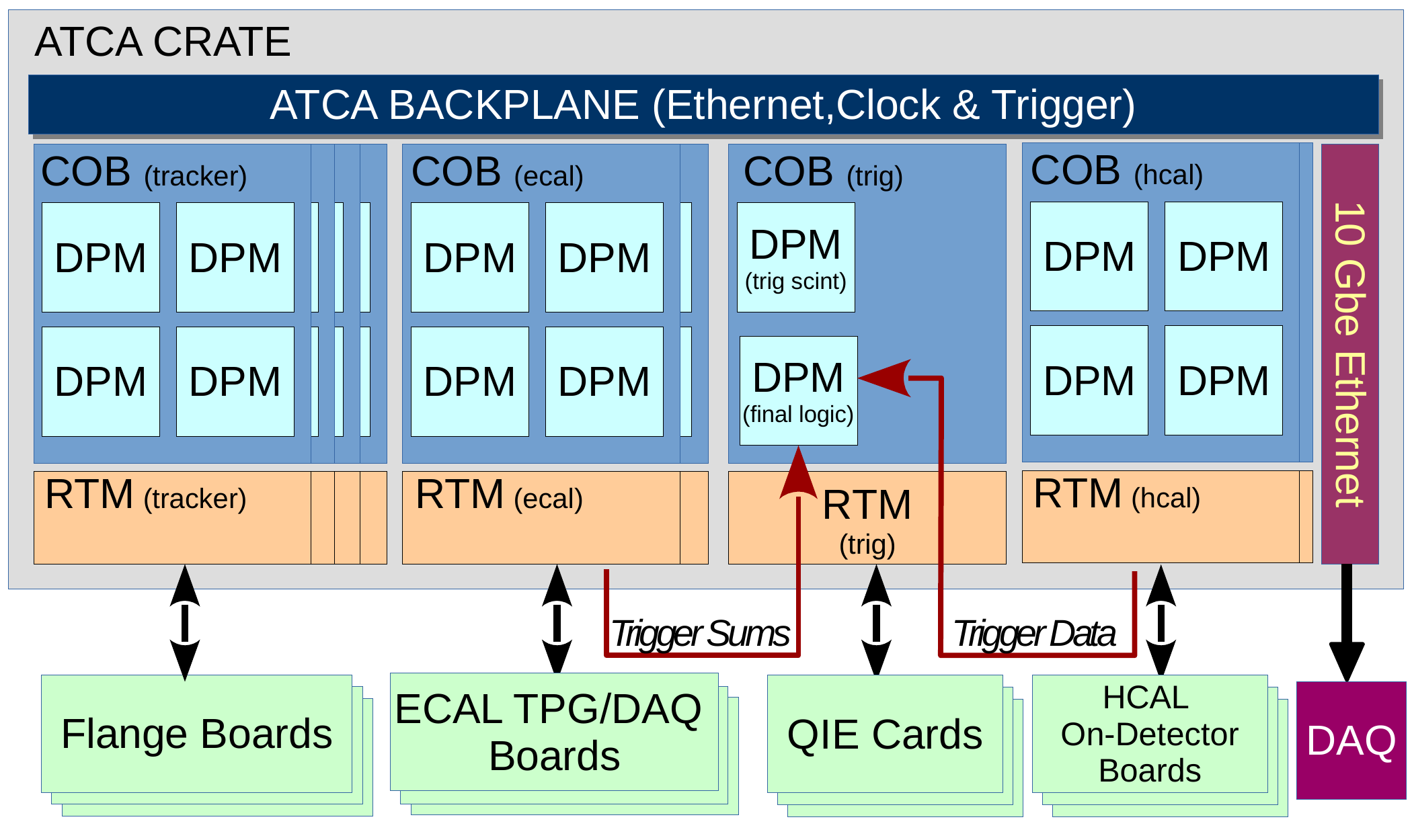}
    \caption{\small{Schematic block diagram of the RCE-based DAQ.}}
    \label{fig:rce_daq}
\end{figure}

\subparagraph*{Overall DAQ Architecture}

One of the DPMs functions as the trigger interface, which receives
trigger signals from the optical fiber module on the RTM, distributes
clock and trigger signals, and handles communication with the software
trigger supervisor.  Specific interfaces to beamline and
infrastructure services are provided through a custom RTM.  Final DAQ
is carried out through the 10~Gbps Ethernet readout path available
separately from each COB.  The COBs in a single ATCA crate are also
interconnected by 10~Gbps readout and contain ARM processors that can
be used to execute high-level trigger algorithms.

To establish a conservative design for the DAQ, we consider a target
DAQ bandwidth of 25~kHz (trigger rejection factor of $5\times
10^{-4}$).  This provides a factor of five safety compared with the
combined expectation for the performance of the hardware trigger and
for the data volume in triggered events.  The breakdown of data
volumes by subsystem is shown in Table~\ref{table:daqSize} and sums to
an estimated total of 3.1 kilobytes per event, for a total DAQ
bandwidth of 77.5~MBps.  This bandwidth is within the
capability of the 10~Gbps Ethernet bandwidth with a comfortable safety factor of
$\approx 10$.  The 10~Gbps ethernet connections from the COBs will be connected to
one or more data storage and data-quality-monitoring PCs.
To estimate the potential total data volume, we assume there is
no high-level trigger, in which case the full sample of $2\times 10^{11}$
events would require 500~TB of storage.

\begin{table}[!th]
  \caption{Estimated DAQ data volume per channel.\label{table:daqSize}}
  \begin{tabular}{|l|r|p{4.5in}|} \hline
    \textbf{Subsystem} & \textbf{kbytes/event} & \textbf{Estimate Notes} \\ \hline
    Trigger & 0.5 & Includes event accounting information (32B), trigger sums from each module of the \ecal (210 B), trigger-counter readout (144B), and readout of trigger information from the \hcal (128B). \\
    Tracker & 1.0 & Each raw hit requires 20B for id and data, and we expect 50 hits/event, including noise hits. \\
    \ecal & 1.2 & Each hit with TDC is 10B including 4B for data and 6B for addressing and overhead, and each (low-amplitude) hit without TDC is 8B.  For $\mu=1$, we expect an average of 125 hits.  We assume are all TDC hits for the estimate. \\
    \hcal & 0.4 & Each channel's readout is 20B for id and data, and we budget for 20 channels in a typical event. \\
    \hline
    {\bf Total} & 3.1 & \\ \hline
    \end{tabular}
\end{table}    

\subparagraph*{Tracking DAQ}

The data acquisition system for the tracking detectors supports the
continuous 46~MHz readout and processing of signals from the silicon-strip sensors. 
It also selects and transfers those events that were
identified by the trigger system to the back-end DAQ for further event
processing at rates approaching 50~kHz. The system adopted is an
evolution of the DAQ used for the HPS SVT and consists of APV25
readout ASICs hosted on hybrid circuit boards integrated into the
tracking modules, a set of Front End Boards (FEBs) inside the magnet bore for power distribution and signal digitization, flange boards
that transmit power and data through the wall of the SVT environmental enclosure, and the remotely located ATCA-based data acquisition hardware and power supplies.

\begin{figure}[htp]
    \centering
    \includegraphics[width=\textwidth]{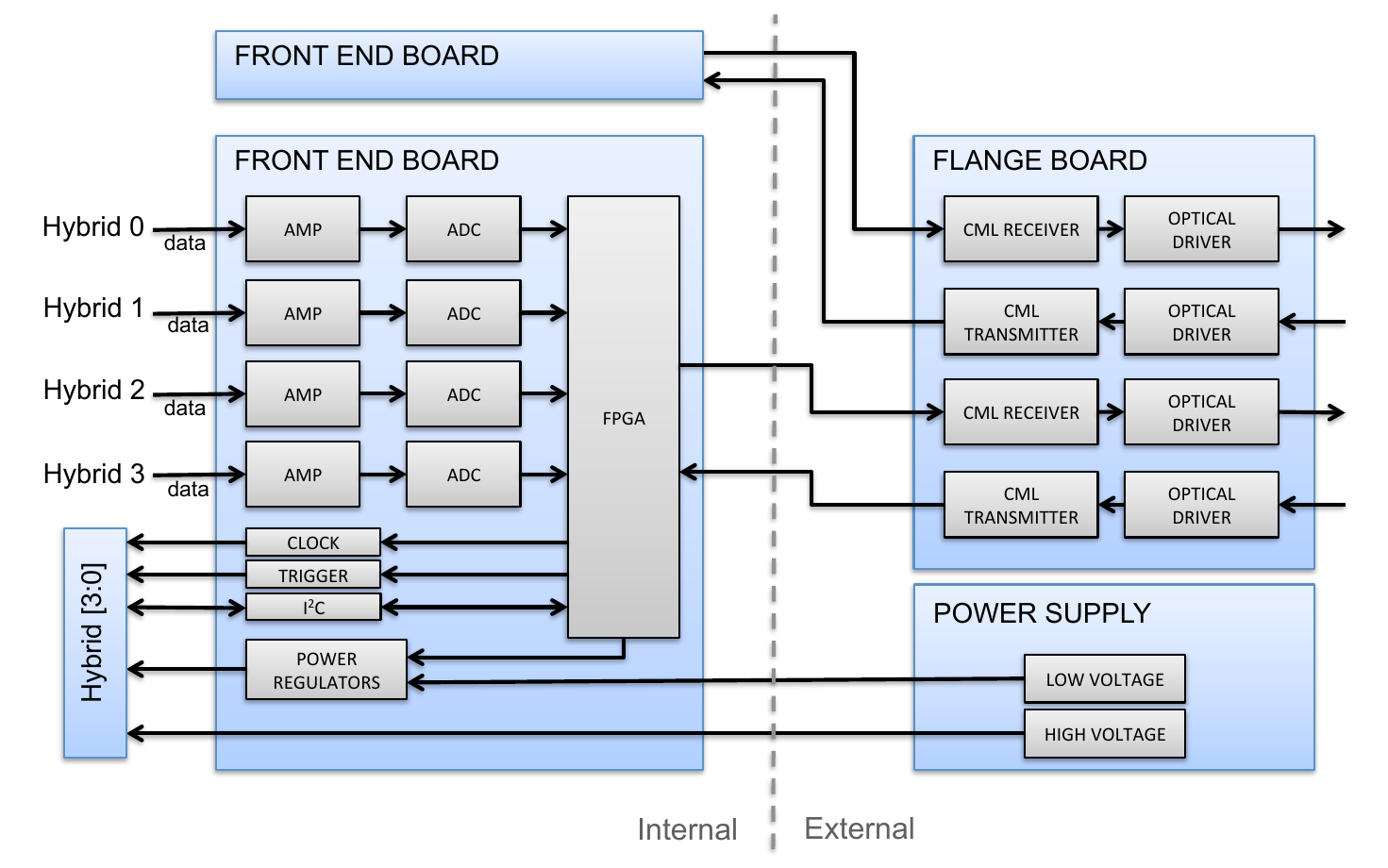}
    \caption{\small{Schematic view of the downstream part of the tracker DAQ.}}
    \label{fig:trackerdaq_internal}
\end{figure}

The APV25 ASIC, initially developed for the Compact Muon Solenoid
silicon tracker at the Large Hadron Collider at CERN, was chosen
because it provides excellent signal to noise, analog output for
optimal spatial resolution, and signal pulse-shape sampling capability
for good hit time resolution. Each hybrid board has five or six analog
output lines (one for each of the APV25 ASICs) that are sent to an FEB
using LVDS signals over about 50~cm of
twisted-pair magnet wire. At the front-end readout board, a
preamplifier scales the APV25 differential current output to match the
range of a 14-bit Analog to Digital Converter (ADC). Each front-end
board services four hybrid boards. The ADC operates at the system clock
frequency of 46~MHz. The digitized output from the front-end board is
sent through compact 8-pair mini-SAS cables to flanges on the SVT enclosure to
connect to the external DAQ, which resides many meters away. The front-end readout board houses an FPGA and buffers signals to
allow for control of the distribution of clock, trigger and I2C
communication with the APV25 ASICs. To further simplify the services
and minimize cabling that enter through the flanges, it
contains linear regulators to distribute and regulate three low
voltage power lines to each of the APV25 ASICs, as well as passing
through the high voltage bias. Figure~\ref{fig:trackerdaq_internal}
shows a schematic layout of this part of the readout chain.

The digitized signals are converted to optical signals just outside
the flanges on custom-built flange boards. Each flange board
houses optical drivers to handle the electrical-optical conversion
and to transmit the optical signals over fibers to the ATCA crate. The
flange board also interfaces the low and high voltage power
transmission from the Wiener MPOD power supplies to the front-end
boards located inside the SVT enclosure.

The optical signals from four hybrid boards on one third of a flange board are
received at one of four sections of the Rear Transition Module (RTM)
board in the ATCA crate.
Each section of the RTM
connects to one of four DPMs.  In order to minimize the complexity of
the system inside the SVT enclosure, all signal processing is done at
the DPM level.  Each data DPM receives the digitized signals from the RTM,
applies thresholds for data reduction and organizes the sample data
into Ethernet frames.  Four COBs are sufficient to handle all of the
hybrid boards of the trackers. The maximum readout rate of the tracking system is
approximately 50~kHz, limited by the APV25 readout rate for six-sample, multi-peak readout.

\subparagraph*{\ecal DAQ}\label{sec:ecaldaq}

\begin{figure}[htp]
    \centering
    \includegraphics[width=\textwidth]{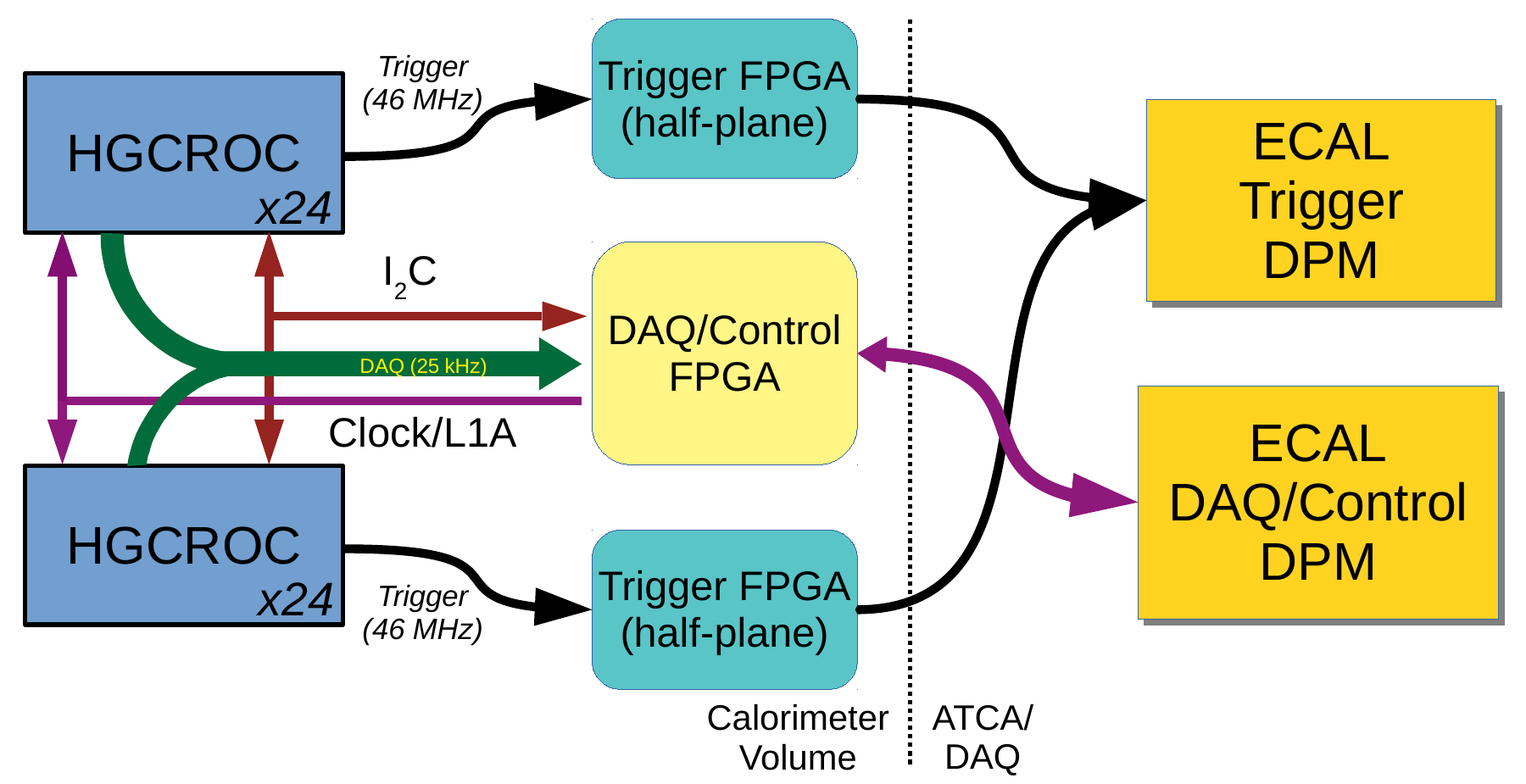}
    \caption{\small{Schematic view of the trigger and readout for the \ecal.}}
    \label{fig:ecaldaq_internal}
\end{figure}

As described above, signals in the Si/W \ecal are digitized in the
HGCROC and transferred on 1.28~Gbps DC-coupled copper data links to
the periphery of the calorimeter.  At the edge of the detector, the
copper links are received by FPGAs and processed before transmission
on optical links off the detector.  The FPGAs also provide the
fast-control and slow-control interfaces to the off-detector
electronics.  The architecture of the trigger primitive and readout
electronics of the \ecal is shown in Fig.~\ref{fig:ecaldaq_internal}.

For trigger purposes, groups of nine cells are summed inside the
HGCROC to form a trigger cell and the energy for each trigger cell is
transferred on the electrical links to an FPGA at the periphery of the
\ecal.  Two links are required for the trigger readout of each HGCROC.
An additional copper data link is used for readout of triggered event
data.  Each module requires six HGCROC chips, resulting in a total of
eighteen high-speed links for each module.

The \ecal layout includes seven modules on each layer, with the central
core module surrounded by six ``peripheral'' modules.  All the modules
of a layer are connected to a motherboard PCB using stacking
connectors.  This PCB also houses three data processing FPGAs, two for
trigger and one for DAQ and controls.

For trigger purposes, the plane is divided into two ``half-planes'' each
containing 21 HGCROC chips.  The trigger data for a half-plane is
transported to a single FPGA on the periphery where summing and other
initial local processing is carried out.  The trigger data is then
transferred by optical link to the ATCA off-detector card where the
final trigger calculation is performed.  A total of 68 optical links
is required for the trigger data, so that the full calculation can be
performed in a single COB card.

The 42 DAQ links from the HGCROC on a plane are connected to the DAQ
FPGA on the motherboard.  The DAQ FPGA is also responsible for
controls and uses a bi-directional data link to communicate with the
electronics in the ATCA crate, where a second COB will be used for
this purpose.

\subparagraph*{Trigger and \hcal Scintillator DAQ}

The target scintillators are read out using SiPM phototransducers and
QIE11 ASICs developed as part of the CMS \hcal Phase~1
upgrade~\cite{cmshcaltdr}.  The QIE11 is a deadtimeless
charge-integrating ADC with an internal TDC capable of 500~ps timing
resolution.  The data from the QIE11 is read out via an Igloo2 FPGA
onto a pair of digital optical data links that operate asynchronously
at 5~Gbps.  The readout of the QIE cards is continuous. For every
46~MHz bunch, the data links will digitize and transmit on the fiber
the complete data for every channel.  Each fiber carries the data for
five channels, so the full trigger scintillator DAQ will require
twelve fibers.  In the ATCA crates, the trigger scintillator readout
and control will occupy a single DPM on one COB, which could be shared
with another portion of the DAQ if required.

%

For the hadron calorimeter, we estimate that 1.25 trigger/readout
fibers will be required for each layer of the calorimeter in the case
of either electronics design under consideration.  In all cases, the
data bandwidth is dominated by that required for triggering
information.  Depending on the number of layers in the \hcal, either
one or two COBs will be required for the \hcal readout, where the
second COB could be shared with the trigger scintillator readout and
the final trigger decision DPM.

\subsection{Computing \people{Omar} \morepeople{Andrew}}
During the design stage, LDMX will be focused on the 
generation of Phase I Monte Carlo samples. The Phase I MC samples will be on 
the scale of 300 TB and will take $\sim$30M CPU hours to generate.  During Phase I
running, LDMX is expected to collect 550 TB of raw data, all of which will
require long-term tape storage.  Reconstruction of a fraction of the raw data
on a daily basis is crucial to monitoring the online performance of the 
detector and will go a long way in mitigating any issues that may arise. Doing
so in a timely manner will require 4k CPU hours per day and ~200 dedicated
nodes.  Reconstruction of the full raw dataset will require 6M CPU hours and 
5 PB of disk space.  For Phase II running these numbers are expected to 
increase by a factor of 2-10.

\clearpage
\section{Simulation and Performance}
%
%
\subsection{Simulation Framework \people{Omar}\morepeople{Andrew}}
\begin{figure}[!ht]
    \centering
    \includegraphics[width=.7\linewidth]{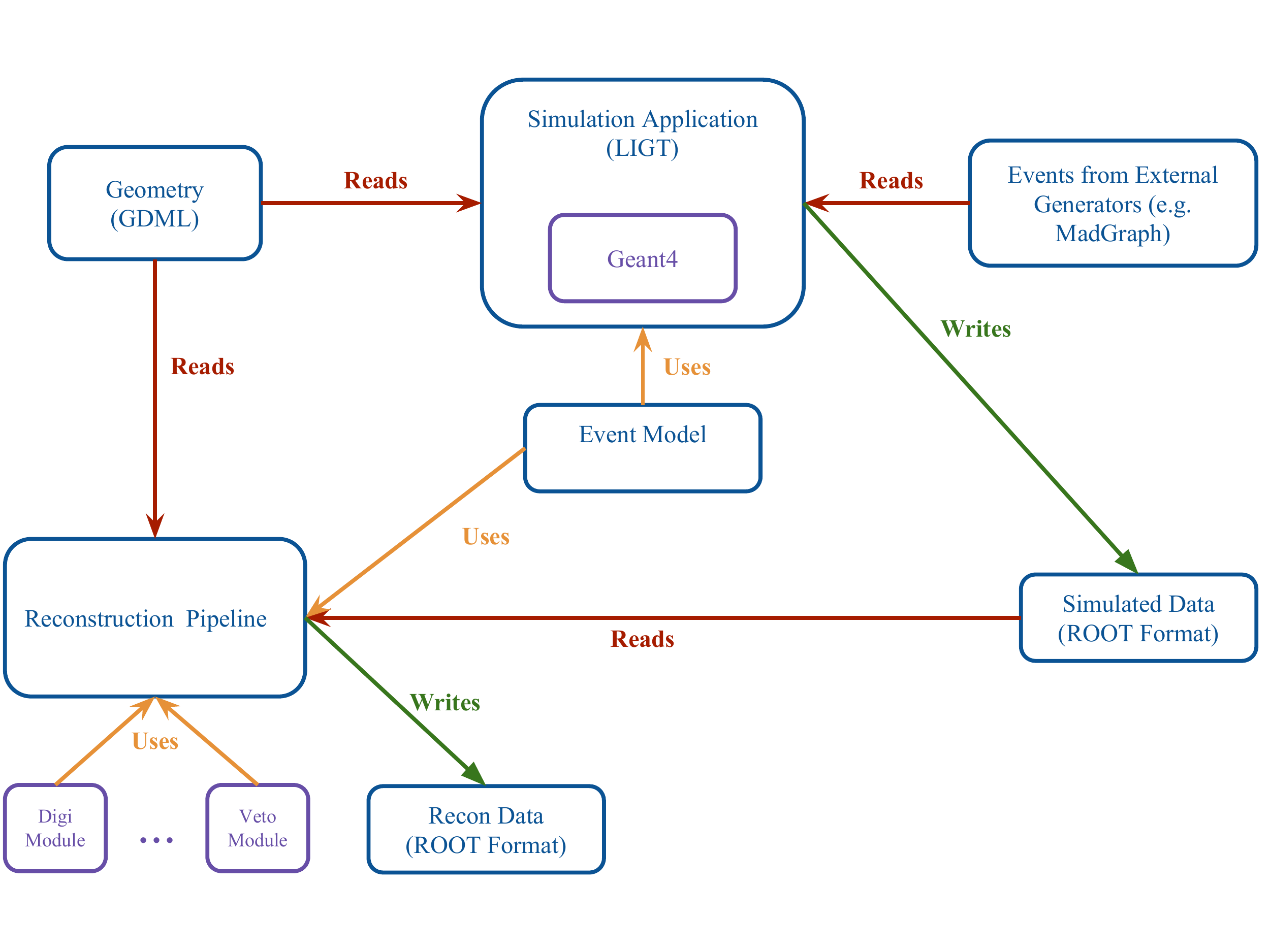}
    \caption{Diagram showing the structure of the LIGT simulation and
             reconstruction application.}
    \label{fig:sim_diagram}
\end{figure}

The propagation of particles through the detector along with their interactions
with material is simulated using the object-oriented LIGT 
(LDMX Interface to \geant Toolkit) framework.  LIGT
wraps a custom version of the \geant~\cite{Agostinelli:2002hh} toolkit
(10.02.p03) and adds several services including dedicated process and 
cross-section biasing capabilities, an event model, ROOT~\cite{Brun:1997pa} 
based persistence, GDML~\cite{1710291} based geometry system as well as an 
analysis/reconstruction pipeline.  
The framework was designed 
with flexibility and ease of use in mind, allowing for the quick study of 
rare physics processes using a wide range of detector concepts. The structure 
of the simulation is described in Figure~\ref{fig:sim_diagram}. 

Event generation is based mainly on sources provided by \geant using the 
physics list FTFP\_BERT augmented by the list 
G4GammaConversionToMuons.  FTFP\_BERT contains all standard 
electromagnetic processes and makes use of the Bertini cascade to model 
hadronic interactions. By default, the Bertini cascade model is used by 
\geant for energies below 3.5 ~GeV.  In order to ensure that the Bertini model
covered our phase space, the threshold was increased to 10~GeV through a direct
modification to \geant. In addition to FTFP\_BERT, the 
G4GammaConversionToMuons is used in order to add the additional reaction
$\gamma \rightarrow \mu \mu$ to the list of physics processes. 

In addition to \geant sources, LIGT can also read events written in the 
Les Houches Event (LHE) format~\cite{Alwall:2006yp} used by the event generator 
MadGraph/MadEvent~\cite{Alwall:2007st}.  An XML parser is used to read the 
event blocks containing the information characterizing each particle.  The 
four-momentum and PDG ID of each of the particles are used in the creation of 
\geant primary particle objects originating from a common vertex. By default, 
the vertex is set to (0, 0, 0) but can be uniformly spread over a volume. These 
particles are then used by the tracking simulation.   

\begin{figure}[!ht]
    \centering
    \includegraphics[width=.7\linewidth]{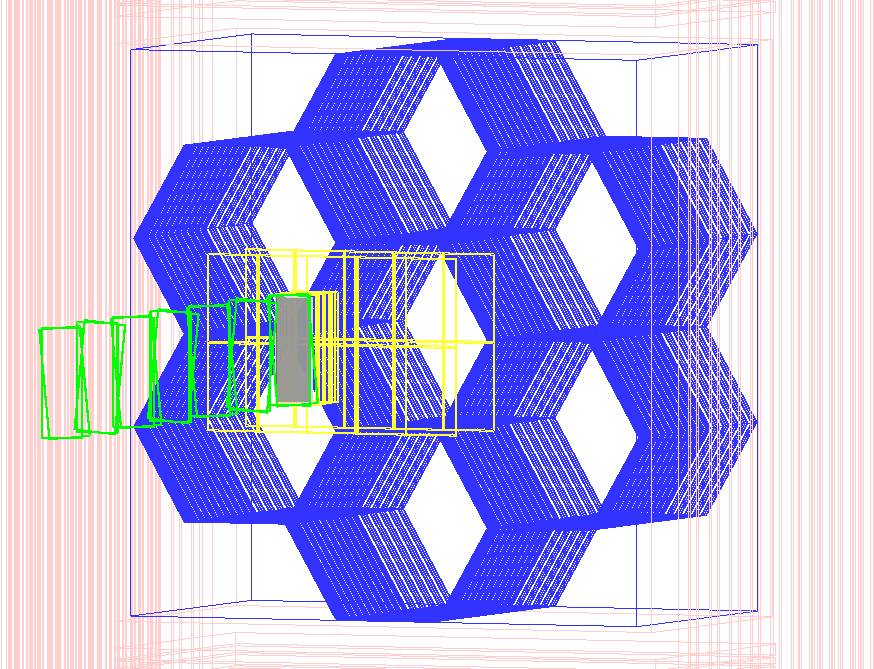}
    \caption{Rendering of the LDMX detector showing the Tagger (green) and 
             Recoil (yellow) trackers, target (grey), \ecal (blue) and \hcal
             (pink) as it appears in the simulation and reconstruction 
             application.  The trigger pads and magnet are also included but 
             are not shown.}
    \label{fig:ldmx_det_rend}
\end{figure}

A description of the geometry is provided to both \geant and the reconstruction
application at runtime using the XML based Geometry Description Markup Language
(GDML). Using GDML, all detectors and their corresponding materials
can be defined using a simple XML tree structure.  Extensions to the GDML format
also allow specifying which detectors are ``sensitive'' as well as the definition of
additional detector properties such as unique identifiers and visualization
parameters.  Furthermore, the
same detector description is used to define geometries that can be used by the
``parallel worlds'' toolkit provided by \geant.  This is commonly used to define
scoring planes which aren't desired to be part of the main geometry.  A 
rendering of the detector being used in the simulation is shown in 
Figure~\ref{fig:ldmx_det_rend}.

To add flexibility to \geant, all user actions (e.g. event,
stepping) are instantiated via an abstract factory.  This allows their 
initialization to occur at runtime which, in turn, makes the simulation 
configurable. An important use case for such a design is in the simulation of
rare physics processes.  A series of ``filter'' modules implementing the 
\geant stepping action interface are used to stop the tracking of events that
don't satisfy specific criteria.  For example, in the case of an electron 
incident on a target, a filter can be used to select events where the electron
bremsstrahlungs and falls below a certain momentum threshold.  Another filter
can then be used to select events where the bremsstrahlung photon undergoes a 
photo-nuclear reaction.  Several sets of these modules are used to quickly
and efficiently simulate several rare processes including photo-/electro-nuclear.

When a subdetector volume is declared sensitive, if a particle enters it 
and takes a ``step'' (i.e. interacts), information associated with that 
interaction (e.g. energy loss, position) is generated.  Subdetector specific 
processors are then used to convert the information associated with a step into
a \geant hit which, in turn, is stored in a container.  Once registered with 
\geant, hit containers can be accessed by the rest of the application.  At the 
end of each event, all available \geant hits are converted to LDMX event model 
objects and persisted to files using the ROOT data format. 

The ROOT based output is processed using a configurable modular reconstruction
pipeline with each stage of the pipeline instantiated via an abstract factory.
A stage is used to either process the event model hit objects read from the 
input file or to process physics objects generated by another stage.  If a 
stage generates a collection of physics objects (e.g. tracks), it is made 
available to all subsequent modules in the stage.  Currently, the 
reconstruction pipeline includes stages that digitize all event model hits,
adds noise, performs basic track finding and applies various vetoes 
optimized to reject previously described backgrounds.  Once all modules have
been executed, the resulting collection of objects is persisted to files
using the ROOT data format.

\subsection{Dark Matter Signal Generation \people{Natalia,Omar} \morepeople{Ruth}}
The new physics of dark matter production is modeled using an external event
generator based on MadGraph/MadEvent4.  Here we describe
this generator, its validation, and the interface with \geant.

MadGraph is an automated tool for calculating the tree-level amplitudes for 
arbitrary physics processes, which allows users to define Feynman rules for
new physics models; MadEvent is a Monte Carlo event generator based on MadGraph.
MadGraph/MadEvent4 (MG/ME) was designed for the study of high-energy collider 
reactions, but minor modifications to the code (e.g. introducing non-zero masses
for incident particles and for the electron, and an electromagnetic form factor
as described in \cite{Tsai:1986tx} for the nucleus) allow for its application
to fixed-target processes.  These modifications and a new-physics model that 
introduces a dark photon with arbitrary mass and kinetic mixing with
the photon has been used for  the APEX test run \cite{Abrahamyan:2011gv} and 
HPS experiment \cite{HPS_proposal_2014}.  For LDMX, we have added LDM particles
(either fermions or scalars) that couple to the dark photon with an arbitrary
interaction strength $g_D$.  This allows us to simulate the signal process of
DM particle pair-production via either decay of an on-shell $A'$ or off-shell
$A'$ exchange.   This report focuses on the on-shell production process, though
the kinematics of the two are very similar.  

Within MG/ME, we generate events for the DM production process 
$e^- W \rightarrow e^- WA^\prime, A^\prime \rightarrow \chi \bar\chi$ where
$W$ represents the tungsten nucleus, $\chi$ represents the dark matter 
particle and $\bar\chi$ its antiparticle. Events are generated assuming a 4 
GeV incident electron and tungsten nucleus at rest as the initial state.  
MG/ME computes a Monte Carlo approximation of the inclusive cross-section for 
this process, and generates a sample of unweighted events in the LHE format.
The inclusive cross-section computed by MadGraph is stable within $1\%$ and 
is consistent within $\sim 30\%$ with independent calculations of the
 cross-section in the Weizsacker-Williams (WW) approximation 
from \cite{Andreas:2012mt}.  The deviations from the WW inclusive cross-section
are largest at high and low masses, and compatible with the size of errors 
expected in the WW approximation. 

Simulating the detector response of these starts by smearing the vertex 
position of the recoiling electron and $W$ nucleus uniformly over the thickness
of the target and a transverse region spanning  $\pm 1$ cm in the $x$ direction 
and $\pm 2$ cm in the $y$ direction about the nominal center of the target.  
Both the recoiling electron and nucleus are then tracked through the detector. 
Although the events also include the \aprime, it is currently ignored by the 
tracking simulation.  For the signal efficiency studies described in 
Section~\ref{sec:sig_eff}, events are generated assuming different 
\aprime masses.



\subsection{Background Sample Generation \people{Natalia,Omar} \morepeople{Nhan}}
\label{sec:bkggen}

A standard set of background samples has been used for the background rejection
studies in Section \ref{sec:bkgrej} and signal efficiency studies in 
Section \ref{sec:sig_eff}.  Unlike the signal events, most of these are 
generated directly in Geant4 10.02.p03, with modifications to the 
photo-nuclear, electro-nuclear, and photon-conversion processes described below
to achieve the accuracy needed for LDMX studies.  

As discussed in Section \ref{sec:proc_bkg}, the leading backgrounds for LDMX 
arise from either a rare interaction of a hard bremsstrahlung photon or 
(more infrequently by a factor of $\mathcal{O}(\alpha/\pi)$) a rare hard interaction of 
the primary electron.  
It is neither necessary nor feasible to generate Monte Carlo events with the full Phase I statistics of incident electrons.  Rather, we use a combination of bremsstrahlung preselection and the Geant4 occurence biasing toolkit to simulate the smaller number of events expected for each of these rare interaction types. 

\noindent
\subparagraph{Hard Bremsstrahlung Filter Pre-Selection}
Many of the important background processes identified in 
Section~\ref{sec:proc_bkg} begin with a hard bremsstrahlung reaction in the 
target, where the resulting high-energy photon subsequently undergoes a rare
interaction in the target, recoil tracker, or \ecal.  Events surviving the 
tracking and \ecal selections are those where the electron recoiling from the bremsstrahlung reaction has an energy below 1.2 GeV, which we call ``hard brem'' events.  For a 10\% $X_0$ target, ``hard brem'' events comprise only about 3\% of all electron interactions.  To simulate them efficiently, we implement a Geant4 filter that propagates the primary electron through the target, putting all secondaries on a waiting stack to minimize the computational expense of simulating non-selected events.  
To maintain a filter efficient for signals within detector resolutions, events where the electron leaves the target with an energy exceeding 1.5 GeV, does not undergo bremsstrahlung, or stops in the target are killed after minimal simulation time.  
For surviving events --- those where the electron undergoes at least one bremsstrahlung reaction and leaves the target with at most 1.5 GeV energy --- the electron track is placed in a waiting stack until other interactions have been simulated.  This allows additional selections to be applied on the bremsstrahlung photon and its subsequent interactions, before incurring the computational cost of simulating an electron shower.

\noindent
\subparagraph{Cross-Section Biasing for photo-nuclear Processes}
The rare photon interactions relevant for LDMX have cross-sections 
$10^{-3}-10^{-5}$ times the conversion cross-section. Thus, even with the
above optimizations, further techniques to enhance the simulated rate of the 
rare process are required.  

Starting in version 10.0, Geant4 has introduced an occurrence biasing toolkit 
that facilitates simulating particle interactions with a biased interaction 
law, e.g.~an enhanced cross-section for a rare process such as photo-nuclear 
interactions or muon conversions.  Each event is assigned a weight to account for
the effect of cross-section biasing on both the occurrence probability of the
biased interactions (a factor of  
$\sigma_{\text{biased}}/\sigma_{\text{physical}}$) and the survival probability
of the interacting particle $z$ (a factor of
$e^{z/\ell_{\text{physical}}-z/\ell_{\text{biased}}}$, where $\ell$ is the 
particle's interaction length).  

For LDMX \ecal photo-nuclear studies, we adopt the following biasing 
prescription:
\begin{equation}
    \sigma_{\text{PN}}^{\text{biased}} 
        = B \sigma_{\text{PN}}^{\text{physical}} \qquad \sigma_{\text{conv}}^{\text{biased}} 
        = \sigma_{\text{conv}}^{\text{physical}} - (B-1) \sigma_{\text{PN}}^{\text{physical}}.
\end{equation}
This biasing scheme guarantees that the \emph{total} interaction cross-section, 
and therefore the photon's non-interaction probability is unaffected by the
biasing.  Thus all photo-nuclear interactions have a uniform weight $1/B$, and 
their distribution through the detector volume matches the expected physical
distribution. For photo-nuclear reactions, a biasing factor 
$B_{\text{\ecal-PN}}=450$ is chosen so that the biased conversion 
cross-section is positive in all detector materials while bringing the biased 
cross-section in Tungsten within an order of magnitude of the physical 
conversion cross-section for efficient simulation.  To obtain an exclusive 
photo-nuclear sample corresponding to $N_e$ electrons on target, we generate 
events corresponding to $N_e/B_{\text{PN}}$ electrons on target with the hard 
bremsstrahlung filter, cross-section biasing, and a second filter that 
\emph{requires} a hard photon to undergo a photo-nuclear reaction.   To generate
MC statistics corresponding to $N_{\text{EOT}} = 1.2\times 10^{14}$ electrons
on target, we generate biased MC events for $2.6 \times 10^{11}$ incident 
electrons, corresponding to a sample of $2.7 \times 10^{9}$ photo-nuclear events.

For the generation of photo-nuclear events originating from the target, the 
biasing description described above is also used with a biasing factor 
$B_{\text{target-PN}} = 450$. By generating events for $1.3 \times 10^{11}$
incident electrons, we can model the target-area PN background corresponding 
to $N_{\text{EOT}} = 2.3 \times 10^{14}$ electrons on target. 

\noindent
\subparagraph{Cross-Section Biasing for Electro-nuclear Processes}

In generating target-area electro-nuclear events, event yields are lower than 
the photo-nuclear counterpart and accurate modeling of the $z$ distribution is
less critical.  Therefore, to increase generation efficiency, we adopt the 
simpler biasing scheme
\begin{equation}
    \sigma_{\text{EN}}^{\text{biased}} 
    = B_{\text{target-EN}} \sigma_{\text{EN}}^{\text{physical}} 
\end{equation}
with a biasing factor $B_\text{target-EN}=5 \times 10^4$.   
By generating events for $5.0 \times 10^9$ incident electrons, we can model the 
target-area EN background corresponding to $N_\text{EOT} = 3.8 \times 10^{14}$ 
electrons on target.

\subsection{Background Modeling \people{Natalia,Omar} \morepeople{Nhan}}
\label{sec:bkgmod}
\paragraph{Photo-nuclear and Electro-nuclear Modeling}
As discussed in Sec.~\ref{sec:proc_strategy}, background processes of particular concern for LDMX are those where the vast majority of the energy of the incident photon is carried by a small number of particles.  It is therefore important to understand not just the overall photo-nuclear rate, but the yields and kinematics of exclusive final states that are particularly difficult to veto.

This motivates using a photo-nuclear model based on nuclear interaction data at energies near the $3-4$ GeV energy scale crucial for LDMX.  At this energy, the interactions of secondaries in a large nucleus can also be significant; these are typically modeled using an intranuclear cascade.  

These considerations strongly motivate using the Bertini Cascade model in \geant~\cite{BertiniCascade}. This model is based on measured cross-sections for photon-hadron, hadron-hadron, and photon-dihadron reactions within the nucleus, and includes 2-to-9-body final states as well as kaon production.  
Final-state phase space distributions for $2\rightarrow 2$ processes are taken from data at low energies, with extrapolations or parameterizations at multi-GeV energies, while multi-body phase space distributions are modeled as uniform.  
The Bertini cascade model has been broadly validated and is the default model for photo-nuclear reactions in Geant with  $<3.5$ GeV incident photons (with higher-energy reactions modeled using a high-energy string model) and for electro-nuclear reactions with virtual photon energies $<10$ GeV.  

In \geant, the Bertini cascade model is the default model for photo-nuclear reactions initiated by $<3.5$ GeV incident photons (with higher-energy reactions modeled using a high-energy string model).  The model is believed to be accurate up to $\sim 10$ GeV incident energies \cite{DennisPrivateCommunication}, and indeed is used in \geant for this purpose for other primary particles.  Given the importance for LDMX of modeling the full physics of the photo-nuclear final state properly, all simulations have been done using a modified FTFP\_BERT physics list, where photo-nuclear processes with $<10$ GeV photons are always modeled using the Bertini cascade.  

Electro-nuclear interactions are closely related to photo-nuclear reactions since they proceed at these energies through a virtual photon.  They are simulated in \geant using the equivalent-photon approximation, with Bertini cascade as the default generator for equivalent photon energies below 10 GeV. In the following discussion, we focus on photo-nuclear reactions for concreteness but the same considerations (for both physics and simulation) apply to electro-nuclear processes. 

\subparagraph{Rare but Important Final States}
Notwithstanding the validation of the Bertini model, the sensitivity of LDMX depends on the rates of sub-dominant processes in small corners of phase space.  These have not received as much scrutiny in the validation of the Bertini model as inclusive distributions, and we have found that the inputs to the Bertini model that govern their rates are, in several cases, significantly overestimated relative to data and/or physical considerations.  
  Therefore, we have investigated in detail three classes of event that are design drivers for LDMX: single forward neutrons (which drive the depth needed for the \hcal), moderate-angle neutron pairs (which drive the width needed for the back \hcal), and hard backscattered hadrons in the so-called ``cumulative'' region --- i.e. kinematics that is only accessible by scattering off a multi-nucleon initial state (which set a sensitivity floor, since they are quite difficult to reject). 

We have found all three of these event types to be significantly over-populated by the Bertini model, relative to data-driven expectations; by contrast, single-pion final states are under-populated. We have identified the origin of these discrepancies and they all appear straightforward to correct by means of minor corrections to the input data and/or algorithms used in the Bertini cascade.   Discussions with the \geant authors and code modifications to this effect are underway.  

Comparisons of the default \geant Bertini Cascade output to relevant data are elaborated in Appendix \ref{sec:PNappendix}, as are the origins for these discrepancies in the Bertini Cascade model.  Here, we briefly summarize our findings and the implementation of \emph{post hoc} corrections to the photo-nuclear model in analyses presented here (presently, several of these corrections are applied only to the small post-veto samples considered in Secs.~\ref{sec:bkgpn} and \ref{sec:bkgen}).  

In the case of single- and di-neutron final states, \geant vastly overestimates (by a factor of $1000$) the rate of two-body dinucleon dissociation by 3--4 GeV photons.  This leads to an overestimate of the inclusive single- and di-neutron event yields by factors of 100 and 500, respectively.  Although the code modifications required to remedy this are straightforward, improving the accuracy of this simulation requires re-generating the entire event sample with new nuclear cross-section models, and is not currently feasible.  Therefore, we have not attempted to correct any studies for this effect, except that when working with sufficiently small post-veto samples, we have down-weighted single- and di-neutron events by factors of 100 and 500 respectively.

\begin{figure}
\includegraphics[height=0.5\textwidth]{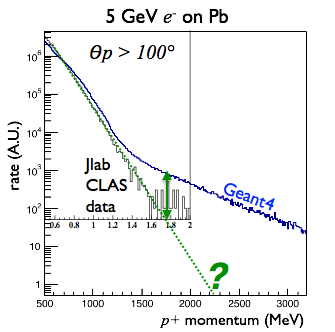}
\caption{\label{fig:bertiniCumulative1} Distribution of proton momenta at $\theta_p > 100^\circ$ in 5 GeV electron scattering off $^{208}Pb$.  The black histogram shows the measured yield in a skim of CLAS eg2 data \cite{StepanPrivateCommunication}, for events with the electron in the CLAS acceptance.  The blue histogram shows the results of a \geant simulation before final-state down-weighting.}
\end{figure}

The over-population of the ``cumulative'' region (very high energy backscattered hadrons) is illustrated, for example, in Figure \ref{fig:bertiniCumulative1} ---  the momentum distribution of protons backscattered beyond $100^\circ$ in 5 GeV $e$-Pb scattering, as extracted from CLAS data \cite{StepanPrivateCommunication} is compatible with the theoretically expected exponential distribution, while an unphysical tail is evident in the Bertini Cascade model.
This tail has been traced to an unphysical model of the transitions between regions of the nucleus with different densities --- highly energetic hadrons at grazing incidence to the boundary between regions are taken to reflect off this boundary, when they should escape the nucleus in a more physical model.  These events have been removed from our sample by a two-fold approach: First, in inclusive studies, events with hard nucleons at $\theta>100^\circ$ and  $w \equiv (p+T)/2 (\sqrt{1.25} - 0.5 \cos\theta) > 1400 $ MeV are down-weighted to a uniform exponential, removing the ``ankle'' in the inclusive distribution and softening this feature in the more backwards phase space (this procedure was adopted before the origin of the tail in Fig.~\ref{fig:bertiniCumulative1} and related distributions was known; the form of the weighting function $w$ is motivated by the phenomenological model of \cite{Degtyarenko:1994tt}). This weighting does not remove all of the unphysical tail events --- for example, those with backscattered pions or wide-angle neutrons are not weighted by this procedure.  
In the analysis of smaller post-\hcal-veto samples in Secs.~\ref{sec:bkgpn} and \ref{sec:bkgen}, we disregard the $w$-dependent event weights but instead have resimulated all events that survived the vetoes from their seeds, using a modified \geant version that eliminates the unphysical reflections from the outset.  Essentially all of these events, once resimulated, involve forward rather than backward-going hadrons and are easily vetoed by the HCal.  These events are, therefore, not included in the event counts. 

We also caution the reader that the single- and di-neutron final state yields (and potentially multi-neutron final states, which are likely dominated by cascades of related reactions) are over-estimated in the samples used in this whitepaper.  It is expected that the inefficiency of the \ecal veto and the size required of the \hcal veto to achieve low background will both be lower when more realistic samples are used.

\paragraph{Modeling of Processes with Muons}

The dominant source of muon pairs in LDMX is photon conversion to $\mu^+\mu^-$ 
via coherent scattering off a nucleus (in the target or \ecal).  This process is
included in \geant but was found to have an unphysical form factor.  
In addition, approximations to the phase-space distribution of outgoing muons
were found to be inaccurate on the tails.  LDMX dimuon samples were therefore 
generated with a modified version of the G4PhotonConversionToMuons class, which
uses the full dimuon phase space distribution from \cite{Tsai:1974} eqn (2.3) 
(assuming elastic recoil of the nucleus and keeping only the $W_2$ component 
of the result).  

\begin{figure}[tbh!]
\begin{center}
    \includegraphics[width=8cm]{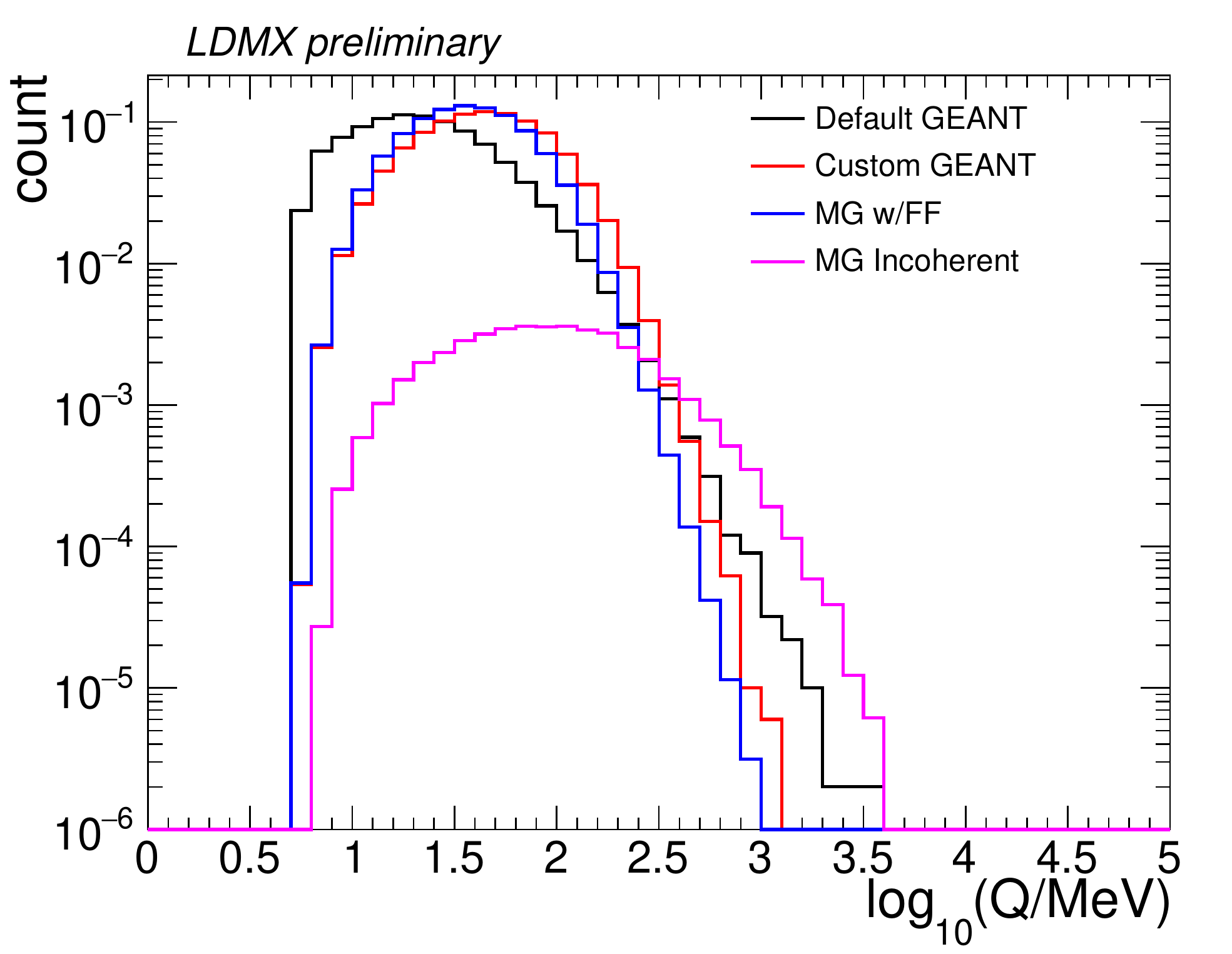}
    \includegraphics[width=8cm]{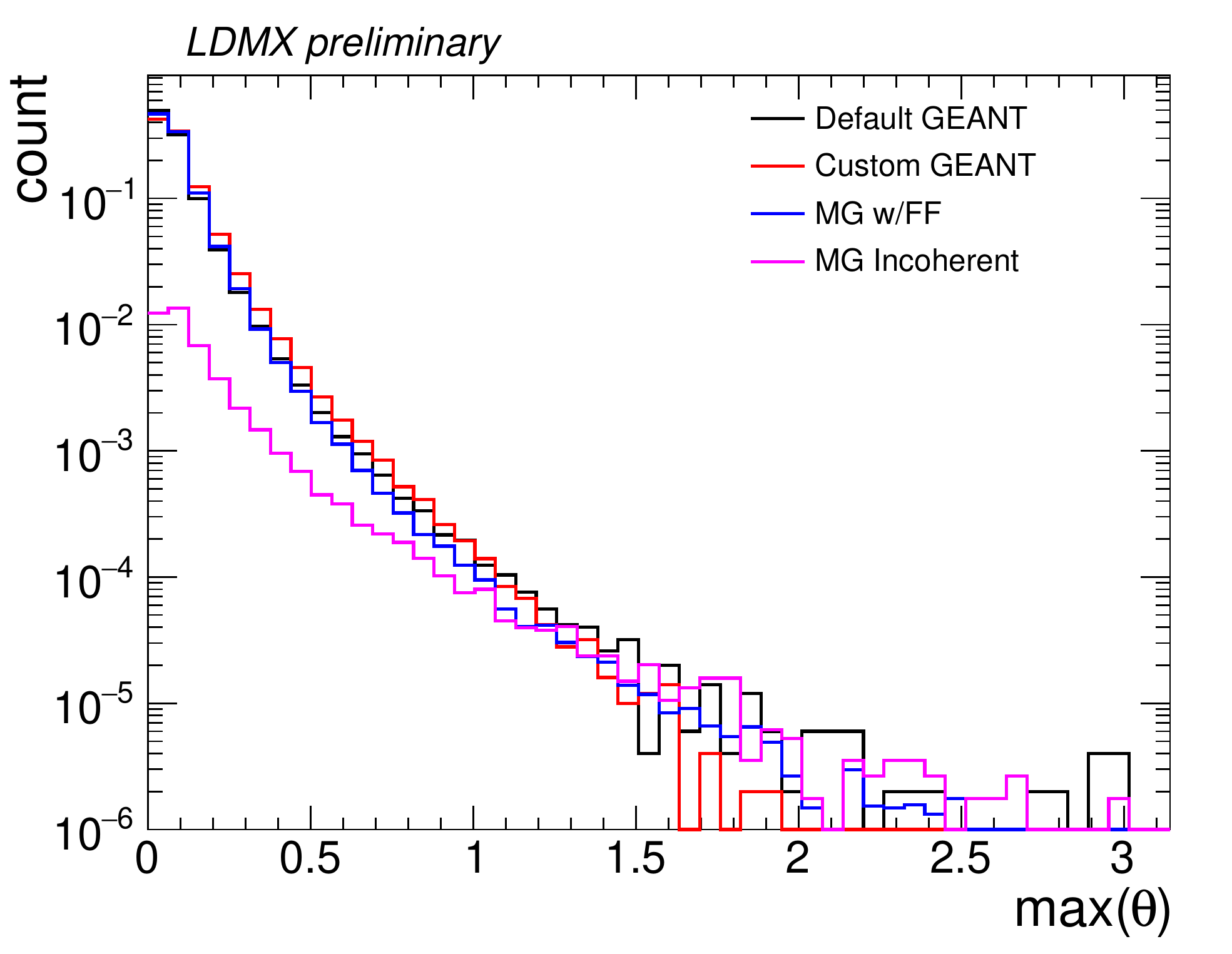}
    \caption{Kinematics of background coherent photoconversion to muons using default GEANT, custom GEANT, {\textsc MadGraph} with a form factor as given in ~\cite{Tsai:1974}, and the incoherent process as well.  On the left is $q$, the momentum transfer and on the right is the maximum polar angle of the two muons.
    }
 \label{fig:gmm}
 \end{center}
\end{figure}

To illustrate the effect of the modified photon conversion process, we compare 
kinematics of photon conversion to muons in \geant, for a monoenergetic 4~\GeV~photon beam incident on a thin target, to a matrix element calculation using {\textsc MadGraph} and {\textsc MadEvent}.
In Fig.~\ref{fig:gmm} (left), we show the momentum transfer $q$ of the photon conversion process where $Q = p_{\mu +} + p_{\mu -} - p_\gamma$ 
and $p$ denotes the three-momentum of the incoming photon and outgoing muons.  
Different colored lines are shown for the three different models of the coherent process: default GEANT, our custom version of GEANT with an improved form factor and phase space integration, and the expectation from MadGraph.  
At high momentum transfer, $Q$, the muons tend to decay at higher angle in the lab frame (with less Lorentz boost) thus making their detection more challenging.  
This can be seen in Fig.~\ref{fig:gmm} (right) which illustrates the polar angle of the muon.  

We also show the expectation from the incoherent process (also simulated using {\textsc MadGraph} and {\textsc MadEvent}) in magenta which has an overall smaller cross-section than the coherent process and therefore, its normalization is plotted relative to the coherent process.  The incoherent-conversion process dominates at large $Q$, where the nuclear form factor suppresses the coherent process.  It can contribute an $\mathcal{O}(1)$ correction to the phase space involving wide-angle muons. 

\subsection{Tracking System Simulation and Performance \people{Omar,Tim} \morepeople{Robert}} 
\label{sec:trk_perf}
\subsubsection{Tagging Tracker}

Two methods are used to benchmark the performance of the tagging tracker and better understand the constraints it provides in preventing mis-tags.  First, the momentum and impact parameter resolutions at the target are determined using an analytic model of the tagging tracker that includes the effect of intrinsic resolutions and multiple scattering in the tracker planes.  Second, full simulation is used to confirm these resolutions and understand reconstruction efficiencies and susceptibility to backgrounds.

\begin{figure}[!htb]
\begin{minipage}{0.48\linewidth}  
    \centering
    \includegraphics[width=\linewidth]{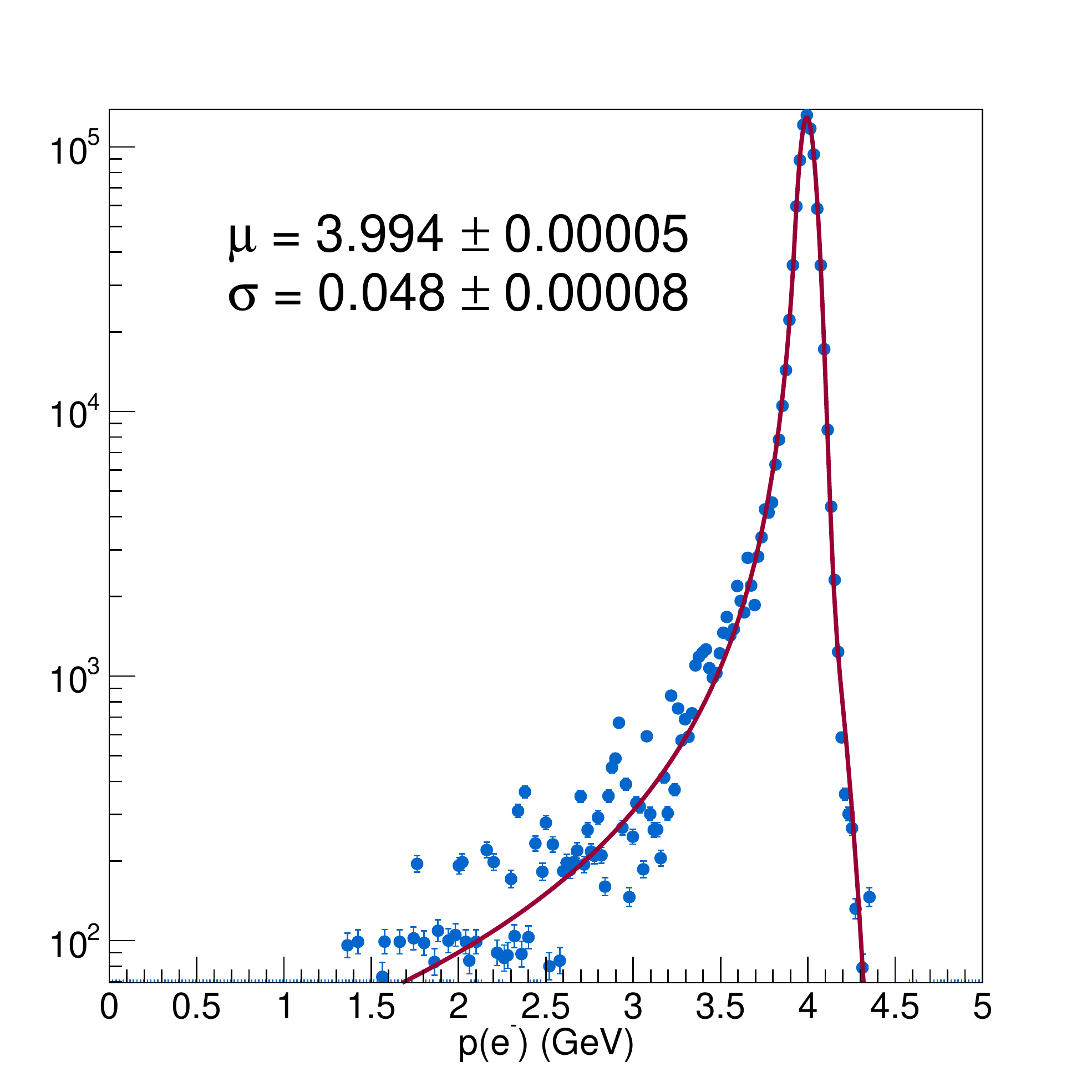}
    \caption{The longitudinal momentum reconstructed by the tagging tracker for 
             a sample of 4 GeV beam electrons. Excellent momentum resolution 
             allows tight selection against any off-energy component of the 
             beam.}
    \label{fig:tagger_4gev_p}
\end{minipage}
\hfill
\begin{minipage}{0.48\linewidth}  
    \centering
    \includegraphics[width=\linewidth]{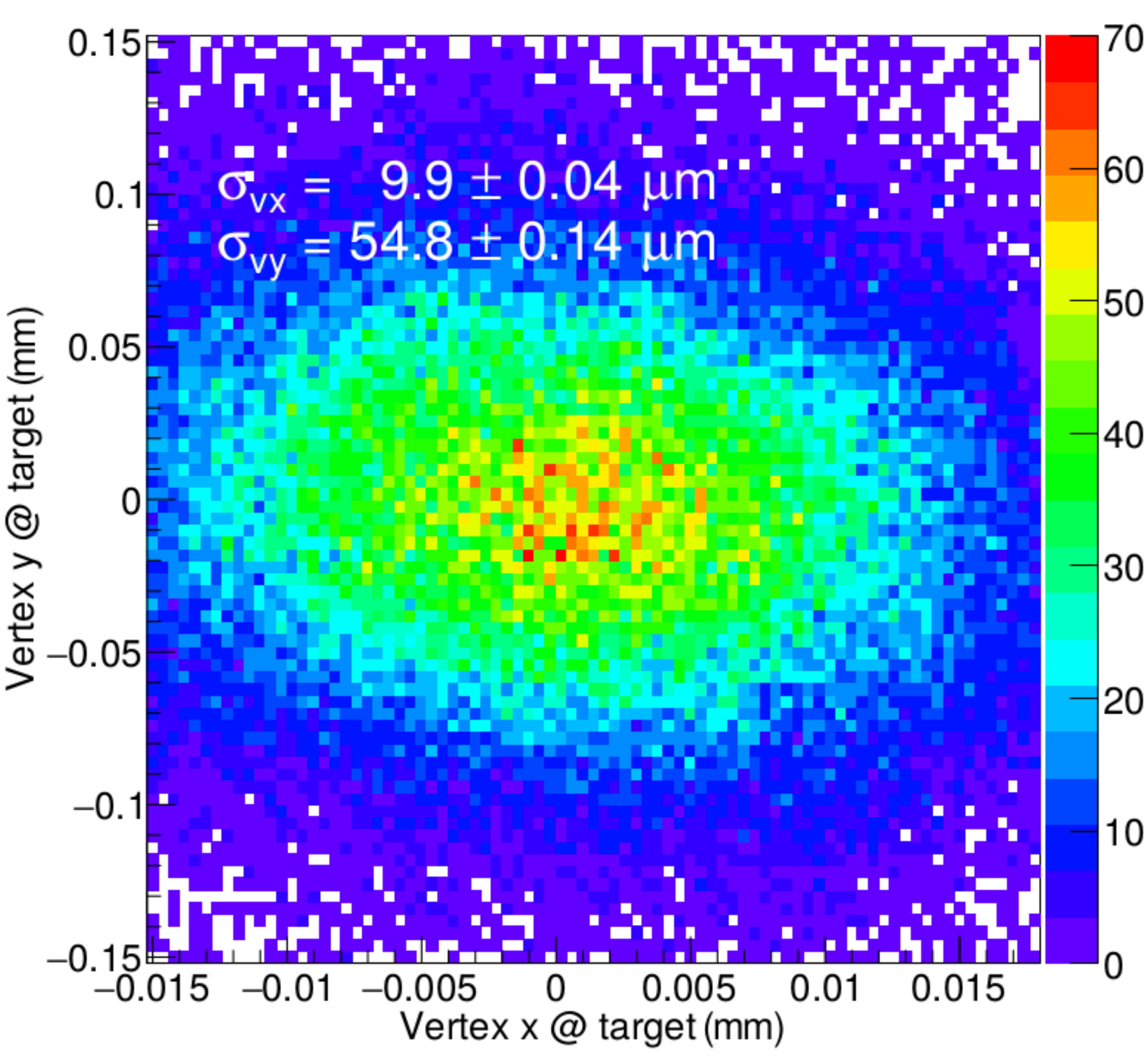}
    \caption{The $x-y$ position at the target reconstructed by the tagging 
             tracker for a sample of 4 GeV beam electrons. Excellent impact 
             parameter resolution provides a strong constraint on a matching 
             recoil track.}
    \label{fig:tagger_4gev_vxvy}
\end{minipage}
\end{figure}

For incoming 4~GeV electrons, the presence of seven three-dimensional  measurements of the trajectories ensures near perfect tracking efficiency and a very low rate of fake tracks. The ability of the tagging tracker to reject beam impurities is discussed further in Section~\ref{section:bkgrejection}. The analytic model finds a longitudinal momentum resolution of approximately 1\% and the corresponding full simulation results show good general agreement, as shown in Fig.~\ref{fig:tagger_4gev_p}. The transverse momentum resolutions are found to be 1.0~MeV and 1.4~MeV in the horizontal and vertical directions, respectively, which are small compared to the 4~MeV smearing in transverse momentum from multiple scattering in the 10\%~$X_0$ target.  Meanwhile, the impact parameter resolution for 4~GeV electrons is expected to be approximately 7~$\mu$m (48~$\mu$m) in the horizontal (vertical) direction. Again, the full simulation shows good general agreement, as shown in Fig.~\ref{fig:tagger_4gev_vxvy}. These results indicate that tight requirements can be made in both the energy and trajectory at the target, which respectively serve to clearly identify signal events and prevent accidental mis-association between incoming particles and tracks reconstructed in the recoil tracker.

In summary, the design of the tagging tracker appears robust enough to provide unambiguous tagging of incoming electrons with the nominal beam energy for Phase I of LDMX. While the system can clearly deal with higher occupancies, further study will be required to find the beam intensity limits for Phase II.

\subsubsection{Recoil Tracker Performance} \label{subsec:recoil_tracker}

The recoil tracker must have a large acceptance for recoiling electrons characteristic of signal events with good resolution for transverse momentum and impact position at the target, both of which are critical for unambiguously associating those recoils with incoming electrons identified by the tagging tracker.  While good reconstruction efficiency for signal recoils is important, it is even more important to have good efficiency for charged tracks over the largest possible acceptance to help the calorimeter veto background events with additional charged particles in the final state.  In addition, the tracker should have sufficient momentum resolution that it can assist the \ecal in identifying events in which an incoming beam electron passes through the target and tracker without significant energy loss. Finally, the recoil tracker can help identify events in which a bremsstrahlung photon generated after the end of the tagging tracker undergoes a \pn reaction in the target or tracker material.  As with the tagging tracker, an analytic model of the recoil tracker is used to estimate momentum and impact parameter resolutions, while full simulation and reconstruction are used to confirm those resolutions, to estimate signal acceptances and tracking efficiencies, and to test the ability of the recoil tracker to reject backgrounds. Resolutions will be discussed further here while acceptance and efficiency for signal events will be discussed in Section~\ref{section:signaleff} and background rejection will be discussed in Section~\ref{section:bkgrejection}.

Measurement of the position and transverse recoil momentum at the target requires precise determination of the track angle at the target.  For the transverse momentum, a good curvature measurement is also required in order to set the overall momentum scale.  At least two 3-d measurements directly downstream of the target are needed to determine the recoil angle, and at least one additional bend-plane measurement is needed for curvature.  For low-momentum tracks, it is sufficient for the third measurement to be in the first four closely-spaced layers, but for high momentum tracks that are nearly straight, hits in both of the downstream axial layers are necessary for good momentum resolution.  

%
%

\begin{figure}[!htbp]
    \centering
    \includegraphics[width=\textwidth]{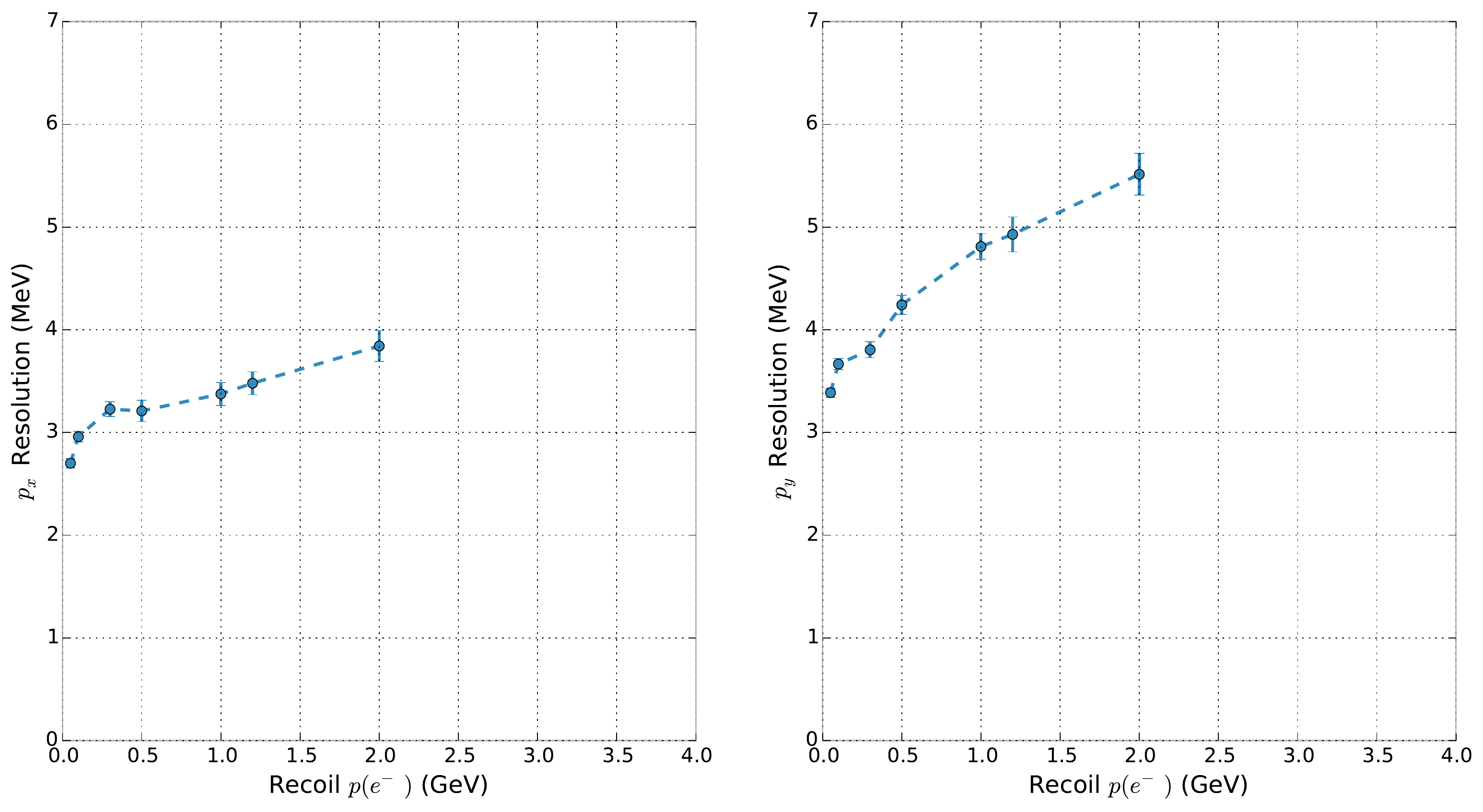}
    \caption{
Resolutions on the components of the momentum transverse to the target for signal recoils. Simulated recoils are assumed to originate at a random depth in the target so that the average resolution is less than the 4~MeV smearing from multiple scattering in the full target thickness. Only the vertical component of the recoil momentum shows significant degradation due to detector resolution at the highest recoil momenta considered.}
    \label{fig:px_py_res}
\end{figure}

\begin{figure}[!htbp]
    \centering
    \includegraphics[width=\textwidth]{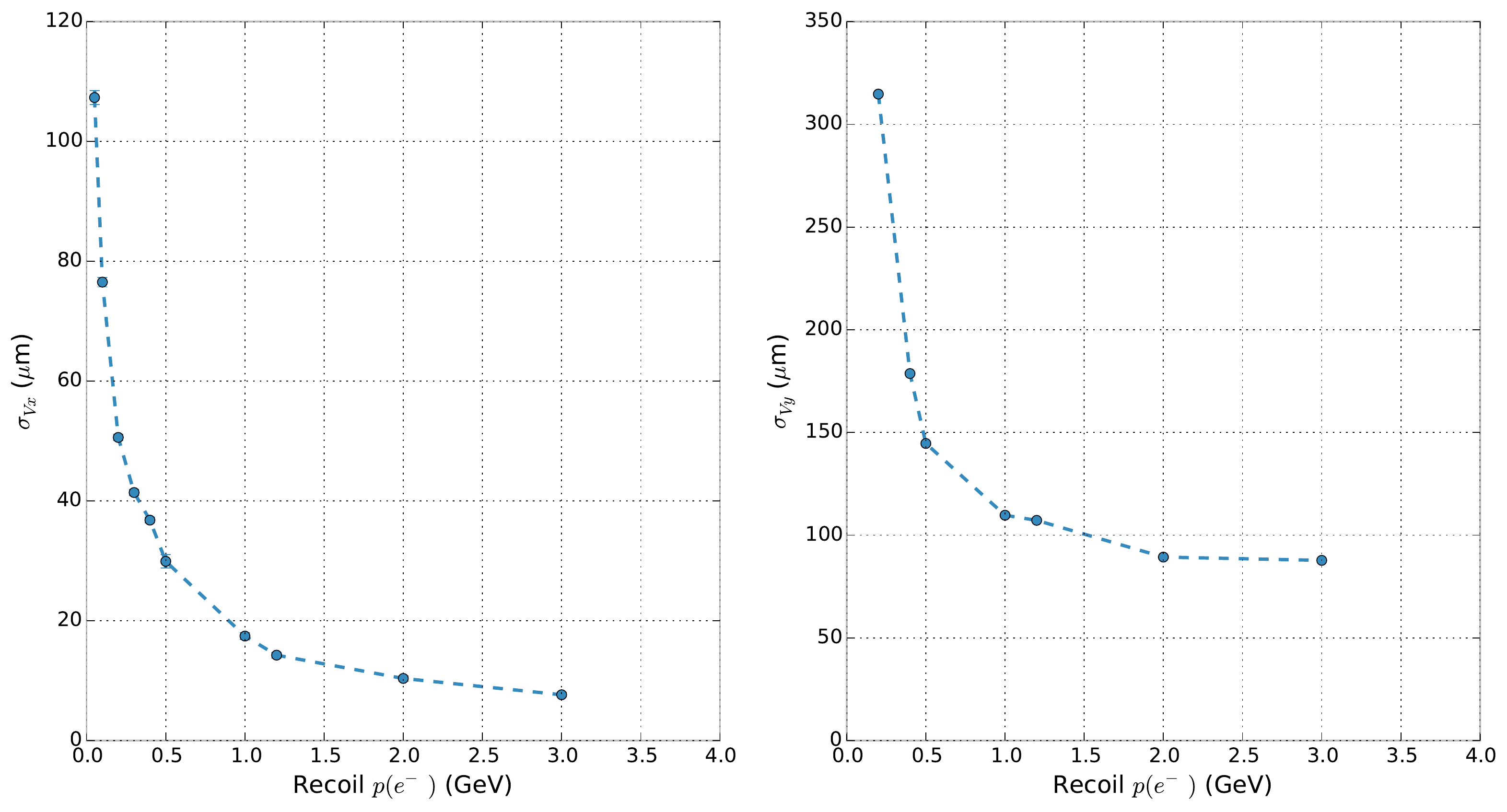}
    \caption{Recoil tracker impact parameter resolutions. Requiring a common impact position at the target with both the tagging and recoil tracker strongly selects against associating mis-reconstructed recoils with a tagged incoming electron.}
    \label{fig:impact_param}
\end{figure}

The ability to distinguish signal from background using the recoil transverse momentum is obviously limited by the multiple scattering in the target, where multiple scattering in a 10\% $X_0$ target results in a 4~MeV smearing in transverse momentum.  Using the analytic model of the recoil tracker, the material budget and single-hit resolutions were designed so that the transverse momentum resolution is limited by multiple scattering in the target over the momentum range for signal recoils. This has been verified in full simulation, as shown in Fig.~\ref{fig:px_py_res}.  Meanwhile, the impact parameter resolution, shown in Fig.~\ref{fig:impact_param}, strongly constrains the phase space for mis-reconstructed tracks to point to the same location in the target as the incoming track reconstructed in the tagging tracker.

\subsection{ECAL Simulation and Performace \people{Joe} \morepeople{Ruth,Mans}}
\label{sec:ecal_perf}
The simulation of the \ecal uses the \geant~framework coupled with an electronics simulation that converts the energy deposits in the silicon into simulated digital hits. Noise is introduced at the digitization stage using a realistic model based on tests performed with prototypes of the CMS HGCROC front end readout chip. The noise RMS is mostly driven by the sensor readout pad capacitance, which in turn depends on the pad geometry. For a 500 $\mu$m thick sensor with pad area of $\sim$0.5~cm$^2$ the capacitance is calculated to be $\sim$25 pF. Assuming a small additional capacitance of $\sim$5~pF for the traces connecting the pads to the readout chip and including the constant term in the HGCROC noise, the RMS is estimated to be approximately 1500 electrons. Under the assumptions that a MIP at normal incidence to the sensor produces $\sim$33000 electrons in $\sim 500$~$\mu$m of silicon, for which the energy-equivalent response in an individual pad is 0.13 MeV, this translates to a noise RMS of approximately 0.006 MeV. A readout threshold of 4 times this noise RMS is found to be reasonably conservative, resulting in sufficiently few noise hits being recorded while also not adversely affecting the probability of reading out genuine hits.

\begin{figure}[ht]
\centering
\includegraphics[width=7cm]{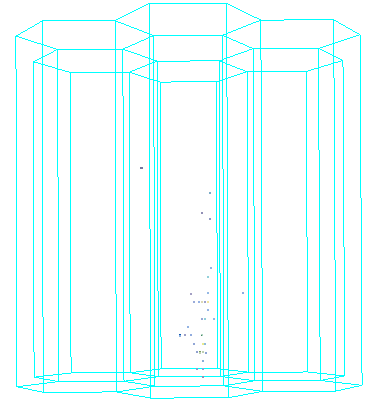}
\includegraphics[width=9.2cm]{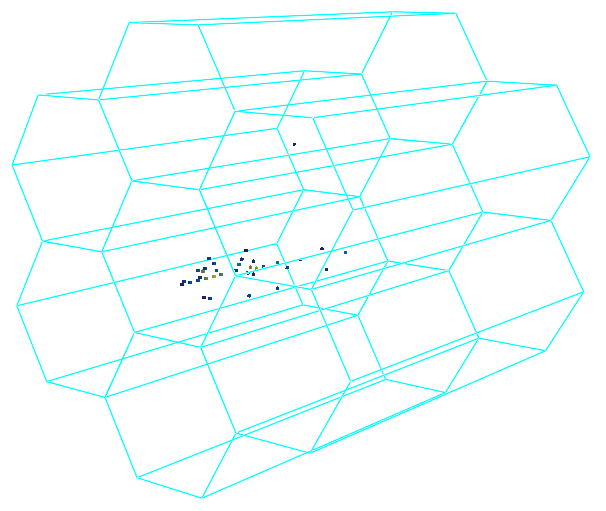}
\begin{flushleft}
\caption{\label{fig:geom} An event display for a \geant simulation of a 4~GeV electron in the LDMX detector.}
\end{flushleft}
\end{figure}

A simulated interaction of a 4 GeV electron in LDMX is shown in Fig.~\ref{fig:geom}. A  4 GeV electron event, on average, results in charge deposition equivalent to $\sim$800 MIP signals observed in the calorimeter cells.

The \ecal is responsible for measuring electrons and photons with extremely high efficiency and good energy resolution.
This majority of events involve easily detected electromagnetic showers, but there are also rare photon and electron interactions such as photon conversion to a muon pair or hadronic, photo-nuclear processes that are more challenging to observe.  The power and versatility of the \ecal to detect this wide range of signatures is a strength of the proposed detector technology.

\paragraph{Inclusive shower}

Reconstructed electromagnetic shower energy is the most powerful discriminant between signal and the most common, electromagnetic bremsstrahlung background.  As shown in Fig.~\ref{fig:LDMspectra}, applying an upper bound on the energy deposited in the \ecal is sufficient to drastically reduce this background.

\paragraph{Photo-nuclear interactions}

Rarer background processes can cause small, irregular, or negligible energy depositions in the \ecal that can, on rare occasions, produce an event structure that is very similar to that expected for signal.  In particular, photo-nuclear processes, in which the photon interaction results in the production of hadrons in the \ecal, have the potential to generate a fake signal.

\begin{figure}[h]
\centering
\includegraphics[width=8.1 cm]{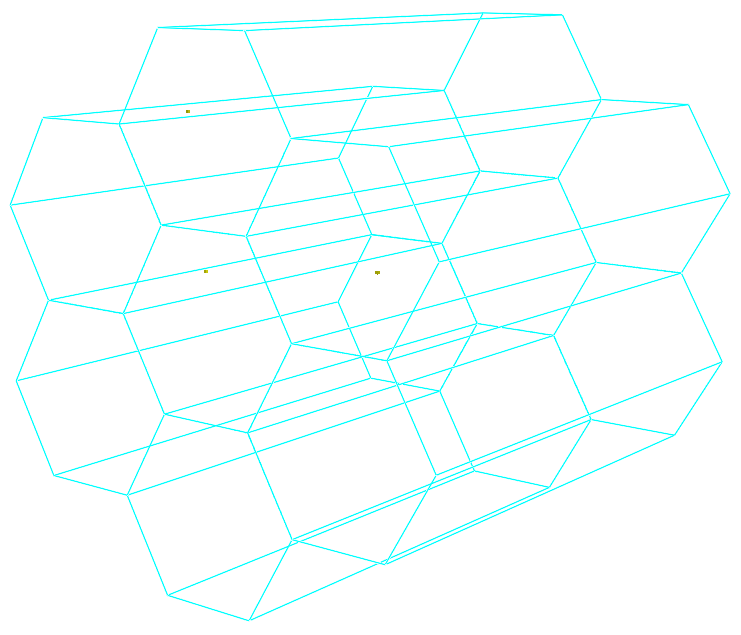}
\includegraphics[width=8.1 cm]{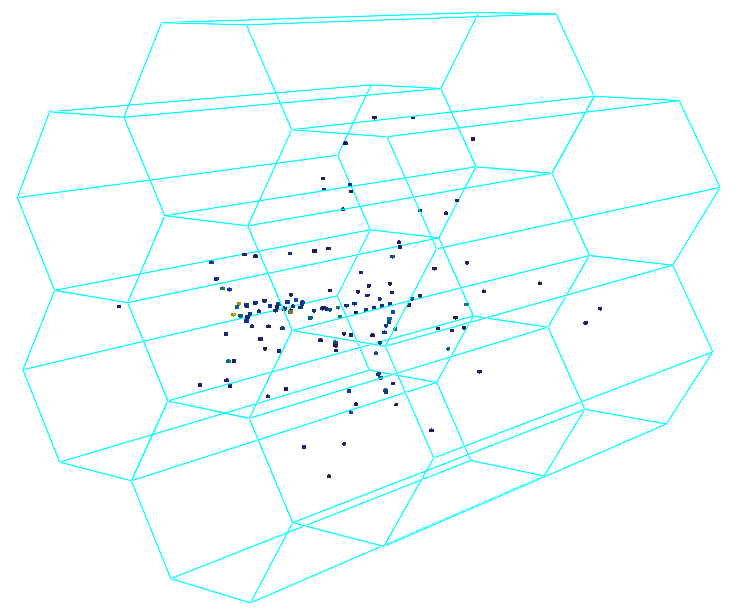}
\begin{flushleft}
\caption{\label{fig:miss} Event display for a $M_{A} = 0.1$~GeV signal event (left) and a photo-nuclear event (right) where in both cases the recoil electron misses the \ecal.}
\end{flushleft}
\end{figure}

\begin{figure}[ht]
\centering
\includegraphics[width=8.1 cm]{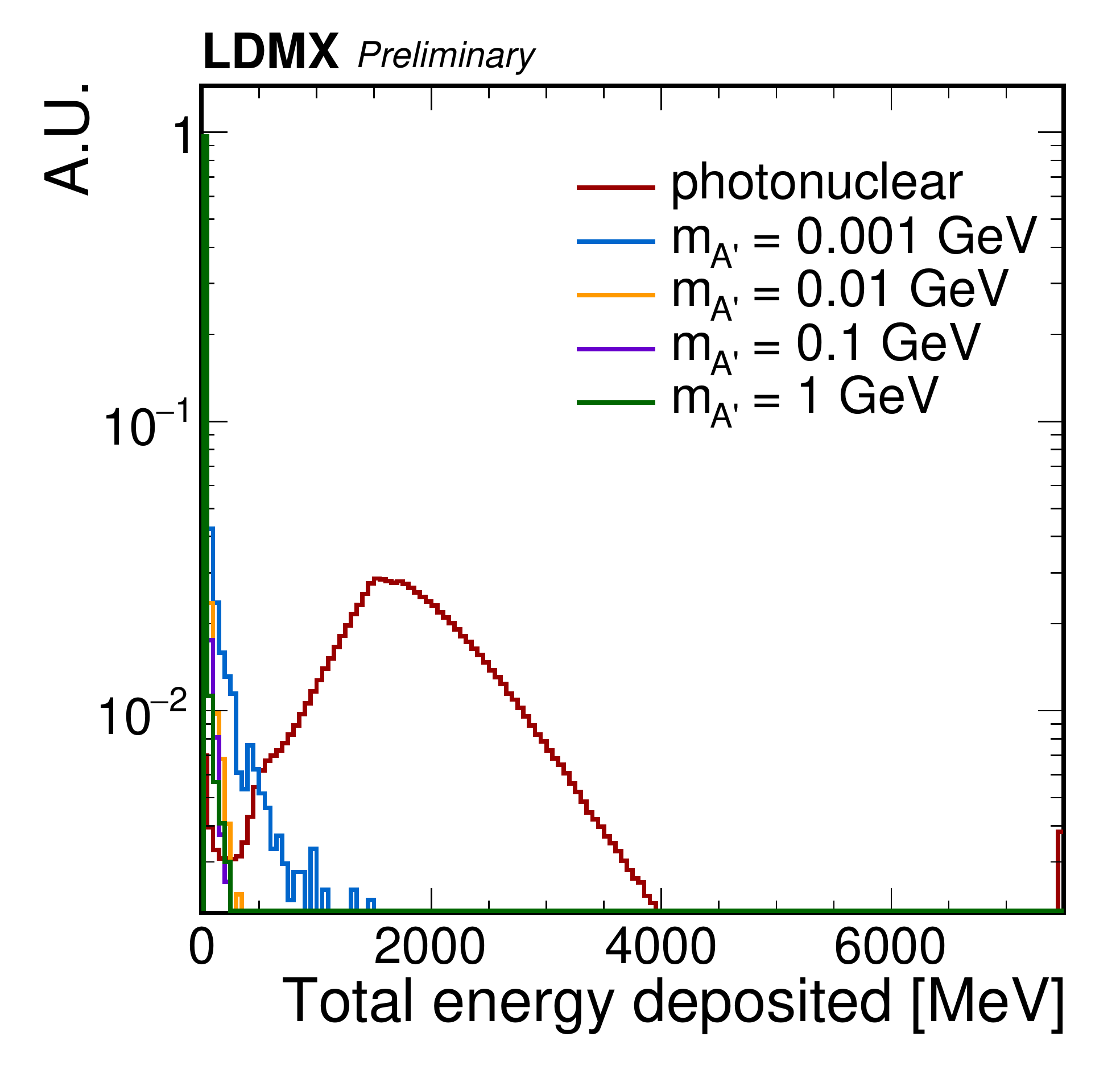}
\includegraphics[width=8.1 cm]{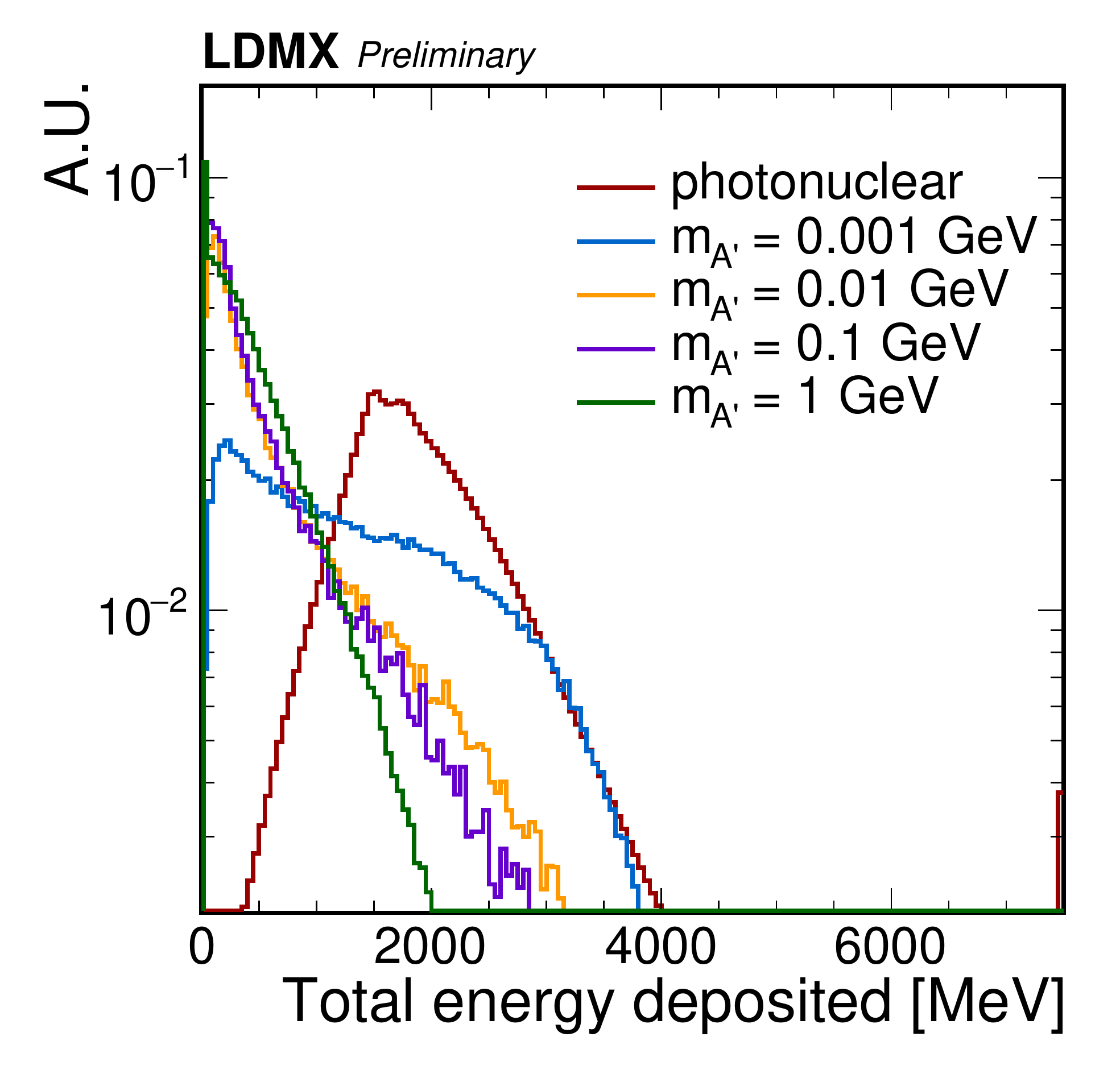}
\begin{flushleft}
\caption{\label{fig:energydep} Total energy deposited in the \ecal for events in which the recoil electron misses the \ecal (left) or is within the \ecal (right), for the photo-nuclear backgrounds and for signal processes with four different mediator masses. All distributions are normalized to unit area for comparison. For all distributions shown in this subsection, the last bin includes the overflow.}
\end{flushleft}
\end{figure}

The optimization strategy discussed later in this section primarily focuses on events in which the recoil electron interacts in the \ecal. However, events in which the recoil electron is emitted at large angles and misses the \ecal represent an interesting category with the distinct experimental signature of very little or no energy appearing in the \ecal. The little that might appear comes from pure noise or from occasional soft photons that have been radiated off of the recoil electron as it is accelerated in the fringe field of the magnet while passing through the recoil tracker. In contrast, for backgrounds, even those involving photo-nuclear processes, a substantial amount of energy usually appears in the \ecal. Event displays for a signal event with $m_{\rm A'} = 0.1$ GeV (left) and a photo-nuclear event (right) for which the recoil electron misses the \ecal in both cases are shown in Fig.~\ref{fig:miss} to illustrate how significant this difference can be. Total energy, computed as the energy sum of all hits above the readout threshold  for events with the recoil electron missing or hitting the \ecal, respectively, are illustrated in the two plots shown in  Fig.~\ref{fig:energydep}. For the case in which the recoil electron misses the \ecal, one sees a much more significant separation between the total energy of photo-nuclear backgrounds and signal events.

\begin{figure}[ht]
\centering
\includegraphics[width=6.5 cm]{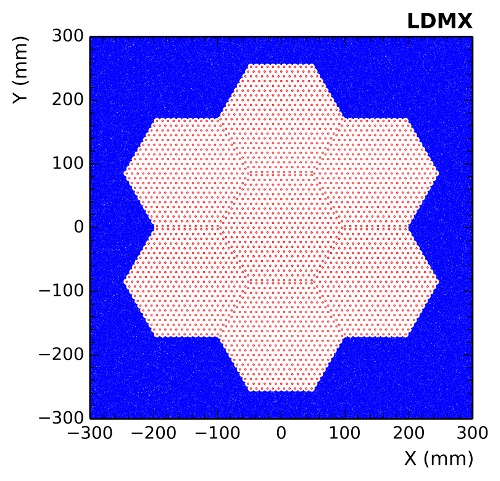}
\begin{flushleft}
\caption{\label{fig:fid} Two million points outside of the fiducial region were generated from a uniform distribution and are shown in blue. The hex cell centers are shown in red, with any areas that are white or red corresponding to the fiducial region.}
\end{flushleft}
\end{figure}

The \ecal fiducial region is illustrated in Fig.~\ref{fig:fid}. Tracking information is used to determine where the recoil electron is expected to hit the \ecal face. If that position is within 5 mm of any \ecal cell center, with cell centers marked by red dots in the figure, the recoil electron is defined to be within the fiducial region. This radius was found to be the smallest that can be used to produce a fiducial region that contains all of the electrons that enter active \ecal regions.

The recoil electron is found to miss the \ecal in a significant fraction of signal events. The fractions of events for which the recoil electron misses the \ecal for signal processes with four different mediator masses, and for photo-nuclear processes, are shown in Table~\ref{tab:table1} for events passing the trigger. One sees that the probability for the recoil electron to miss the \ecal is correlated with the mediator mass. As the mediator mass decreases, the recoil electron is likely to carry more energy and undergo a smaller  transverse deflection.

\begingroup
\begin{table*}[!bh]
\caption{\label{tab:table1}
Fraction of events after applying the trigger requirement in which the recoil electron misses the \ecal for signal processes involving four different mediator masses, and for photo-nuclear backgrounds.}
\begin{ruledtabular}
\begin{tabular}{ l @{\qquad} ccccccc }
\textrm{1.0 GeV A'}&
\textrm{0.1 GeV A'}&
\textrm{0.01 GeV A'}&
\textrm{0.001 GeV A'}&
\textrm{Photo-nuclear}\\
\colrule 
0.38 & 0.36 & 0.25 & 0.07 & 0.07 & \\
\end{tabular}
\end{ruledtabular}
\end{table*}
\endgroup

Several variables, constructed from information read out from the \ecal, have been found to be powerful in distinguishing photo-nuclear background from signal processes. The most powerful discriminating variables are related to energy. In the case of photo-nuclear backgrounds, the photon and recoil electron can both produce hits in the \ecal. This motivates the use of the variables that capture increases in energy and hit count as listed below.
   \begin{itemize}
     \item {\bf Number of readout hits:} the total number of hits in the \ecal above the readout threshold.
     \item {\bf Total energy deposited:} the energy sum of all hits above the readout threshold.
     \item {\bf Total tight isolated energy deposited:} the energy sum of all isolated hits in the inner ring; where a hit is isolated if none of its neighbor cells in the same plane are above the minimum readout threshold, and the inner ring is defined as those cells in the plane that neighbor the cell that contains the shower centroid. The latter is defined as the energy weighted average (x,y) position of all hits in an event.
     \item{\bf Highest energy in a single cell:} the highest amount of energy deposited in a single cell.
   \end{itemize}
   
\begin{figure}[ht]
\centering
\includegraphics[width=8.1 cm]{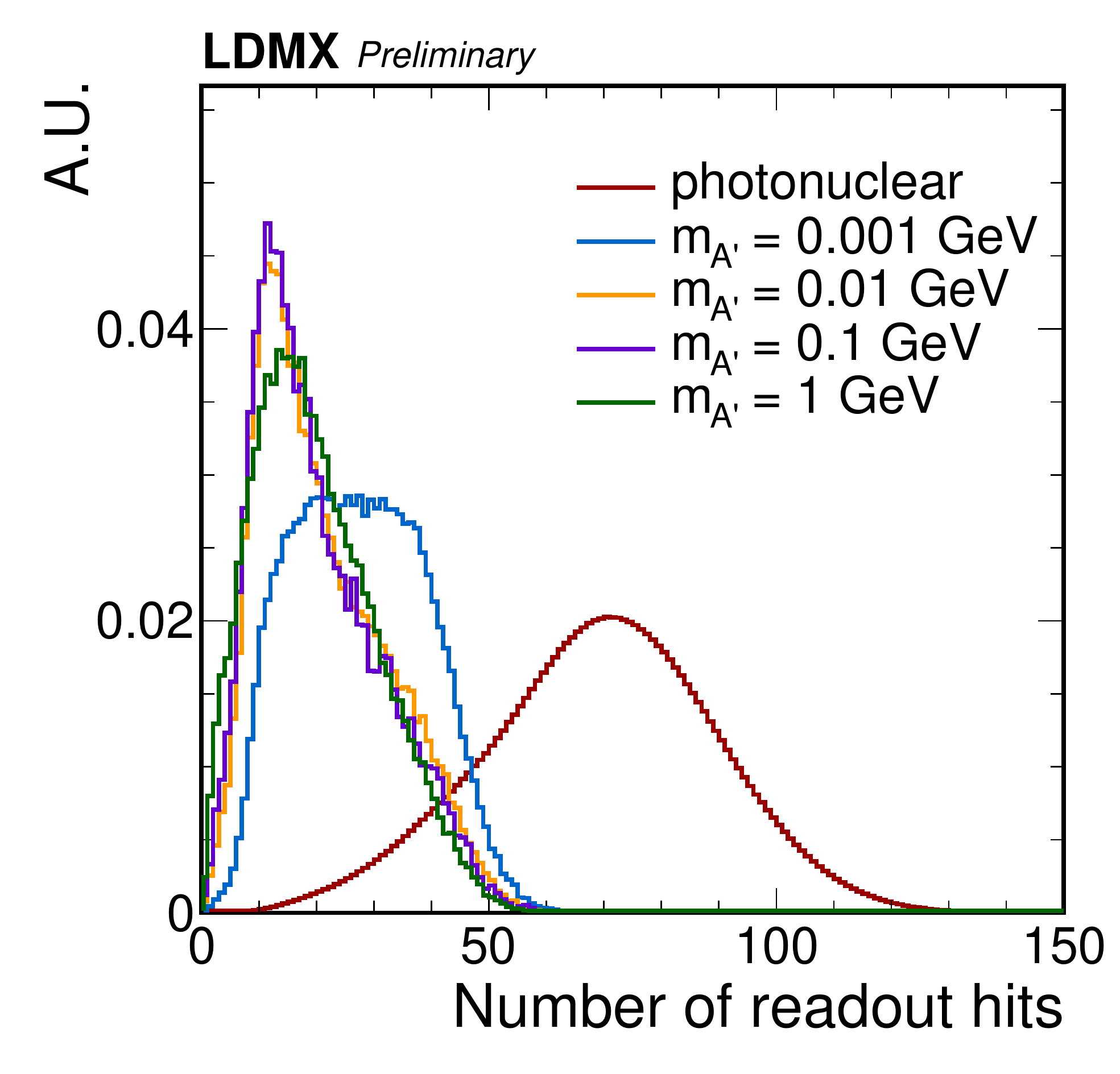}
\includegraphics[width=8.1 cm]{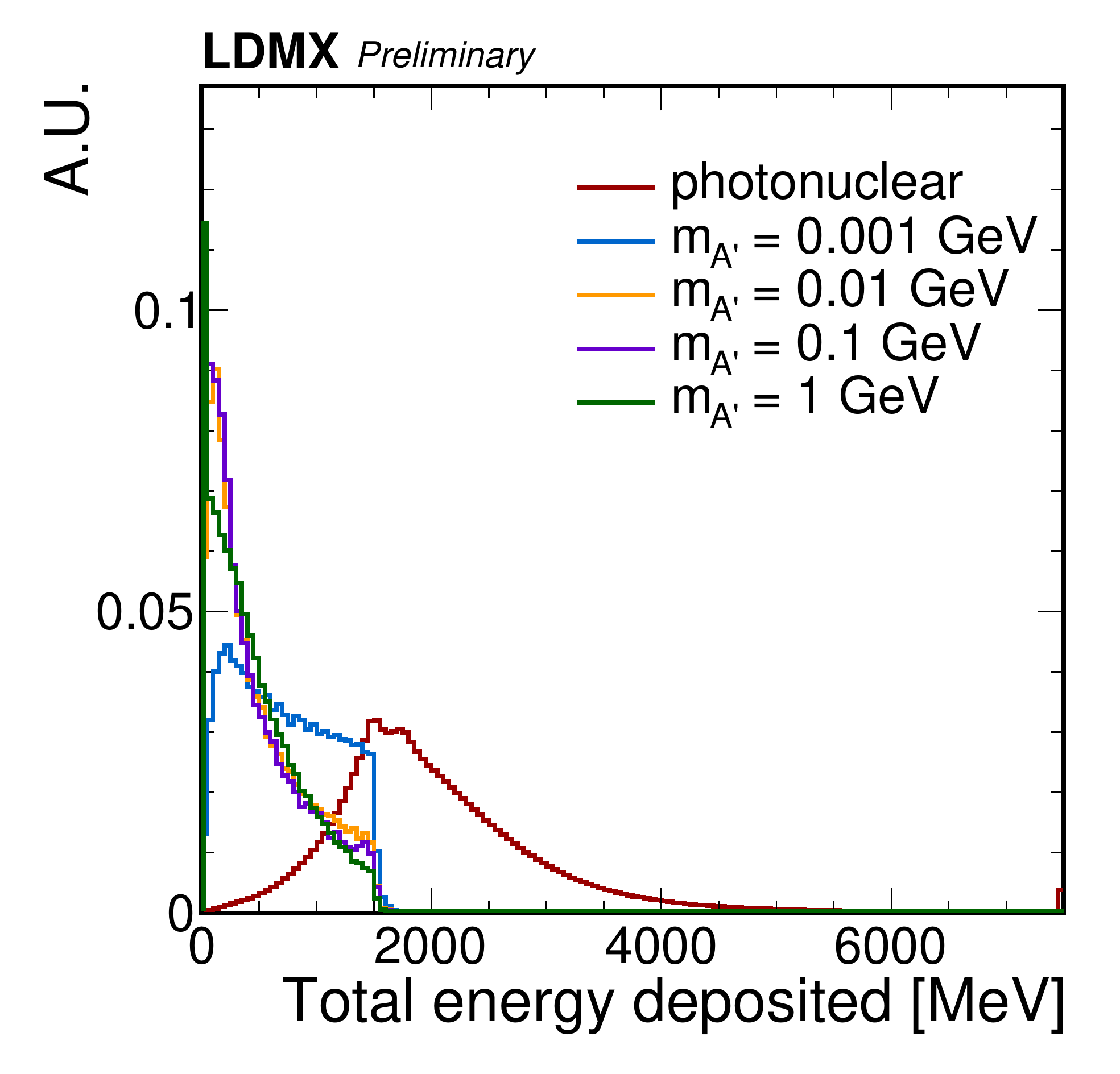} \\
\includegraphics[width=8.1 cm]{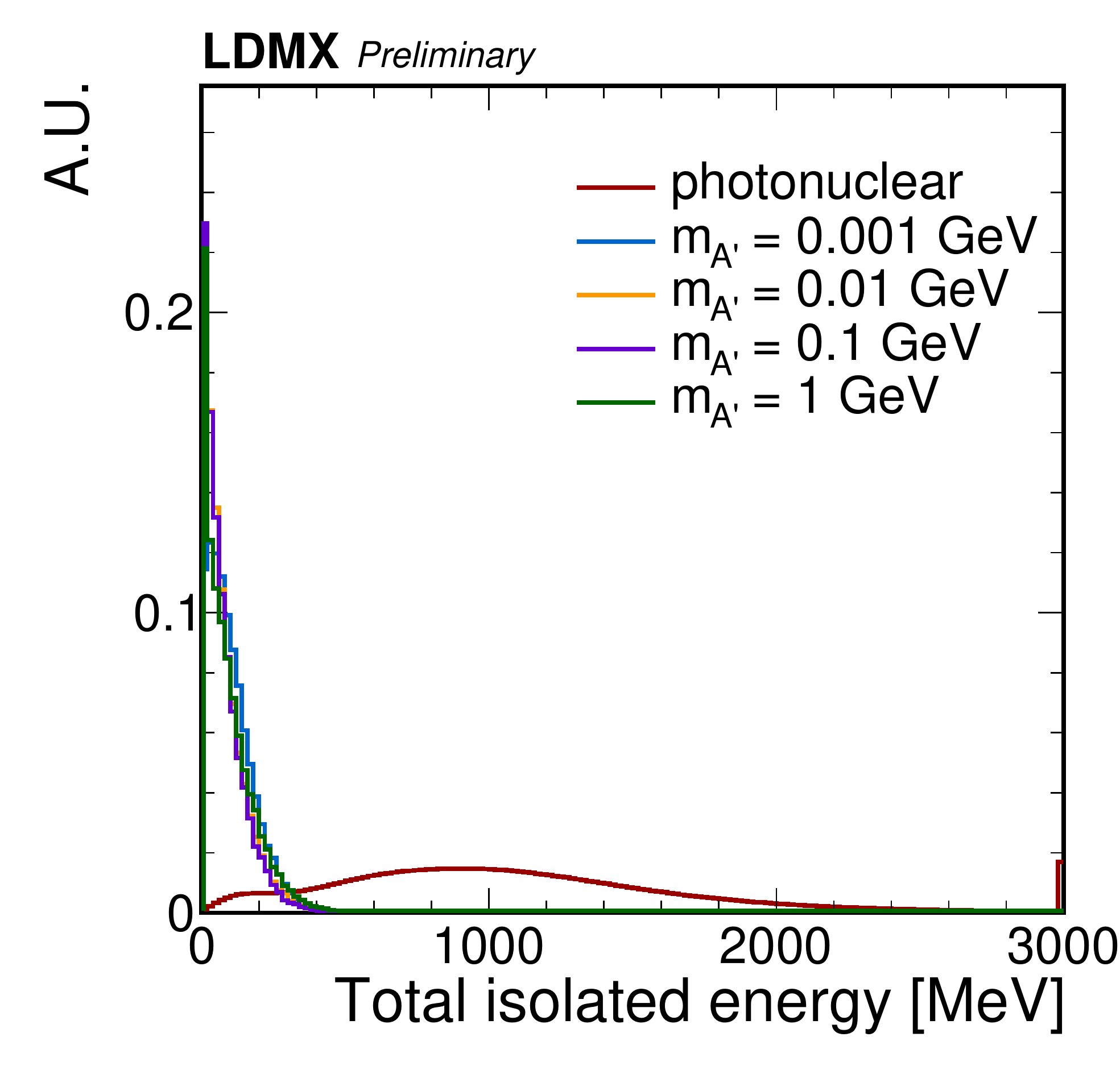}
\includegraphics[width=8.1 cm]{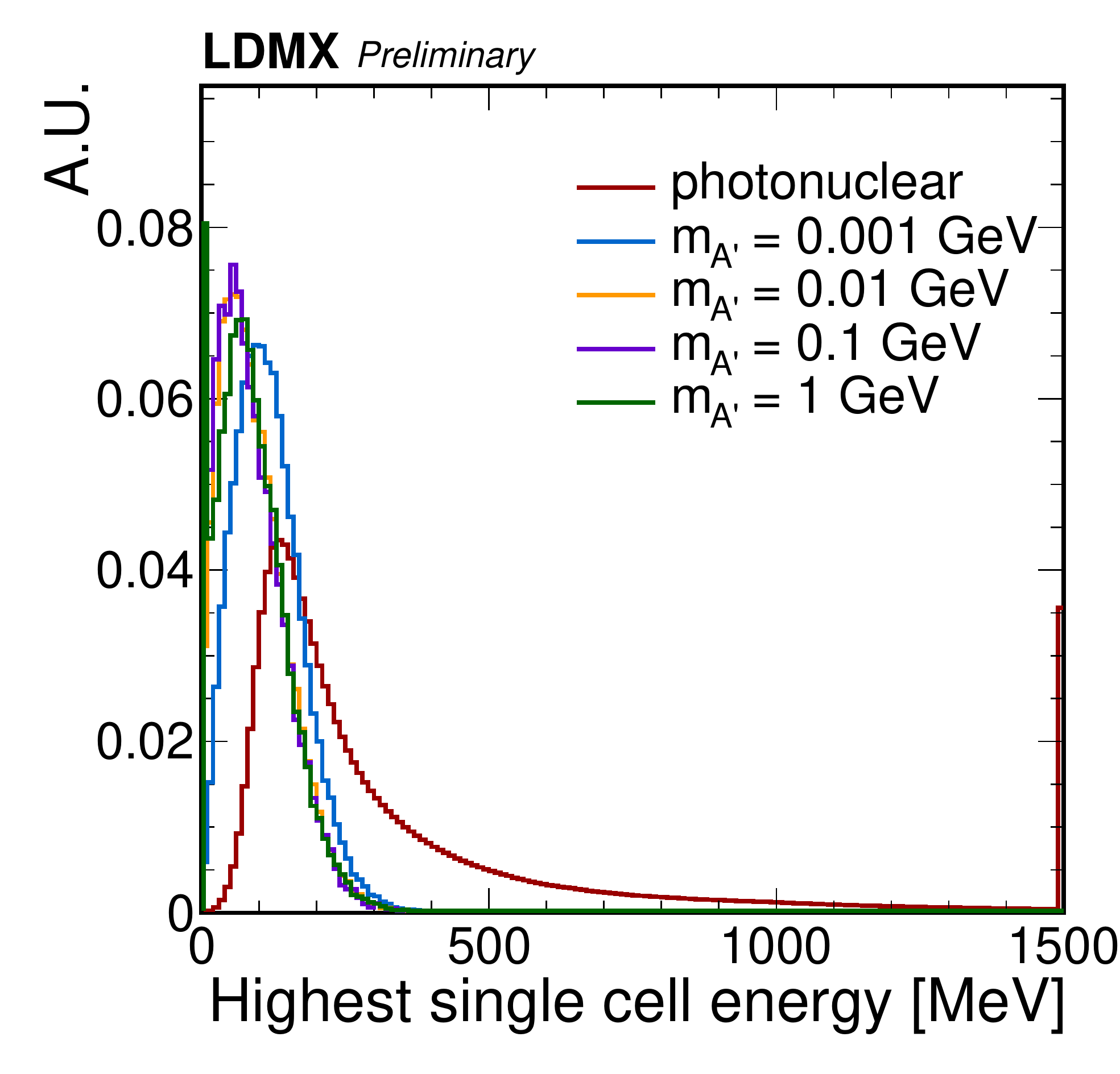}
\begin{flushleft}
\caption{\label{fig:energyvars} Distributions of quantities related to the energy deposited in the \ecal for photo-nuclear and signal processes in events passing the trigger in which the recoil electron is within the \ecal fiducial region. From top left to bottom right: number of readout hits, total energy deposited, total tight isolated energy deposited, and highest energy in a single cell. All distributions are normalized to unit area. The sharp cutoff at 1.5 GeV that is a consequence of the trigger definition as discussed in the text.}
\end{flushleft}
\end{figure}

Distributions of these variables for photo-nuclear and signal processes are shown in Fig.~\ref{fig:energyvars}. Only events passing the trigger for which the recoil electron is within the \ecal fiducial region are plotted. The sharp cutoff seen in the total energy at 1500 MeV for signal events is due to the trigger requirement. A cutoff is not seen for photo-nuclear backgrounds because the trigger uses only the first 20 layers of the \ecal. These are adequate for containment of the vast majority of showers from electron recoils but photo-nuclear interactions in background events often occur in, or extend to, the other layers of the \ecal.  

We defined a series of cuts on this set of variables that are aimed at preserving very high signal efficiency while rejecting a substantial fraction of photo-nuclear background events. Table~\ref{tab:cuteff} lists the signal and background efficiencies corresponding to these cuts, applied in sequence.

\begin{table}
\caption{\label{tab:cuteff}
Signal and background efficiencies corresponding to cuts applied on energy-related \ecal observables, following the trigger requirement. The selection requirement listed on each row is applied in sequence to those listed in previous rows. Only events in which the recoil electron is within the \ecal fiducial region are considered.
}
\centering
\begin{tabular}{l | c|c|c|c|c}
\hline
Selection & \textrm{1.0 GeV A'}& \textrm{0.1 GeV A'}& \textrm{0.01 GeV A'}& \textrm{0.001 GeV A'}& \textrm{Photo-nuclear}\\
\hline
Number of readout hits $<$ 60 & 1.00 & 1.00 & 1.00 & 0.999 & 0.311 \\
Total energy deposited $<$ 1500 MeV & 0.997 & 0.994 & 0.994 & 0.984 & 0.208 \\
Total isolated energy $<$ 400 MeV & 0.990 & 0.991 & 0.990 & 0.979 & 0.094 \\
Highest single cell energy $<$ 300 MeV & 0.986 & 0.988 & 0.986 & 0.972 & 0.090 \\
\hline
\end{tabular}
\end{table}

Further discriminating power between signal and photo-nuclear backgrounds can be obtained by exploiting information about the shower profile in the longitudinal and transverse directions within the \ecal. For these variables, the separation between signal and background distributions is not sufficiently sharp to allow us to place a single cut with high signal efficiency and high background rejection. However, the signal and background distributions are sufficiently distinct to motivate the use of these variables, in combination with the energy-related variables described above, in a multivariate analysis based on a boosted decision tree (BDT). The quantities related to the longitudinal and transverse shower profiles that are included in the BDT are described and motivated below.

\begin{enumerate}
   \item{\bf Longitudinal shower distributions:} In photo-nuclear interactions, a substantial amount of energy can be found in the back layers of the \ecal, leading to a deeper longitudinal profile for photo-nuclear events.  Together with the energy associated with the recoil electron, which is mostly in the front part of the \ecal, photo-nuclear events tend to have a more extensive longitudinal profile than signal. The variables listed below characterize the effects of this longitudinal profile.
   \begin{itemize}
     \item{\bf Deepest layer hit:} the layer number of the deepest \ecal layer that has a readout hit. The \ecal layers are assigned integers in the range [0,32], where the layer 0 is at the front, and layer 32 is at the back of the \ecal.
     \item{\bf Average layer hit:} the energy-weighted average of the layer numbers corresponding to all readout hits in an event.
     \item{\bf Standard deviation of layers hit:} the standard deviation of energy weighted layer numbers for all hits in an event.
   \end{itemize}
   
   \item{\bf Transverse shower distributions:} In photo-nuclear interactions, the distribution of energy  arising from the photon is found to also have a broader transverse profile than the recoil electron. The transverse separation of the photon and recoil electron by the magnetic field also contributes to a wider transverse profile of photo-nuclear events. The variables listed below characterize transverse energy profiles.
   \begin{itemize}
     \item{\bf Transverse RMS:} a two dimensional, energy weighted RMS centered on the shower centroid.
     \item{\bf Standard deviations of x and y positions:} the energy-weighted standard deviations for the x and y positions of all hits in an event. As before, the x and y positions of each hit are individually weighted by the energy of the hit.
   \end{itemize}
\end{enumerate}

\begin{figure}[ht]
\centering
\includegraphics[width=8.1 cm]{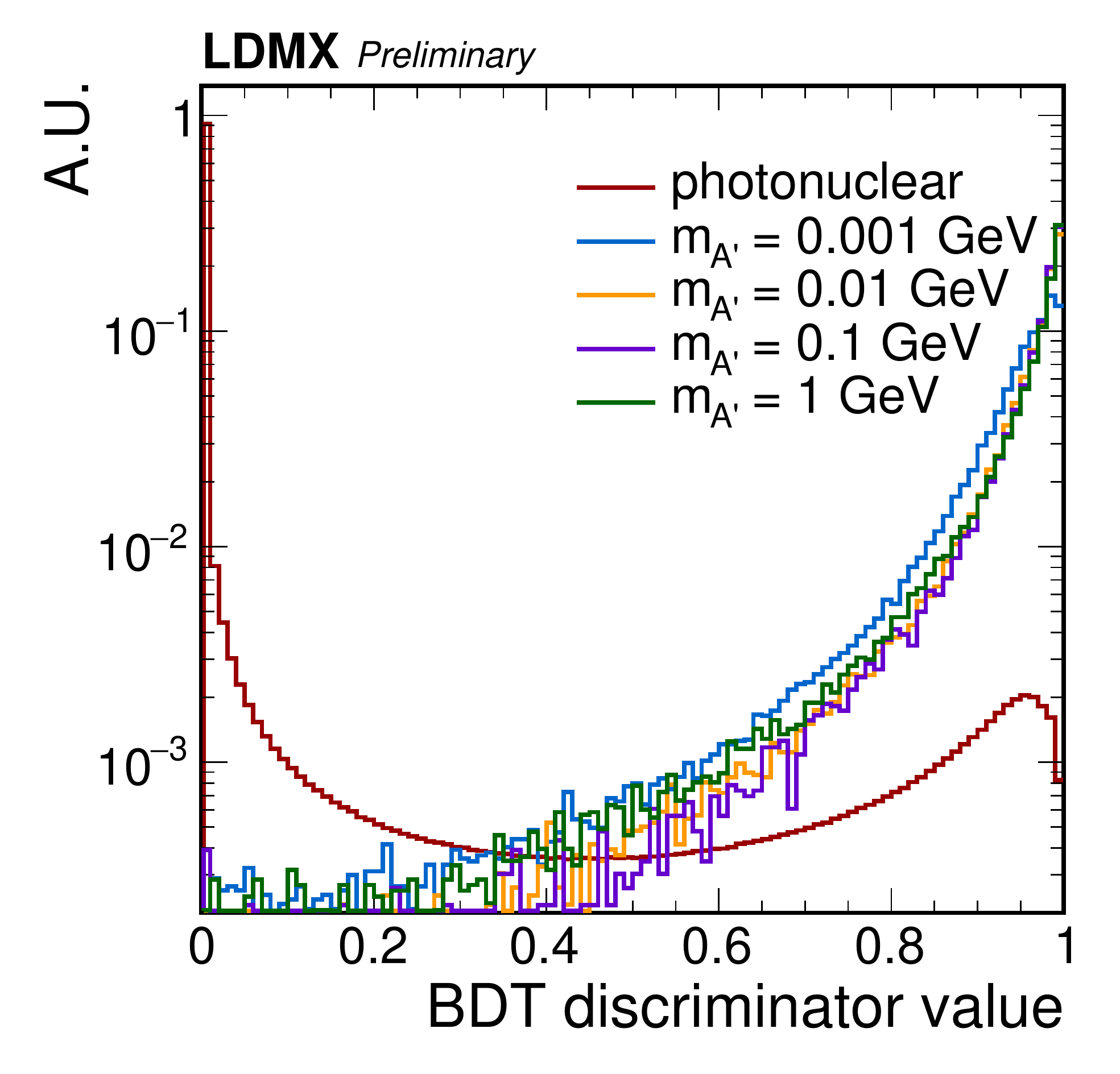}
\begin{flushleft}
\caption{\label{fig:bdtdist} Distributions of the \ecal BDT discriminator value for signal and photo-nuclear events passing the trigger in which the recoil electron is within the \ecal fiducial region. All distributions are normalized to unit area.}
\end{flushleft}
\end{figure}

\begin{figure}[tbp]
\centering
\includegraphics[width=8.1 cm]{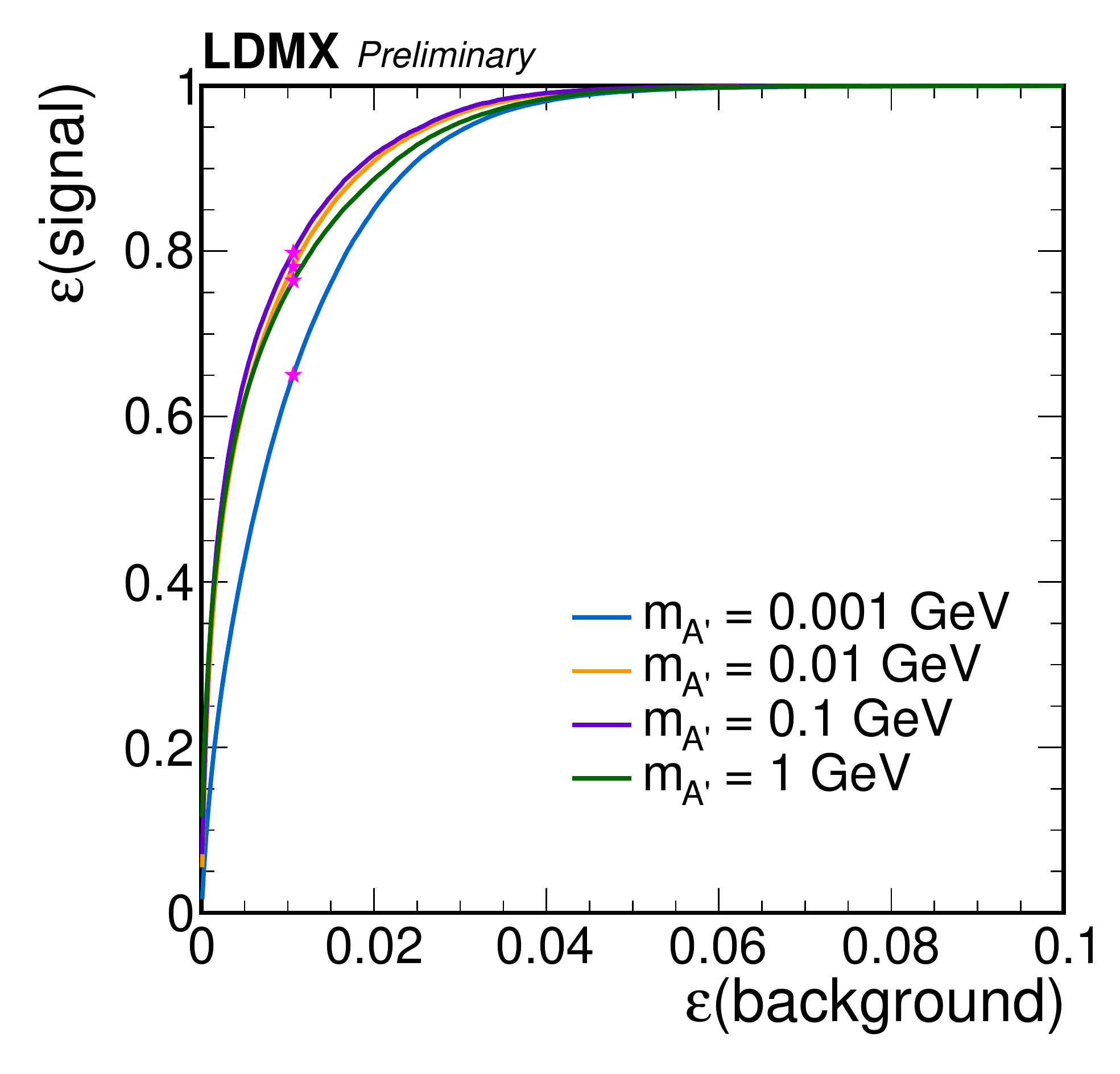}
\begin{flushleft}
\caption{\label{fig:roc} ROC curves for the \ecal BDT evaluated for signal and photo-nuclear events passing the trigger in which the recoil electron is within the \ecal fiducial region.}
\end{flushleft}
\end{figure}

The BDT is trained against photo-nuclear events while the signal sample used for training corresponds to a mixture of events simulated with four different mediator masses (0.001 GeV, 0.01 GeV, 0.1 GeV, and 1 GeV). Figure~\ref{fig:bdtdist} shows the distribution of the BDT discriminator value for signal and background events after requiring that they pass the trigger and have a recoil electron that is within the fiducial region of the \ecal. The ROC curves showing the signal efficiencies for different mediator masses as a function of the background efficiency corresponding to different BDT thresholds are plotted in Fig.~\ref{fig:roc}. A BDT threshold of 0.94  as indicated by magenta dots on the ROC curves in the figure corresponds to a rejection of $99\%$ of photo-nuclear events in the fiducial sample, and signal efficiencies ranging between $\sim 70$ and $\sim 80\%$. By applying a more stringent BDT cut, we can achieve background efficiencies at the sub-percent level, while still retaining reasonably high signal efficiencies.


As discussed in Section~\ref{sec:bkgmod}, the simulated photo-nuclear background sample is affected by an unphysical model of reflections that results in a significant population of events with backward-going hadrons that are hard to veto. The nominal \ecal design and performance studies were based on this sample, and are consequently sub-optimal. In order to gauge the true potential performance we can achieve with the \ecal, we have re-evaluated the distribution of the discriminator value and the ROC curves for the nominal BDT using a small sample of photo-nuclear events that are simulated with a modified \geant version that eliminates these unphysical reflections as described in Appendix~\ref{sec:PNappendix} (Fig.~\ref{fig:g4fix}). Once again, only events passing the trigger that have a recoil electron within the \ecal fiducial region are shown. The removal of the unphysical tail events significantly improves the performance of the BDT, since these events constituted the overwhelming majority of events with high BDT discriminator values in the nominal photo-nuclear sample, as seen in Fig.~\ref{fig:g4fix} (left). The expected background rejection corresponding to a BDT threshold of 0.94 is improved by a factor of $\sim$50 as shown in Fig.~\ref{fig:g4fix} (right). 

The improvements seen here are likely to represent lower bounds on what can be achieved, since the nominal \ecal BDT was trained using a photo-nuclear sample containing the unphysical tail events. In the future we will employ a new version of the BDT trained on a sufficiently large sample in which these unphysical effects are eliminated from the simulation, which could further improve the expected performance of the \ecal. The removal of these unphysical events from consideration may also affect the optimization of the design of the \ecal, with a  corresponding enhancement in performance.  

\begin{figure}[tbp]
\centering
\includegraphics[width=8.1 cm]{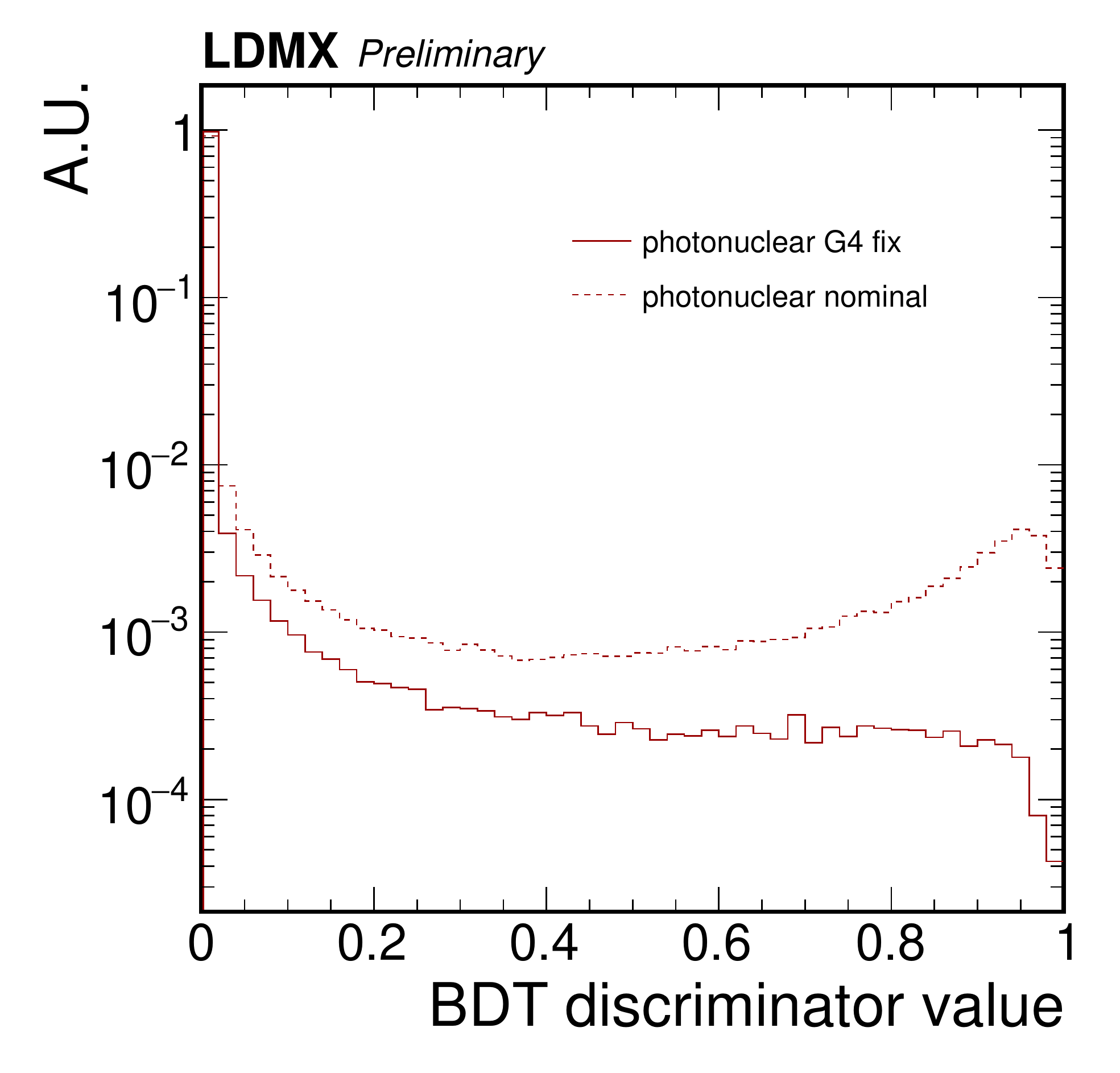}
\includegraphics[width=8.1 cm]{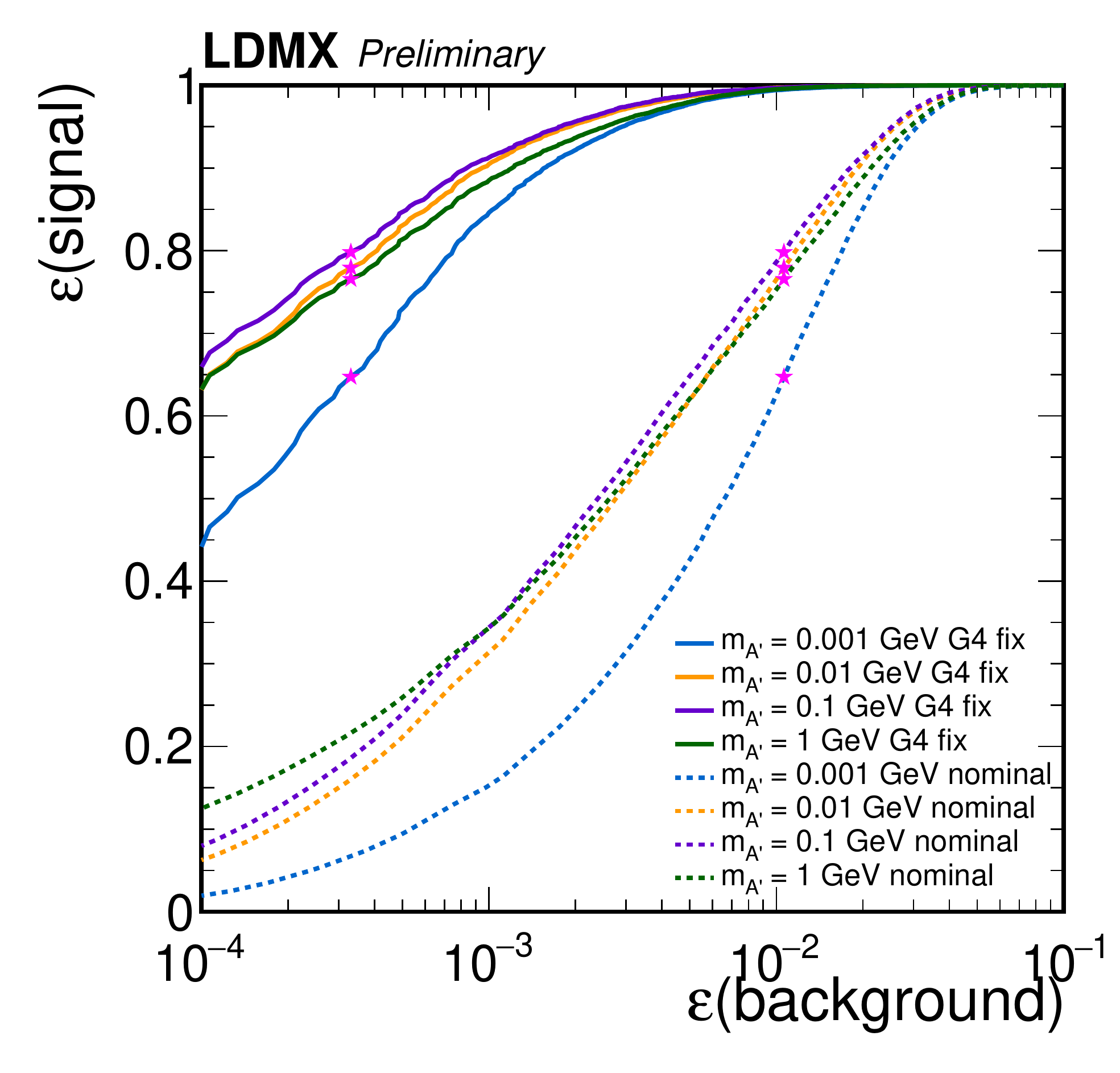}
\begin{flushleft}
\caption{\label{fig:g4fix} Left: Comparison of the nominal BDT distribution for photo-nuclear events with the distribution obtained in a small sample simulated with the modified \geant version that eliminates the unphysical population of events with backward-going hadrons. Right: ROC curves for the \ecal BDT evaluated using the nominal and modified photo-nuclear samples.}
\end{flushleft}
\end{figure}

\subparagraph{Muons and minimum ionizing particles}

\begin{figure}[ht]
\centering
\includegraphics[width=7 cm]{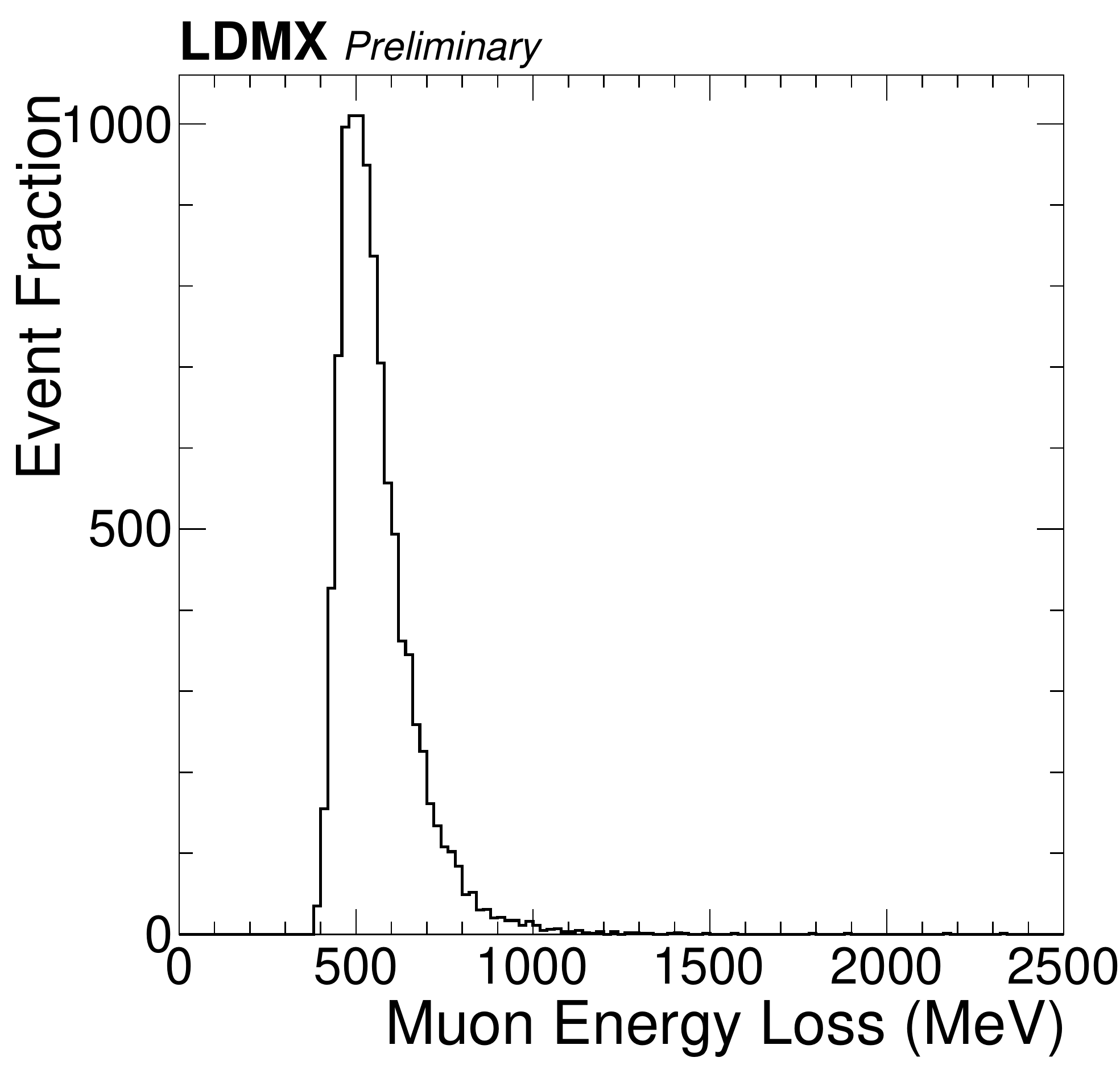}
\begin{flushleft}
\caption{\label{fig:mipecal}Energy loss in the \ecal of muons with an incident energy of 2.5~GeV.}
\end{flushleft}
\end{figure}

Another important class of signals that could be distinguished in the \ecal are those of minimum ionizing particles (MIPs).  
Long tracks from MIPs can be observed in the \ecal due to its high degree of transverse and longitudinal segmentation.  
Efficiently identifying such MIP tracks can provide a strong handle for vetoing backgrounds that would be complementary to vetoes from the \hcal system which does not have adequate segmentation to track MIPs. This is a particularly important \ecal  capability for identifying muons which will frequently, but not always penetrate into the \hcal. 
In order to understand the energy range for muons that do not reach the \hcal, we perform a simple study to understand the MIP energy loss of muons in the \ecal.  
This is shown in Fig.~\ref{fig:mipecal} where it is seen that muons typically lose $\sim$500~MeV or more in the \ecal.
Therefore, for most muons at or below $\sim$500~MeV, we would rely solely on the \ecal information for muon background rejection.

Of course, it is also possible for a muon to decay in flight in the \ecal;  $\mu^- \to e^- + \nu_e + \bar{\nu}_\mu$, so that it produces no veto signal in the \hcal.  This is a much more rare occurrence and since the muons are pair-produced, this rarer background is found to be negligible relative to the dominant backgrounds.

\subsection{HCAL Simulation and Performace \people{David, Bertrand, Nhan, Andrew} \morepeople{Torsten}}
\label{sec:hcalperf}
The HCAL performance is studied using simulations based on the \geant framework, combined with a model of the scintillator and readout 
responses to convert the energy deposited inside the active medium into photoelectrons (PEs). As described previously (see section~\ref{sec:detector.hcal}), 
the scintillator layers are longitudinally segmented in an alternate ($x,\ y$) configuration to allow ambiguity resolution. In the current simulation, each bar is 20 mm thick and 50 mm wide. 

Single neutron penetration studies to optimize the \hcal depth and transverse dimensions have been done with steel absorber thickness between 2 mm and 100 mm. The transverse dimensions were 3 $\times$ 3 meters, with smaller sizes studied using dimensional cuts. The bulk of the full background studies have been done using 50 mm plates and a total \hcal depth of 13~$\lambda_A$.

In both cases, the scintillator response was simulated by generating a Poisson-distributed number of photo-electrons based on the energy deposited in each scintillating bar, and adding the noise. An early estimate of 10 PE / MeV of energy deposited in the bar was used; future simulations will be based on the photo-electron yield measured by the Mu2e experiment and the updated bar geometry (see section ~\ref{sec:detector.hcal}). These values include quenching effects on the scintillation light yields (Birks' law), parametrized from earlier measurements of plastic scintillators~\cite{NIM80.2.239}. The noise is generated from a Poisson process with a mean of 0.004 PE / bar.  The amount of simulated noise is based on the consideration that we plan to read out both ends of a bar with SiPMs and thus the coincidence noise is relatively small.  We consider an event to be vetoed if {\it{any}} bar contains at least 3 PEs.

\begin{figure}[hbtp]
\begin{center}
    \includegraphics[width=0.48\textwidth]{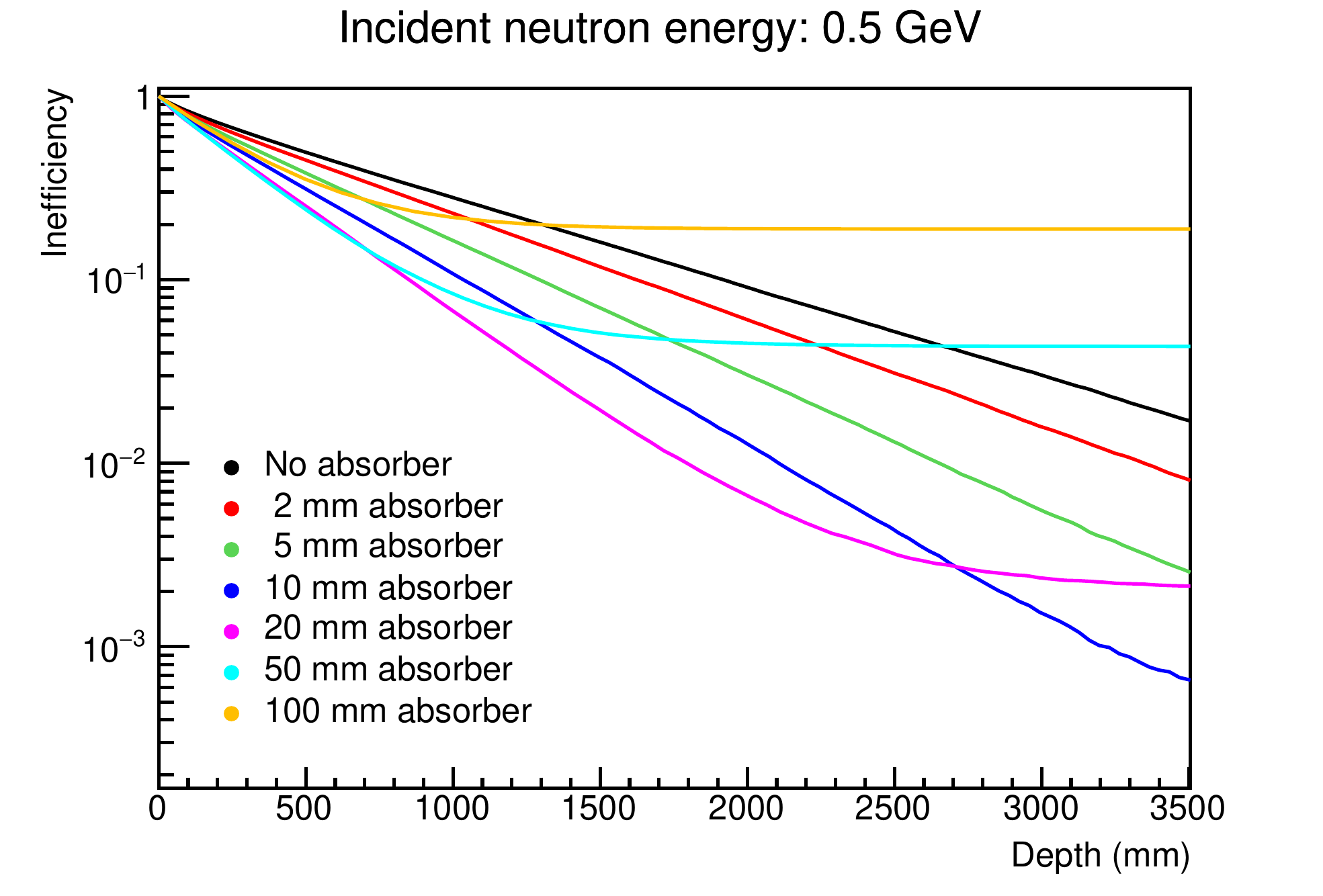}
    \includegraphics[width=0.48\textwidth]{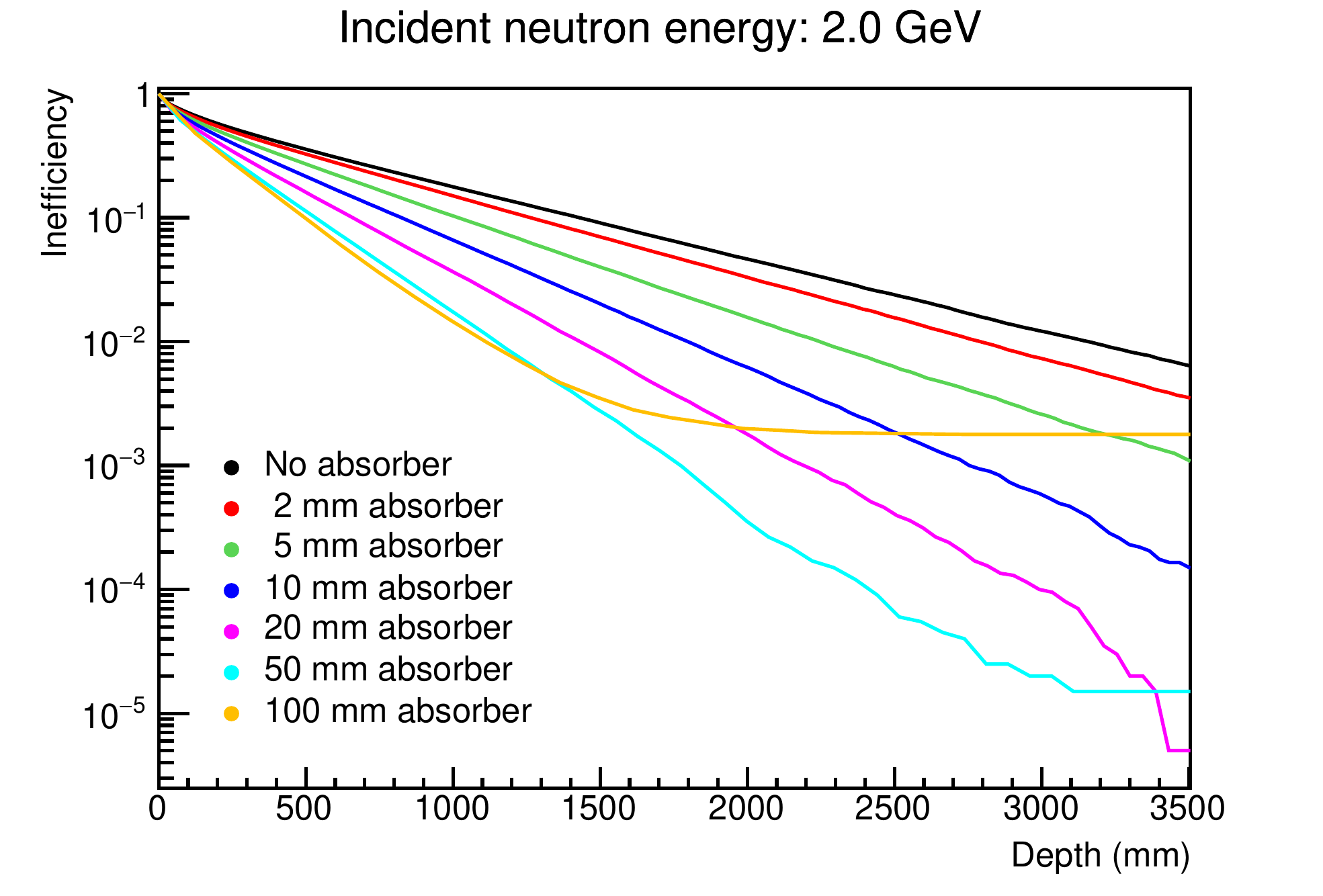}
    \includegraphics[width=0.48\textwidth]{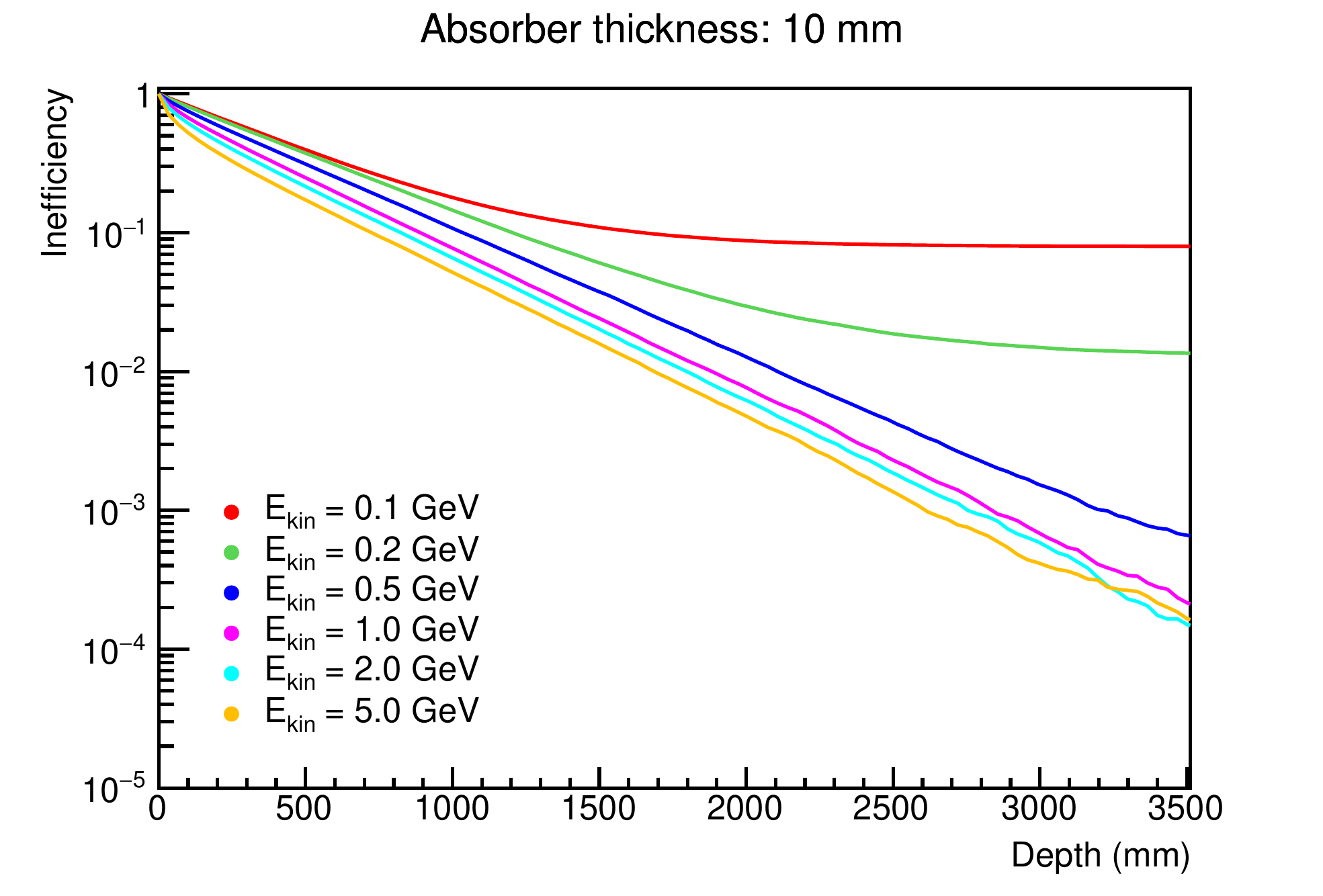}
    \includegraphics[width=0.48\textwidth]{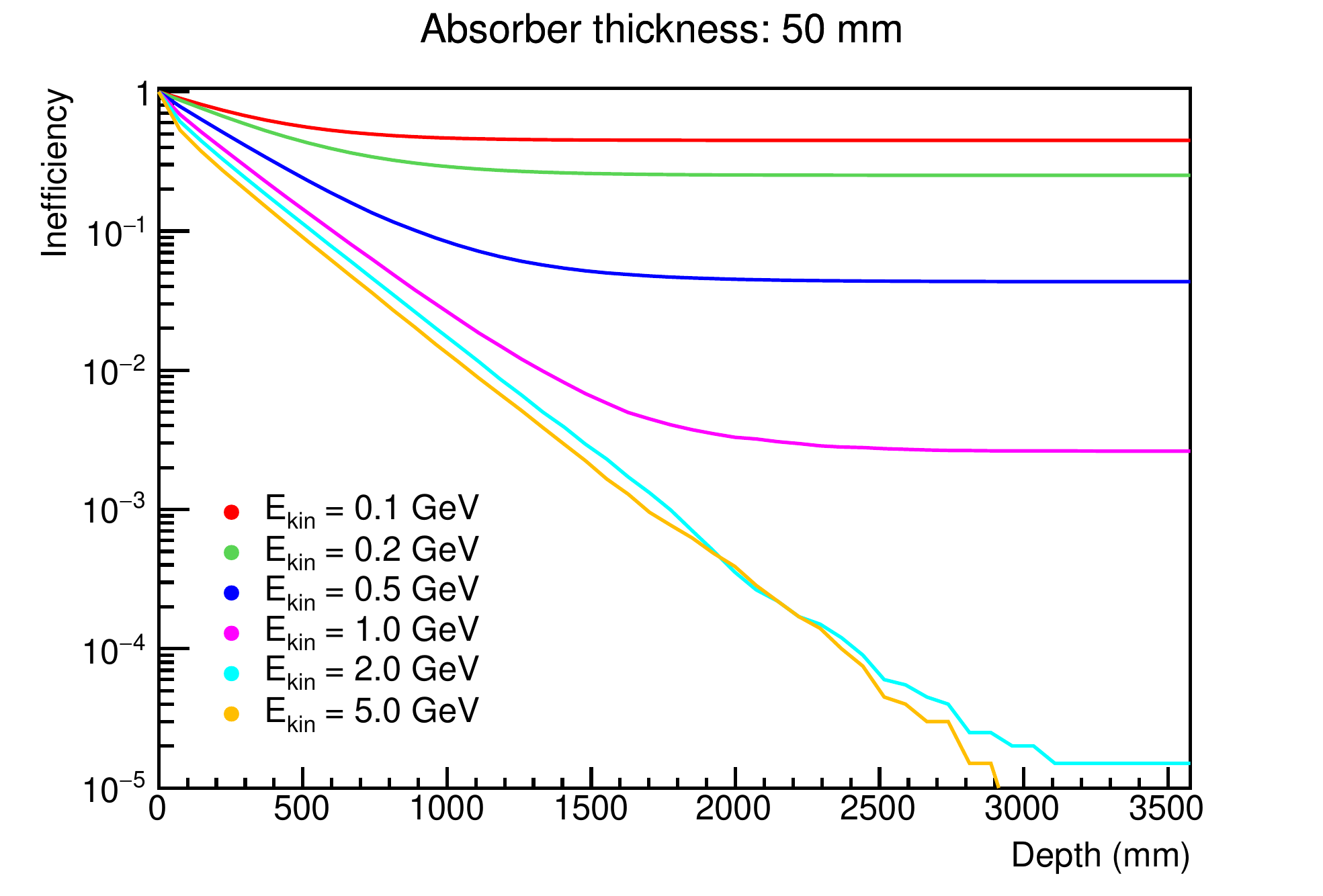}
    \caption{Top: Single neutron veto inefficiency as a function of the sampling fraction for (left) 500 MeV and (right) 2 GeV incident neutrons. 
             Bottom: Single neutron veto inefficiency as a function of the incident neutron energy for (left) 10 mm and (right) 50 mm absorber thickness.}
 \label{fig:hcalneutrons}
 \end{center}
\end{figure}

\begin{figure}[hbtp]
\begin{center}
    \includegraphics[width=0.75\textwidth]{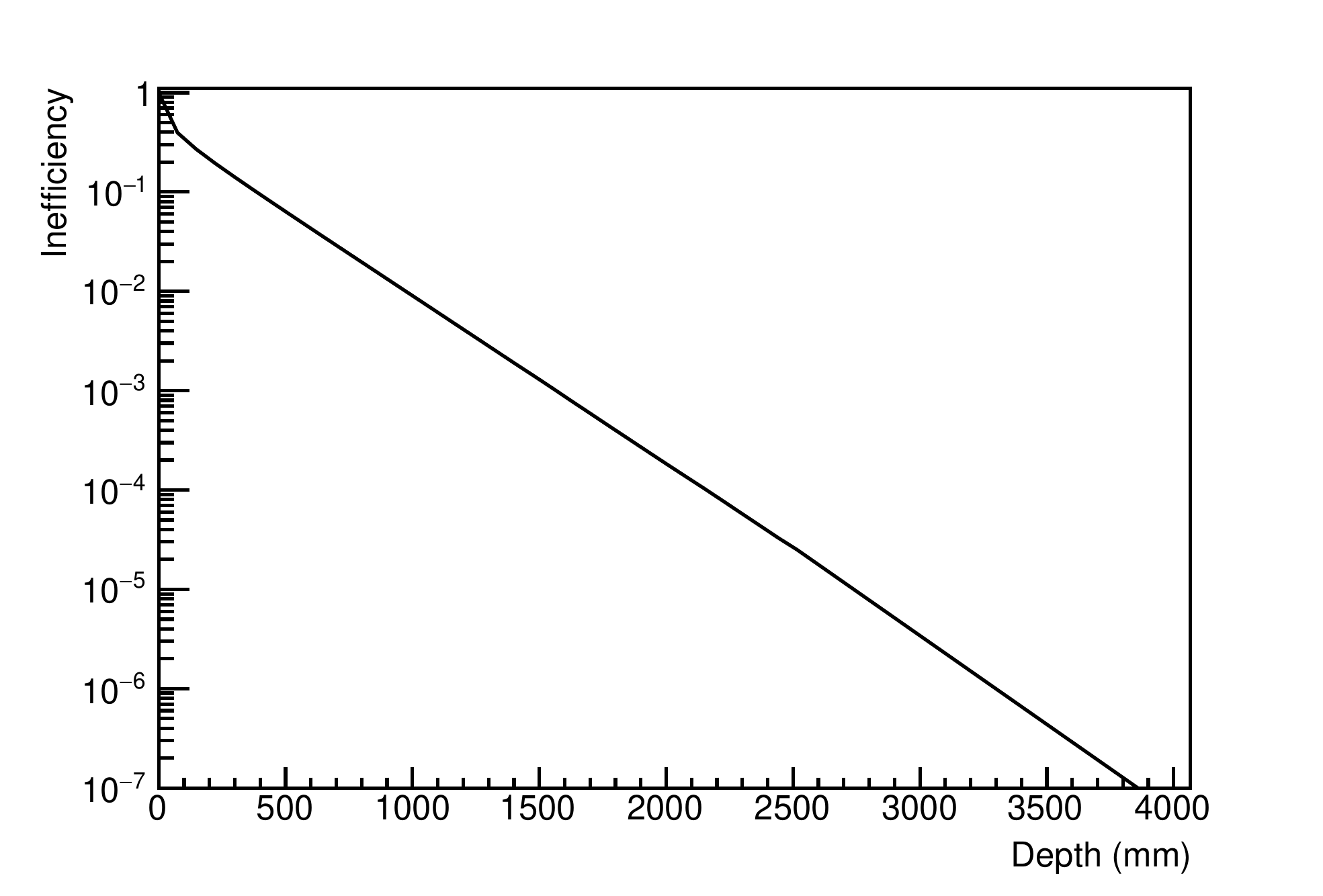}
    \caption{Single neutron inefficiency as a function of depth of the \hcal for a neutron of incident energy of $3~\GeV$.}
 \label{fig:hcalneutronsdeep}
 \end{center}
\end{figure}

In the most challenging cases, the hadronic veto system must detect a few neutrons having energies ranging from 100 MeV to a several GeV. To guide 
the design of the detector, we perform studies to characterize the detection efficiency of a single, normally incident neutron. The veto inefficiency 
as a function of the \hcal depth is shown in Fig.~\ref{fig:hcalneutrons} for different incident neutron energies and absorber thicknesses. As 
expected, thinner absorbers are less ineffective at vetoing energetic neutrons, while the veto capabilities of thicker absorbers decrease
for low-energy neutrons as they are fully absorbed in the steel plates. We also study a very deep sampling hadronic calorimeter with 50~mm 
thick absorber to understand the rate of single energetic (few GeV) neutrons punching through the \hcal as a function of depth. The results are 
displayed in Fig.~\ref{fig:hcalneutronsdeep}. After a few layers, the inefficiency decreases approximately by an order of magnitude per 0.6~m of \hcal length. A depth of $\sim 3$~m should be sufficient to reach the required single neutron inefficiency (see Sec.~\ref{sec:bkgmod}).


\begin{figure}[hbtp]
\begin{center}
    \includegraphics[width=0.75\textwidth]{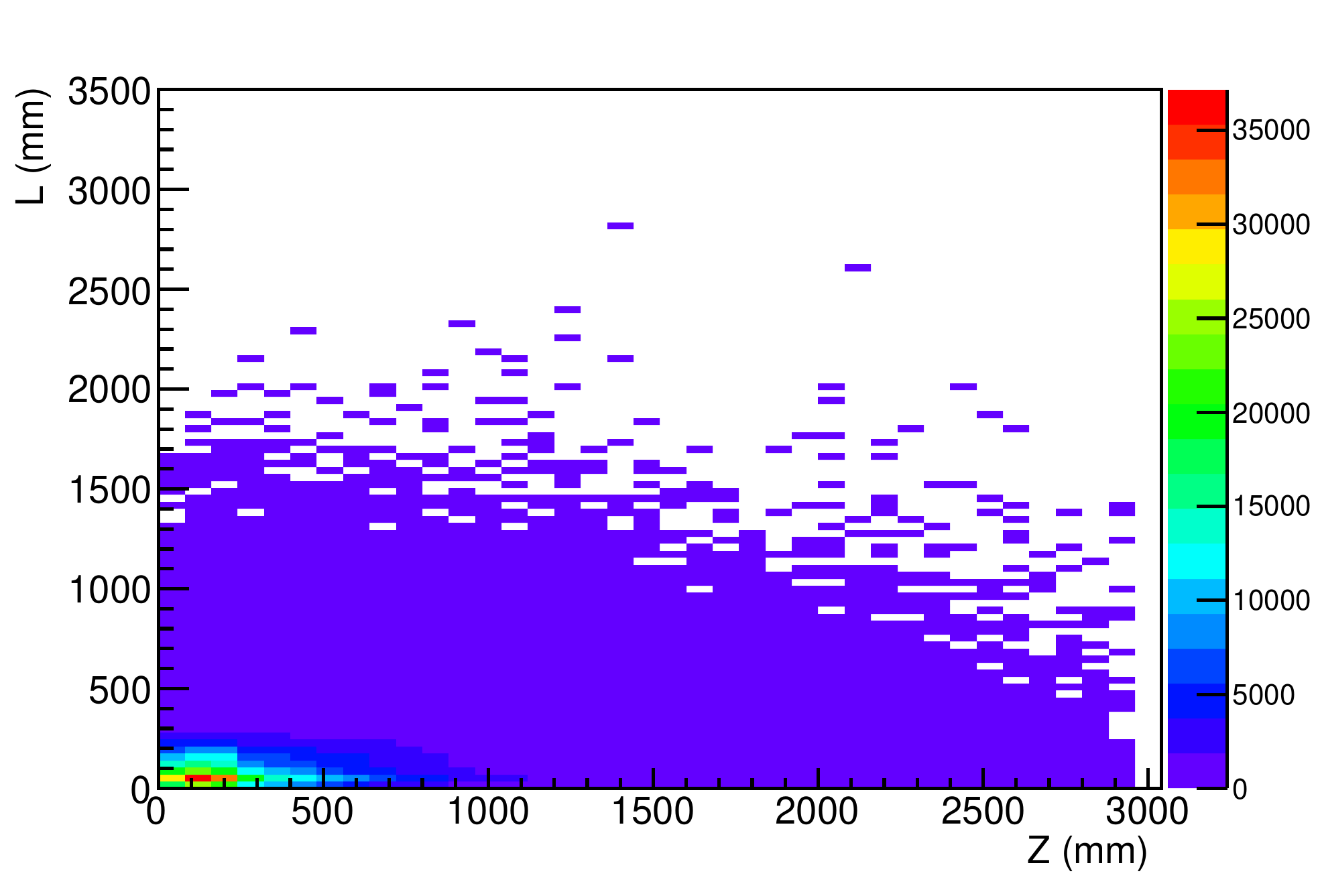}
    \caption{The transverse ($L$) and longitudinal position ($Z$) of the veto hit minimizing the transverse \hcal dimension. A \hcal size of 3m~$\times$~3m$\times$~3m is selected as a starting point to study the required dimensions.}
 \label{fig:hcaldimopti}
 \end{center}
\end{figure}

The Main \hcal transverse dimensions are studied using a  sample of photo-nuclear events corresponding to $10^{14}$ EOT (see Sec.~\ref{sec:bkggen}). We select events containing a single reconstructed track, a value 
of the BDT output compatible with the signal hypothesis, and no veto hit found in the Side \hcal. These PN events typically contain a 
few energetic neutrons emitted forward and must be vetoed by the Main \hcal. For each event, we determine the position of the veto hit minimizing the transverse dimension. The distribution of the corresponding transverse and longitudinal positions 
is displayed in Fig.~\ref{fig:hcaldimopti}. 
In Fig.~\ref{fig:hcaldimopti}, the longitudinal size $L$ is defined as $L = 2 \times {\rm min} ( {\rm max} (|x|,|y|) )$ where $x$ and $y$ are the coordinates of each hit in the \hcal.
While these preliminary results point towards a transverse dimension of the order of 2.5--3~m, additional studies are required to improve our understanding
of the rate and kinematics of photo-nuclear interaction at moderate to high angles.
For more discussion, see Sec.~\ref{sec:bkgmod}.
 
\begin{figure}[hbtp]
\begin{center}
    \includegraphics[width=0.48\textwidth]{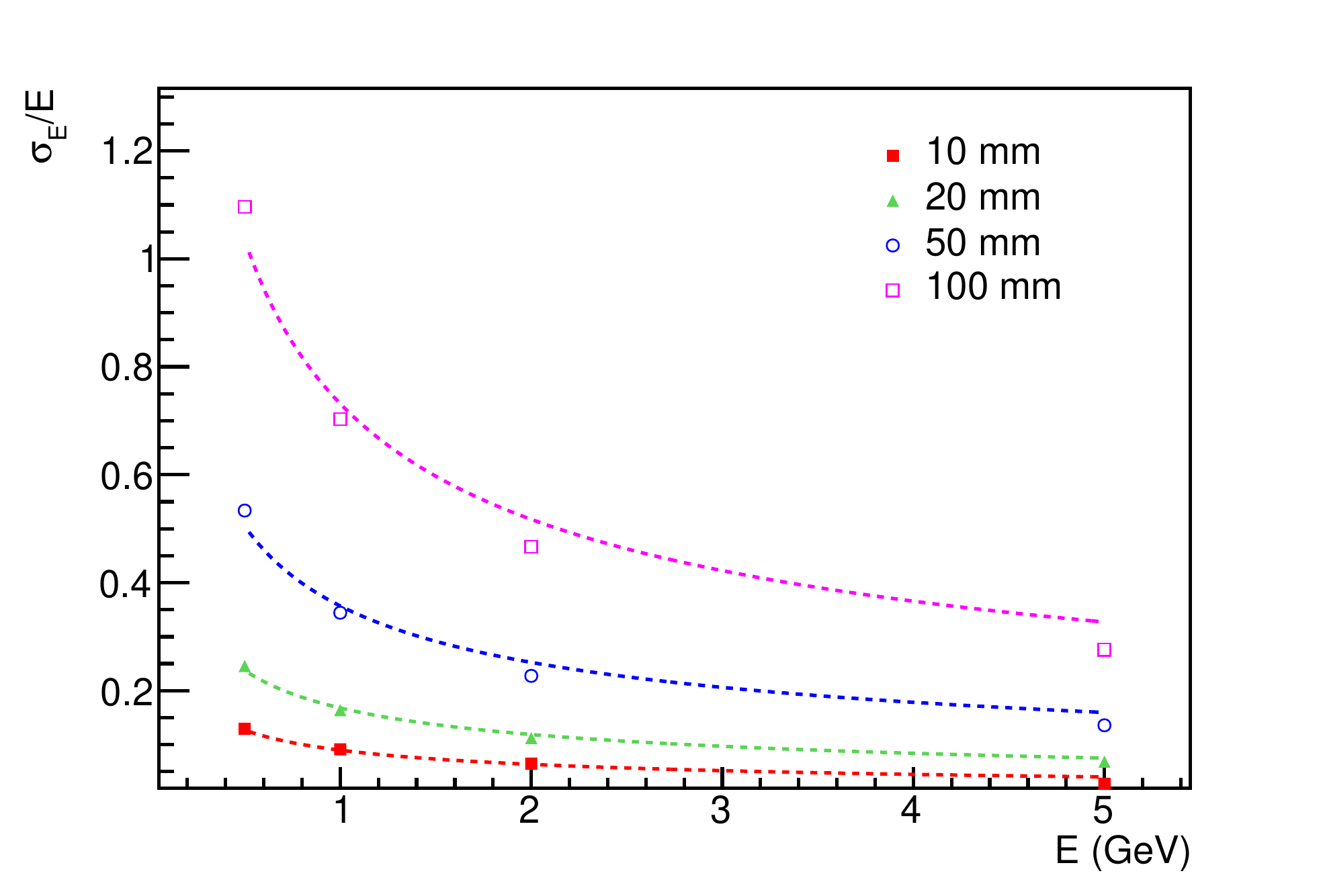}
    \includegraphics[width=0.48\textwidth]{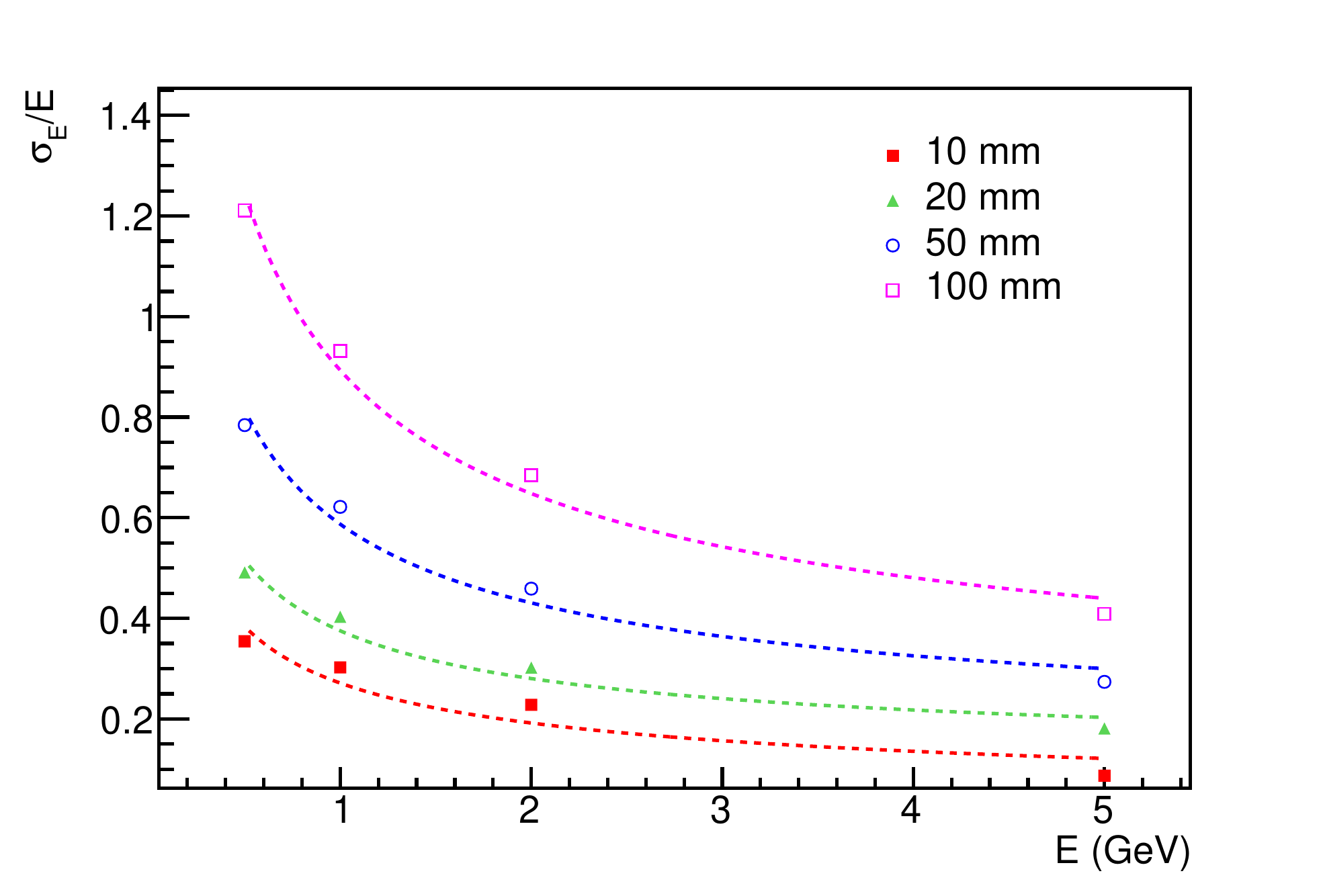}
    \includegraphics[width=0.48\textwidth]{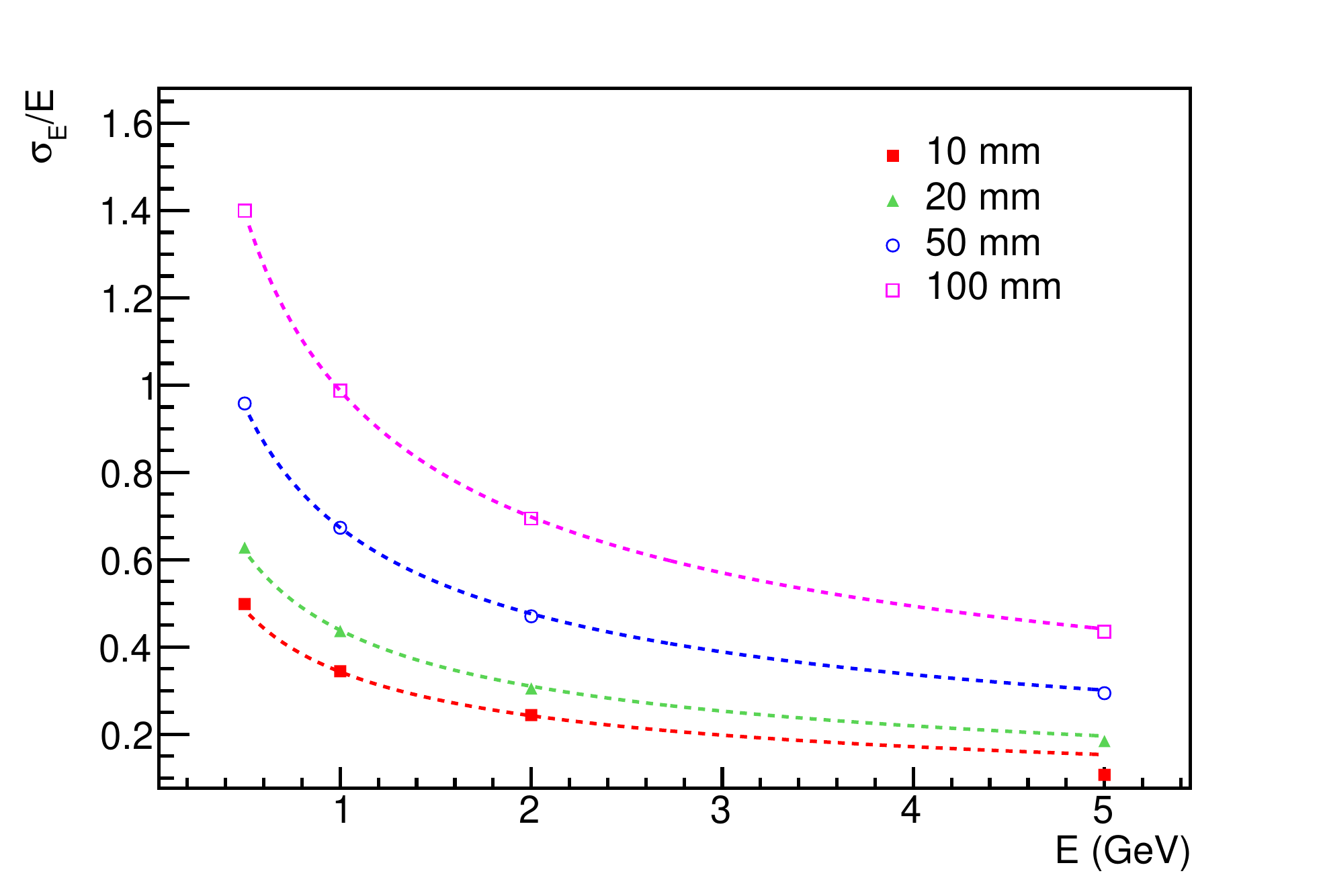}
    \caption{The energy resolution as a function of the incident particle energy for (top left) electrons,  (top right) pions and (bottom) 
             neutrons. The data are fit with a function of the form $a \oplus b/\sqrt{E}$.}
 \label{fig:hcalresol}
 \end{center}
\end{figure}

The energy resolution is estimated using the RMS of the sum of all PEs inside the \hcal (normalized to the mean value). The resolution as a function of the incident particle energy is displayed in Fig.~\ref{fig:hcalresol} for electrons, pions, and neutrons. The data are well 
approximated by a function of the form $a \oplus b/\sqrt{E}$. The energy resolution for $1~\GeV$ neutrons is around 45\% for 20 mm thick absorber, 
increasing to $\sim 70\%$ for a thickness of 50~mm.

\clearpage

\subsection{Trigger System Performance \people{Mans} \morepeople{Bertrand, Nhan}}
\label{ssec:performance-trigger}

As described above, the primary physics trigger is based on the total
energy observed in the calorimeter, combined with a requirement to
identify the number of incoming electrons in the trigger
scintillators.  The performance of this trigger for signal is
discussed in detail in Section~\ref{ssec:signaleff-trigger}.

Besides the primary physics trigger, the LDMX trigger system will also
allow the selection of events for calibration, alignment, and
background studies.  Each event will be marked with the set of
triggers that fired.  The trigger will include input from the
\hcal to allow selection of events with hadrons or
muons.  Additional samples of events will also be acquired with a
zero-bias requirement, using only a requirement on energy deposition
in the trigger scintillators, and with a veto on energy in the trigger
scintillator.

An initial draft trigger menu is shown in
Table~\ref{tab:trigger_menu}.  The table is built using a target
trigger rate of 5~kHz, which provides a safety factor of ten compared
with the limits expected from the tracker DAQ, which provides the
strongest limit on trigger rate.  As a result, the trigger and DAQ
systems have sufficient capacity to add additional monitoring
triggers, to increase the bandwidth assigned to these triggers, or to
define new physics triggers, e.g. triggers appropriate for nuclear
physics studies.  The signal trigger bandwidth can also be increased
if required, for example for a larger average number of incoming beam
electrons per bucket.

\begin{table}
  \caption{Draft trigger menu for LDMX, showing the primary
    contributions to the trigger budget for a $\mu_e=1.0$ 46~MHz beam
    rate, with no corrections for any
    overlaps.}\label{tab:trigger_menu} \begin{tabular}{|l|r|r|} \hline
    Trigger & Prescale factor & Rate (Hz) \\ \hline \em{Physics
    Trigger} & 1 & {\em 4000} \\ \hline
    \em{Background-Measurement Triggers} &  & \em{500} \\
    \hspace{0.2in} ECAL Missing-Energy $>$ 1~GeV & 5000 & 100 \\
    \hspace{0.2in} HCAL hit $>$ 2 MIP & 1000 & 100 \\
    \hspace{0.2in} HCAL hit $>$ 20 MIP & 1 & 100 \\
    \hspace{0.2in} HCAL MIP track &  & 200 \\
    
    \em{Detector-Monitoring Triggers} &  & \em{500} \\
    \hspace{0.2in} Zero-bias (trigger scintillator ignored) & $4.6 \times 10^{5}$ & 100 \\    
    \hspace{0.2in} Beam-arrival (trigger scintillator) & $1.5 \times 10^{5}$ & 300 \\    
    \hspace{0.2in} Empty-detector (trigger scintillator veto) & & 100 \\
    \hline
    {\bf Total Trigger Budget} & & 5000 \\ \hline
  \end{tabular}
  \\
\end{table}

\clearpage
\section{Background Rejection Studies}
\label{sec:bkgrej}
\label{section:bkgrejection}

\subsection{Rejection of Beam Backgrounds \people{Tim} \morepeople{Robert,David}}
An incoming charged particle within the signal recoil acceptance that
is wrongly reconstructed as a good incoming beam electron in the tagging tracker would be an irreducible background. However, incoming 4~GeV beam electrons have a precisely-defined trajectory through the tagging tracker which - together with the design of the tracker - makes the likelihood of such reconstruction errors vanishingly small.

In order for an incoming low momentum particle to fake a beam-energy electron in the tagging tracker, a number of conditions must simultaneously be met. First, the incoming particle must reach the first layer of the tagging tracker or it will not intersect any other material before hitting the magnet. Because the first layer of the tagger is well inside the field region of the magnet, most low momentum particles are swept away, and only particles within a small region of energy-angle phase space will reach the first tagging layer. Second, that particle must undergo a hard scatter in the first layer, of order 100 mrad, but within a window approximately 1 mrad wide both vertically and horizontally, that puts it on the correct trajectory to make a hit in the correct location in the next layer.  Finally, the particle must repeat a similarly unlikely scatter 5 more times in each subsequent layer so that it makes on-trajectory hits in all the correct locations in all of the layers of the tagging tracker before reaching the target and passing into the recoil tracker.

\begin{figure}[!hb]
    \centering
    \includegraphics[width=10cm]{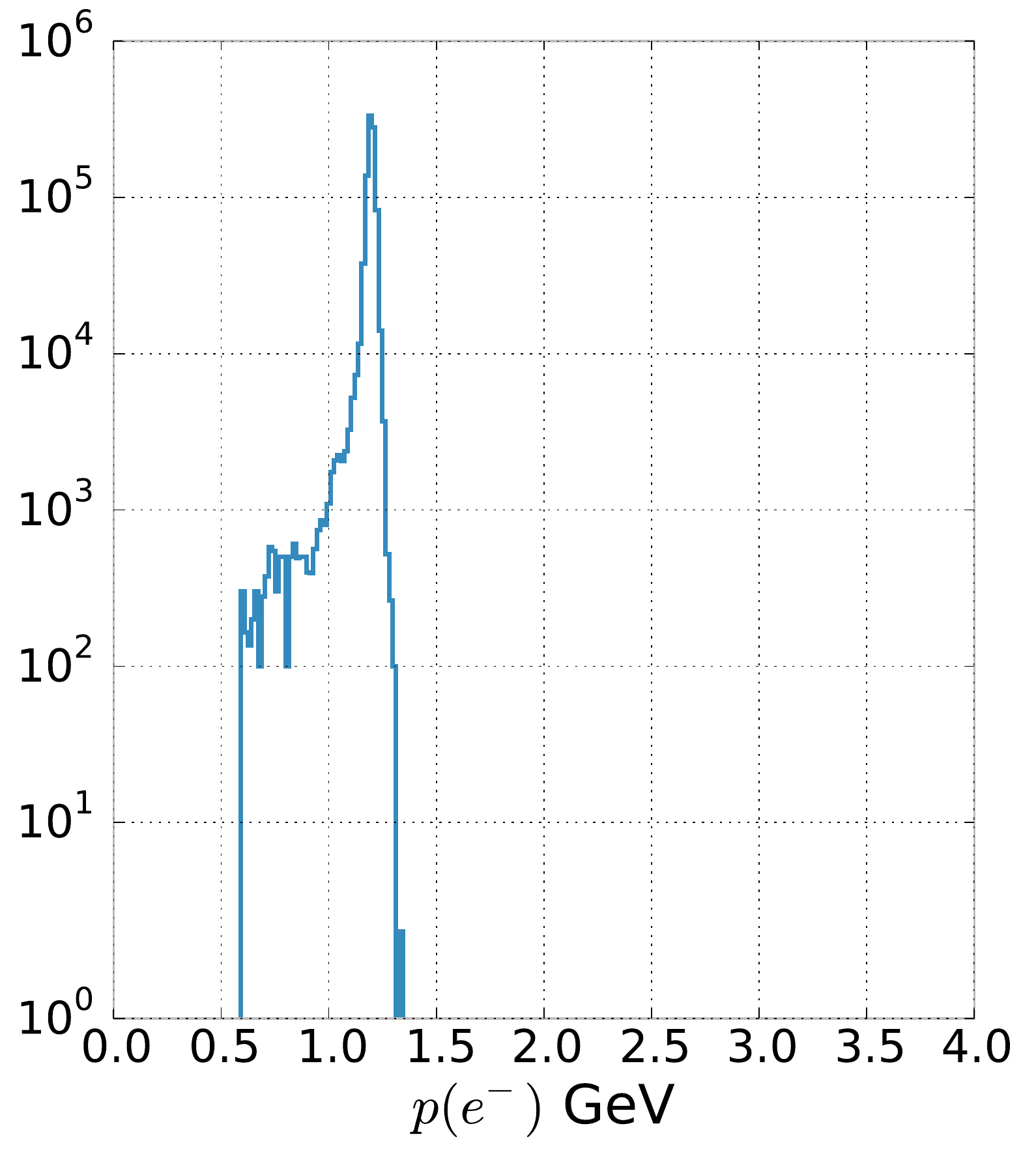}
    \caption{\small{Reconstructed total momentum for a sample of 1.2 GeV beam
                    electrons in the tagging tracker. Of $10^6$ incident particles, none is reconstructed with momentum greater than 1.4 GeV.}}
    \label{fig:tracking_1pt2gev}
\end{figure}

Taking into consideration the design of the tagging tracker, these 
requirements place a very heavy penalty on any off-energy component in the
beam.  First, incoming particles with momentum less than approximately 500 MeV 
will not hit the first layer of the tagger unless they are significantly 
off-trajectory as well.  Furthermore, even at 500 MeV, scatters of more 
than 10$^\circ$ are needed in order for the incoming particle to appear to be on
the correct trajectory. It is clear then that the most challenging scenario is
large contamination with incoming charged particles at the top of the momentum 
range for signal recoils, nominally 1.2 GeV.  Such particles have the highest 
likelihood of reaching the first layer of the tagging tracker without being 
bent away by the magnetic field and require much smaller scatters and/or track 
reconstruction errors to result in fake tags. In order for a 1.2 GeV particle to
make a trajectory through the tracker consistent with a 4 GeV track, six 
successive scatters of approximately 50 milliradians must occur, each suppressing the rate by several orders of magnitude. It is therefore expected that even relatively impure beam will not present a significant background to the experiment.

As a preliminary test of the robustness of the tagging tracker against mis-tagging off-energy incoming electrons, a sample of simulated 1.2~GeV electrons on a trajectory that allows them to pass through all seven layers of the tagging tracker was used to estimate how often they could be mis-reconstructed as full-energy beam electrons. The results, including noise and pattern recognition effects, are shown in Figure~\ref{fig:tracking_1pt2gev} and confirm that such particles cannot be mistaken for 4~GeV electrons at a level of less than one part in 10$^6$, limited here by simulation statistics.

\subsection{Rejection of Non-Interacting Electrons \people{Tim,Mans} \morepeople{Joe, Ruth}}

The LDMX target is only a small fraction of a radiation length, so the leading order process for an incoming 4~GeV electron is to pass through the target and both trackers without any significant energy loss and shower electromagnetically in the \ecal. Such events can constitute a background if these electrons are measured by the recoil tracker and \ecal as having energy consistent with signal, less than 1.2~GeV. Therefore, the combination of the recoil tracker and \ecal must be able to reject 4~GeV electrons as recoils within the energy range for signal at the level of $4\times10^{14}$ and $10^{16}$ for Phase I and Phase II, respectively.

\begin{figure}[htbp]
    \centering
    \includegraphics[width=\textwidth]{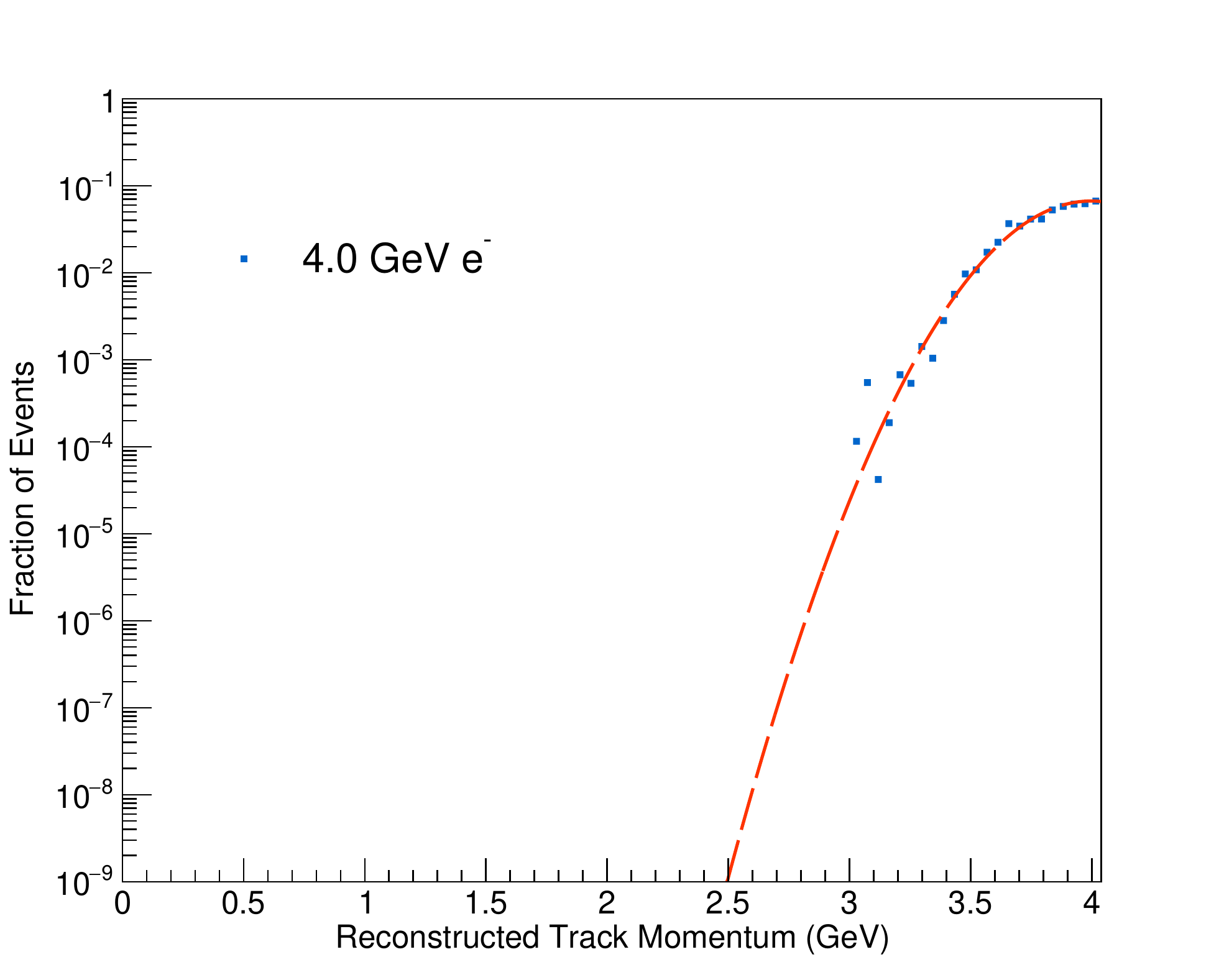}
    \caption{The reconstructed recoil track total momentum for a sample of 4~GeV
             beam electrons that experience very little energy loss through 
             the target and trackers. We find that the tracker is able to reject 
             non-interacting electrons at level of at least $10^{-5}$, limited only by Monte Carlo statistics for this study.
             } 
    \label{fig:recoil_p}
\end{figure}

As will be shown in Section~\ref{section:EM_showering_background}, the resolution of the \ecal alone is sufficient to suppress backgrounds from 4~GeV electromagnetic deposits so that they will not be reconstructed as having less than 1.2~GeV to the level of less than 1 event in $10^{15}$. The recoil tracker, although optimized for low-energy recoils, adds similar, if not better, discriminating power for such events. As can be seen from Figure~\ref{fig:recoil_p}, the resolution of the recoil tracker for 4~GeV electrons that pass through the target and trackers without significant energy loss provides at least five more orders of magnitude of rejection of this background, where much larger statistics would be required to find the point where fake tracks due to noise and pattern recognition errors could contribute.  In combination with an energy measurement from the \ecal this background will be insignificant.

\subsection{Rejection of Backgrounds with a Hard Bremsstrahlung Photon}

A class of background processes which presents varying challenges on the 
detector are the cases when the incoming electron undergoes a hard 
bremsstrahlung in the target and loses a majority of its energy. Then, based 
on how the photon interacts, a signal-like signature can be mimicked.  For 
example, if the electron has less than 1.2~\GeV~of energy and the photon 
leaves no detectable signature in the detector, this would be a signal-like 
signature.  

The basic detector signature for the nominal dark matter candidate is an 
incident 4~\GeV~electron, measured by the tagging tracker, which interacts in 
the target and leaves a less than 1.2~\GeV~electron track in the recoil 
tracker matched to a shower in the \ecal and no other detector signals.
This would pass the trigger requirement discussed in 
Sec.~\ref{ssec:performance-trigger}. In the following subsections, we study 
the cases where the photon from hard bremsstrahlung could mimic this signal-like
signature. Specifically, we explore the following scenarios in detail:
\begin{itemize}
    \item The photon undergoes a normal EM shower in the \ecal but sampling 
          calorimeter fluctuations result in a significant mismeasurement of 
          the photon
    \item The photon converts, through a number of different processes, 
          into a $\mu^+\mu^-$ pair
    \item The photon undergoes a \pn or \en interaction resulting in a final
          state with hadrons
\end{itemize}
The samples used to to study these processes are detailed in 
Tab.~\ref{tab:bkg_samples}.  

\begin{table}[htbp]
    \begin{tabular}{ r | c | c } 
        \hline
        \hline
        \textbf{Simulated Sample} & \textbf{Number of events} & \textbf{EoT equivalent} \\
        \hline
        Target $\gamma^\ast \to \mu \mu$    & $3.3 \times 10^7$ & $1.1 \times 10^{15}$ \\
        \ecal $\gamma^\ast \to \mu \mu$     & $7.3 \times 10^8$ & $6.6 \times 10^{14}$ \\
        Target \pn                          & $5.2 \times 10^7$ & $1 \times 10^{14}$ \\
        \ecal \pn                           & $2.0 \times 10^9$ & $1.17 \times 10^{14}$ \\
        Target \en                          & $3.2 \times 10^7$ & $3.8 \times 10^{14}$ \\
        \hline
        \hline
    \end{tabular}
    \caption{ Dedicated simulation samples for background investigation. }
    \label{tab:bkg_samples}
\end{table}

\subsubsection{Photons that Shower Electromagnetically \people{Mans} }
\label{section:EM_showering_background}

As the \ecal is a sampling calorimeter, natural electromagnetic shower fluctuations can produce showers which have a low charge deposited in the silicon detector compared with the total energy of the shower.  For such an event to appear as a signal event, it would be necessary for the shower in the calorimeter to not only match the recoil electron in energy, but for the shower to appear in the calorimeter at a position consistent with the low-momentum electron's bending in the dipole field. The most-challenging region of phase space to reject is therefore where the momentum of the electron is near the upper acceptance limit for the analysis -- at 1.2 GeV.

Accordingly, we have studied the behavior of the \ecal for events where a recoil electron of 1.2 GeV is produced at the target along with a 2.8 GeV photon.  The electron and photon originated at the same position in the target and were propagated collinearly into the \ecal. We disabled all photo-nuclear, electro-nuclear, and muon-conversion processes in \geant to allow a factorized study of this background.

Based on the extrapolated fit, the expected number of events of this type to yield less than 1500 MeV in the \ecal--assuming $4\times10^{14}$ EoT--is $\sim 10^{-2}$ events. Here we have also taken into account the fact that only $\sim6\times 10^{-2}$ of 4 GeV electrons on target result in a recoil electron with less than 1.2 GeV energy. Thus, pure electromagnetic shower fluctuations are not expected to be a significant source of background for $4\times10^{14}$ EoT or even upwards of $10^{15}$ EoT.

\begin{figure}[tb]
\centering
\includegraphics[width=14 cm]{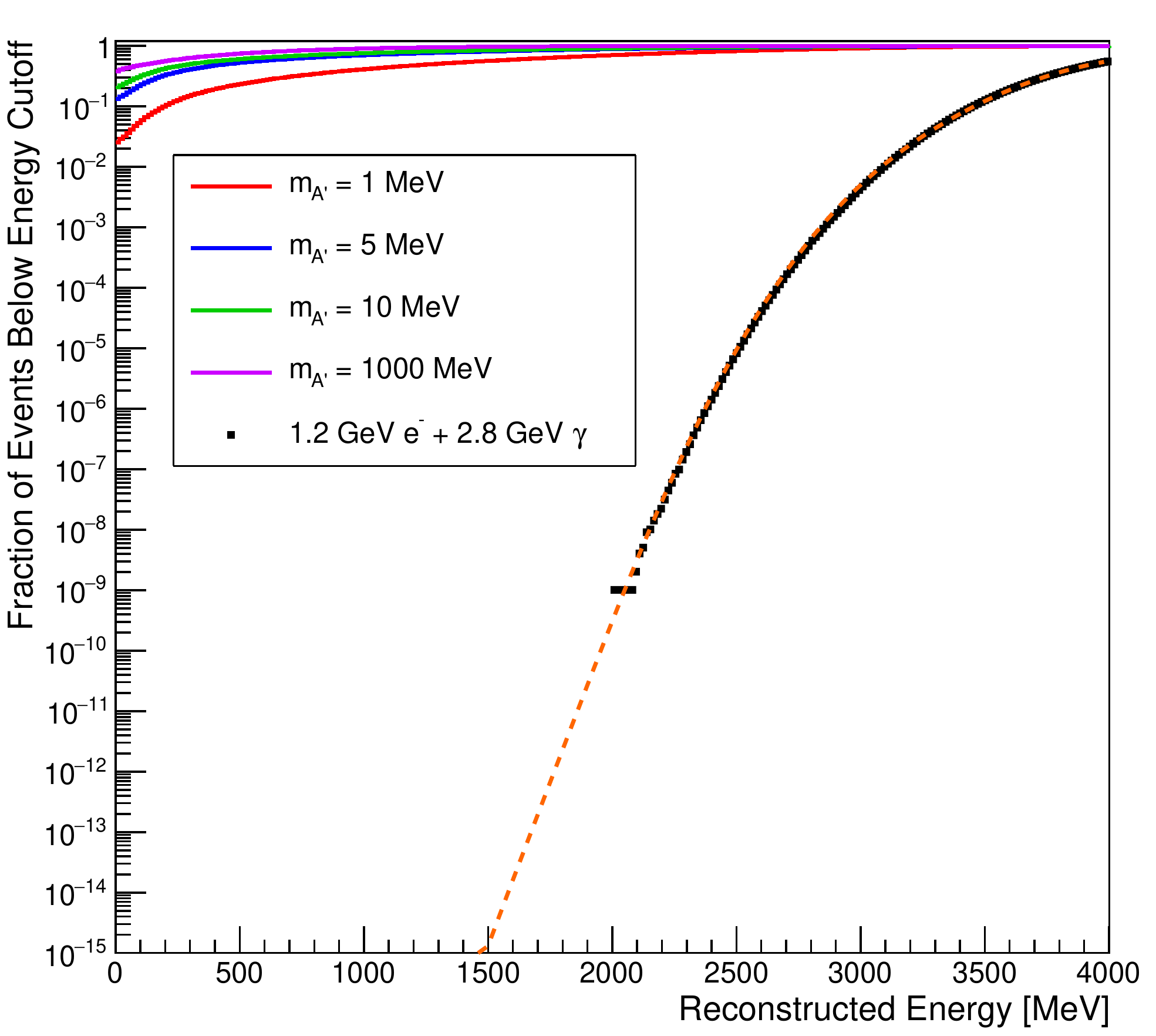}
\caption{The \ecal reconstructed-energy distribution for 1.2 GeV electron plus 2.8 GeV photon events and signal events for various $A'$ masses are plotted over a range extending to 4 GeV. By extrapolating the fit of the electron-plus-photon events, we find that the fraction of these events with a measured energy of less than 1500 MeV is on the order of $10^{-15}$.}
\label{fig:em_shower_egamma}
\end{figure}

\subsubsection{Photons Undergoing a Muon Conversion \people{Andrew,Nhan} \morepeople{Bertrand}}
Another challenging background for the LDMX detector which is qualitatively 
different from many of the other backgrounds are processes that involve the production of muons.  An overview of the dominant photoconversion and 
trident processes which produce muons is given in Sec.~\ref{sec:proc_bkg}.
Here we consider strategies to reject these muon background produced both in the target and in the \ecal.
The primary sample of muons produced through photoconversion is simulated in \geant with the cross section scaled up by a large factor as described in Sec.~\ref{sec:bkggen}.  
The kinematics are simulated with the modified differential cross-section based 
on an improved nuclear form-factor described in Sec.~\ref{sec:bkgmod}.

The main rejection handles for muon backgrounds, depending on where they are produced, are:
\begin{itemize} 
	\item Trigger information: The energy of the recoiling electron should be lower than the energy threshold such that the event passes the nominal trigger.
	\item \hcal information: This is the primary rejection handle for muon backgrounds as muons typically travel through the tracking and \ecal systems and leave a MIP track through the \hcal with many signals which would be vetoed by the baseline \hcal selection.  The most challenging scenarios to veto are often when the muon decays in the \ecal producing neutrinos.  
	\item Tracking information: If the muons are produced in the target, the recoil tracker system may identify more than one track.  In this case, the event should be vetoed.  
	\item \ecal information: In the event that tracking and \hcal systems do not veto the event, then the muons have often decayed in the \ecal.  The muons often leave a MIP track in the \ecal before decaying or ranging out and this, in addition to a large energy deposit, is the primary handle for rejecting muons in the \ecal.
\end{itemize} 
Because of the distinction between muon events produced in the target or \ecal, we will discuss them separately.

\paragraph{Muons from the target}

For events where muons are produced in the target, a sample equivalent to $1.1\times 10^{15}$~EoT is produced.
For a sample of $1.1\times 10^{15}$~EoT, we expect the electron to undergo a hard bremsstrahlung at a rate of 0.0329 and of those resulting photons, we expect a muon pair conversion rate of $9.1 \times 10^{-7}$.  
Therefore we generate a $3.3\times 10^{7}$ event sample with this dedicated process where muons are produced in the target.

Of the sample of dedicated muon events produced in the target, using the hadronic veto described in Sec.~\ref{sec:hcalperf}, we find only 9 muon pair events remain which have not been vetoed by the \hcal.  These remaining events are generically characterized by one hard forward muon, one soft muon, and one soft electron.
Then the high energy muon decays in the \ecal volume, the soft muon and electrons range out or decay in the \ecal as well.
Events of this type are easily vetoed by the tracking system which can individually identify the 3 charged particle tracks.  
In addition to the tracking information, the \ecal information can provide complementary information to further veto this event.

\begin{figure}[!bh]
\centering
\includegraphics[width=14cm]{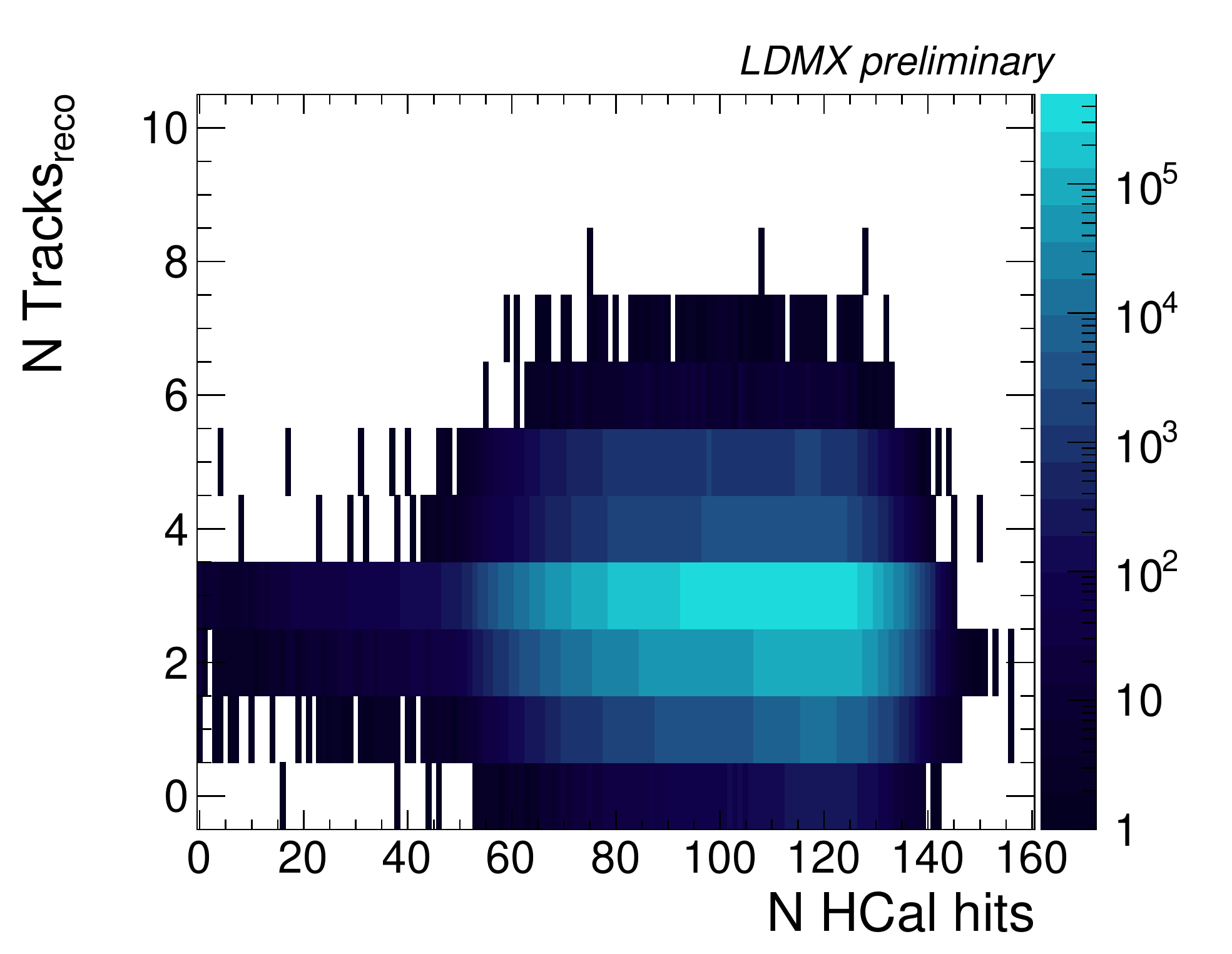}
\begin{flushleft}
\caption{\label{fig:muonTarget} Number of \hcal hits versus reconstructed tracks for muon pair conversion background.}
\end{flushleft}
\end{figure}

To illustrate the rejection power of muon pairs produced in photoconversion in the target region, we select those events which pass the trigger selection.
Then, of those events, we characterize the event by how many tracks are reconstructed and the number of hits found in the \hcal system.
This is plotted in Fig.~\ref{fig:muonTarget} where on the $x$-axis is the number of \hcal hits and the $y$-axis is the number of reconstructed tracks. 
We see that though there are several events which are not vetoed by the \hcal, but most are vetoed by the tracking requirement.
There is also the opposite scenario where the event is not vetoed by the tracking requirement but is vetoed by the \hcal veto.  
This typically occurs when the recoiling electron and one of the muons is very soft and does not result in reconstructed tracks.  However, the remaining hard muon is identified by the \hcal veto. 

The signal region is where there are {\it exactly} zero \hcal hits and 1 reconstructed track.
In Fig.~\ref{fig:muonTarget}, we find that there are 2 events which remain after the reconstructed track and \hcal hit selection.
Of those 2 events, neither event passes the \ecal BDT which is described in Sec.~\ref{sec:ecal_perf}.
Therefore, we find no background events are contained in the signal region.
This is summarized in Table.~\ref{tab:mumu_rejection}.

\paragraph{Muons produced in the \ecal}

We also study the case where the muon pair is produced from photoconversion in the \ecal.  
Simulation is performed using \geant, again with the modified modeling of the photon conversion to a muon pair.
In order to simulate a high statistics sample of muon pair background events, we perform final-state reweighting of this process
as described in Sec.~\ref{sec:bkggen}.
We generate a sample that corresponds to $6.6 \times 10^{14}$~EoT.  

\begin{figure}[!tbh]
\centering
\includegraphics[width=14cm]{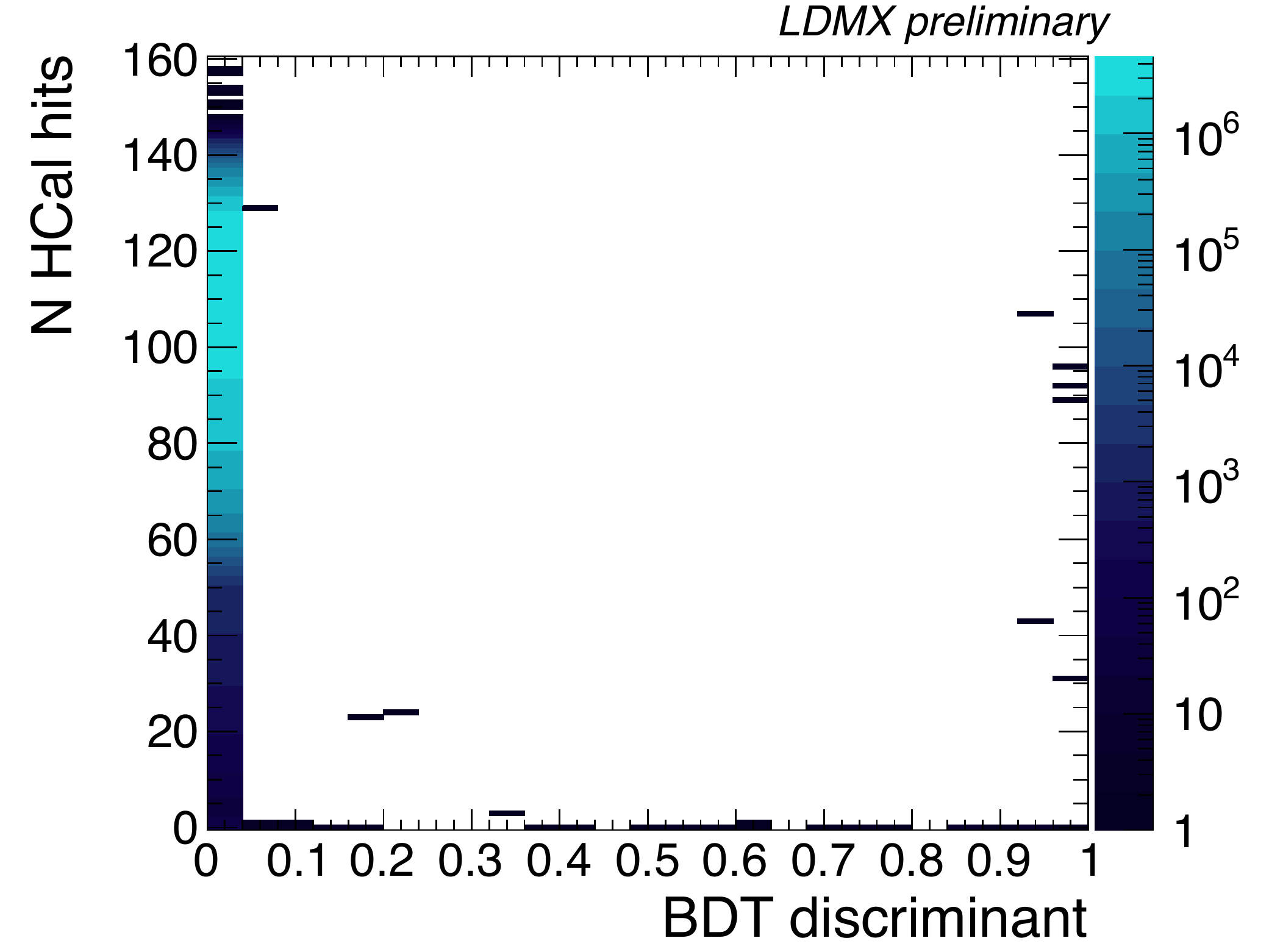}
\begin{flushleft}
\caption{\label{fig:muonEcal} Number of \hcal hits versus the BDT score for muon pairs produced in the \ecal}
\end{flushleft}
\end{figure}

\begin{figure}[!tbh]
\centering
\includegraphics[width=12cm]{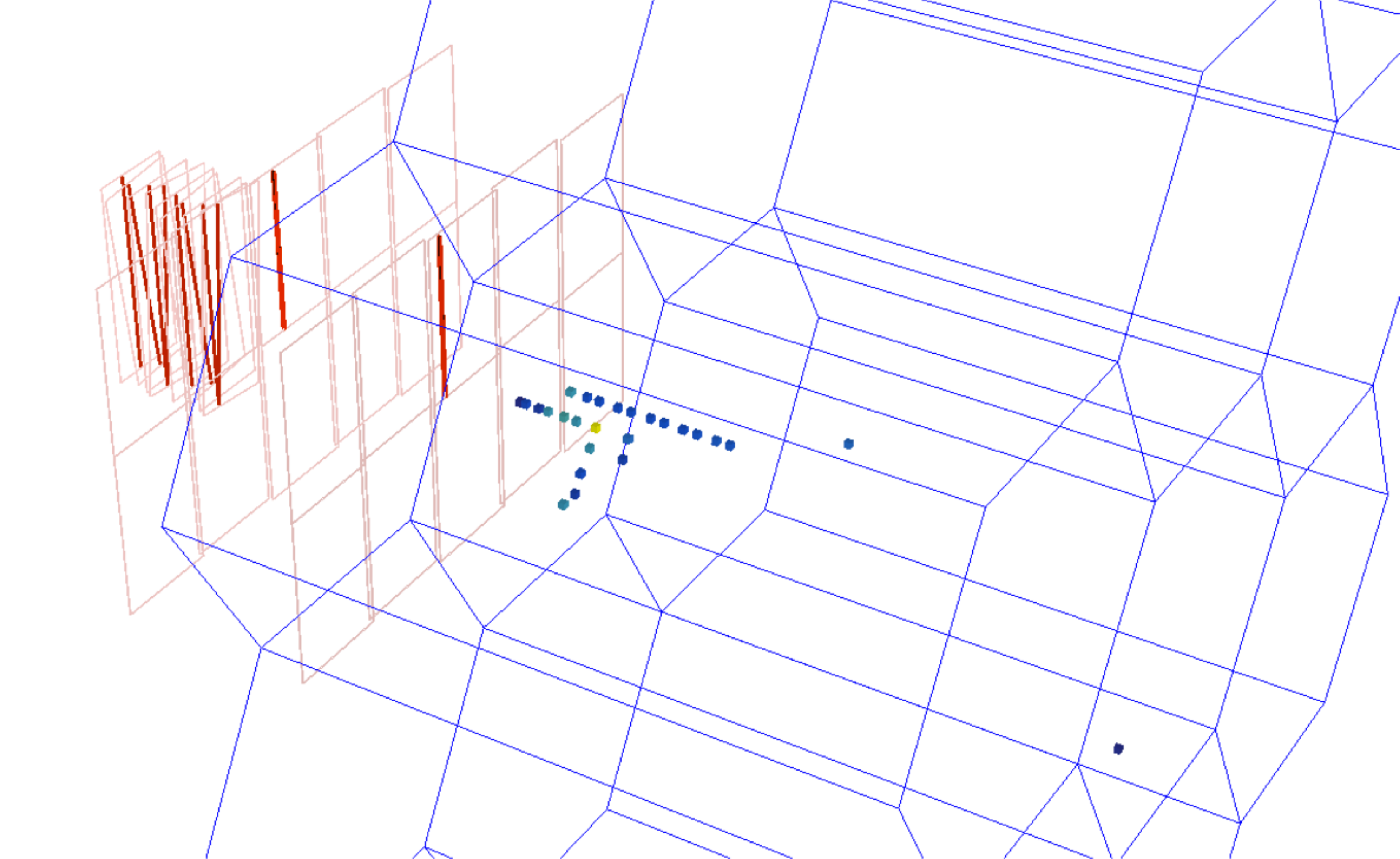}
\begin{flushleft}
\caption{\label{fig:muonEcalEv} Event display of an $\gamma^\ast \to \mu \mu$ interaction in the \ecal which does not pass the \hcal veto.}
\end{flushleft}
\end{figure}

In the case where photon conversion occurs in the \ecal, we no longer have a handle from the recoil tracking system.
To illustrate the rejection power of muon pairs produced in the \ecal, we select those events which pass the trigger selection.
We can first attempt to veto the event based on the number of hits in the \hcal system.
MIP tracks typically leave many hits in the \hcal and this is the primary mode for vetoing these types of events. 
After the \hcal veto, there are still $\mathcal{O}(100)$ events remaining.
These are typically events where one soft muon ranges out in the \ecal and one hard muon decays into an electron and neutrinos.  
Therefore, we can use further \ecal information to reject these remaining events which survive the \hcal veto.  
As a simple means of veto, we use the same BDT that is used in to identify photo-nuclear events and these events are fairly easily vetoed by the \ecal BDT.  
To illustrate the rejection power of these types of events, we show the number of \hcal hits versus the \ecal BDT output in Fig.~\ref{fig:muonEcal}.
We find that after requiring zero \hcal hits, of those remaining events only 1 event also passes the \ecal BDT veto, a BDT value greater than 0.94.
For example, the event with the highest BDT score surviving the \hcal veto is shown in Fig.~\ref{fig:muonEcalEv}.
Beyond the usual photo-nuclear shower handles, there is a clear muon MIP track in the \ecal which would be identified easily with dedicated algorithms. 
In future work, we note that it is reasonable to design an even more efficient veto for muon MIPs in the \ecal and this remaining event would be vetoed.  
The results for the rejection of muons originating in the \ecal are also presented in Table~\ref{tab:mumu_rejection}.

\begin{table}[!t]
    \centering
    \begin{tabular}{ l | c c  } 
        
        \multicolumn{3}{c}{Cutflow and event yields} \\ [0.5ex]
        \hline\hline
        & \textbf{Target-area $\mu\mu$} & \textbf{\ecal $\mu\mu$}  \\ [0.5ex]
        \hline
        EoT equivalent & $1.1 \times 10^{15}$ & $6.6 \times 10^{14}$ \\ 
        \hline
        \hline
        Events passing trigger & $2.14 \times 10^7$ & $1.50 \times 10^8$ \\
        Passing \hcal veto & 36 & 169 \\
        Passing $N_{\rm track}$ veto & 2 & 169 \\
        Passing \ecal BDT veto & 0 & 1 \\
        \hline
    \end{tabular}
    \caption{The estimated levels of dimuon background events after successive 
             background rejection cuts.}
    \label{tab:mumu_rejection}
\end{table}

\subsubsection{Photons Undergoing a Hard Photonuclear Interaction \people{Nhan, Omar} \morepeople{Ruth}}
\label{sec:bkgpn}
A background process that is difficult to distinguish from signal is the case
where a hard photon undergoes a \pn reaction in either the target or
in the \ecal absorber layer resulting in the conversion of the photon's energy
into outgoing hadrons.  The resulting final states can vary greatly because both 
nuclear disintegration and/or hadron production can occur.  Since neutrons do
not lose energy to ionization, and protons follow a very steep ionization curve,
the detector response also greatly varies. Vetoing of such challenging 
processes requires all sub-detector systems to work in complementarity. 

Several handles have been studied in order to understand how to best veto
these events. For example, as shown in Fig.~\ref{fig:energyvars}, \pn interactions 
have distinctly different \ecal shower signatures when compared to more common 
electromagnetic processes that serve as a powerful discriminator which can be 
used to veto these events.
In processes involving neutrons that do not interact in the \ecal, the \hcal
is relied on to identify them. 
Finally, the recoil tracking system can be used to identify events with
additional charged hadrons produced in these \pn processes, either forward going 
if the \pn process occurs in the target or via back-scattering 
particles if the \pn interaction occurs in the \ecal.

With this in mind, the \pn background rejection strategy is roughly broken into 
three main tasks:
\begin{enumerate}
	\item shower profile rejection using the \ecal
	\item hadronic tails with a high efficiency, large angular coverage \hcal veto
	\item track and hit multiplicity vetoes using the tagging and recoil trackers
\end{enumerate}
Development of the background rejection strategy was done using a sample of 
events where a hard bremsstrahlung occurs in target and the resulting photon
undergoes a \pn reaction.  As shown in Tab.~\ref{tab:bkg_samples}, an \ecal \pn 
(target-area \pn) sample
of $2.0\times10^{9}$  ($5.2\times10^{7}$) events was used which is what is
expected assuming $1.17\times 10^{14}$ ($1\times 10^{14}$) EoT. The details
of the sample generation are discussed in Sec.~\ref{sec:bkggen}.  Only the 
subset of events that pass the
trigger requirements as described in Sec.~\ref{ssec:performance-trigger} are 
considered. A summary of the final rejection strategy is given in 
Table~\ref{tab:pn_rejection}, the details of which are discussed below. 

\begin{table}[!t]
    \centering
    \begin{tabular}{ l | c c  } 
        
        \multicolumn{3}{c}{Cutflow and event yields} \\ [0.5ex]
        \hline\hline
        & \textbf{\ecal PN} & \textbf{Target-area PN} \\ [0.5ex]
        \hline
        EoT equivalent & $4 \times 10^{14}$ & $4 \times 10^{14}$ \\
        \hline\hline
        $E_{recoil}< 1.5$ GeV, trigger requirement & $2.7 \times 10^8$ & $2.2 \times 10^7$ \\ 
        \ecal Shower-Profile BDT         & $2\times 10^6$ & $8.2\times 10^5$  \\ 
        \hcal Max PE $<3$                & 0.55 & 28             \\ 
        Single track with $p < 1.2$ \GeV & 0.51 & 23             \\ 
        Recoil activity Cut              & 0.41 & 23             \\ 
        \ecal activity cut               & 0.24 & 23             \\ 
        Tagging tracker activity cut      & 0.24 & 0              \\ 
        \hline
    \end{tabular}
    \caption{The estimated levels of background \pn events, scaled to 
             $4\times10^{14}$ EoT, after successive 
             background rejection cuts. Note that all single- and di-nucleon 
             events have been re-weighted in accordance with the value on 
             Tab.~\ref{tab:event_weight}.
             }
    \label{tab:pn_rejection}
\end{table}

\paragraph{Shower profile rejection}

As detailed in Sec.~\ref{sec:ecal_perf}, there is a distinct difference 
in the \ecal signatures between dark matter signal-like processes and \pn 
interactions. The variables with the strongest discrimination power are 
exploited to distinguish \pn reactions from normal electromagnetic showers 
in the \ecal. Specifically, the information from these variables are combined 
using a gradient boosted decision tree (BDT) in order to further reject
\pn reactions while maintaining a high signal efficiency.  

Based on the BDT performance metrics discussed in Sec.~\ref{sec:ecal_perf}, a 
cut of 0.94 on the BDT discriminant was determine to best balance the rejection
of \pn background and the preservation of signal. Applying this requirement, 
roughly 98\% and 96\% of \ecal \pn and target-area \pn backgrounds, 
respectively, are rejected while preserving 65-85\% of the signal,
depending on the mediator mass. Those events that remain typically fall into
one of two categories: 
\begin{itemize}
    \item \pn events that contain backwards going charged hadrons
    \item \pn events that produce forward energetic neutrons
\end{itemize}
A subset of the first class of events can be rejected by making use of activity
in the trackers while it is the task of the \hcal to efficiently veto the second 
class of events. 

\paragraph{Hadronic tails}

A subset of events that the \ecal has difficulty vetoing are typically 
characterized by neutral hadrons which escape the \ecal without leaving any 
significant energy deposition. The \hcal is left to detect these events with
high efficiency.

\begin{figure}[!t] 
    \centering
    \includegraphics[width=.8\textwidth]{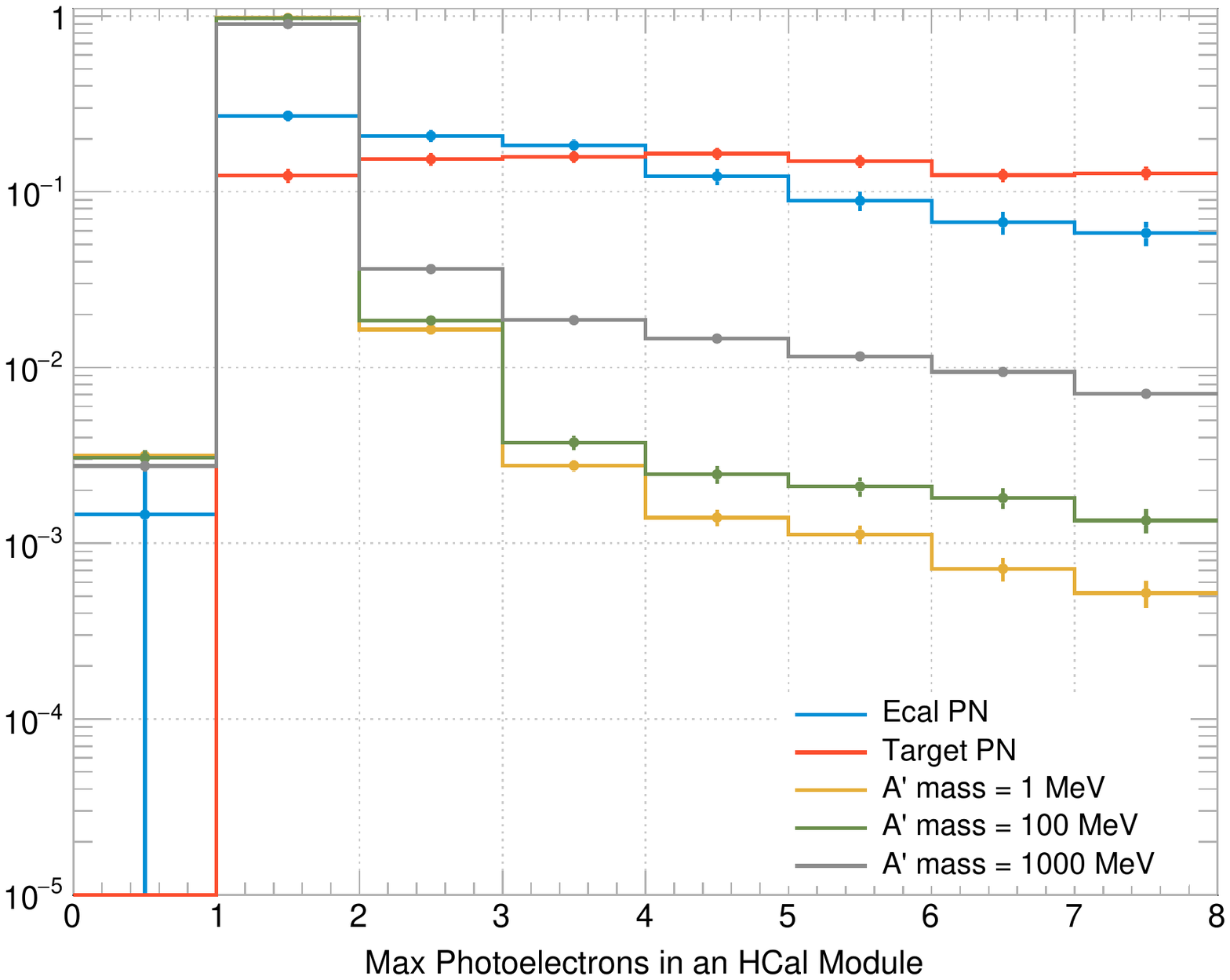}
    \caption{The maximum photoelectrons in a layer of the \hcal per event observed
             for 1 (yellow), 100 (green) and 1000 (grey) MeV signal, 
             target-area (orange) and \ecal (blue) \pn events.
             On average, noise will contribute 1 PE per event which results in 
             the distributions peaking at 1 PE. Requiring the max PE $<3$ is 
             found to reject 99.99\% of background events that are failed to be
             rejected by the \ecal BDT while preserving 99\% of the signal.
    }
    \label{fig:max_pe_pn_bkg}
\end{figure}

As shown in Fig.~\ref{fig:max_pe_pn_bkg}, a strong discriminator between signal
and \pn background is the maximum number of PE measured in the scintillating bars of the \hcal (max PE). Typically, the recoil electron in a signal event will enter the
\ecal and shower with very little energy escaping into the \hcal.  As 
discussed in Sec.~\ref{sec:hcalperf}, noise will contribute on average 1 PE 
per event. Given that signal events will typical see very little activity in 
the \hcal, most events are expected to have a max PE of 1.  In contrast, 
\pn backgrounds will leave multiple hits in the \hcal resulting in a larger value of max PE. This is shown in Fig.~\ref{fig:max_pe_pn_bkg} which overlays the maximum PE 
per event observed in signal and \pn background events.  With this
in mind, requiring an event to have a max PE $<3$, will reject 99.99\% of the
remaining background while still preserving 99\% of the remaining signal.

\begin{figure}[!t]
    \centering
    \includegraphics[width=.75\linewidth]{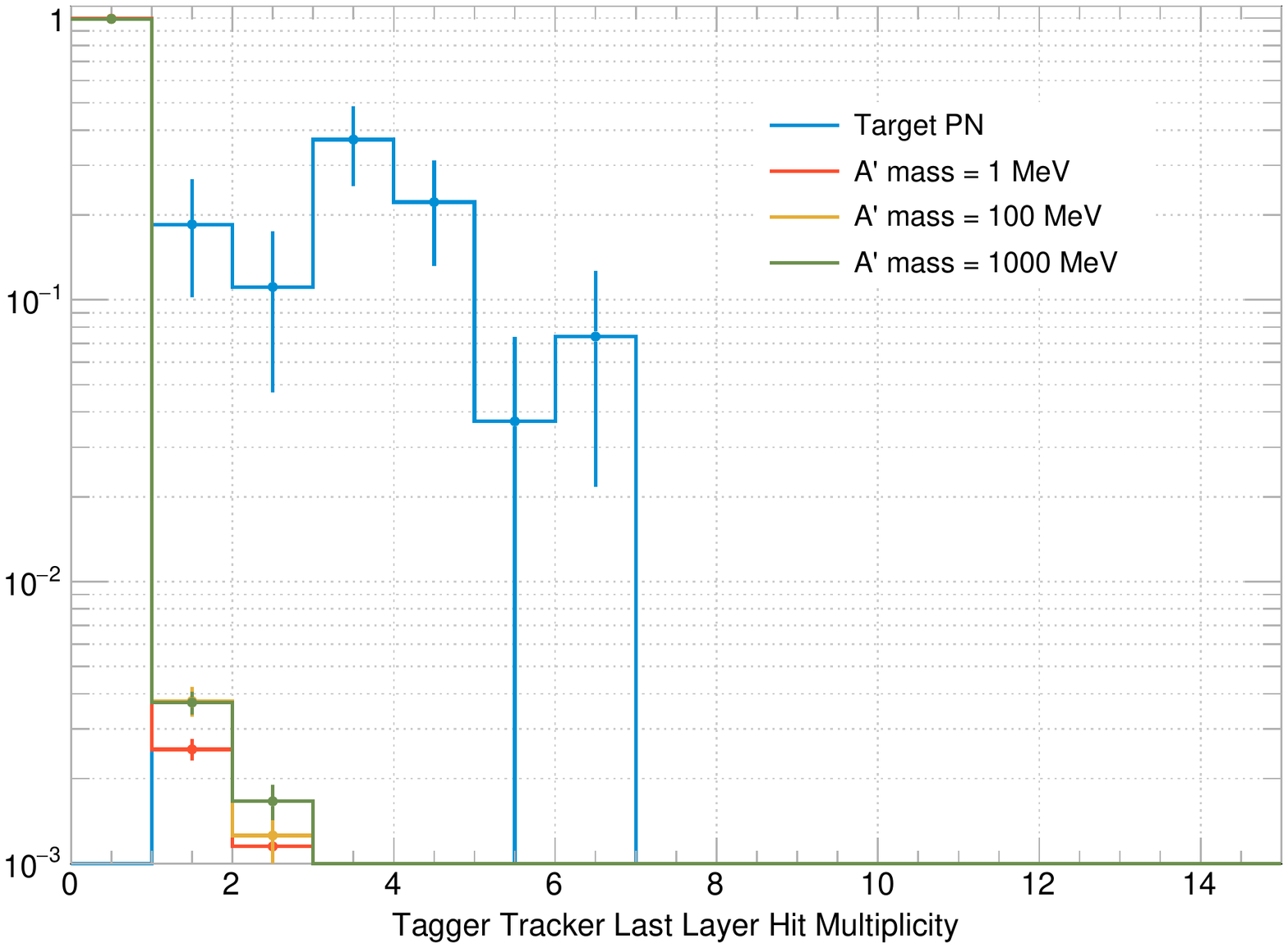}
    \caption{The number of hits in the last layer of the Tagging tracker, 
             excluding the contribution from the incident beam electron, for
             target-area (blue) \pn events after the application of calorimeter 
             and recoil tracker vetoes. For comparison, $1~\MeV$ (orange), 
             $100~\MeV$ (yellow), $1000~\MeV$ (green) $A'$ signal are also shown. \\} 
    \label{fig:tagger_hit_mult}
    \includegraphics[width=.8\linewidth]{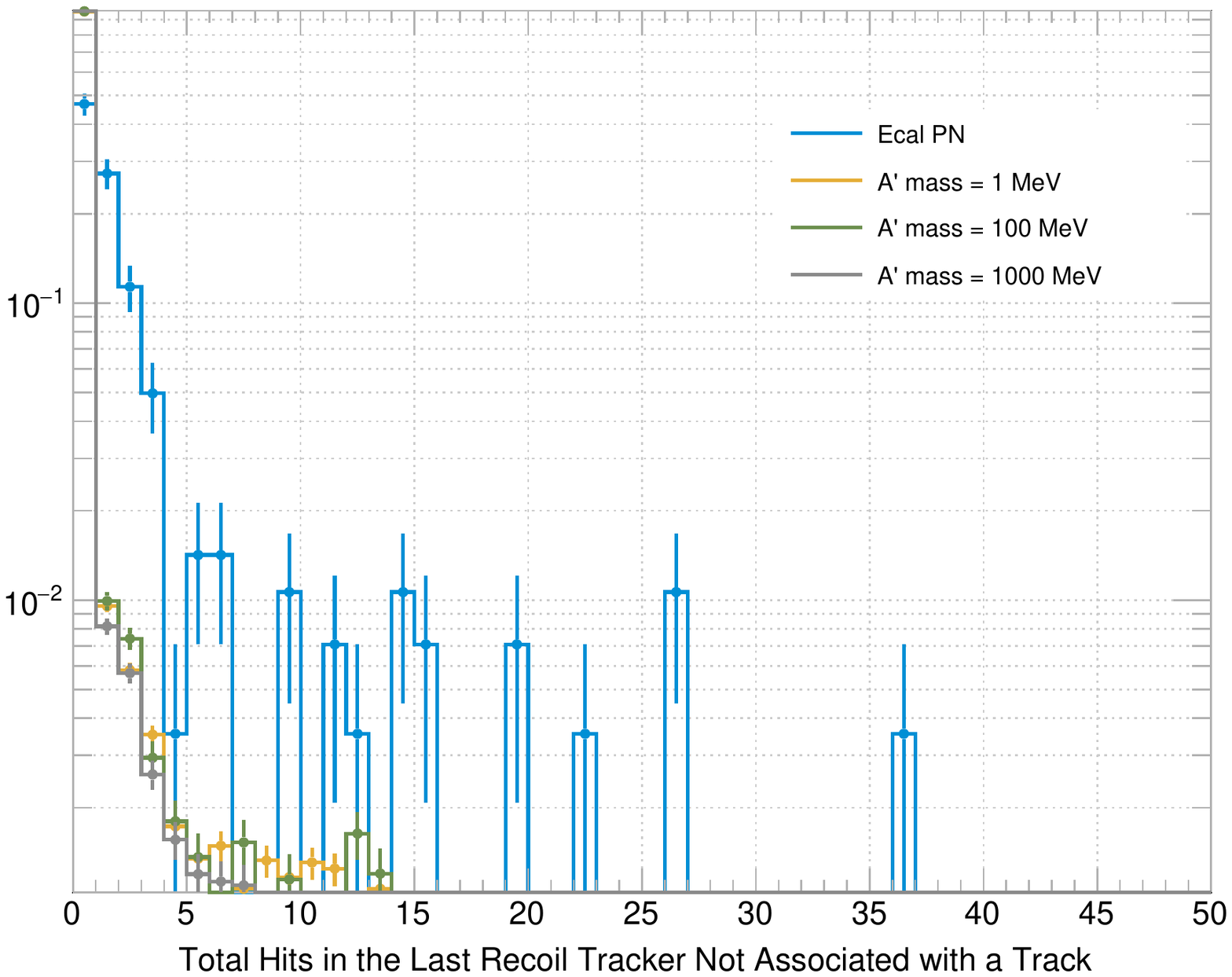}
    \caption{The number of hits in the last layer of the Recoil tracker, 
             excluding the contribution from hits associated with tracks, for
             \ecal (blue) \pn events after the application of calorimeter 
             and recoil tracker vetoes. For comparison, $1~\MeV$ (yellow), 
             $100~\MeV$ (green), $1000~\MeV$ (grey) $A'$ signal are also shown.} 
    \label{fig:recoil_hit_mult}
\end{figure}


\paragraph{Tracker Veto}

After exhausting all calorimeter handles, there still remains a subset of events
with forward or backward going charged hadrons that can be rejected by 
the trackers. 
A significant number of these remaining events have multiple reconstructed 
tracks in the Recoil tracker.  Furthermore, as discussed in 
Sec~\ref{sec:proc_sig}, only signal events where the reconstructed momentum of
the recoiling $e^{-}$ is below 1.2~\GeV{} are considered to encompass the 
signal region.  Therefore, simply requiring that the Recoil tracker has only a
single reconstructed track with a momentum $<1.2$~\GeV{} vetoes an additional 
18\% and 6\% of the remaining target and \ecal \pn events, respectively.

For events that see a \pn reaction occur in the target, the Tagging tracker 
can be used to further reject these events. Given that the 
noise occupancy per tracker sensor is expected to be at the 10$^{-4}$ level,   
the last layer of the Tagging tracker will typically only see a single hit
from the incident beam electron.  For \pn events, the backward going charged 
hadrons will leave multiple hits in the last layer.  This can be seen in 
Fig.~\ref{fig:tagger_hit_mult} which shows the hit multiplicity in the last
Tagging tracker layer excluding the hit due to the incident beam electron.  
Requiring that no  hits other than that due to the incident beam electron 
are observed in the last layer of the Tagger tracker rejects the remaining
target area \pn events while still preserving  99\% of the remaining 
signal.

For \ecal \pn events, the last layer of the Recoil tracker can be used in 
the same way the last layer of the Tagging tracker was used to reject events
with backwards going charged hadrons.  Typically, hits observed in the last
layer of the Recoil tracker will be associated with a track.  However, given the
coverage of the last Recoil tracker layer when compared to the rest of the 
layers, a subset of \ecal \pn events will see backwards going charged hadrons
produced at angles such that they only hit the last layer. This can be seen in 
Fig~\ref{fig:recoil_hit_mult}, which shows the hit multiplicity in the last
Recoil layer after excluding hits associated with tracks. Requiring that all
the hits in the last Recoil layer have an associated track rejects an additional
25\% of the remaining events while preserving 97\% of the remaining signal.

\paragraph{Rejection of High-Energy Backscatters}

In large-scale simulations of both \pn and \en reactions, we found that \geant
was over-populating the wide-angle/backwards high energy tails. As discussed
in Section~\ref{sec:bkgmod} and Appendix~\ref{sec:PNappendix}, almost all of these ``ghost'' events 
populating these regions of phase-space were found to be due to an unphysical 
mechanism.  Therefore, ghost events are not considered when estimating the 
background yields surviving all of the vetoes. The parameterization used to 
determine if an event is unphysical is discussed in the 
Appendix~\ref{sec:PNappendix}.

\paragraph{Re-weighting of Single and Di-nucleon Final States}

As discussed in Section~\ref{sec:bkgmod} and Appendix~\ref{sec:PNappendix}, the single- and di-nucleon final
state yields were found to be overproduced by \geant by at least a factor of 
100.  In order to get a better estimate of the expected number of background 
events that survive all of the vetoes, events with single and di-nucleon final
states were given a conservative weight as listed on 
Table~\ref{tab:event_weight}. The estimated levels of background events shown in 
Tables~\ref{tab:pn_rejection} and~\ref{tab:en_rejection} reflect the use of these
weights.

\begin{table}[!ht]
    \centering
    \begin{tabular}{ l | c } 
        
        Final State & Weight \\ [0.5ex]
        \hline\hline
        Single neutron & 0.01  \\
        Di-neutron     & 0.002 \\
        Di-proton      & 0.01  \\
        proton-neutron & 0.02  \\
        \hline
    \end{tabular}
    \caption{
            The event weights applied to single- and di-nucleon final states.
            The weights are used to account for the overproduction of such
            event types by \geant. 
            }
    \label{tab:event_weight}
\end{table}

\subsection{Rejection of Electronuclear Interaction Backgrounds \people{Omar} \morepeople{Natalia}}
\label{sec:bkgen}

Electro-nuclear interactions are a background component which are similar to 
photo-nuclear interactions, though they have a soft recoil electron momentum 
and a lower overall cross-section.  Therefore, the rejection strategy is 
similar to that used on photo-nuclear interactions.  Since we can track the
momentum of an electron with the beam energy, here we focus on \en interactions
which occur in the target rather than in the \ecal. 

\begin{table}[!t]
    \centering
    \begin{tabular}{ l | c  } 
        
        \multicolumn{2}{c}{Cutflow and event yields} \\ [0.5ex]
        \hline\hline
        & \textbf{Target-area EN} \\ [0.5ex]
        \hline
        EoT equivalent & $4 \times 10^{14}$ \\
        $E_{recoil}< 1.5$ GeV, trigger requirement & $1.4 \times 10^7$ \\ 
        \ecal Shower-Profile BDT          & $6.2 \times 10^5$   \\ 
        \hcal Max PE $<3$                 & 26 \\ 
        Single track with $p < 1.2$ \GeV  & 13 \\ 
        Tagging tracker activity cut       & 0.0127 \\ 
        \hline
    \end{tabular}
    \caption{The estimated levels of background \en events, 
        scaled to $4\times10^{14}$, after successive background rejection cuts.
        Note that all 
        single- and di-nucleon events have been re-weighted in accordance with
        the value on Tab.~\ref{tab:event_weight}. 
         }
    \label{tab:en_rejection}
\end{table}

\begin{figure}[!tbh]
\centering
\includegraphics[width=.7\textwidth]{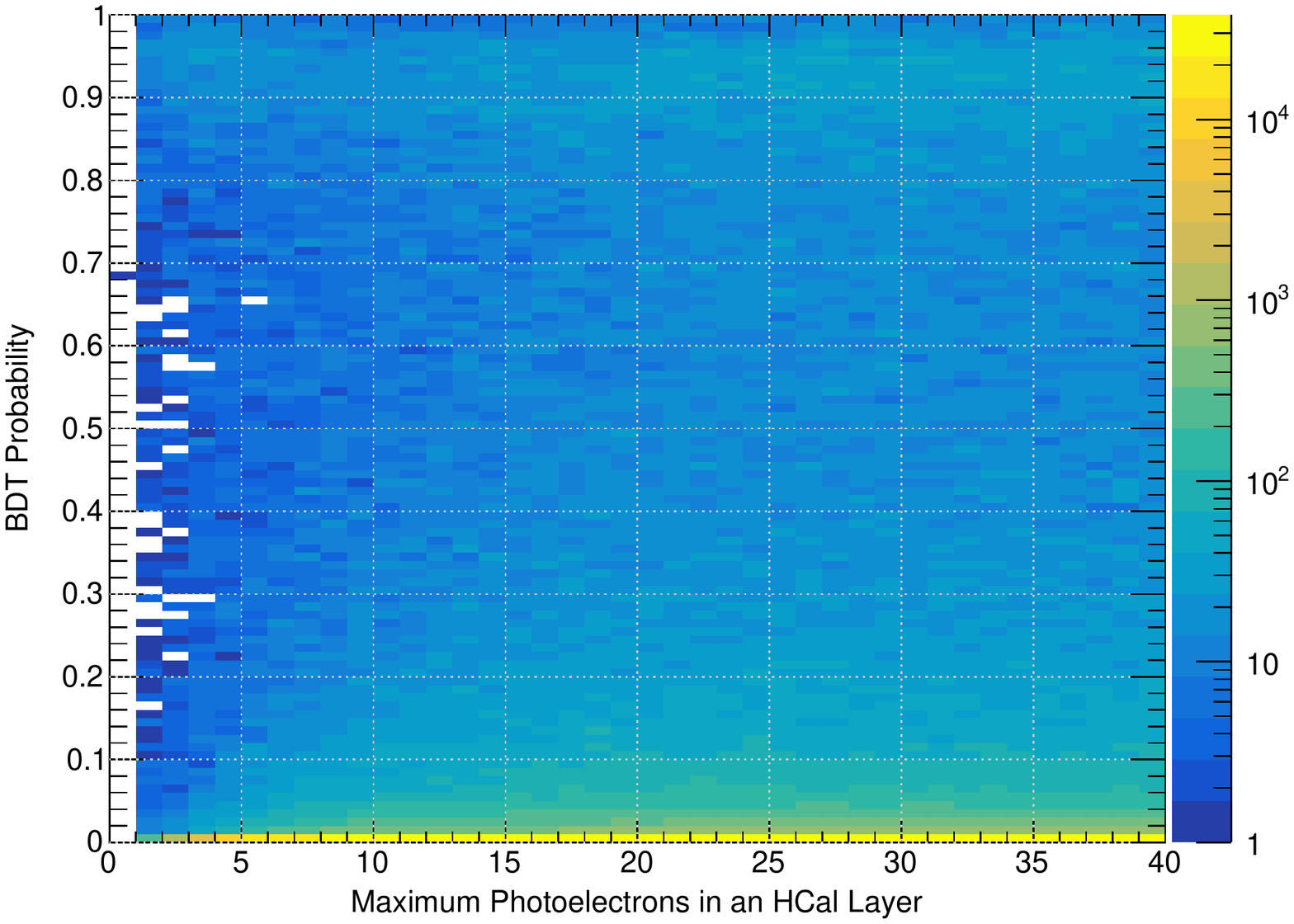}
\begin{flushleft}
\caption{\label{fig:en_hcal_bdt} The maximum number of photoelectrons versus 
         the \ecal BDT score for \en events passing the trigger.}
\end{flushleft}
\end{figure}

\begin{figure}[!bh]
\centering
\includegraphics[width=.7\textwidth]{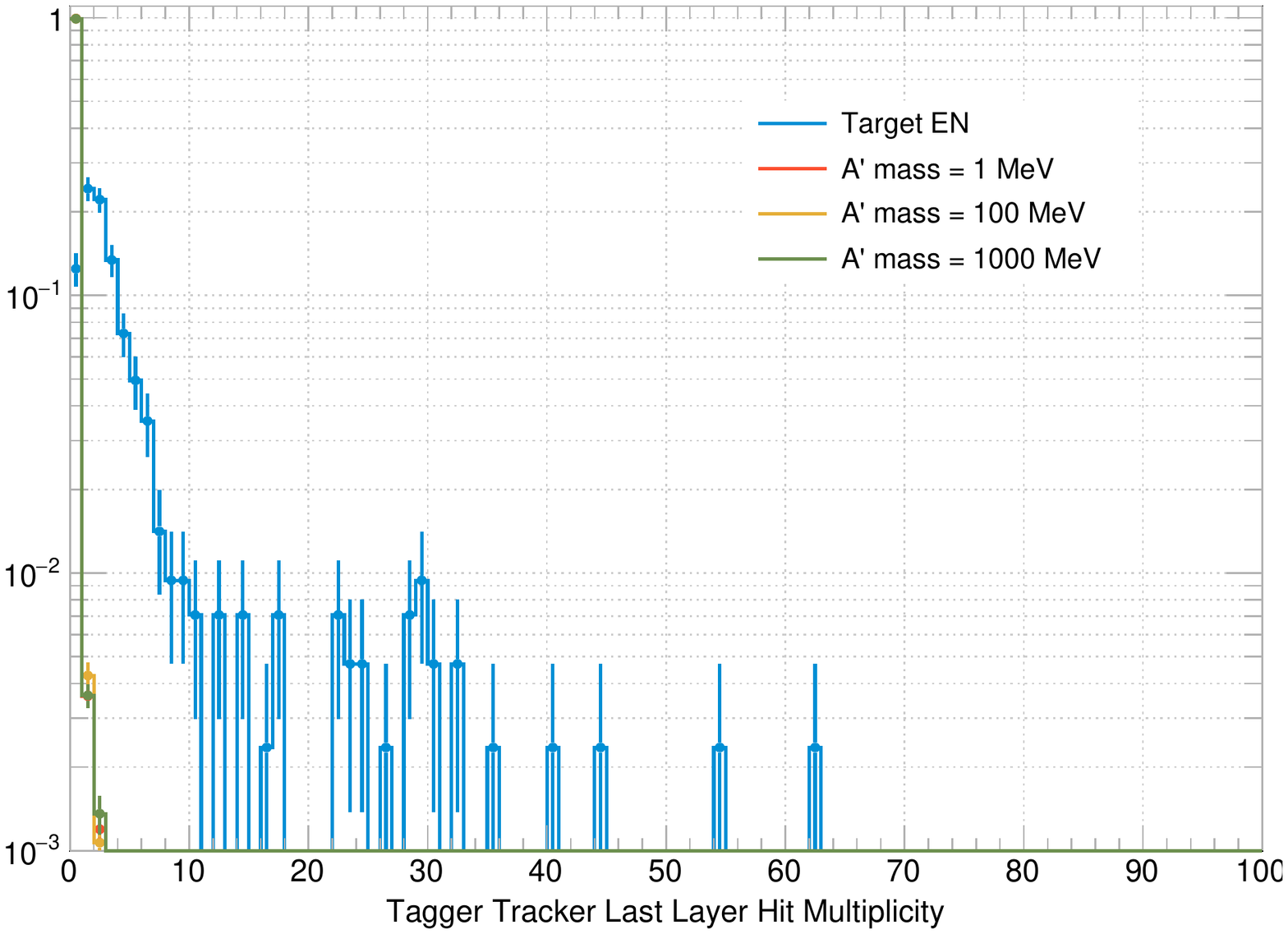}
\begin{flushleft}
    \caption{The number of hits in the last layer of the Tagging tracker, 
             excluding the contribution from the incident beam electron, for
             target-area (blue) \en events after the application of calorimeter 
             and recoil tracker vetoes. For comparison, $1~\MeV$ (orange), 
             $100~\MeV$ (yellow), $1000~\MeV$ (green) $A'$ signal are also shown.} 
    \label{fig:en_tagger_hit_mult}
\end{flushleft}
\end{figure}

We simulated a sample of \en interactions in the target corresponding to 
$3.8 \times 10^{14}$~EoT. Of those events which pass the trigger criteria, we 
determine which events pass both the \ecal BDT and \hcal max PE selection. 
As shown in Fig.~\ref{fig:en_hcal_bdt}, 26 events lie within the signal region
which is defined as the region where the maximum PE is less than 3 and the BDT 
discriminant value is greater than 0.94.  Applying the recoil tracker 
requirements described above (single track plus 1.2 GeV momentum requirement)
rejects an additional 50\% of events. Finally, for those events that pass the
recoil tracker veto, we plot the hit multiplicity in the last layer of the 
tagging tracker. As shown in Fig.~\ref{fig:en_tagger_hit_mult}, requiring that 
no other hits other than that due to the incident beam electron are observed in
the last layer of the Tagging tracker, rejects all but a fraction on an event
while still preserving 99\% of the remaining signal. 

\clearpage
\section{Signal Efficiency Studies} 
\label{sec:sig_eff}
\label{section:signaleff}

\subsection{Trigger Efficiency \people{Mans}}
\label{ssec:signaleff-trigger}

As described above, the primary physics trigger is based on the missing energy in the event, based on the total
energy observed in the calorimeter combined with the incoming energy as determined by the number
of incoming electrons in the trigger scintillators.  While the average number of incoming 
beam electrons is 1, the number in any given bunch crossing will vary from zero to ten or more.  
For each number of incoming beam electrons $n$, a different threshold is set to reject events 
consistent with $n$ full-energy electrons impacting the calorimeter while retaining events with $n-1$ full-energy electrons and possibly one low-energy electron impacting the calorimeter.  These thresholds are tuned
to maximize the efficiency for signal, while keeping the calorimeter trigger bandwidth within
its targeted allocation of 4~kHz.  The resolution of the calorimeter implies that the rate for a fixed missing-energy threshold 
will increase with increasing numbers of incoming electrons.

\begin{figure}[b]
  \begin{center}
    \includegraphics[width=0.7\linewidth]{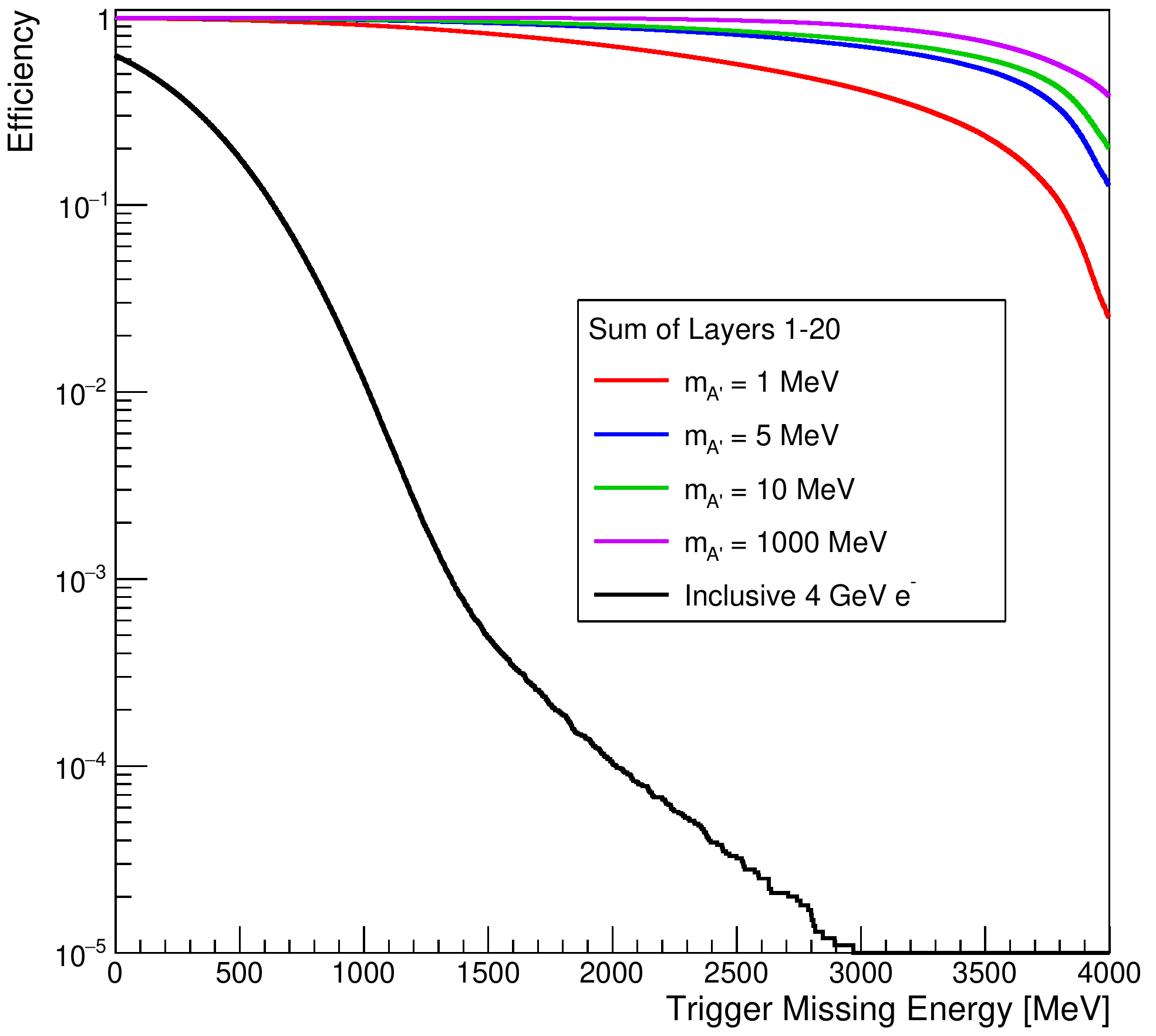}
    \end{center}
  \caption{Performance of the primary physics trigger for LDMX for a single incoming beam electron.  The
    efficiency for signal electrons for different $A'$ masses and the trigger
    rate for all backgrounds induced by beam electrons are shown as a
    function of the trigger missing-energy threshold.}\label{fig:trigger_rejection}
\end{figure}

Figure~\ref{fig:trigger_rejection} shows the simulated performance of
the primary physics trigger for a single incoming beam electron for signal and background.

The performance for the current optimization of the trigger threshold table for $\mu=1$ is shown in 
Table~\ref{tab:trigger_mu1}.  The signal inefficiency column takes into account that, for example, there are four opportunities for a signal event to be produced in an $n_\mathrm{beam}=4$ event.

\begin{table}[t]
\caption{Trigger thresholds, rates, and total inefficiency contribution as function of number of incoming beam
electrons for an average number of one electron per bunch.}
\label{tab:trigger_mu1}
\begin{center}
\begin{tabular}{|ccccccc|}\hline
   & Fraction of & Trigger Scintillator& Missing Energy & Calorimeter Trigger & Rate & Signal \\
$n_\mathrm{beam}$ & Bunches (Signal) & Efficiency & Threshold [GeV] &  Efficiency & [Hz] & Inefficiency \\ \hline
1 & 36.8\% (36.8\%) & 100\%  &  2.50  & 99.2\%  &  588  & 0.3\% \\
2 & 18.4\% (36.8\%) & 97.4\%  &  2.35  & 98.0\%  & 1937  & 1.7\% \\
3 &  6.1\% (18.4\%) & 92.4\%  &  2.70  & 91.6\%  & 1238  & 2.8\% \\ 
4 &  1.5\% (6.1\%) & 84.3\%  & 3.20  & 77.2\%  &  268  & 1.6\% \\
\hline
Total &   &   &  &   & 4000 & 8.8\% \\ \hline
\end{tabular}
\end{center}
\end{table}

\subsection{Recoil Acceptance and Efficiency \people{Omar}}

\begin{figure}[!thbp]

\begin{minipage}{0.48\linewidth}  
\centering
    \includegraphics[width=7cm]{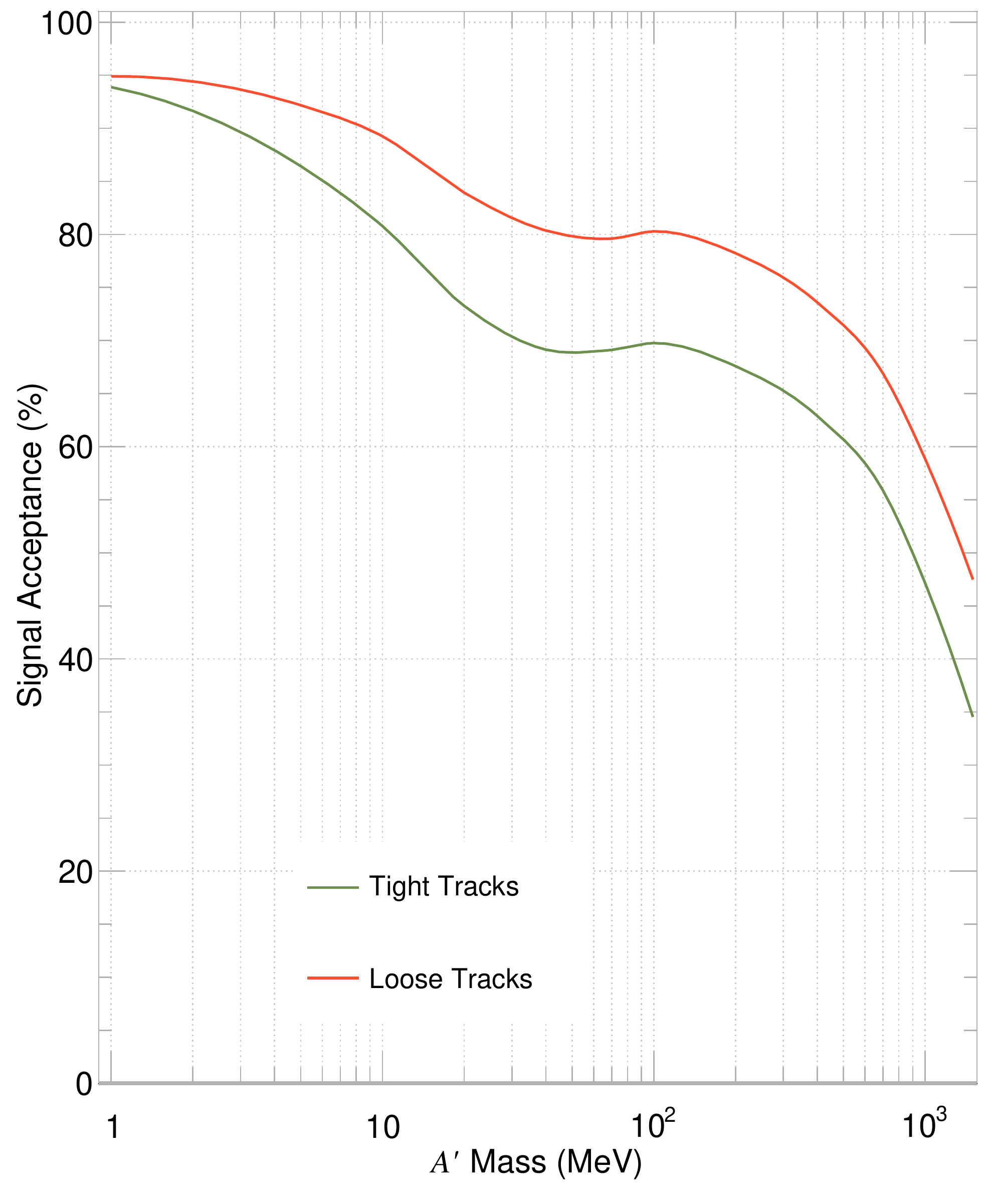}
    \caption{The recoil tracker acceptance to signal recoils as a function of 
    \aprime mass. The acceptance drops with \aprime mass as the polar angle of 
    the recoils increases.}
    \label{fig:tracking_accept}
\end{minipage}\hfill
\begin{minipage}{0.48\linewidth}
  \centering
    \includegraphics[width=7.5cm]{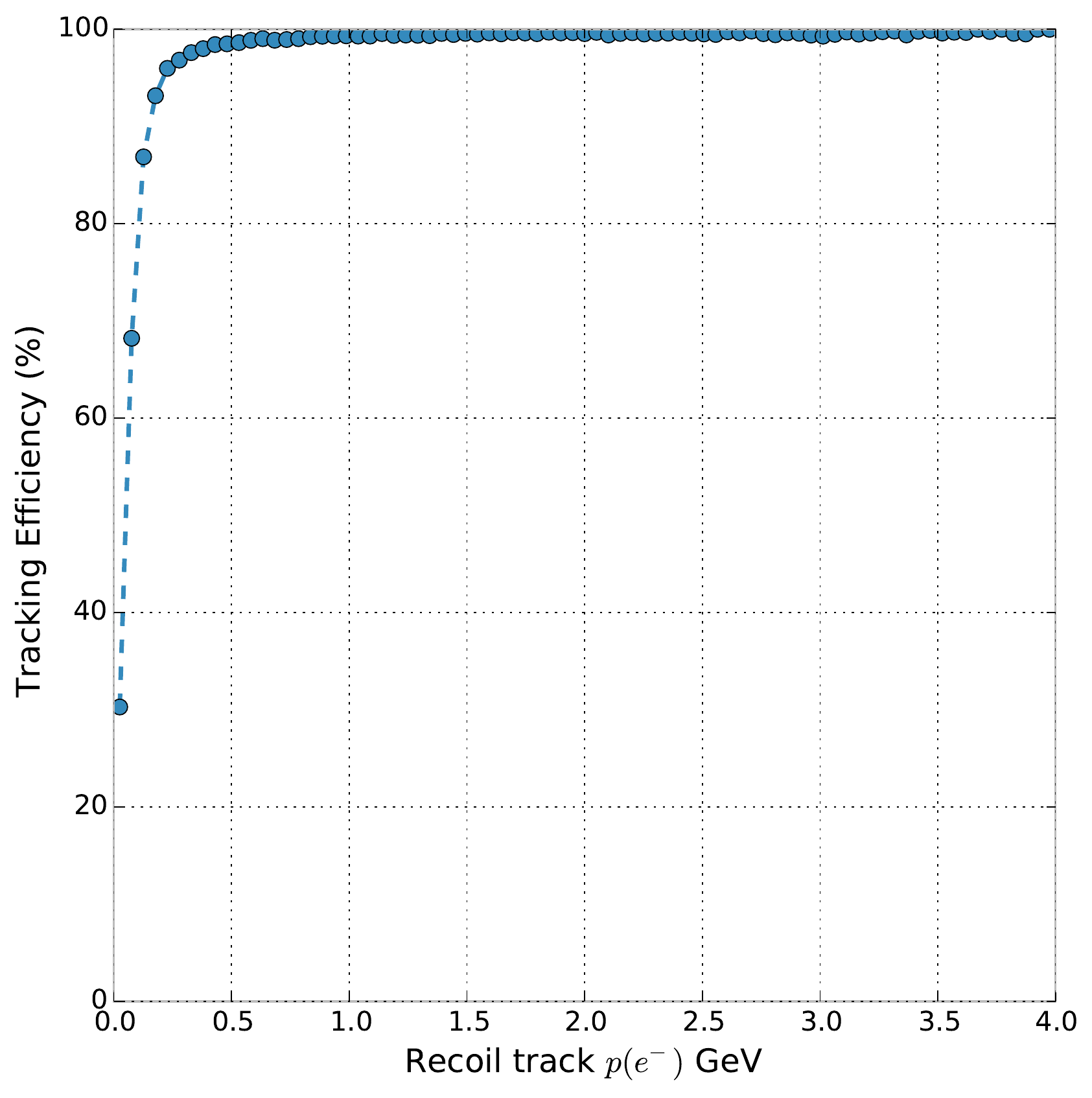}
    \caption{The tracking efficiency for signal recoils as a function of the
    momentum for electrons that pass the acceptance criteria described in the 
    text. Since data from all \aprime mass points are used, these efficiencies
    are averaged with respect to different recoil polar angle distributions 
    for each sample.}
    \label{fig:tracking_eff}
\end{minipage}

\end{figure}

As discussed in Sec. \ref{subsec:recoil_tracker}, the recoil tracker must not 
only have good acceptance for signal recoils but also for charged tracks over 
the largest possible acceptance.  Since high-momentum signal recoils will 
nearly always pass through all six layers, the acceptance near the top of the 
energy range for signal recoils is near unity, only reduced by the small 
single-hit inefficiency in the last two layers. However, at low momentum a 
large number of tracks can escape detection. Therefore, in order to estimate
the signal acceptance using simulated signal events, we apply both 
``loose'' and ``tight'' track requirements.  A loose track requires that the 
recoiling electrons leave hits in the first two 3-d layers which is 
sufficient for pattern recognition and angle estimation. For those events where
the best vertex and $p_T$ resolution is desired, the recoiling electrons are
required to leave hits in at least two of the 3-d layers and at least four
hits total. The resulting acceptance for the recoil tracker as a 
function of mediator mass using the loose (orange) and tight (green) track 
requirements is shown in  Fig.~\ref{fig:tracking_accept} while the average of 
the track finding efficiencies for accepted tracks in those 
samples as a function of momentum is shown in Fig.~\ref{fig:tracking_eff}.

\begin{figure}[!hb] 
    \centering
    \includegraphics[width=.73\textwidth]{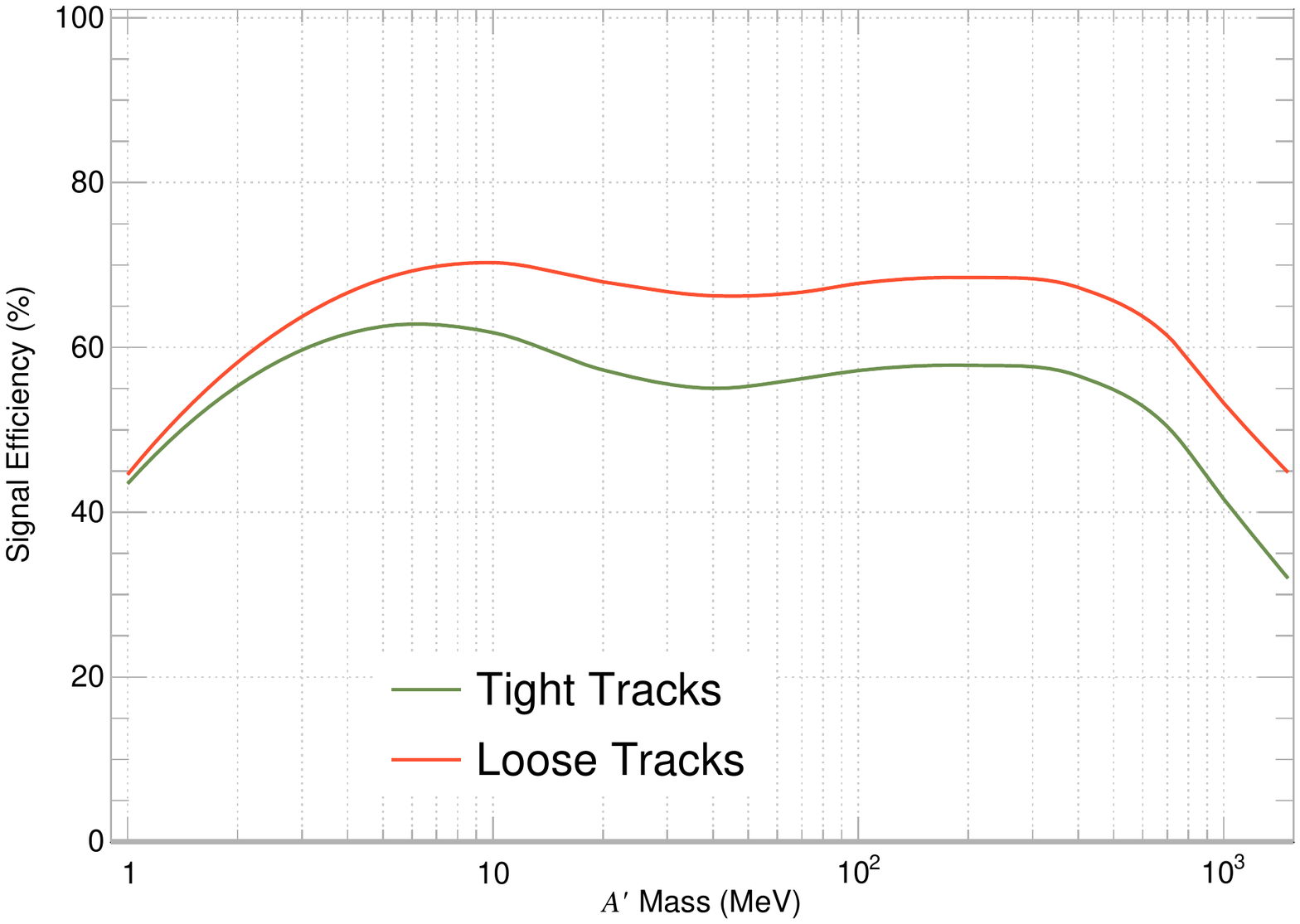}
    \caption{The signal efficiency as a function of mediator mass after requiring
             an event to have only a single track below 1.2 GeV.  The signal 
             efficiency assuming a loose track requirement is shown in orange
             while the efficiency after imposing a tight track requirement is
             in green.}
    \label{fig:tracker_sig_eff}
\end{figure}

The recoil tracker will play a leading role in rejecting beam, \pn, \en and 
muon conversion backgrounds that originate in the target.  Assuming the average 
number of incident beam electrons is 1, most signal events will see a single
track in the recoil tracker.  Furthermore, as discussed in 
Sec.~\ref{sec:proc_sig}, only events where the recoil electron has a momentum 
below 1.2 GeV are considered.  Therefore, requiring an event to contain a single
track in the recoil tracker with reconstructed momentum $<1.2$ GeV, aides in the
rejection of background while maintaining a high signal efficiency. The impact
of these cuts on the  signal efficiency using both the loose and tight track
requirement is shown in Fig.~\ref{fig:tracker_sig_eff}.

\subsection{Signal Efficiency of Calorimeter Vetoes \people{Omar}}

\begin{figure}[!hb] 
    \centering
    \includegraphics[width=.7\textwidth]{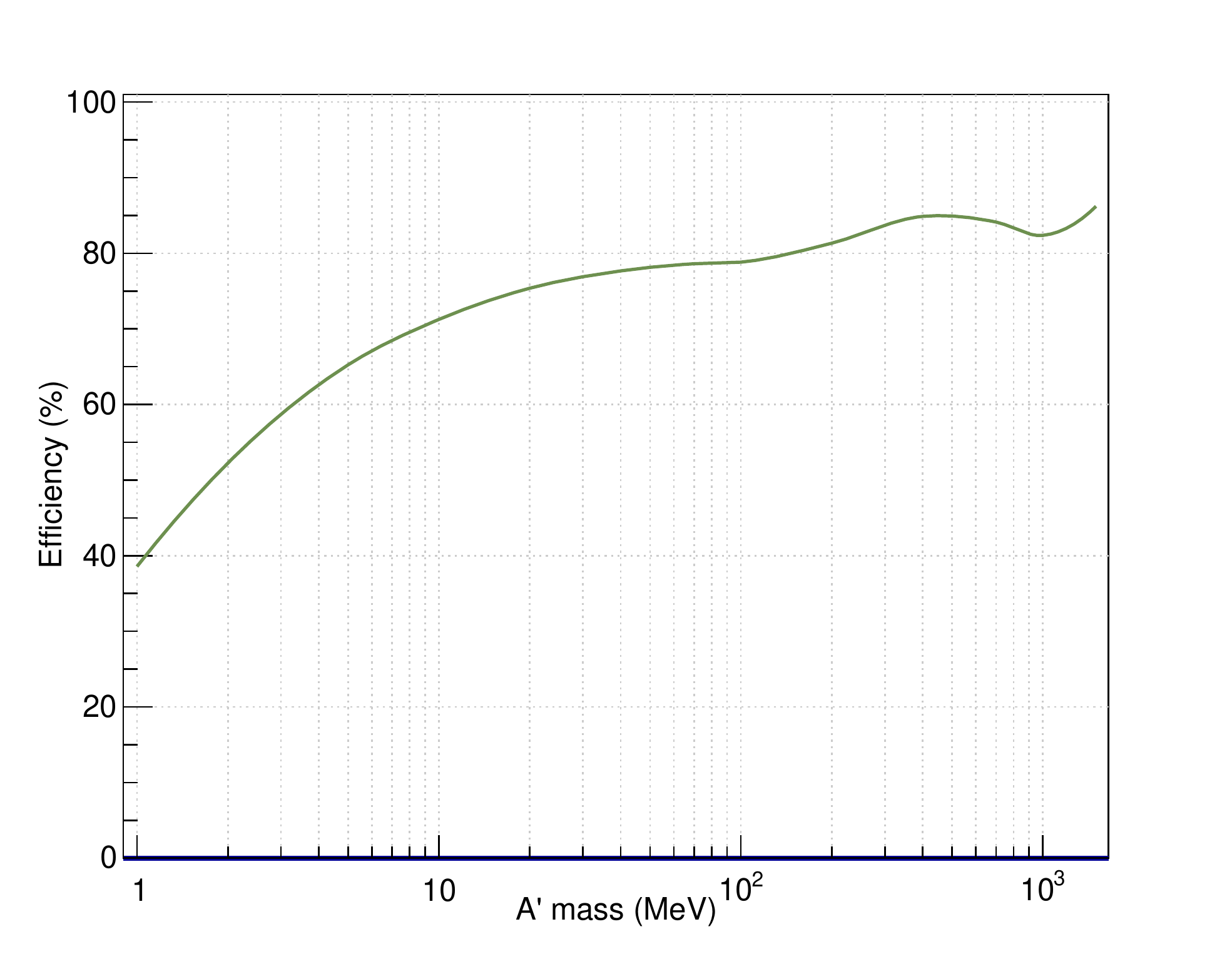}
    \caption{The signal efficiency as a function of mediator mass after applying
             a BDT threshold of 0.94 on an inclusive signal sample.}
    \label{fig:ecal_sig_eff}
\end{figure}

As detailed in Sec.~\ref{sec:detector.ecal}, the difference in hit multiplicity
and energy deposited in the \ecal as well as the transverse and 
longitudinal shower shapes are combined into a BDT and used to discriminate 
between signal and \pn background.  Based on the performance of the BDT, a 
threshold of 0.94 was found to balance the rejection of background versus the 
preservation of signal.  The impact of applying such a requirement on an inclusive 
signal 
sample is shown in Fig.~\ref{fig:ecal_sig_eff} as a function of mediator mass.

\begin{figure}[!thbp]

\begin{minipage}{0.48\linewidth}  
\centering
    \includegraphics[width=8.2cm]{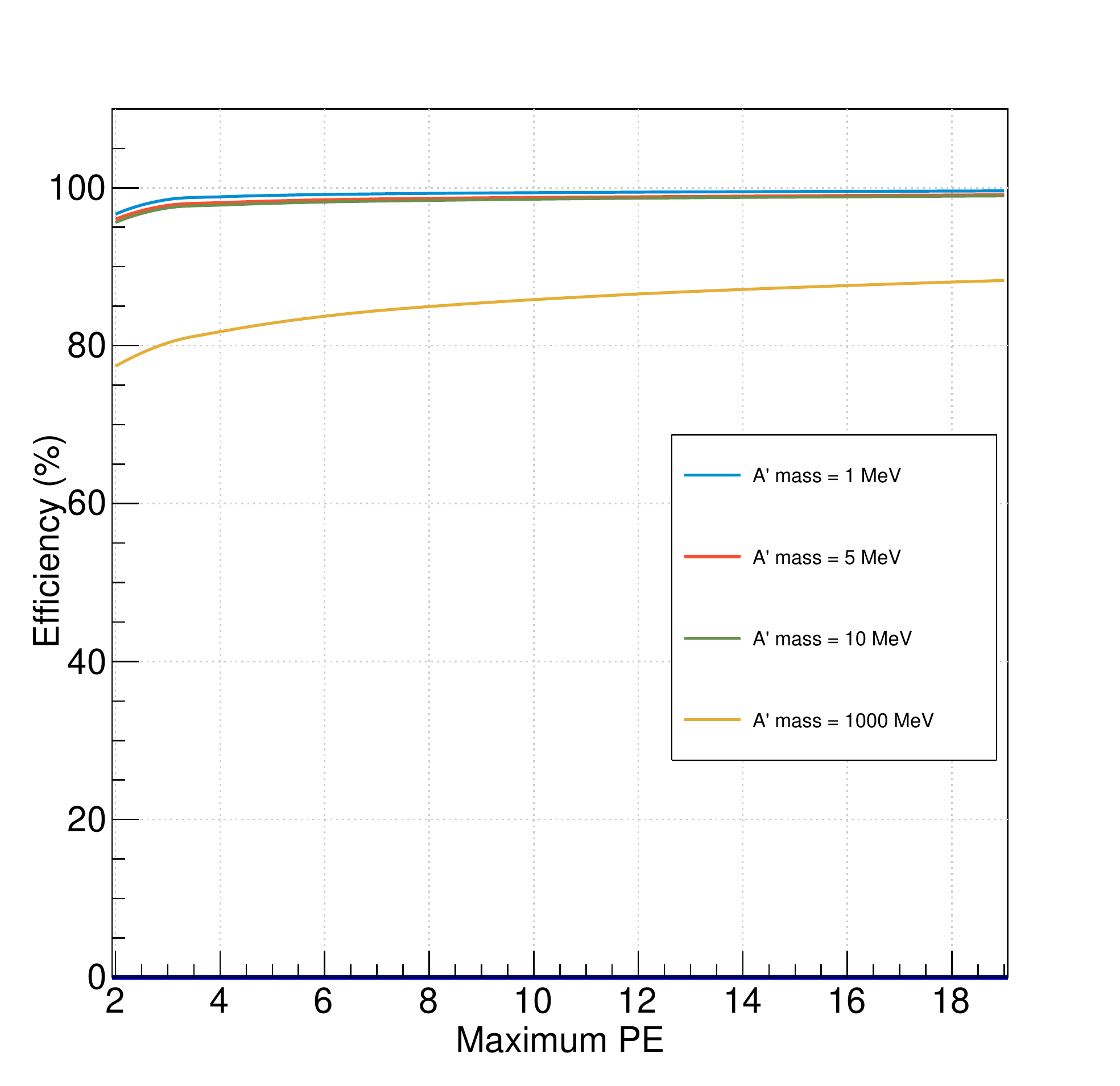}
    \caption{The signal efficiency as a function of the mediator mass and the largest PE yield measured in the \hcal.}
    \label{fig:hcal_sig_eff_pe}
\end{minipage}\hfill
\begin{minipage}{0.48\linewidth}
  \centering
    \includegraphics[width=8cm]{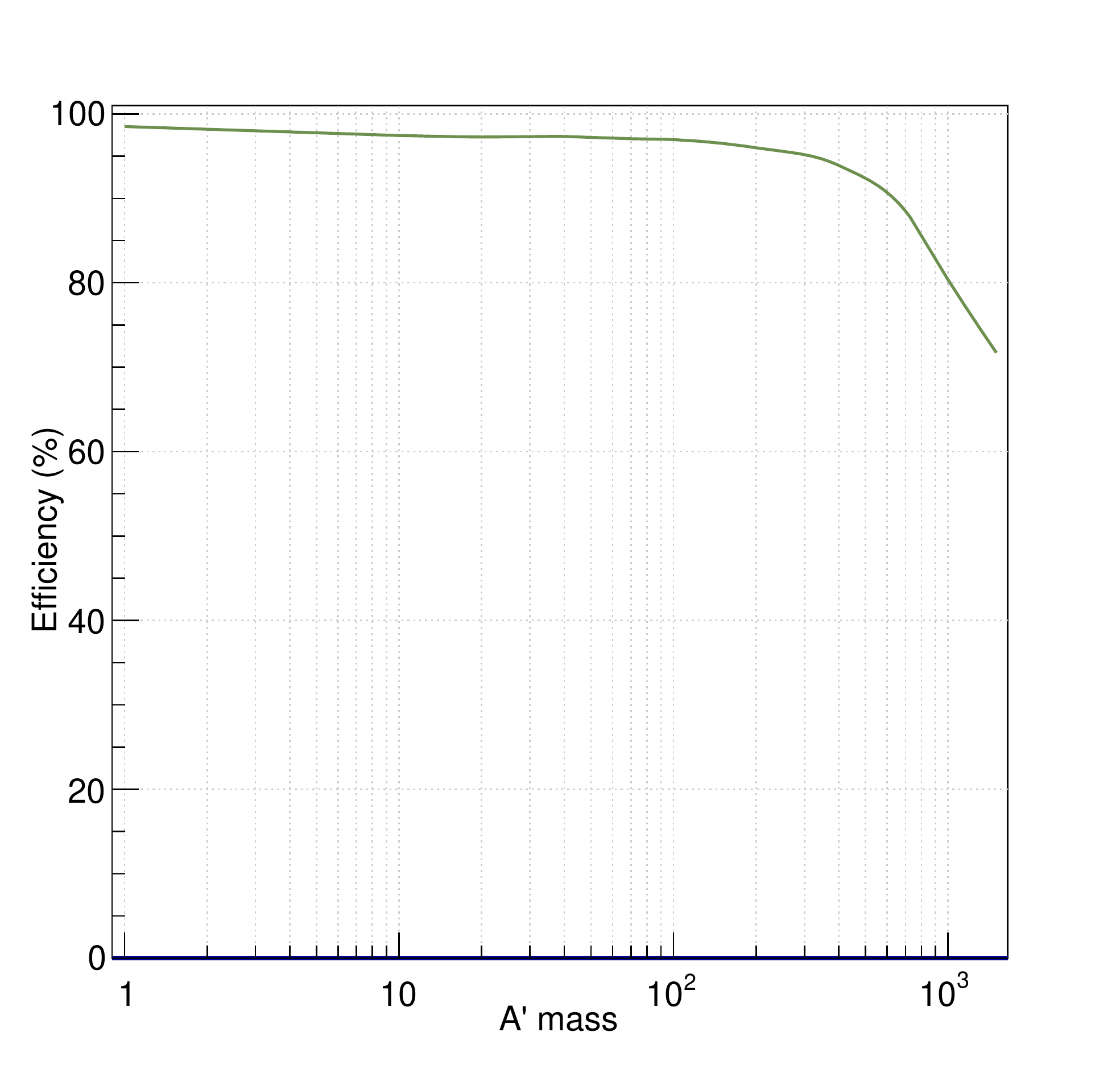}
    \caption{The signal efficiency as a function of the mediator mass after 
            requiring a max PE less than 3 in the \hcal. The drop in efficiency at high 
            mass is a result of signal recoils entering the \hcal and showering
            resulting in multiple PE's.}
    \label{fig:hcal_sig_eff}
\end{minipage}

\end{figure}

A typical signal event will see the recoil electron enter the \ecal and shower
with very little energy escaping into the \hcal.  In fact, assuming the average
number of incident beam electrons is 1, most signal events will see a maximum PE
in the \hcal of 1 PE due to the expected noise contribution 
(see Sec.~\ref{sec:hcalperf}).  In contrast, backgrounds will leave multiple 
hits in the \hcal resulting in multiple PE's per layer.  This is shown 
in Fig.~\ref{fig:hcal_sig_eff_pe}, which shows the signal efficiency for 
different \aprime masses along with the \pn background efficiency as a function of
maximum PE.  By requiring that the maximum PE observed in an event is $<3$, the
signal efficiency is maximized while also rejecting a large portion of the
background.  The impact of this cut on an inclusive signal sample is shown in 
Fig.~\ref{fig:hcal_sig_eff}.  The drop in 
efficiency at high mass, is a result of signal recoils entering the \hcal and
showering, resulting in multiple PE's.  Future iterations of the veto will 
mitigate the inefficiency by excluding the \hcal region where the shower 
occurs when determining the maximum PE. 

\subsection{Signal Efficiency Summary}

\begin{figure}[!ht] 
    \centering
    \includegraphics[width=.7\textwidth]{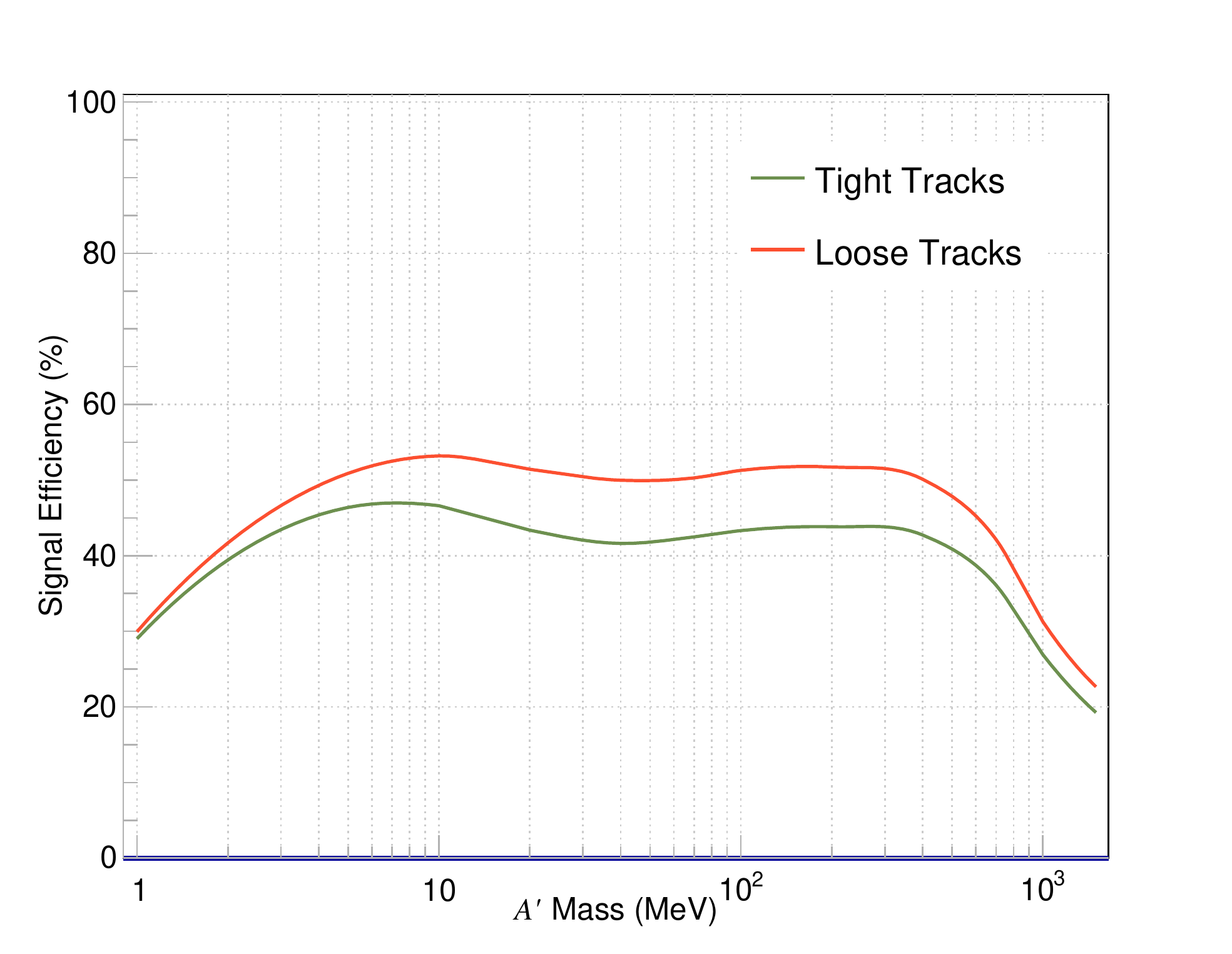}
    \caption{The signal efficiency after applying all background rejection 
             cuts.} 
    \label{fig:sig_eff}
\end{figure}

The signal efficiency after applying all background rejection cuts discussed 
above is shown in Fig.~\ref{fig:sig_eff} and summarized for three \aprime
mediators in Tab.~\ref{tab:sig_eff}.  After all cuts, the signal efficiency
is approximately 50\%.  This efficiency is used in the calculation of the 
expected sensitivity described in the next section.

\begin{table}[!b]
    \centering
    \begin{tabular}{ l | c c c  } 
        
        \multicolumn{4}{c}{Signal Cutflow} \\ [0.5ex]
        \hline\hline
        & \textbf{1 MeV} & \textbf{100 MeV} & \textbf{1000 MeV} \\ [0.5ex]
        \hline
        & \multicolumn{3}{c}{$\epsilon$ (\%)} \\
        \hline
        Trigger requirement & 57 & 91 & 97 \\ 
        \ecal Shower-Profile BDT         & 38 & 79 & 82 \\ 
        \hcal Max PE $<3$                & 37 & 76 & 63 \\ 
        Single track with $p < 1.2$ \GeV & 31 & 56 & 34 \\ 
        Recoil activity Cut              & 30 & 54 & 33 \\ 
        \ecal activity cut               & 29 & 53 & 32 \\ 
        Tagger tracker activity cut      & 29 & 51 & 31 \\ 
        \hline
    \end{tabular}
    \caption{Signal efficiency for 1, 100 and 1000 MeV after successive 
             background rejection cuts.}
    \label{tab:sig_eff}
\end{table}

\clearpage
\section{Sensitivity and Background Considerations}\label{sec:phase1}
In this section we evaluate the expected sensitivity of a  $4 \times 10^{14}$ EOT LDMX run, based on the  the preceding discussions of background rejection, signal efficiency, and dark matter production cross-sections.  We highlight the value of electron $p_T$ measurement: in the event that a significant deviation from the expected background rate is observed, the recoil electron $p_T$ spectrum will be a powerful tool for characterizing events as signal- or background-like and can even be used to constrain the mass-scale of the mediator and/or dark matter particles being produced. 

For the purpose of this discussion, we focus attention on one prominent benchmark model: production of mediators decaying into dark matter.  The breadth of sensitivity of the LDMX missing momentum search to other dark matter models is discussed in Section \ref{sec:breadth}.

\begin{figure}[htbp]
\includegraphics[width=0.7\textwidth]{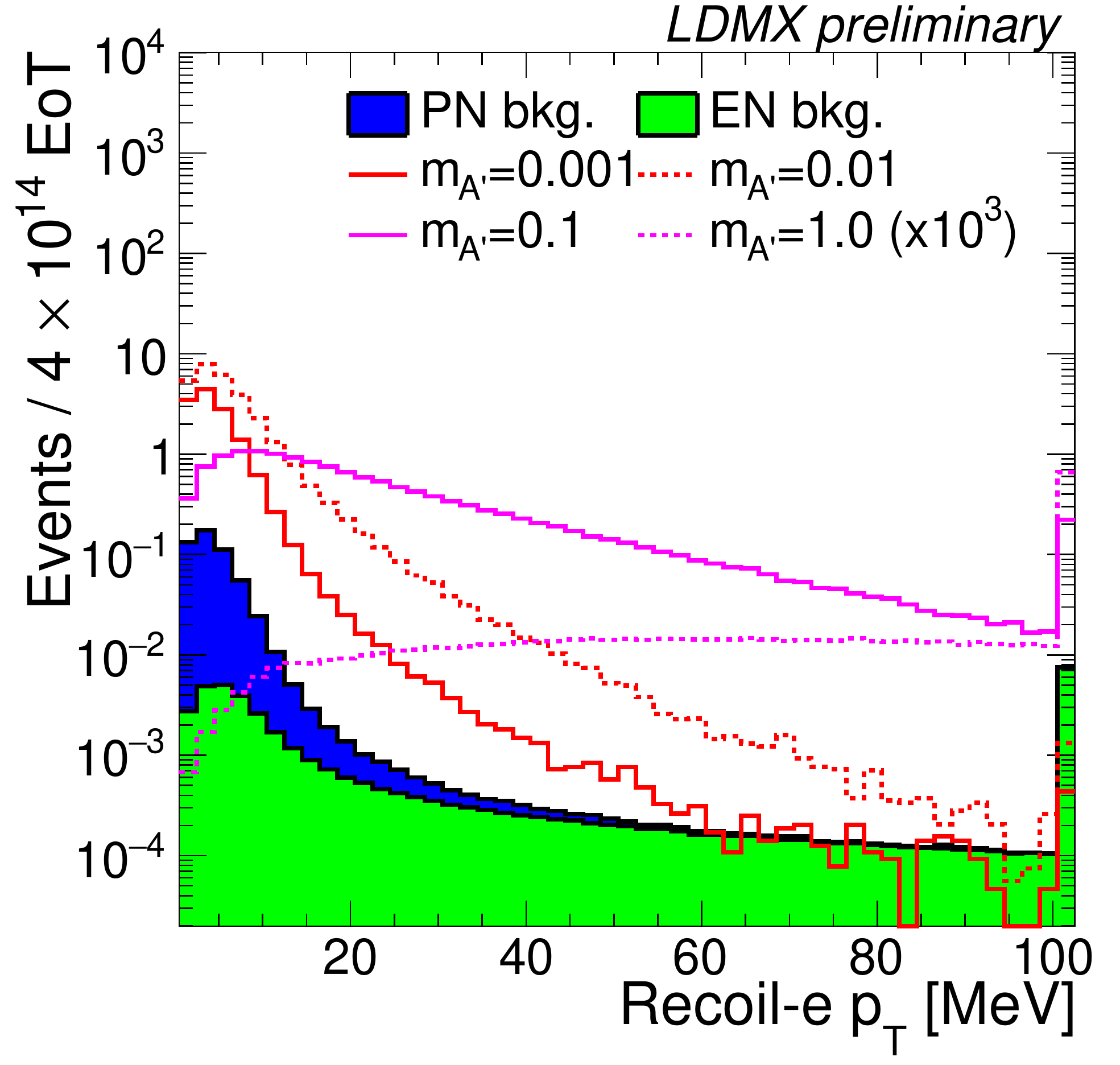}
\caption{\label{fig:recoil_pt_stack} Distribution of recoil electron transverse momentum $p_T$ for backgrounds (solid histograms) and dark matter signals with mediator masses of 0.001, 0.01, 0.1, and 1 GeV after all analysis selections.  Signal yields are scaled to the thermal freeze-out elastic scalar dark matter model, assuming $\alpha_D=0.5$ and $m_{\chi}/m_{A'}=1/3$. Among other kinematic measurements, both recoil electron transverse momentum and missing momentum will provide considerable kinematic discrimination between background and signal, as well as sensitivity to the mediator and dark matter mass.}
\end{figure}

The simulation studies presented here indicate an expected background of $\leq 0.5$ events (and this is expected to improve considerably as our analysis studies develop).  The expected recoil electron $p_{T}$ distribution for dark matter signal and backgrounds, normalized to this expectation, are shown in Fig.~\ref{fig:recoil_pt_stack}.  The signal distributions are normalized to the cross section expected from the thermal freeze-out elastic scalar dark matter model, assuming $\alpha_D=0.5$ and $m_{\chi}/m_{A'}=1/3$. The background $p_T$ distributions are essentially independent of the \ecal BDT discriminator and \hcal signal, and so can be reliably estimated from control samples where one of the \hcal or \ecal veto is inverted or loosened.  As illustrated, the signal $p_T$ distributions are readily distinguished from those of the background in most of the mass range of interest, $m_{A'}\gtrsim 5$ MeV.   With enough signal events, the $p_T$ distribution could also be used to estimate the mediator mass, assuming on-shell mediator production as the signal.  
 
\begin{figure}[htbp]
\includegraphics[width=0.7\textwidth]{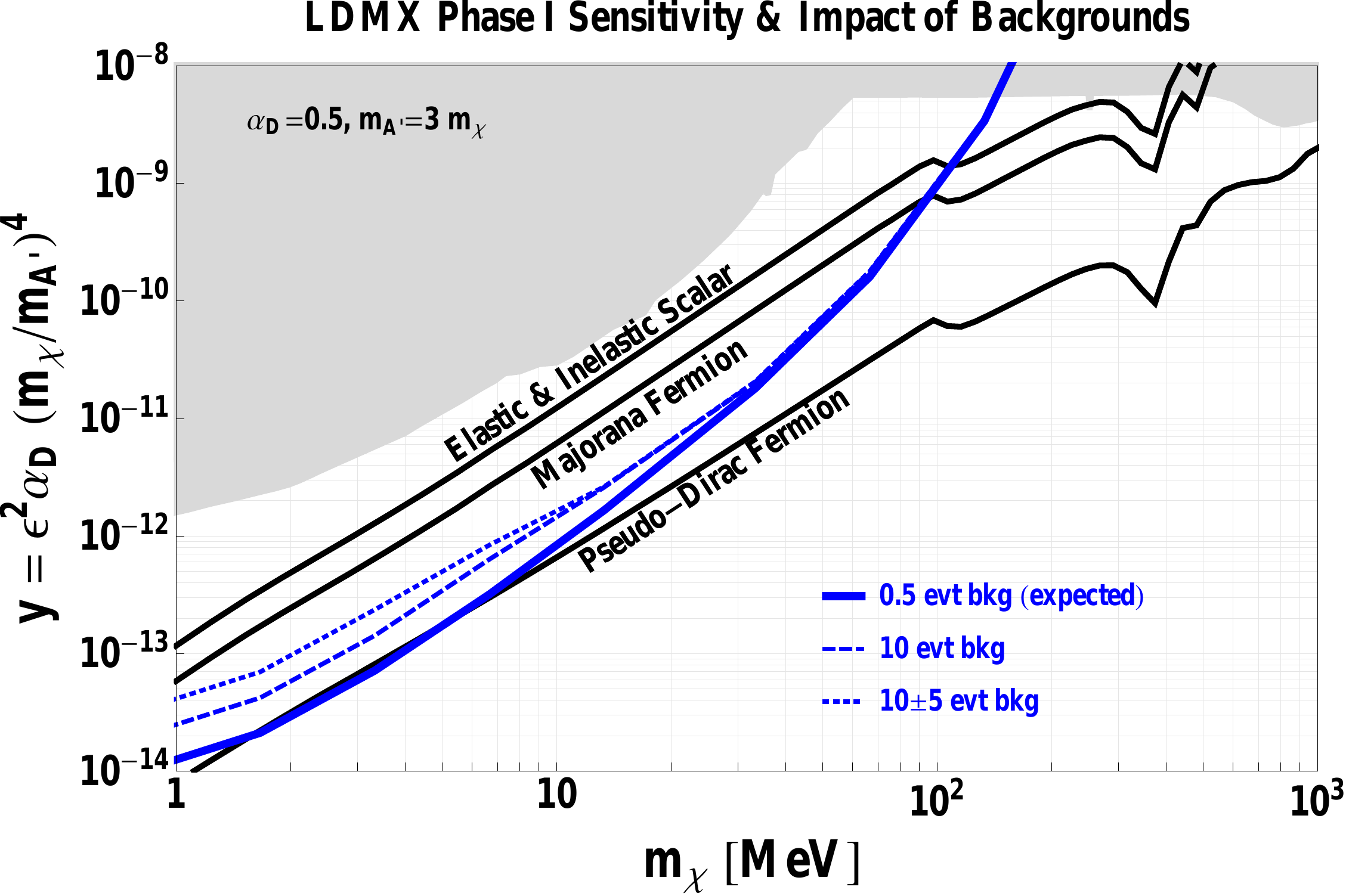}
\caption{\label{fig:reach1} Projected sensitivity in the $y$ vs. $m_{\chi}$ plane for a $4 \times 10^{14}$ EOT $4 \ \GeV$ beam energy LDMX run (solid blue curve), for the case of on-shell mediator production and decay into dark matter. Thermal relic targets are shown as black lines. Grey regions are (model-dependent) constraints from beam dump experiments and \babar. The dashed and dotted curves illustrate the robustness of this search to any unexpected photo-nuclear backgrounds at the 10-event level.  In this case, a mass-dependent optimized $p_T$ cut can be used to reduce the background level, recovering nearly the same sensitivity at high dark matter masses.  The dotted line further assumes, pessimistically, that such background can only be normalized to within a 50\% systematic uncertainty using veto sidebands as control regions.}
\end{figure}

Based on the total expected yields, assuming 0.5 background events and with no additional selections, we obtain the 90\% C.L. sensitivity illustrated by the thick blue line in Figure \ref{fig:reach1} for the model of on-shell mediator decay into dark matter.  We note that, given the low expected background rates, the reach shown here may be considered conservative. For example, thickening the target by a factor of few could boost sensitivity by a similar factor, and loosening the track requirements would additionally enhance the high-mass signal acceptance by 30--50\% above what is shown. It is likely that either of these directions could be pursued without compromising the low-background expectations --- though determining to what degree requires further study.   At this level of background rejection, there is no benefit to a further selection on electron $p_T$.  However, if unexpectedly larger backgrounds are encountered, electron $p_T$ serves as a crucial discriminator and, over much of the mass range of interest, allows us to recover the low-background sensitivity.  This is illustrated by the dashed line in Fig.~\ref{fig:reach1}, where we have assumed a total background of 10 events with a bremsstrahlung-like electron $p_T$ distribution (accounting for the electrons' scattering in the target).  For dark matter masses above $\sim 5$ MeV, the $p_T$ distribution of signal is substantially broader than the background $p_T$ distribution. A simple mass-dependent $p_T$ cut can be used to recover excellent sensitivity to dark matter in the presence of unexpected backgrounds. The dotted line further illustrates the effect of systematic uncertainties in the background rate (for the sake of illustration, a pessimistic 50\% uncertainty is shown). The robustness of our sensitivity is encouraging, as is the overall sensitivity with respect to direct thermal freeze-out predictions.

\clearpage
\section{Extending the reach of LDMX}
\label{sec:ExtendedReach}

In the preceding section, we have demonstrated the potential for LDMX to achieve sensitivity to dark photons decaying to dark matter in the mass range below 1~GeV (see \cite{LDMXSciencePaper} for other applications). We assumed $ 4\times10^{14}$ electrons incident upon a 0.1$X_0$ tungsten target at an average rate of one 4~GeV electron per 20~ns. This could be achieved in $\sim$1~year of operation with a live time of $\sim 10^{7}$ seconds. With this Phase I of the experiment, LDMX would be able to probe much of the thermal freeze-out parameter space for scalar and Majorana dark matter. In this section we discuss ways to extend the reach of LDMX to cover additional milestones such as more decisively exploring the challenging pseudo-Dirac fermion dark matter parameter space, wherever possible. To this end, we consider what might realistically be achieved by means of small changes to the Phase I configuration and estimate how our reach can be extended by extrapolating from our previous results. We then discuss more substantial changes to the experiment that could potentially provide significant enhancements, but whose impacts cannot be as easily extrapolated from our Phase I studies. These enhancements can be taken conservatively to represent ways of shoring up the results of the earlier extrapolations, but they may also enable significant extension of the reach of LDMX. A more detailed study is required to fully quantify LDMX Phase II performance. This will be the subject of a future note.
\vskip 0.2cm
{\bf Extrapolating from Phase I.} 
\vskip 0.2cm
For the thermal-relic scenarios we are considering, the production of dark photons, as well as the final state kinematics of both signal and background events, vary with mass. With this in mind, and in anticipation of other mass-related considerations, we break the full mass range of interest into four regions of $m_{\chi}$ defined as follows: $(0.01,20]$, $(20,75]$, $(75,150]$ and $(150,300]$~MeV, corresponding to $0.03 < m_{A'} \le 900$~MeV for $m_{A'}/m_{\chi}=3$. The choice $m_{A'}/m_{\chi}=3$ and $\alpha_D=0.5$ is motivated by the desire to consider regions of parameter space that are challenging for experiments to reach, but not tuned to the especially challenging resonance region centered on $m_{A'}/m_{\chi}=2$. We now show that most of the remaining pseudo-Dirac fermion parameter space may be reached for the first three mass ranges by means of modest and or easily accessible extensions of the Phase~I experiment, while for the highest mass range, more significant changes may be required.  The former include: (i) increased thickness of the target from 0.1$X_0$ to values ranging from 0.15$X_0$ to 0.4$X_0$, (ii) increased electron beam energy of 8 GeV as currently being planned for the LCLS-II high energy upgrade (LCLS-II HE), and (iii) an increase in the average number of electrons per 20~ns to $\mu_e=2$. For the two highest mass ranges we also consider: (iv) increasing the beam energy to 16~GeV, as would be possible in a CERN implementation of LDMX, and (v) changing the target material from tungsten to aluminum. For the highest mass range we consider (vi) a multiplicity of 5-10 electrons per 20~ns. 

The impact of (i) is straightforward to estimate since the hard bremsstrahlung rate is linearly proportional to the target thickness.  A potential drawback of a thicker target is increased multiple scattering of the recoil electron. This adds uncertainty to the recoil electron $p_T$ - an important quantity that can help discriminate signal from background, or provide an estimate of an $A'$ mass associated with candidate signal events. However, since the $p_T$ generated by multiple scattering goes as the square root of target thickness, the change in resolution for small changes in thickness is modest.
We can clearly increase the target thickness in the two highest mass ranges where the recoil electron $p_T$ for signal events is significantly greater than the level of dispersion introduced by multiple scattering in even a target of 0.4$X_0$. For the lower two mass ranges, we consider increasing target thickness to 0.15$X_0$.

\begin{figure}[htbp]
\includegraphics[width=0.8\textwidth]{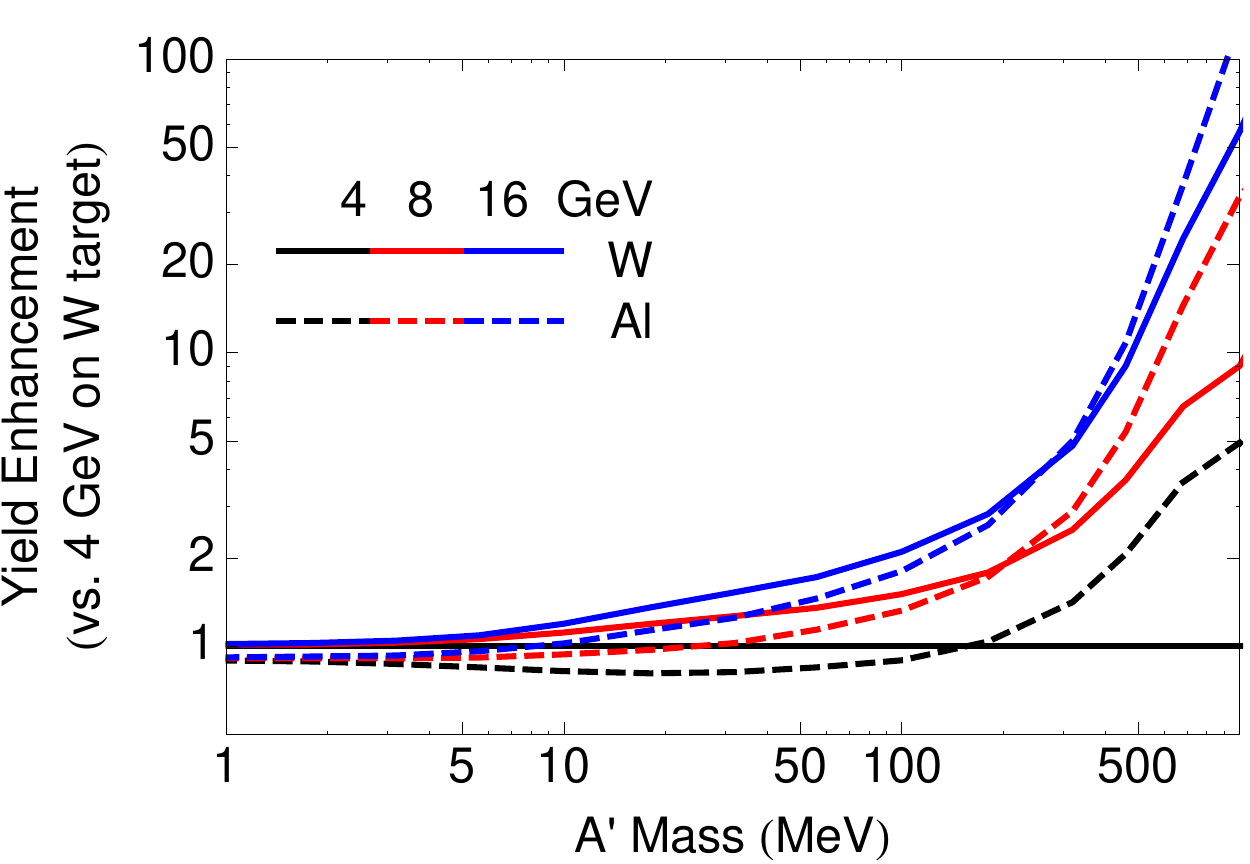}
\caption{\label{fig:xsecratio} Both beam energy and target material affect the dark photon production cross-section, with especially large effects at high masses.  This figure illustrates how increasing the LDMX beam energy to 8 or 16 GeV, and/or switching from Tungsten to Aluminum targets (at 10\% $X_0$ in each case), impacts the signal production cross-section for different dark photon masses.  In each case, we assume the kinematic selection $E_{recoil}<0.3 E_{beam}$ as was used in 4 GeV studies. This is conservative for higher-energy beams.}
\end{figure}

The impact of (ii), raising the beam energy, is also relatively straightforward to estimate since it can directly impact the signal production cross section. The increase is negligible in the lowest mass range, but becomes increasingly significant at higher masses, as illustrated in Figure \ref{fig:xsecratio}. 

One consequence of (iii), doubling the mean number of electrons on target per 20~ns sampling time, is a somewhat more challenging environment for triggering and reconstruction. It should not present any show-stoppers. Nevertheless, it is not as straightforward to extrapolate the impact of this change. The most common signal-like event type would be one in which there is a potential signal, as defined for the case of one electron on target, but now accompanied by another beam electron that loses very little energy in the target and tracker, and showers in the calorimeter. The final state would contain an electron at beam energy and either a soft recoil electron and noise in the case of signal, or noise and the remnants of a photo-nuclear interaction in the case of background. Our Phase I configuration assumes a beam distribution of $\Delta x \times \Delta y = 2\times 8$~cm$^2$ at the target. It would be necessary to isolate and identify all products of the two electrons in this region. The electromagnetic shower produced in the \ecal by the electron at or near full beam energy could overlap with the photon or recoil electron from the electron that interacted in the target. If the photon undergoes a photo-nuclear interaction yielding hadrons that do not produce a large signal in the \ecal, then the potential to miss it is increased. One must either reject all events in which a beam electron in the \ecal is near to where a hard photon is expected to be, or the \hcal must be relied upon to veto such events. The latter is feasible, given the performance of the \hcal. 

It follows that an increase in the average number of electrons per sample period will mainly reduce acceptance. For exactly two electrons randomly distributed over a $2\times 8$~cm$^2$ target area, simple acceptance studies indicate that overlaps may occur $\sim 10$\% of the time for the  detector used in our Phase~I study.  An additional new source of acceptance loss is punch-through from the additional EM shower into the \hcal that would enhance the probability of an \hcal veto. For 2 electrons on target, we determine $\sim1.7~(3.0)\%$ loss of signal for a veto on total energy in the \hcal above a threshold of 8 PE for 4~(8)~GeV electrons.  

The impact of (v), changing the target from W to a lower Z material such as Al, is an increase in the production cross section for more massive dark photons. As mass increases, the minimum momentum transfer $q_{min}=m_{A'}^2/(2 E_{beam})$ increases. When the wavelength associated with momentum transfer becomes comparable to or smaller than the size of the nucleus, coherence in the scattering begins to degrade, resulting in a suppression of the cross section. For smaller (lower Z) nuclei, the onset of this behavior is pushed to higher $A'$ mass. We calculate that a switch to Al from W for a 4~GeV electron beam will produce a factor of two or more increase in cross section for $A'$ masses above $\sim$400~MeV as seen in Figure \ref{fig:xsecratio}. For an 8~(16)~GeV electron beam, the cross section is doubled for $A'$ masses above $\sim$500~(900)~MeV.  A lower-$Z$ target does have the side-effect of increasing the photo-nuclear and electro-nuclear yield for given target thickness (which scales roughly as $A/Z^2$).  But we expect that this increased background can be readily rejected, especially for higher beam energies where photo-nuclear products are more collimated and the cross-sections for low-multiplicity final states are suppressed.

The impact of (vi), increasing the electron multiplicity to an average of 5-10 electrons per 20~ns, is only considered for an implementation where the beam could be spread over a much larger area. This is discussed in more detail below. 

For illustration purposes, we now present some operating scenarios for each of the four mass ranges as summarized in Table \ref{table:ExtendedLDMX}. For the lowest mass range, $0.01\le m_{\chi}<20$~MeV, the signal production rate is relatively high and the pseudo-Dirac fermion target can be reached with a factor of $\sim$2 increase in the luminosity relative to Phase~I. This could be achieved with $\sim$2 years of additional running or one could reduce the required run time by increasing the thickness of the target and/or doubling the mean number of electrons on target. There is no real gain from increased beam energy. Several options for achieving the desired increase in less time are shown in Table \ref{table:ExtendedLDMX}.  For $20\le m_{\chi}<75$~MeV the pseudo-Dirac fermion target can be reached with a factor of $\sim$6 increase in luminosity relative to Phase-I. There is a small impact from increased beam energy on production of dark photons. We can also increase the W target thickness slightly to 0.15~$X_0$ and/or double the number of electrons on target. 

The $75\le m_{\chi}<150$~MeV mass region is more challenging, with nearly two orders of magnitude increase in sensitivity relative to Phase~I needed to reach the pseudo-Dirac fermion target. Fortunately, in this mass range we gain as much as a factor of $\sim$4 in the $A'$ production cross section by switching to an 8 GeV beam. Another factor of 4 is obtained by using a 0.4~$X_0$ W target.  This leaves us to find a factor of $\sim$6, which can be achieved in several ways. One could double the number of electrons on target and run for $\sim$3 years. Alternatively the tungsten target could be replaced by Aluminum to gain a factor of 1.5, cutting the run time to 2 years. Shifting to the CERN implementation of LDMX, one could run at 16 GeV to gain a factor of 4 relative to 8 GeV for a Tungsten target. One could then cut back to an average of one electron per sample period and run 2 years.

The $150\le m_{\chi}<300$~MeV mass region is the most challenging, with a factor of $6\times 10^3$ increase in sensitivity needed to reach the pseudo-Dirac fermion target at $m_{\chi}=300$~MeV. This cannot be achieved even with 4 years of running with 16 GeV electrons, a 0.4$X_0$ W target, and a $\mu_e =2$ electrons per 20~ns sample. However, with an Aluminum target, the pseudo-Dirac target fermion could be reached with pileup of 5-10 electrons on target. In this case, one would need to make use of a broader beam of order 20-30~cm~$\times$~4-8~cm. The average electron density at the target would be about one-third  that of the Phase I experiment but it would nevertheless be a significantly more complicated environment, where it is harder to maintain high signal efficiencies and high background veto efficiencies. For instance, for  $\mu_e =5$ electrons per 20~ns sample, the rate at which punch through would induce an \hcal veto for events with more than 8 PE is 8~(14)~\% for 4~(8)~GeV beams. 

Table \ref{table:ExtendedLDMX} lists all of the illustrative scenarios discussed above. Taken at face value, the scenarios imply that the lowest two mass regions could be probed to the level of the pseudo-Dirac fermion target in 2 years of running with a 0.15$X_0$ W target, and an average of one 8~GeV electron per 20~ns sample period. The third highest mass range could be probed to the pseudo-Dirac fermion target with an 8~GeV electron beam in 2 years of running with a 0.4$X_0$ Al target and an average of 2 electrons per sample period, or 1 year of running with a 16 GeV beam, a 0.4$X_0$ W or Al target, and an average of one electron per 20~ns. Assuming one could achieve efficient discrimination of signal and noise with 2 electrons per sample period, the first 3 mass ranges can be covered with LDMX at SLAC in as little as $\sim~2$ years of running with the upgraded LCLS II-HE accelerator. 

For the highest mass range, 150 $\le M_{\chi}<$ 300 MeV, the running scenarios described above will cover substantial new ground, albeit leaving gaps in the pseudo-Dirac fermion parameter space. Although Belle II may provide good coverage above $\approx$100 MeV, it would be prudent to consider how LDMX might also cover this region. This may be achievable with 16~GeV electrons at an average multiplicity $\mu_e =5-10$ electrons per 20~ns sample on an Al target, spread over $30\times 4$~cm$^2$. For 10 electrons on target this would require a $\sim$2 year run.


\begin{table}[ht]
\begin{center}
\caption{\label{table:ExtendedLDMX}
Extrapolating to Phase~II. The ``Factor needed" is the increase in luminosity relative to Phase I that is needed to reach the pseudo-Dirac fermion thermal relic sensitivity for $m_{A^\prime}/m_{\chi}=3$ and $\alpha_D=0.5$ for masses in the range indicated. This is by no means the only useful figure of merit for improvement, but it nonetheless provides a concrete starting point for design and performance considerations. The "Factor achieved" is the approximate increase for a combination of changes to the Phase I experiment as indicated. The scenario denoted by ``*'' is reflected in Figure \ref{fig:extendedLDMX}}
\vskip 0.2cm
\small
\begin{tabular}{|c|c|cc|c|c|c|c|c|c|}
\hline             
Mass Range & Factor && E$_e$ & E$_e$ & Target & Target &$\mu_e$& Years & Factor\\
$[$MeV$]$  & needed   && $[$GeV$]$ & Factor &$[X_0]$& Factor & & running &achieved\\ \hline 
\hline
   		         &    && 4 & 1 & 0.15 W & 1.5   & 1.5 & 1  &         \\ 
   $0.01\le M_{\chi}<20$ & 2 && 4 & 1 & 0.1 W  & 1   & 1.5 & 1.5 & $\sim$2 \\ 
  		         &   && 4 & 1 & 0.15 W & 1.5 & 1 & 1.5   &          \\ 
		\hline
   		       &    && 8 & 2 & 0.1 W  & 1   & 2 & 1.5 &        \\ 
  $20\le M_{\chi}<75$  & 6  && 8 & 2 & 0.15 W & 1.5 & 1 & 2   & $\sim$6\\ 
                       &    && 4 & 1   & 0.15 W & 1.5 & 2 & 2   &         \\ 
         \hline
   		               &     && 8  & 4  & 0.4 W  & 4 & 2 & 3 &  \\ 
$75\le M_{\chi}< 150$ & 80 && 8 & 4  & 0.4 Al  & 6 & 2 & 2 &$\sim 80$\\
  		      &              & &16 & 8 & 0.4 W  & 4 & 1.5 & 1.5 & \\ 
   	                   &         && 16 & 8 & 0.4 Al & 4 & 1 & 2 & \\ 
         \hline
                       &               &*& 8 & 8  & 0.4  Al & 13 & 2 & 4 & $\sim 8\times10^2$\\ 
$150\le M_{\chi}<300$ & $6\times 10^3$&& 16   & 45 & 0.4 W & 4& 2 & 4 & $\sim 1\times 10^3$ \\ 
  	                  &  		       && 16 & 45 & 0.4 Al& 8& 5 & 4& $\sim 7\times 10^3$\\ 
  	                  &  		       &&16 & 45 & 0.4 Al& 8& 10 & 2& $\sim 7\times 10^3$\\ 

         \hline\hline
\end{tabular} 
\normalsize
\end{center}
\end{table}
\vskip0.2cm
{\bf More possibilities }
\vskip0.2cm

The preceding discussion highlighted a number of number of straightforward ways to extend the reach of LDMX. In contrast to these simple enhancements, increasing the electron multiplicity per 20~ns requires more significant changes to the experiment. As mentioned previously, the simplest way to deal with higher multiplicities is to spread the beam over a larger target area, of order $\Delta x \times \Delta y = (20-30)$~cm$\times (4-8)$~cm. This would require an apparatus with a larger spectrometer and ECal, while the HCal could be largely the same. 

In addition to increasing the acceptance of the trackers and ECal to accommodate a larger beam spot, there are other changes that can mitigate the effects of pileup. First, the beam can also be spread out in time. In all of the accelerator scenarios under consideration, a 5~ns bunch structure is envisioned, so that 4 bunches would be contained in each 20~ns sampling period. For the cases where the average electron multiplicity is two or more, one could imagine placing half of the mean number of electrons in each of the second and third bunches where charge collection efficiency is highest for 20~ns front-end readout. The tracker and calorimeters can easily distinguish a 5~ns difference in signal times, adding another dimension to help resolve overlapping objects. Second, increasing the granularity of the ECal - conceivably by a factor of two - can provide better $e/\gamma$ discrimination and improve the identification of minimum ionizing tracks from photonuclear reactions at high multiplicities.  Third, a stronger and/or longer magnet would provide better separation of recoil electrons and bremsstrahlung photons, which would improve the performance of the ECal at high multiplicities. Last, thick targets could be segmented to incorporate intermediate planes of scintillator or silicon to better image the topologies of multi-electron events.

Finally, there are other changes that would provide large improvements in reach for the highest mediator masses. For example, the very broad angular distribution for recoils results in very low recoil tracking efficiencies at high mass. Providing recoil detection that surrounds the target area could improve the signal acceptance at high mass by a factor of two. Given the rarity of such events, these detectors could be relatively simple in comparison to the trackers, such as overlapping layers of scintillator that surround the tracking volume.

From a conservative viewpoint, these changes to the Phase I detector presented earlier would enhance the ability of LDMX to achieve excellent coverage to the pseudo-Dirac fermion scenario in all four mass ranges in Table \ref{table:ExtendedLDMX}. From a more ambitious viewpoint, they could enable LDMX coverage to be extended to higher dark matter and mediator masses, and to explore far beyond the sensitivity milestones considered in this note. Projected Phase II sensitivity for conservative and optimistic scenarios are considered in the next section. A quantitative analysis of these possibilities will be the subject of a future note.

\vskip 0.2cm
{\bf Phase II Projected Sensitivity} 
\vskip 0.2cm

While several run configurations are given in Table~\ref{table:ExtendedLDMX} for illustrative purposes, in practice only one or two such configurations would be used for a ``Phase II'' LDMX run. For concreteness, we show the sensitivity of just one representative future scenario, denoted by the ``*'' line in Table \ref{table:ExtendedLDMX}. This corresponds to an $8 \ \GeV$ beam energy, $0.4 X_0$ Al target, and $\mu_e=2$, with a total run time of 4 years (assuming $\sim 30\%$ duty cycle). The sensitivity of this ``extended'' configuration is compared to the sensitivity of a ``Phase I'' configuration in most of the figures shown in the next section, as well as in Figure~\ref{fig:extendedLDMX} below. 

\begin{figure}[htbp]
\includegraphics[width=0.8\textwidth]{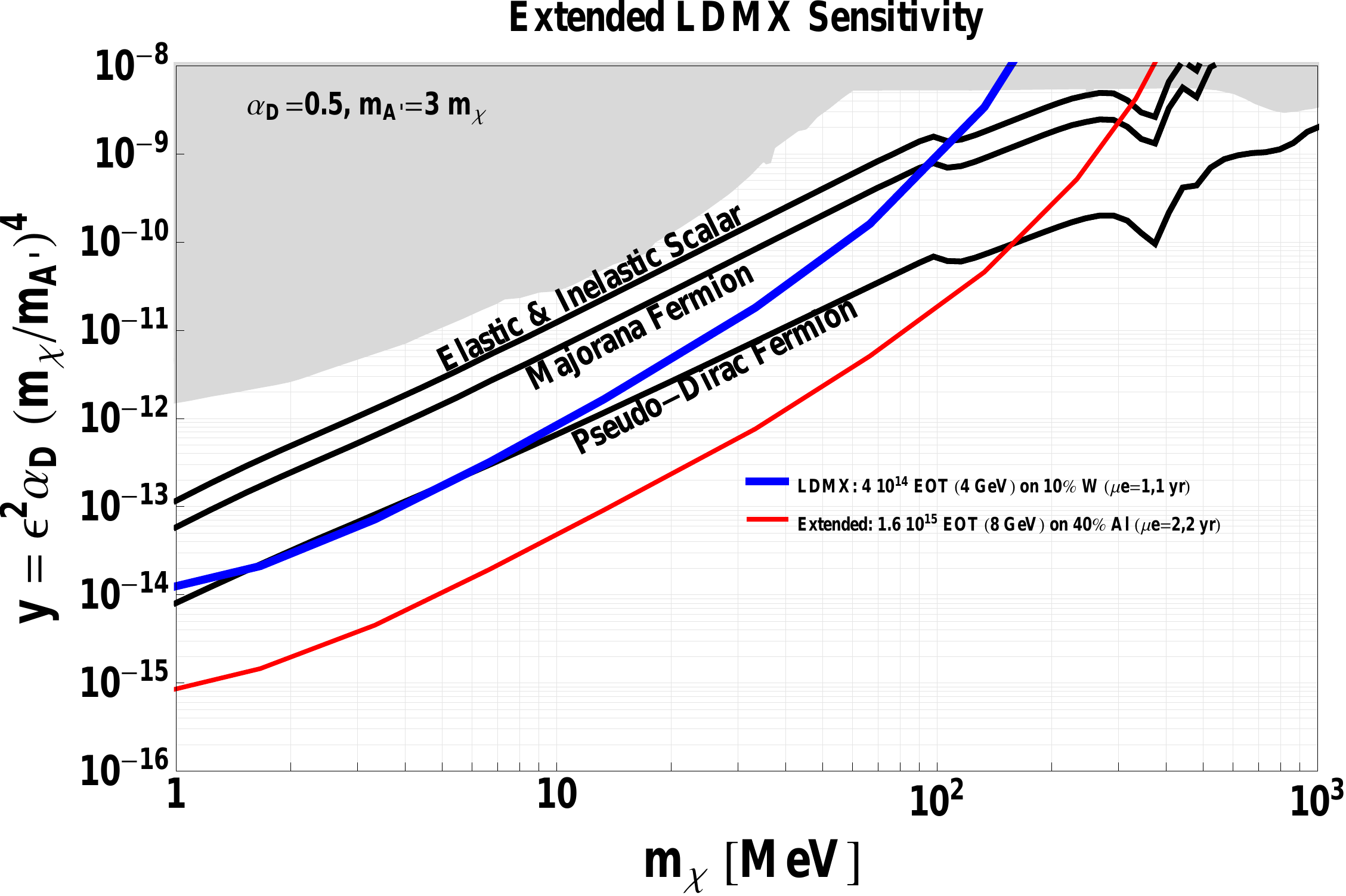}
\caption{\label{fig:extendedLDMX}The blue line is the sensitivity of the ``Phase I'' LDMX discussed throughout this whitepaper, conservatively assuming 0.5 background events.  A scaling estimate of the sensitivity of the scenario denoted by the ``*'' line in Table \ref{table:ExtendedLDMX} is illustrated by the red line.  We have again assumed low background, which is consistent with the expected reductions (relative to our $4 \ \GeV$ study) in both the yield of potential background, and improved rejection power at higher energies.}
\end{figure}

\clearpage
\section{Breadth of the LDMX Physics Program}\label{sec:breadth}
A measurement of missing momentum provides broad sensitivity to light dark matter and other dark sector physics. The reason is simple -- if the new particles couple to electrons (which is well-motivated), then LDMX can produce them up to a mass of $\sim\GeV$. For dark matter and other invisible particles at colliders, this will result in significant missing momentum.   This section expands on the sensitivity of LDMX (both a near-term run as studied in this document and a longer-term, higher-luminosity run as discussed in the preceding section) to some of these physics possibilities.   Here, we will explicitly quantify LDMX sensitivity to scalar and fermion dark matter produced through an on- or off-shell mediator (including the near-resonance region), B-L gauge boson mediated scalar and fermion dark matter, millicharge particles, and invisibly decaying dark photons and B-L gauge bosons, as well as enumerating a number of other scenarios discussed more fully in ~\cite{LDMXSciencePaper}.  In all of the following, we illustrate the sensitivity of both of the ``Phase I'' LDMX run analyzed in detail in this whitepaper ($4 \times 10^{14}$ EOT at 4 GeV impinging on a $10\% \,X_0$ Tungsten target, shown in blue) and an ``extended'' run as envisioned in Sec.~\ref{sec:ExtendedReach} ($3.2\times 10^{15}$ EOT at 8 GeV impinging on a $40\%\,X_0$ Aluminum target) assuming a low background in each case. 

\begin{figure}
\includegraphics[width=16cm]{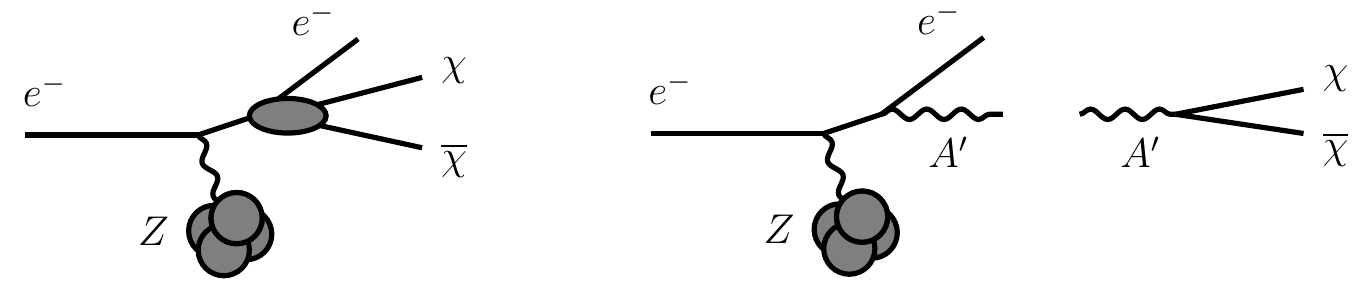}
\caption{\label{fig:BothSignalReactions2}
Left panel: Feynman diagram for direct dark matter particle-antiparticle production. This process yields an irreducible contribution to the LDMX signal even if other channels are present. Right panel: Feynman diagram for radiation of an on-shell mediator particle off a beam electron, which then decays to produce dark matter. Measuring both of these (and similar) reactions is the primary science goal of LDMX, and will provide broad and powerful sensitivity to light dark matter and many other types of dark sector physics.}
\end{figure}

As discussed in the introduction, {\bf both} direct dark matter production and on-shell mediator production can be probed with LDMX. These reactions are illustrated in Figure~\ref{fig:BothSignalReactions2}. 
Direct dark matter production (left panel of Figure~\ref{fig:BothSignalReactions2}) can dominate when the mediator is lighter than $2 m_{\chi}$ or when the DM-SM interaction is a higher dimension operator. In this case, the rate observed in LDMX provides a measure of the dark matter coupling to electrons. Even if other channels dominate dark matter production at LDMX, this direct dark matter production channel is always present. Thus, regardless of the relative ordering of masses between the mediator and the dark matter, dark matter production reactions can be probed in LDMX.
On-shell mediator production (right panel of Figure~\ref{fig:BothSignalReactions2}) can dominate the LDMX signal when the mediator is heavier than $2 m_{\chi}$, in which case dark matter can be produced through the on-shell decay of the mediator. In this case, the rate observed in LDMX primarily measures the dark mediator's coupling to electrons.  
\begin{figure}[htbp]
\includegraphics[width=0.48\textwidth]{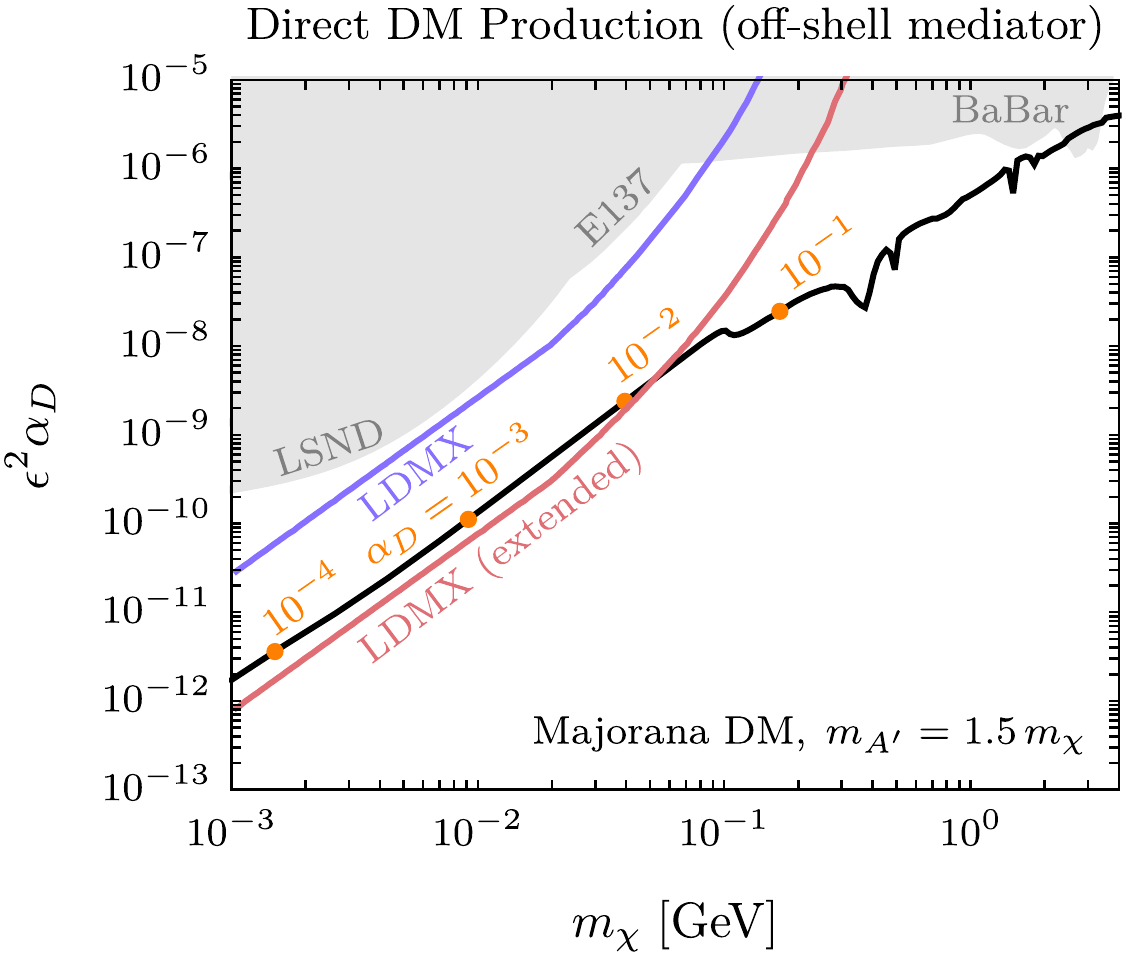}%
\includegraphics[width=0.48\textwidth]{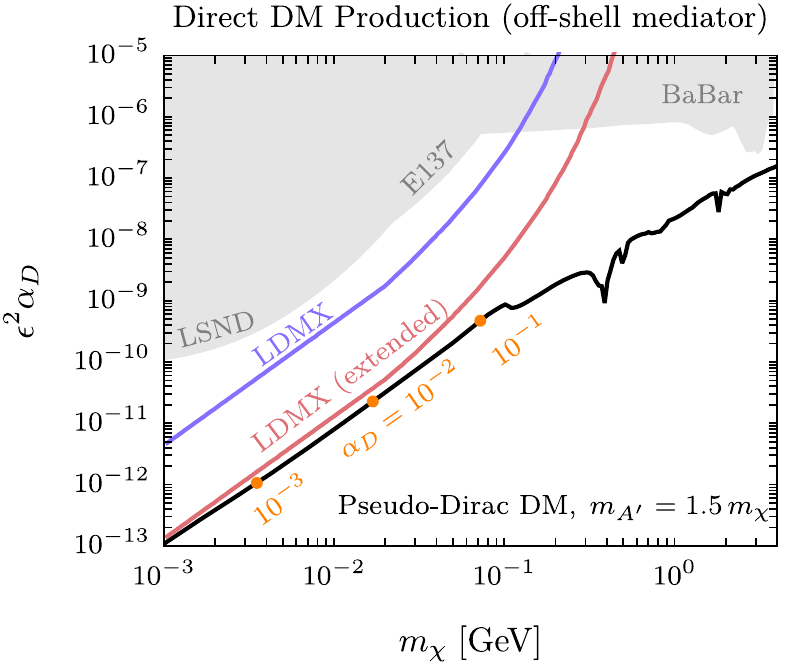}%
\caption{\label{fig:reachOffShell} Projected sensitivity to direct dark matter production (left panel of Figure~\ref{fig:BothSignalReactions}) through off-shell mediator exchange for Majorana dark matter {\bf (left)} and (pseudo-)Dirac dark matter {\bf(right)}. The sensitivity is expressed in terms of the dark matter coupling to electrons, $\epsilon^2\alpha_D$, versus dark matter mass $m_{\chi}$. In both cases, $m_{A^\prime}/m_{\chi}=1.5$ is used. The solid blue curve corresponds to $4 \;\GeV$ beam energy and $4 \times 10^{14}$ EOT. The red curve is the extended LDMX run highlighted in Table~\ref{table:ExtendedLDMX}. Gray regions are existing constraints from beam dumps LSND~\cite{deniverville:2011it} and E137~\cite{batell:2014mga}, and from \babar~\cite{Lees:2017lec}. The black curves are thermal freeze-out targets along which $\chi$ makes up the observed dark matter abundance. Orange dots along the thermal target indicate the range of $\alpha_D$ below which forbidden annihilation channels can be neglected~\cite{DAgnolo:2015ujb,Cline:2017tka}.}
\end{figure}

In Figure~\ref{fig:reachOffShell}, we show the sensitivity to dark matter production through an off-shell mediator, corresponding to the direct dark matter production process of Figure~\ref{fig:BothSignalReactions}. The sensitivity is expressed in terms of the dark matter coupling to electrons, $\epsilon^2\alpha_D$, versus dark matter mass $m_{\chi}$. LDMX yields for (pseudo-)Dirac dark matter is larger than for Majorana because of near threshold velocity suppression in the case of Majorana DM. In each plot of Fig.~\ref{fig:reachOffShell} the solid 
black line shows the thermal freeze-out target along which $\chi$ makes up all of the observed 
DM of the universe. In both cases, the overall sensitivity of LDMX with respect to direct thermal freeze-out predictions is encouraging. We also show existing constraints from beam dumps LSND~\cite{deniverville:2011it} and E137~\cite{batell:2014mga}, and from \babar~\cite{Lees:2017lec} as gray shaded regions. 
Even though $m_\chi < m_{A'}$ and secluded annihilations $\chi \chi \rightarrow A' A'$ are 
forbidden at zero temperature, they may still deplete $\chi$ abundance in the thermal 
bath of the early universe~\cite{DAgnolo:2015ujb,Cline:2017tka}. If the dark sector coupling $\alpha_D$ is large enough, this process can determine the relic abundance, such that there is no longer a sharp freeze-out target for $\epsilon^2 \alpha_D$. The values of $\alpha_D$ above which this ``forbidden'' channel dominates are indicated by orange dots as a function of $m_\chi$ in 
Fig.~\ref{fig:reachOffShell}.

\begin{figure}[htbp]
\includegraphics[width=0.48\textwidth]{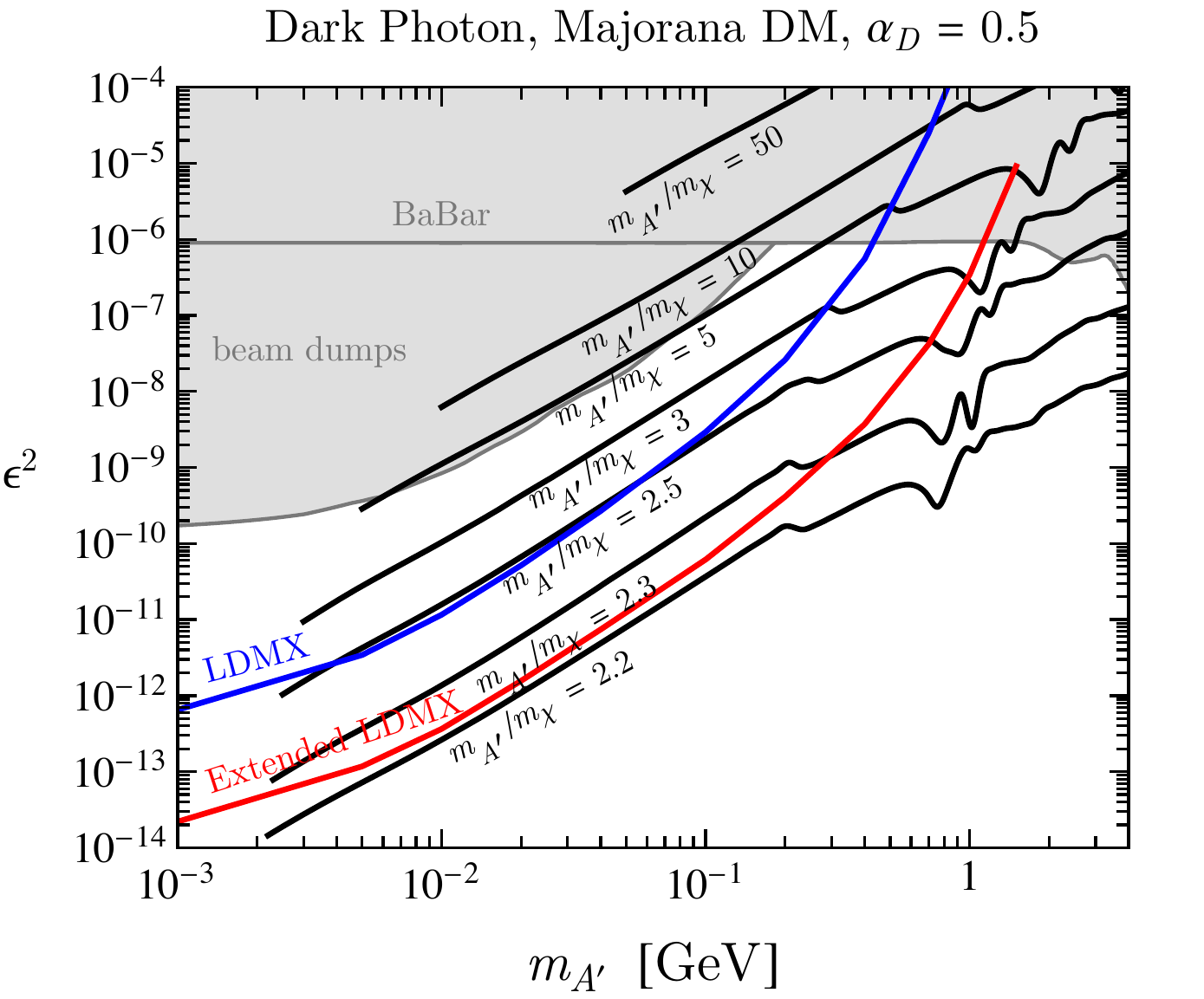}
\includegraphics[width=0.48\textwidth]{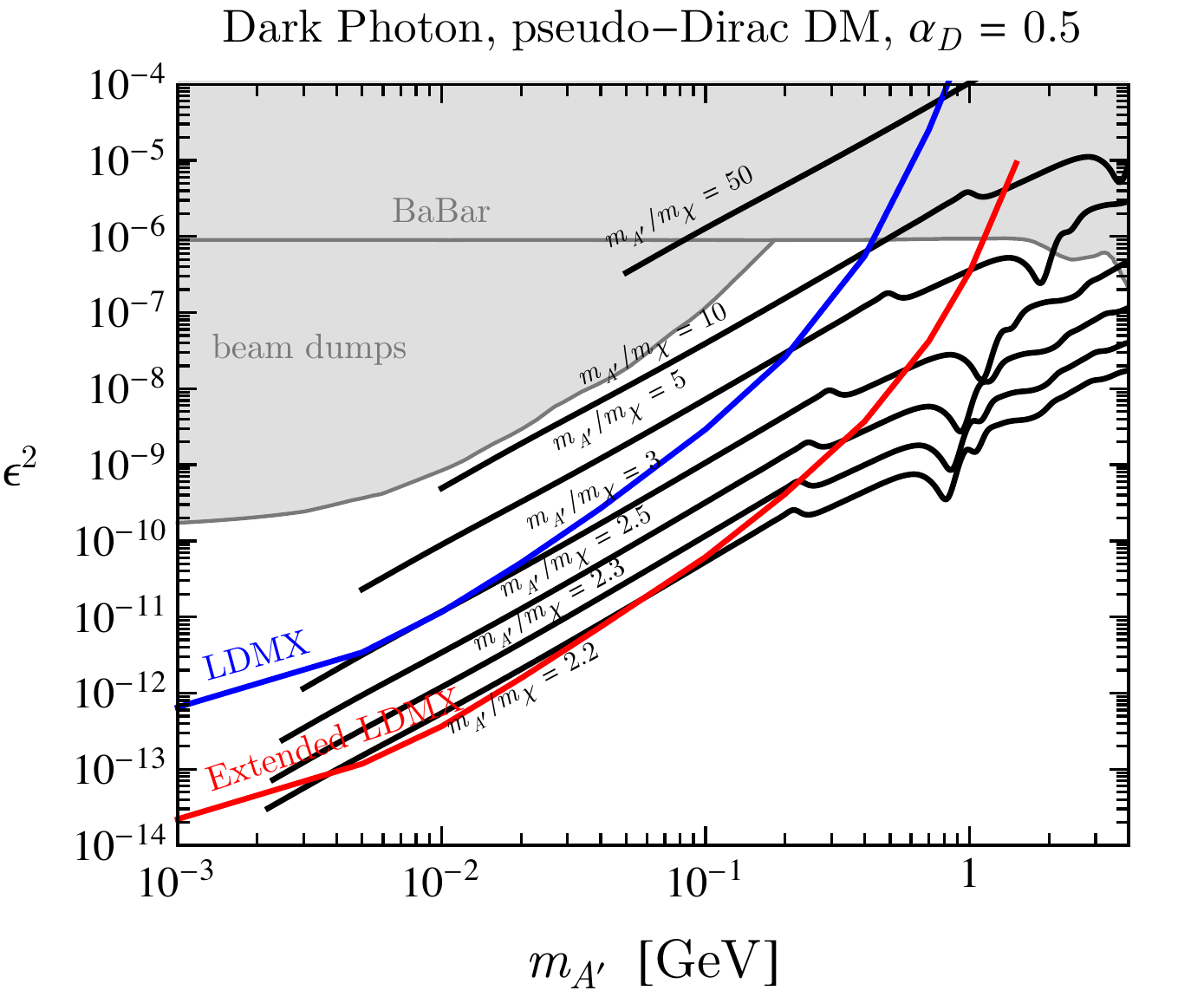}
\caption{\label{fig:NearResonance} Sensitivity in the $\epsilon^2$ vs $m_{A'}$ plane to $A^\prime$ production, with $A^\prime \rightarrow \chi\bar{\chi}$ (on-shell mediator production of dark matter). In much of the viable parameter space for light dark matter, this type of reaction dominates the dark matter yield in accelerator experiments. Shown for comparison are curves corresponding to direct thermal freeze-out reaction accounting for the observed density of dark matter for Majorana {\bf (left)} and Pseudo-Dirac {\bf (right)} dark matter; the gray regions are excluded by existing constraints. The mass ratio $m_{A'}/m_{\chi}$ is varied from $50$ to $2.2$, while $\alpha_D=0.5$. The thermal relic sensitivity is most challenging to reach as the parameters approach the narrow resonance region $m_{A'}/m_{\chi}=2$ with large $\alpha_D$. The blue line is the sensitivity of the ``Phase I'' LDMX run with $4 \ \GeV$ beam energy and $4 \times 10^{14}$ EOT. A scaling estimate of the sensitivity of the extended run scenario highlighted in Table \ref{table:ExtendedLDMX} is illustrated by the red line. Note that as $\alpha_D$ is decreased relative to the reference value shown here, the relic curves and beam dump constraints  shift uniformly upwards in the parameter space, whereas the \babar\ exclusion region is unchanged.}
\end{figure}

The main physics results in Section \ref{sec:goals} emphasize the LDMX sensitivity to dark-matter thermal targets in terms of the variable $y = \alpha_D \epsilon^2 (m_\chi/m_{A^\prime})^4$ as a function of the dark matter mass $m_\chi$.   Although this parameter space is mainly useful for illustrating (and comparing) accelerator based sensitivity in a conservative manner, as was done in the preceding sections, it is also useful to show the sensitivity in the native parameter space of the mediator's coupling to electrons in the $\epsilon$ vs. $m_{A^\prime}$ plane, which is shown in Figure~\ref{fig:NearResonance} for the case of on-shell mediator production and decay into dark matter, where the mass ratio $m_{A'}/m_{\chi}$ is varied from $50$ to $2.2$ to illustrate the resulting variation in the predicted coupling strength for direct thermal freeze-out. At larger mass ratios, the thermal targets move to larger values of $\epsilon$ (scaling as $(m_{A'}/m_\chi)^4$ due to the propagator suppression of the annihilation cross-section; near $m_{A'}/m_{\chi}=2$, the annihilation rate is dominated by the $A'$ resonance and so depends much more dramatically on the mass ratio. As Figure~\ref{fig:NearResonance} illustrates, LDMX can robustly explore deep into the direct thermal freeze-out territory over the sub-GeV mass range, and is especially strong below $\sim 300 \;\MeV$ masses where even few-percent tuning about the resonance region can be explored. 

\begin{figure}[htbp]
\includegraphics[width=0.45\textwidth]{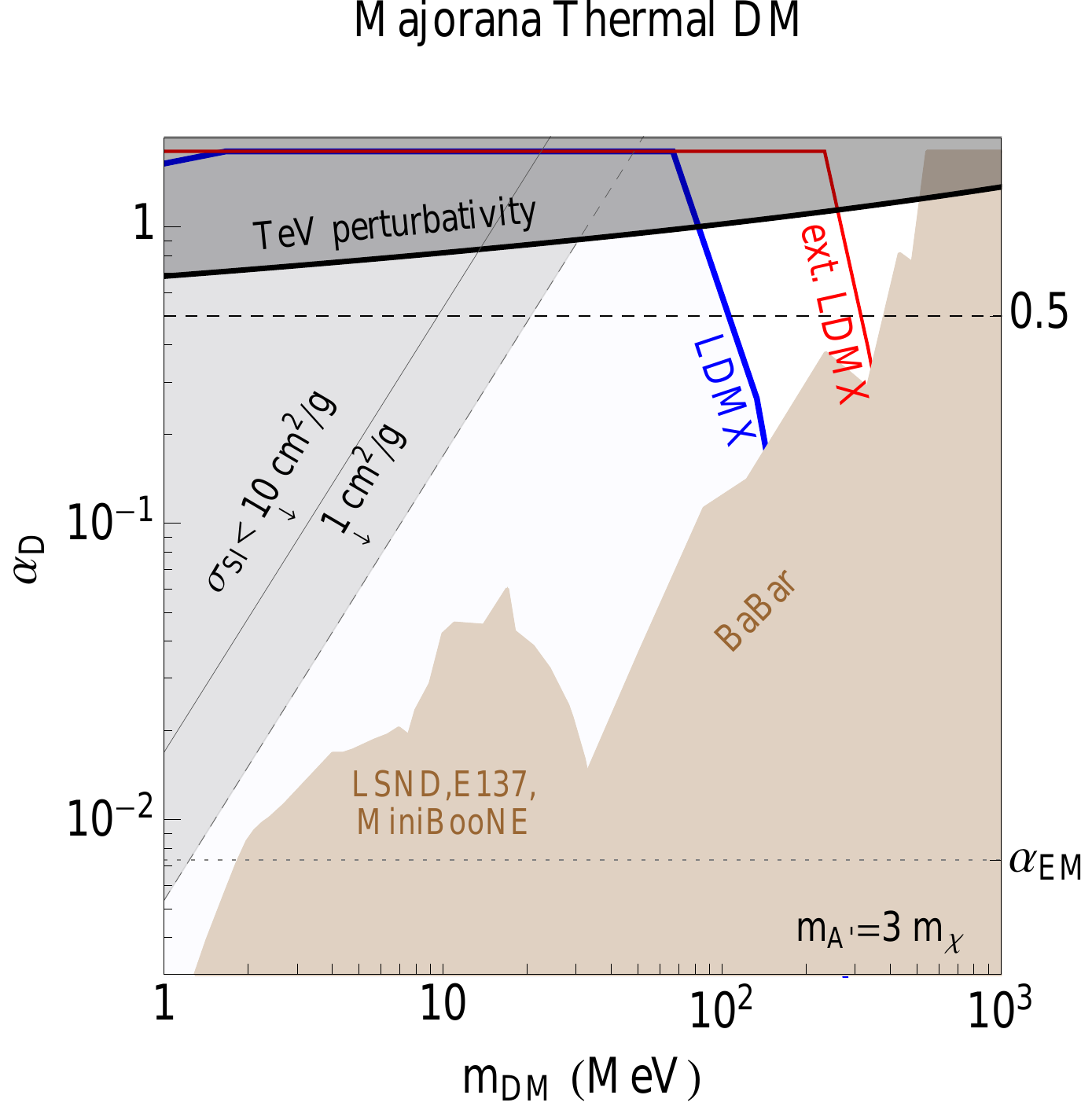}	
\includegraphics[width=0.45\textwidth]{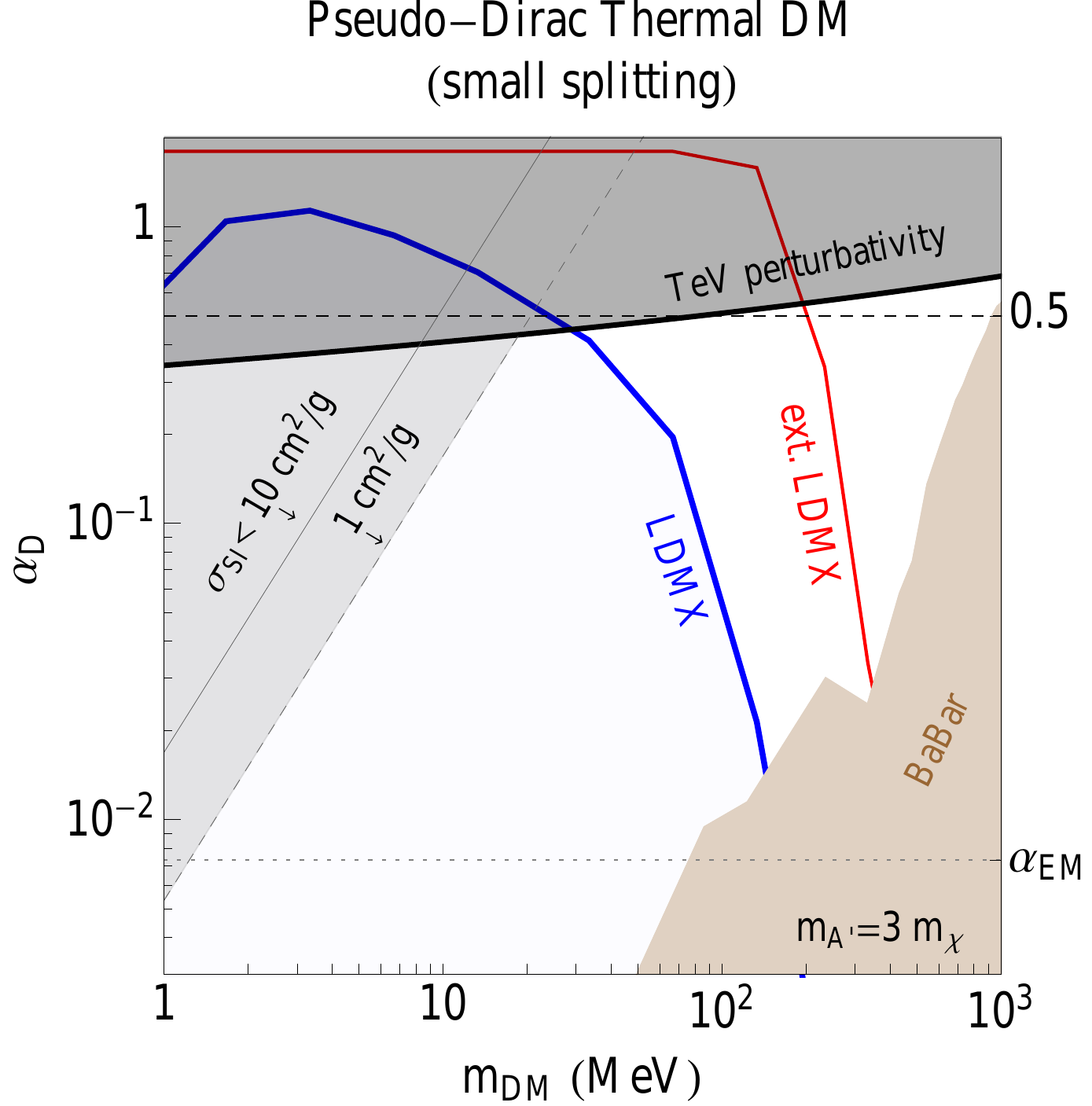}
\caption{\label{fig:alphaD}Projected LDMX lower bounds on $\alpha_D$ vs. $m_\chi$ for thermal relic Majorana (left) and pseudo-Dirac (right) fermions. The region below the blue curves will be explored by LDMX Phase I. The  upper limits on $\alpha_D$ from TeV-scale perturbativity and from dark matter self-interaction bounds of 1 (10) cm$^2$/g are shown in dark and light gray, respectively.  Direct searches lead to \emph{lower bounds} because their yield scales as $y/\alpha_D$ (missing momentum/energy/mass) or $y^2/\alpha_D$ (beam dumps) while the thermal target depends only on $y$. The most significant constraints from existing accelerator-based experiments --- a \babar\ missing mass search \cite{Lees:2017lec} and a reinterpretation of the LSND neutral-current $\nu -e$ scattering measurement as a bound on light DM \cite{deniverville:2011it} --- are indicated by the filled brown region extending from the bottom of the plot (for pseudo-Dirac DM, these and the recent NA64 missing energy search \cite{Banerjee:2017hhz} are also the most sensitive, but their constraints are outside the range of the plot).  LDMX expected sensitivity is indicated by a blue line, with light blue shading and assumes a $4 \times 10^{14}$ EOT run. 
}
\end{figure}

The projected LDMX sensitivity can also be interpreted as an \emph{upper bound} on the coupling $\alpha_D$ of light dark matter to the mediator, as a function of mass, if we assume that the product of couplings $\alpha_D \epsilon^2$ lies precisely at the thermal relic line.  This interpretation of the expected LDMX sensitivity is shown in Figure \ref{fig:alphaD} for both Majorana and pseudo-Dirac dark matter. These figures also show several complementary direct and indirect constraints: direct searches for dark matter production at \babar~\cite{Lees:2017lec} and LSND~\cite{deniverville:2011it} exclude the coupling ranges shaded in brown (in the case of pseudo-Dirac dark matter, the best existing constraints at low mass come from LSND~\cite{deniverville:2011it} and NA64~\cite{Banerjee:2016tad,Banerjee:2017hhz}, but are below the range of the plot). Demanding that the $U(1)_D$ coupling remain perturbative up to 1 TeV (``TeV perturbativity'') leads to a constraint that the running coupling at $m_\chi$ must be below the ``TeV perturbativity'' line. Finally, demanding that the DM self-interaction cross-section lie below $\sigma/m = 1$ cm$^2$/g excludes the light-gray shaded region (a thin gray line illustrates the weaker constraint $\sigma/m < 10$ cm$^2$/g).  

\begin{figure}[htbp]
\includegraphics[width=0.67\textwidth]{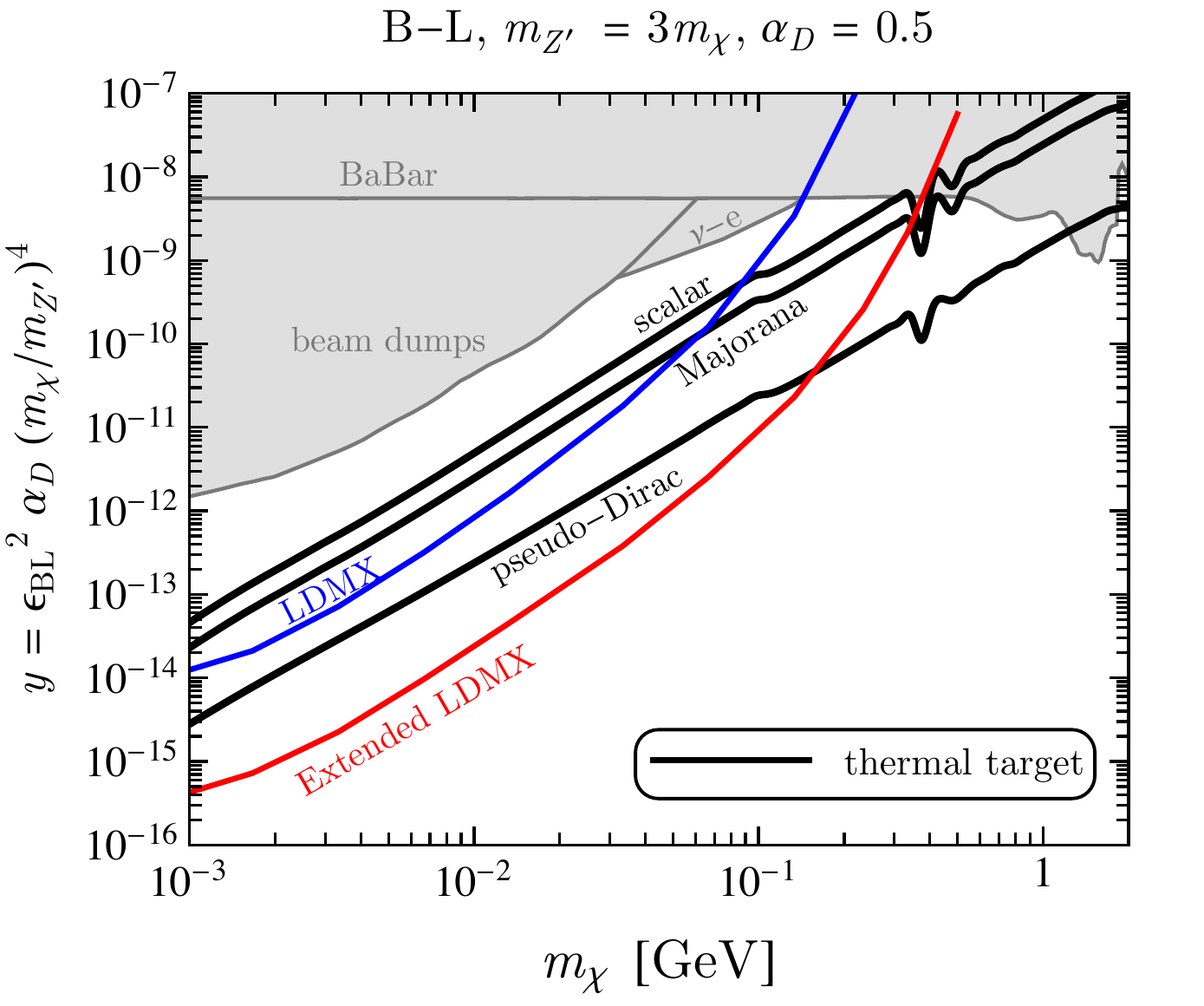}
\caption{\label{fig:BminusLThermal} Projected sensitivity of LDMX in the $y$ vs $m_{\chi}$ plane for light dark matter models coupled to the Standard Model via a B-L gauge boson $Z^\prime$. Thermal relic curves are shown for $\alpha_D=0.5$ and $m_{Z^\prime}/m_{\chi}=3$. The blue line is the sensitivity of the ``Phase I'' LDMX run with $4 \ \GeV$ beam energy and $4 \times 10^{14}$ EOT. A scaling estimate of the sensitivity of the extended run scenario highlighted in Table \ref{table:ExtendedLDMX} is illustrated by the red line.}
\end{figure}

\begin{figure}[htbp]
\includegraphics[width=0.6\textwidth]{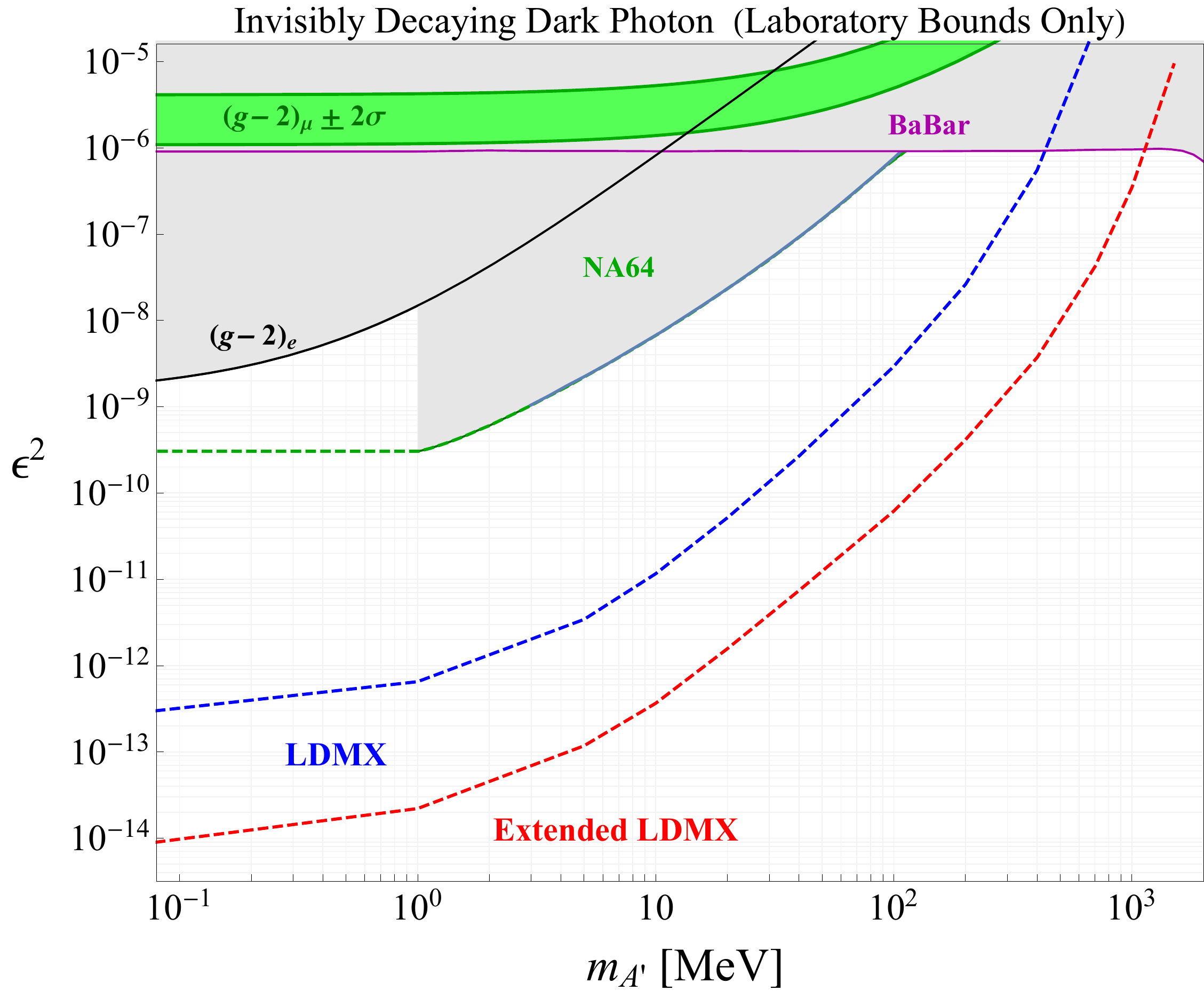}
\includegraphics[width=0.62\textwidth]{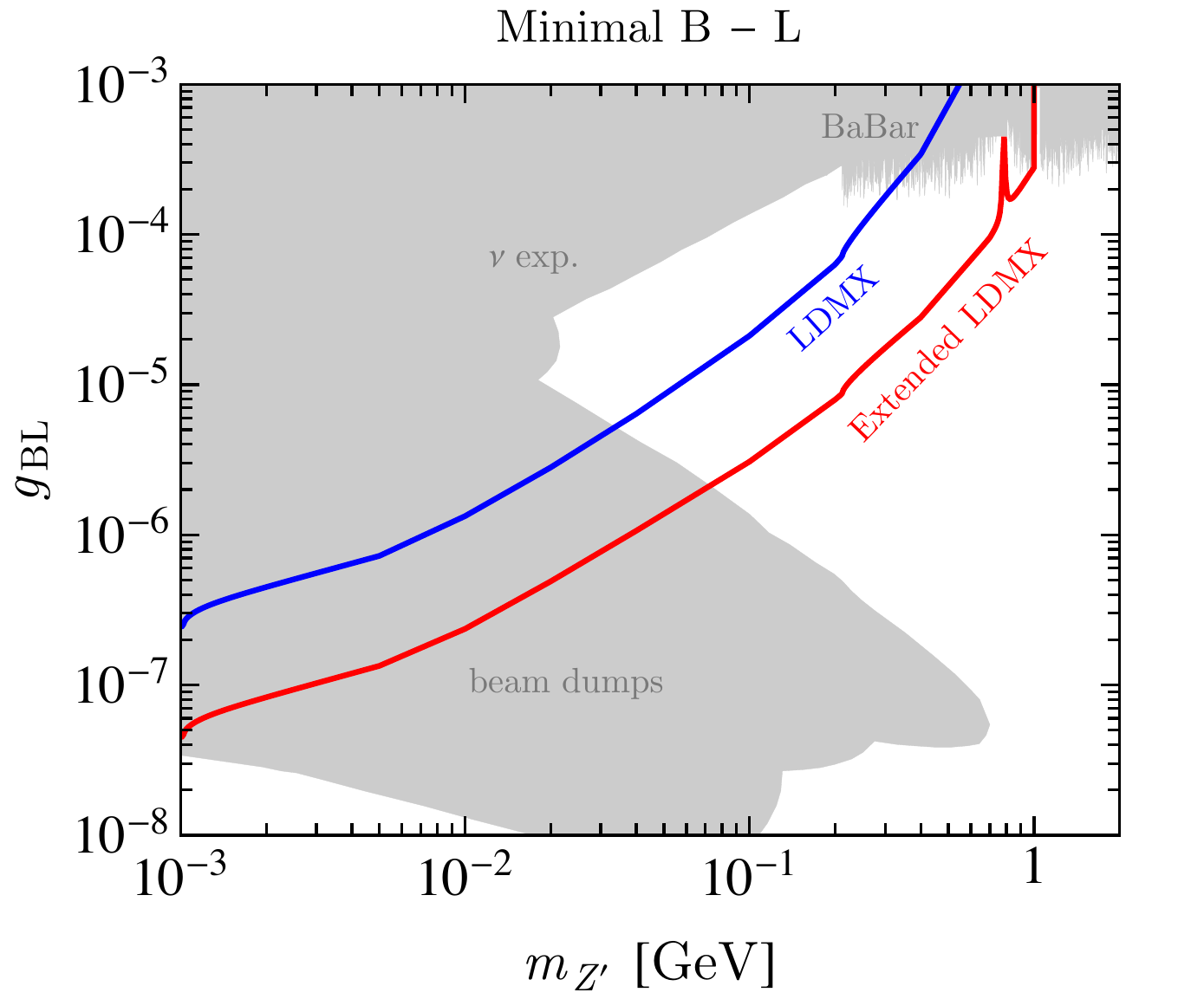}
\caption{\label{fig:InvisibleMediator} Sensitivity to invisibly decaying dark photons (top) and B-L gauge bosons (bottom). The blue line is the sensitivity of the ``Phase I'' LDMX run with $4 \ \GeV$ beam energy and $4 \times 10^{14}$ EOT. A scaling estimate of the sensitivity of the extended run scenario highlighted in Table \ref{table:ExtendedLDMX} is illustrated by the red line.}
\end{figure}

\begin{figure}[htbp]
\includegraphics[width=0.67\textwidth]{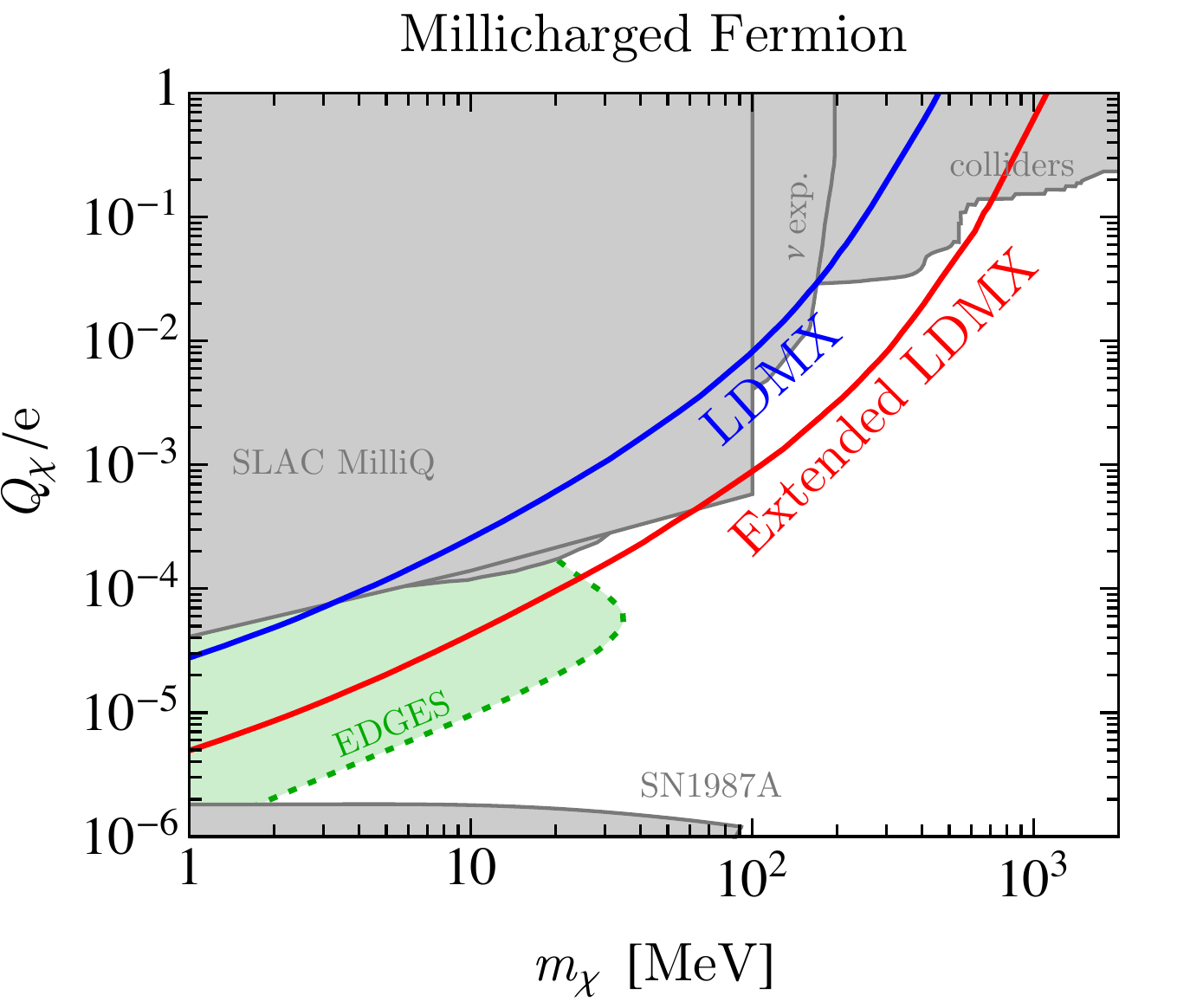}
\caption{\label{fig:millicharge} Sensitivity to millicharge fermion particles with charge $Q_{\chi}/e$ vs mass. Production in LDMX occurs through an off-shell photon. Grey regions are existing constraints. The green shaded region represents parameter
space where a millicharged dark matter subcomponent can 
accommodate the 21 cm absorption anomaly reported by the EDGES 
collaboration \cite{Bowman:2018yin,Barkana:2018lgd,Berlin:2018sjs,Kovetz:2018zan}.
The blue line is the sensitivity of the ``Phase I'' LDMX run with $4 \ \GeV$ beam energy and $4 \times 10^{14}$ EOT. A scaling estimate of the sensitivity of the extended run scenario highlighted in Table \ref{table:ExtendedLDMX} is illustrated by the red line.}
\end{figure}

As discussed in detail in \cite{LDMXSciencePaper}, the range of dark matter and mediator physics probed by fixed-target missing momentum measurements extends far beyond dark photon mediators, and has many other applications to new physics searches as well. For example, Figure~\ref{fig:BminusLThermal} illustrates the LDMX sensitivity to dark matter interacting via a B-L $Z^\prime$ gauge boson.  The thermal targets and other experiments' sensitivity change by $\mathcal{O}(1)$ factors relative to the dark photon mediator case.  Other models feature more dramatically altered constraints and thermal targets, to which LDMX nonetheless has substantial sensitivity, as catalogued in \cite{LDMXSciencePaper}.  For example:
\begin{itemize}
\item Thermal freeze-out through mediators coupled to anomaly-free combinations of lepton numbers, e.g. $L_e - L_\mu$,
\item Thermal freeze-out through mediators coupled to baryon number, where radiatively generated (and hence suppressed) couplings to electrons nonetheless control the thermal abundance of dark matter lighter than the pion ---  LDMX can explore this possibility directly, which to date is only constrained by indirect searches \cite{Dror:2017ehi,Dror:2017nsg},
\item Thermal freeze-out through leptophilic or electrophilic (pseudo)scalar mediators, including parameter space where such mediators can explain the anomalous measurements of the fine structure constant,\cite{Davoudiasl:2018fbb} or muon magnetic moment \cite{Bennett:2006fi}
(the simpler case of direct annihilation through a scalar mediator that mixes with the Higgs boson is already excluded by meson decay data \cite{Krnjaic:2015mbs}),
\item Asymmetric dark matter annihilating directly to SM particles through any of the mediators above,
\item Dark matter that consists of stable mesons from a confined dark sector, in which case new mechanisms \cite{Hochberg:2014kqa,Hochberg:2015vrg,Kuflik:2015isi,Kuflik:2017iqs,Berlin:2018tvf} can dominate its depletion at couplings below the usual scalar thermal target, and
\item Freeze-in of dark matter with keV-scale mass through a $\sim 100$ MeV-mass mediator in models with a low reheating temperature.
\end{itemize}

In addition, a missing momentum search at LDMX is sensitive to a range of other new-physics scenarios, potentially unrelated to dark matter.   Figure~\ref{fig:InvisibleMediator} illustrates the sensitivity of LDMX to invisible dark photons and to minimal B-L $Z^\prime$ gauge bosons, via their invisible decays to neutrino final states. Figure~\ref{fig:millicharge} illustrates the sensitivity of the LDMX missing momentum search to production of millicharged particles.  Millicharge production in LDMX occurs through off-shell photon exchange, and particles with sufficiently small millicharge $Q_{\chi}/e$ have no additional interactions in the detector. 

\begin{figure}[htbp]
\includegraphics[width=0.45\textwidth]{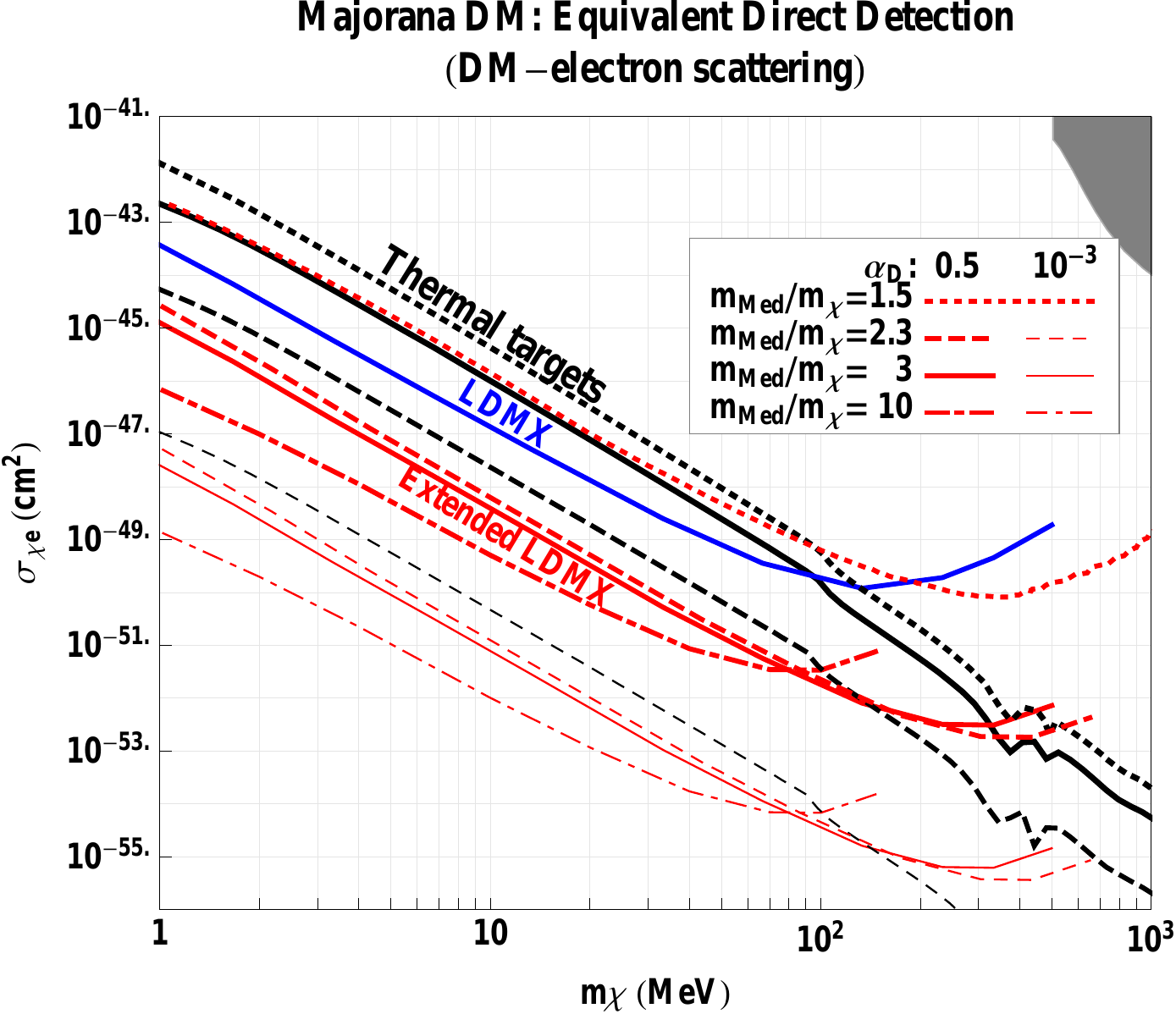}	
\includegraphics[width=0.45\textwidth]{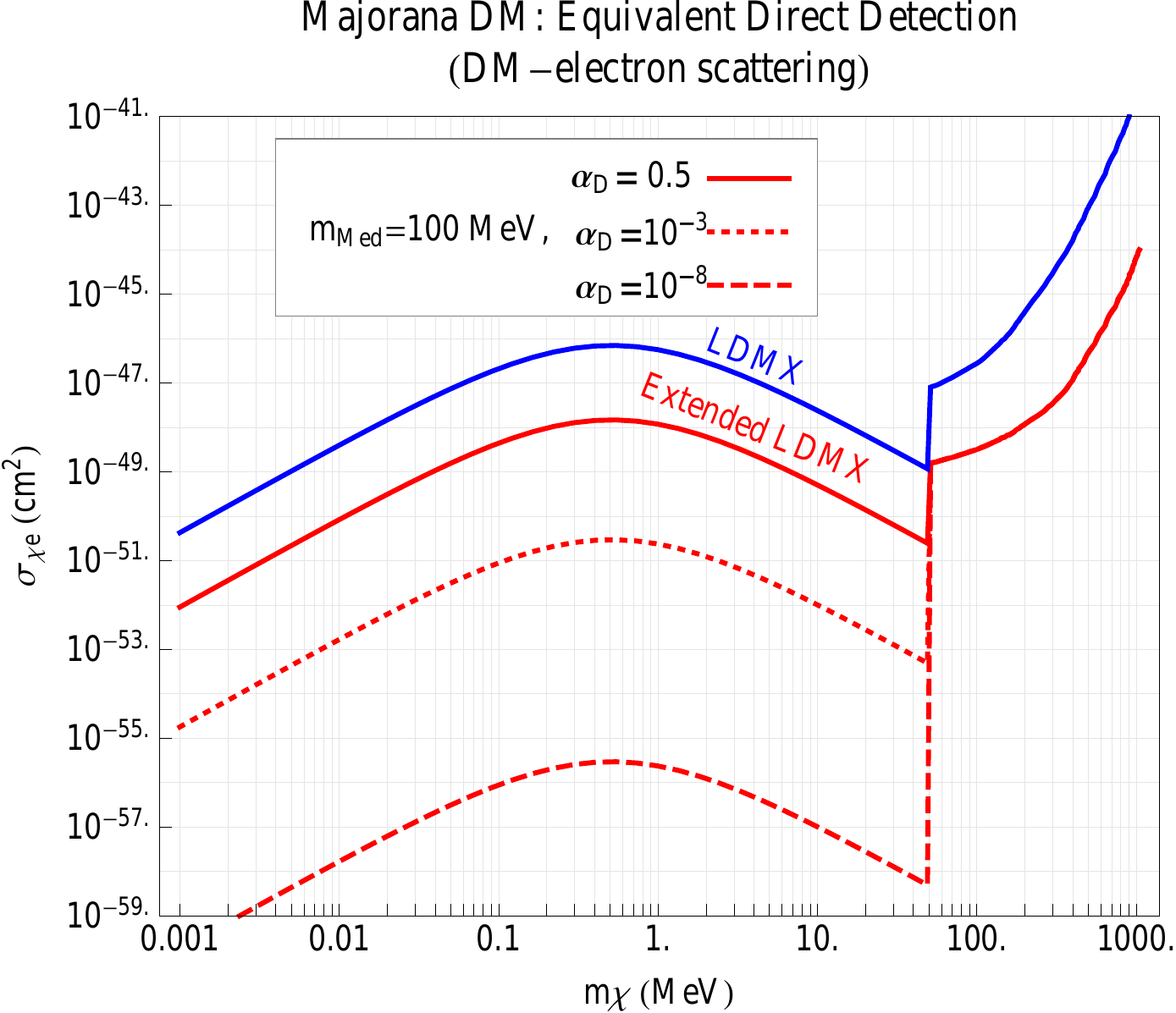}
\caption{\label{fig:directDetectionEquiv} Translation of dark matter and mediator coupling sensitivity into DM-electron scattering rates for various assumed model parameters and Majorana dark matter. Model-dependent constraints (including constraints from past accelerator-based DM searches) are not shown.  In the {\bf left} panel, each sensitivity projection assumes a fixed ratio of mediator to DM particle mass --- ratios of $1.5$ and $2.3$ correspond to off-shell production and resonance-dominated thermal freeze-out respectively; ratios of $3$ and $10$ are two illustrative examples of thermal freeze-out without resonance enhancement, where on-shell mediator production dominates the LDMX signal.  In each case, sensitivities are computed and mapped to equivalent direct detection sensitivities for both $\alpha_D = 0.5$ (thick lines) and $\alpha_D =10^{-3}$. Where only one line is shown ($m_{\rm Med}=1.5 m_{\chi}$) the equivalent direct detection sensitivity is independent of $\alpha_D$. The ``thermal target'' cross-section for each parameter choice is also shown (for resonant annihilation only, this target depends significantly on $\alpha_D$; the thin black dashed line illustrates the thermal target for $m_{\rm Med}=2.3 m_{\chi}$ and $\alpha_D = 10^{-3}$).  The blue solid line corresponds to the initial phase of LDMX as presented in this note; to simplify the plot this is shown only for one benchmark. The red projection lines are scaled to the extended LDMX luminosity denoted by the  ``*'' line in Table \ref{table:ExtendedLDMX}. 
We also show the constraint from CRESST~\cite{Angloher:2015ewa,Petricca:2017zdp} as the gray region.
The {\bf right} panel illustrates equivalent sensitivity for a fixed mediator mass and dark matter masses as light as a keV, for $\alpha_D = 0.5$, $10^{-3}$, and $10^{-8}$, where the latter coupling is suitable for low-reheat-temperature freeze-in as discussed in \cite{LDMXSciencePaper}. The equivalent cross-section sensitivity is greater for pseudo-Dirac and inelastic scalar models, and weaker for scalar elastically scattering DM. 
}
\end{figure}

To close, we comment on how LDMX sensitivity is related to scattering rates in terrestrial direct detection experiments. As underscored, accelerator experiments probe dark matter coupling vs mass, and are therefore complementary to direct detection scattering rates. However, for a given model, one can map coupling sensitivity into scattering rates. In the case of Majorana dark matter, this mapping is straightforward --- direct detection cross sections for electron scattering are 
\begin{eqnarray}
\sigma_{\chi e} = 32 \pi \frac{\alpha \alpha_D \epsilon^2 \mu_{\chi e}^2}{m_A^{\prime 4}} \left(\frac{\mu_{\chi e} v_{rel}}{  m_{\chi} }  \right)^2,
\end{eqnarray}
where $\mu_{\chi e}$ is the $\chi-e$ reduced mass. 
In this case, the LDMX sensitivity (again, assuming a mass ratio $m_{A^\prime}/m_\chi=3$ and $\alpha_D = 0.5$) is shown in Figure \ref{fig:directDetectionEquiv}.  The effective cross-section sensitivity \emph{improves} as $\alpha_D$ is decreased; as the mass ratio is increased the LDMX sensitivity region moves down (greater sensitivity) and to the left (lower masses).  The effective cross-section sensitivity is model-dependent --- for elastic scalar dark matter, LDMX sensitivity maps to higher cross-sections than those shown in the plot, while for inelastic models (scalar or fermion) LDMX sensitivity maps to lower cross-sections.



\begin{acknowledgments}
Support for UCSB involvement in LDMX is made possible by the Joe and Pat 
Yzurdiaga endowed chair in experimental science. RP acknowledges support 
through The L’Or\'{e}al-UNESCO For Women in Science in Sweden Prize with 
support of the Young Academy of Sweden. PS and NT are supported by the U.S. 
Department of Energy under Contract No. DE-AC02-76SF00515. GK, NT, AW are 
supported by the Fermi Research Alliance, LLC under Contract No. 
DE-AC02-07CH11359 with the U.S. Department of Energy, Office of Science, 
Office of High Energy Physics. BE and DH are supported by the US Department of
Energy under grant DE-SC0011925. 
\end{acknowledgments}

\clearpage
\begin{appendices}

\section{Photonuclear Modeling: Few-Neutron Final States and Hard Backscatters}\label{sec:PNappendix}
In Section \ref{sec:bkgmod} we highlighted three classes of event that are design drivers for LDMX, and which were found to be mismodeled in the default \geant Bertini cascade model: single forward neutrons, moderate-angle neutron pairs, and hard backscattered hadrons in the ``cumulative'' region.

Here, we describe in more detail our findings about these classes of events, the data on which we base our estimates for their yields, and the path forward to improve modeling of these final states and phase-space regions in \geant.  

\subsection{Forward Single Neutrons and Neutron Pairs}
The rates of single neutrons and neutron pairs are two design drivers for the \hcal.  When events have only a single forward neutron (accompanied only by hadrons at energies $<100$~MeV and/or angles $>90\deg$), non-interaction of that neutron represents a single-point failure of the veto system.  Requiring that these be rejected to the $<1$ event level determines a lower bound on the depth of the detector.  While single neutrons typically go forward due to energy conservation, pairs of neutrons are dominantly produced at nontrivial opening angle, so that their non-interaction probability is determined by the depth \emph{and} transverse size of the \hcal. 

Neutron pairs can originate from a primary photon reaction on a di-neutron $\gamma \,(nn) \rightarrow n\, n$, or from a sequence of reactions (e.g. $\gamma N \rightarrow \pi n,\quad \pi n \rightarrow n \pi$, where the pion backscatters at low momentum in the 2nd reaction, effectively ``converting'' into a hard neutron). 
A single hard neutron can arise from a highly asymmetric 2-body final state --- for example for $\gamma \,(nn)\rightarrow n n$ [$\gamma n \rightarrow n \pi^0$] with a 3 GeV incident photon, the phase space with $|\cos\theta_{cm}| > 0.7$  [$\cos\theta_{cm} < -0.77$] leads to a hard forward neutron with the other secondary backscattered. They can also arise from a smaller fraction of phase space for 3-body final states.  The rates for these processes \emph{per inclusive photonuclear interaction} for 3 GeV photons impinging on $^{184}$W --- can be inferred from exclusive cross-section data on light nuclei.  These estimates, and the yields according to the default Bertini cascade model, are summarized in Tables \ref{tab:BertiniVsPhysicsTableTwoNeutron} and \ref{tab:BertiniVsPhysicsTableOneNeutron}.  As shown, the rates for these processes in \geant appear to be overestimated by 2 to 3 orders of magnitude!  The origins for these discrepancies in the code have been identified, and work is underway on remedying them in collaboration with the \geant collaboration.  In the following, we summarize the data used to extract our physical estimates for each class of reactions, as well as the origin for the unphysical event yields in the \geant code.


\begin{table}[htbp]
\begin{tabular}{l|l l l l |l }
Primary Process & $N_{T}/\rm{nucleus}$ & $\sigma_{T}^{inclusive}$ & Secondary process & Probability &  Di-neutron Yield  \\ \hline
$\gamma (nn) \rightarrow \pi n $ &  300 {(3900)}& $7$ {(200)} nb & -- & -- & $ 1 \times 10^{-4}$ {(0.16)} \\
$\gamma n \rightarrow \pi^0 n $ & 110 & 0.9 $\mu$b & $\pi^0 n \rightarrow n \pi^0$ 
 & $\sim 0.01$ & $\sim 1 \times 10^{-4}$ \\ 
$\gamma p \rightarrow \pi^+ n $ & 110 & 2.2 $\mu$b & $\pi^+ n \rightarrow n \pi^+$ 
& $\sim 0.01$ & $\sim 1 \times 10^{-4}$ \\ \hline
\multicolumn{5}{l}{\bf Total}  & $\mathbf{\lesssim 3\times 10^{-4}}$ {\textbf{(0.16)}} \\
\end{tabular}
\caption{\label{tab:BertiniVsPhysicsTableTwoNeutron} 
Estimated yield of di-neutron events \emph{as a fraction of inclusive photonuclear reactions} for the exclusive processes discussed in the text, in the benchmark case of a 3 GeV photon impinging on $^{184}$W. Where  we have found discrepancies between the Bertini Cascade model and our calculations, the rate in the Bertini Cascade model is given in parentheses.  The yield for two-neutron final states from di-neutron dissociation is overestimated in \geant by a factor of at least $\sim 500$, due to several factors discussed in the text.  Note that neither our estimates nor the \geant rates listed above account for the survival probability of the final-state neutrons (i.e.~the probability that it exits the nucleus before scattering into a multi-particle final state), which will mildly suppress the di-neutron rate relative to the above.}
\end{table}

\begin{table}[htbp]
\begin{tabular}{l| l l l l  | l  }
Process & $N_{T}/\rm{nucleus}$ & $\sigma_{T}^{inclusive}$ & kin.~requirement  & frac.~in $1n$ kinematics & Yield in $1n$ kinematics \\
\hline
$\gamma (nn) \rightarrow n\, n$  &  300 {(3900)}& 7 {(200)} nb &  $|q|>0.7$ & 0.9 {(0.3)} & $< 1\times10^{-4}$ {(0.05)} \\
$\gamma (pn) \rightarrow p\, n$  &  400 {(2600)}& 7 {(200)} nb &  $q<-0.7$ & 0.3 {(0.15)} & $ 3\times10^{-5}$ {(0.015)} \\
$\gamma p \rightarrow \pi^+\, n$  & 74 & 2 $\mu$b  &  $q<-0.77$ & 0.02  {(1.0)}  & $1\times10^{-4}$ {(0.003)}\\
$\gamma n \rightarrow \pi^0\, n$  & 110 & 1 $\mu$b  &  $q<-0.77$ &  0.04 {(1.0)}  & $2\times10^{-4}$ {(0.005)} \\
\hline
\multicolumn{5}{l}{\bf Total} &  $\mathbf{\sim 4\times10^{-4}}$ {\textbf{(0.07)}}\\
\end{tabular}
\caption{	\label{tab:BertiniVsPhysicsTableOneNeutron} Estimated yield of effective single-neutron final states (with any additional hadrons produced backwards in the lab-frame) for the exclusive processes discussed in the text, in the benchmark case of a 3 GeV photon impinging on $^{184}$W. Where  we have found discrepancies between the Bertini Cascade model and our calculations, the rate in the Bertini Cascade model is given in parentheses.  Again the yield of such events in the Bertini Cascade model is dominated by an un-physically large contribution from di-nucleon dissociation.  The physical contributions of these reactions are obtained by numerically integrating and interpolating deuteron dissociation cross-section data from \cite{Mirazita}; conservative assumptions are used for the $\gamma (nn)$ initial state (see text) but the yield is likely a factor of $\sim 60$ lower.
A distinct (physically dominant) contribution to single-neutron final states arises from $\gamma N\rightarrow \pi N$ reactions with a hard forward neutron and backward pion. While the cross-sections for these reactions in the \geant Bertini model are taken from data, the angular distributions used are unphysical as discussed in the text. 
The yields of two-step processes (e.g.~$\gamma p\rightarrow \pi^+ n$, where a forward pion backscatters off a neutron) are not shown --- estimates analogous to those in Table \ref{tab:BertiniVsPhysicsTableTwoNeutron} indicate that they are subdominant.
These estimates suggest that the total yield of single-neutron-like final states is overestimated by a factor of ${\cal O}(100)$ in the Bertini model, and the physically dominant reaction giving rise to such events is modeled incorrectly.  Note that neither our estimates nor the \geant rates listed above account for the survival probability of the final-state neutron (i.e.~the probability that it exits the nucleus before scattering into a multi-particle final state), which will mildly suppress the single-neutron rate relative to the above.}
\end{table}

\paragraph{Di-nucleon dissociation}
Two-body di-nucleon dissociation yields can be inferred from data, albeit with some sizable uncertainties.  The effective number of quasideuterons $P_{QD}$ in heavy nuclei can be inferred from data via photonuclear interaction cross-sections with $\sim100$ MeV photon energies (see e.g.~\cite{Anghinolfi:1985bi} Figure 1), and is usually parameterized as ${\cal P}_{QD} = L \frac{Z (A-Z)}{A}$, where the ``Levinger factor'' $L$ for large nuclei agrees within $\sim 20\%$ accuracy with the Local Density Approximation (LDA) expectation $L = 10.83 - 9.76 A^{-1/3}$ (for Tungsten, $L\approx 9$) \cite{Benhar:2003xr}.  The effective number of di-neutrons and di-protons are expected to scale similarly with a combinatoric factor, e.g. ${\cal P}_{nn} = L \frac{1}{2} \frac{(A-Z)^2}{A}$.  Multiplying these numbers by a dinucleon dissociation cross-section yields a cross-section \emph{per nucleus}, which can be compared to the inclusive photonuclear cross-section to obtain a yield \emph{as a fraction of inclusive photonuclear reactions}.  

The differential cross-section for deuteron dissociation has been measured with excellent angular coverage up to 3 GeV, allowing a fairly precise estimate of the 
contributions of $\gamma (pn) \rightarrow p\, n$ to the single-neutron final state (when the proton backscatters).  
The differential cross-section for di-proton dissociation (in $^3$He) has been measured at specific kinematics up to 4.7 GeV \cite{Mirazita,Pomerantz:2009gf}.  The 2-body dissociation cross-section for photon energies above 2 ~GeV was found to be a factor of 20 smaller for di-protons than for deuterons, which may be attributable to a cancellation between two scattering amplitudes in the hard re-scattering model (HRM) \cite{SargsianPRC}. We are not aware of any direct measurements of the di-neutron dissociation cross-section, which contributes to both single- and di-neutron final states.  
In the tables, we have assumed a dissociation cross-section of $7$~nb for both di-neutrons and quasideuterons.  This cross-section is obtained by numerically integrating and interpolating deuteron dissociation cross-section data from \cite{Mirazita}.  This is the largest plausible estimate for the dineutron dissociation cross-section, and indeed it appears more reasonable to assume that the dineutron dissociation cross-section is further suppressed by 1/20, like its diproton counterpart.  We do not include this suppression, however, because for LDMX design it is important not to underestimate the yields of difficult final states (this process is, in any case, a subleading contribution to the physical yield). Single-neutron final states receive contributions from the forward and backward phase-space of di-neutron dissociation (the forward peak is, again, based on deuteron data and likely overestimated), and from the backward phase-space of quasi-deuteron dissociation.  

The standard Bertini model dramatically over-estimates di-nucleon dissociation cross-sections, as shown by the numbers in parentheses in  \ref{tab:BertiniVsPhysicsTableTwoNeutron} and \ref{tab:BertiniVsPhysicsTableOneNeutron}.
This large discrepancy can be traced to three errors in the \geant Bertini cascade code:
\begin{itemize}
\item The parameterized cross-section for di-nucleon dissociation exceeds the measured deuterium 2-body dissociation cross-section starting around 500 MeV photon energies, with the discrepancy growing dramatically at multi-GeV energies, where the dissociation cross-section is expected to scale as $\sim E_{\gamma}^{-10}$. For example, the cross-section for two-body photodisintegration of deuterium at 2.8 GeV has been measured at Jefferson Lab to be about 5~nb\cite{Mirazita}; the Bertini cascade code uses a cross-section of 200 nb for 1.8 GeV $<E_\gamma <$ 5.6 GeV.
\item The di-neutron density used in the \geant model is a factor of $\sim7$ larger than the ${\cal P}_{nn}$ given above, because all neutrons within a spherical shell zone are allowed to pair with one another to form di-nucleons.  We note that the quasideuteron number ${\cal P}_{pn}$ has been extracted from data in multiple ways, and from a variety of nuclear models, with mutual agreement within $\sim 20\%$ for all of these models.
\item The selection of targets (proton, neutron, or di-nucleon) for photonuclear interaction in the Bertini Cascade Monte Carlo is not proportional to the \emph{physical} path lengths for these interactions, but to the path lengths when an interaction is forced --- effectively rendering all reactions equally likely irrespective of their specified cross-sections.
\end{itemize}

\paragraph{Single-Pion Photo-Production}
The processes that we expect to dominate the single-neutron final state are $\gamma n \rightarrow \pi^0 n$ and $\gamma p \rightarrow \pi^+ n$ in the region of phase space where the pion backscatters.  The rates for these and other few-body photon-nucleon processes --- inclusively and in the single-neutron phase space --- can be readily extracted from data.  Single-pion photoproduction cross-sections off protons and neutrons and their angular distributions have been measured in e.g.~\cite{Bartholomy:2004uz,Zhu:2004dy} at multi-GeV photon energies; the backscatter kinematics, which according to our estimates dominates the yield of single-neutron final states, exhibits a universal form of $k^3 d\sigma/du$ for $E_\gamma > 4 \GeV$ \cite{Anderson:1969bq} which we extrapolate to $\sim 3$ GeV and use to extract the cross-section for the region of phase space where the nucleon is undetectable (backscattered or below 100 MeV in the lab frame).  Neglecting the $<20\%$ effect of shadowing for 3--4 GeV photons in heavy nuclei \cite{Michalowski:1977ic}, these can simply be multiplied by $A$ to estimate the per-nucleus production cross-section, and again compared to the inclusive photonuclear cross-section to obtain a yield as a fraction of the inclusive photonuclear rate.  As shown in Table \ref{tab:BertiniVsPhysicsTableOneNeutron}, these reactions are expected to dominate the single-nucleon final state.

We have found two relevant mismodelings in the Bertini model for the $\gamma N\rightarrow \pi n$ processes.  The total rates for these processes are correctly modeled, but their angular distributions are not.  In particular, the final-state phase space for $\gamma p \rightarrow n \pi^+$ ($\gamma n \rightarrow n \pi^0$) is parameterized by a single, sharply forward-peaked function for photon energies above 1.77 GeV (2.4 GeV) parameterizing the $t$-channel (forward pion) pole.  The $u$-channel (forward neutron) pole contribution at few-GeV energies (see \cite{Anderson:1969bq}) accounts for only 1--2\% of the total cross-section, but is highly relevant for the LDMX veto since it leads to a forward-going neutron with the more easily detected pion backscattering out of acceptance.  Neglecting the $u$-channel pole would naively \emph{underestimate} the contributions of $\gamma N\rightarrow n\pi$ reactions to the single-neutron final state.  However, in addition the angular distributions of the pion and nucleon in the collision CM frame were reversed in the Bertini cascade code.  The combined effect of these two errors is to overestimate the rate of $\pi n$ final states in the effectively single-neutron phase space by a factor of 25-50.  

A corollary of the above is that the \emph{single-pion} final state, which is expected in approximately 0.1\% of photonuclear reactions, is essentially absent from our \geant samples (see Table \ref{tab:BertiniVsPhysicsTableOnePion} for an incomplete summary).  However, high-energy pions are expected to be much easier to veto than single neutrons: a forward $\pi^0$ simply converts the full energy of the incident photon into two $\gamma$'s, maintaining an essentially electromagnetic shower in the \ecal, while a forward $\pi^+$ leads to a MIP track in both the \ecal and the \hcal (similar to the case of muon pair production).  

\paragraph{Cascade Reactions}
Another class of process that contributes significantly to di-neutron final states is a cascade of reactions on single-nucleon targets: an initial photon-nucleon reaction produces a pion-neutron final state, and the pion subsequently elastically scatters off a neutron. 
Rates given in Table \ref{tab:BertiniVsPhysicsTableTwoNeutron} reflect a simple order-of-magnitude estimate of the probability for the 2nd reaction to occur, given by the ratio of the exclusive cross-section for elastic back-scattering ($\cos\theta_{cm}< -0.75$) to the total pion-nucleon cross-section, at a reference energy of 1.5 GeV.  A more careful calculation of this rate depends on the distribution of energy, angles, and intra-nuclear location of the initial reaction.  Indeed, the need to model such effects carefully is precisely the rationale for using the Bertini Cascade Model (with corrections to address the mismodelings noted above) rather than a stand-alone generator.  Our estimates suggest that cascades of single-nucleon processes have a higher rate than primary di-neutron dissociation, and that 
the total yield of di-neutron final states is overestimated by a factor of $\sim 500$.

\begin{table}[htbp]
\begin{tabular}{l| l l l l  | l  }
Process & $N_{T}/\rm{nucleus}$ & $\sigma_{T}^{inclusive}$ & kin.~requirement  & frac.~in $1\pi^\pm$ kinematics & Yield in $1\pi^+$ kinematics \\
\hline
$\gamma p \rightarrow \pi^+\, n$  & 74 & 2 $\mu$b  &  $q>0.92$ & 0.1  (0)  & $3\cdot 10^{-4}$ (0)\\
$\gamma n \rightarrow \pi^-\, p$  & 110 & 2 $\mu$b  &  $q>0.92$ &  0.1 (0)  & $5\cdot 10^{-4}$ (0)\\
$\gamma N \rightarrow N\pi \pi$ &  184 & 39 $\mu$b & * & ? & ${\cal O}(10^{-4})$ \\
\hline
\multicolumn{5}{l}{\bf Total} &  $\mathbf{\sim 0.001}$  \textbf{(${\cal O}(10^{-4})$)}\\
\end{tabular}
\caption{\label{tab:BertiniVsPhysicsTableOnePion} Estimated yield of effective single-charged-pion final states (with any additional hadrons backscattered or carrying $<100$ MeV kinetic energy) for the exclusive processes discussed in the text, in the benchmark case of a 3 GeV photon impinging on $^{184}$W. Where  we have found discrepancies between the Bertini Cascade model and our calculations, the rate in the Bertini Cascade model is given in parentheses.  The physical yield of such events is dominated by exclusive $\pi N$ final states where the pion is quite forward in the CM frame.  This kinematics is absent from the default Bertini model because the final state kinematics of the pion and neutron are reversed in the simulation.   The yield of such events in the \geant model is presumably dominated by the corners of $N\pi \pi$ phase space where two of the particles are backwards or slow-moving -- we have not estimated the fraction of events that populate this phase-space carefully, but expect it to be roughly percent-level, giving rise to a subdominant but not negligible component of the single-pion final states.  While the single-pion final state is underestimated in the default \geant model, we expect that it is much easier to reject than single-neutron final states, as it leaves distinctive signals in both the \ecal and the \hcal.}
\end{table}

\subsection{High-Energy Backscatters}
In large-scale simulations of photonuclear reactions for LDMX, we identified a surprisingly large number of events where much of the incident photon's energy was carried by a hadron produced at backward angles.  
Conservation of momentum requires that, for such a reaction to occur, the particle must recoil against a multi-nucleon composite which carries multi-GeV forward momentum but negligible energy.  The physics of such ``cumulative'' hadron production is related to intranuclear correlations.  Cumulative hadron production has been studied in many experiments --- it has been found to be well described by a near-exponential distribution.
Published data on cumulative hadrons from photonuclear reactions is limited to Copper and Lead targets and final-state hadron kinetic energies below 0.3 GeV (see e.g. \cite{Alanakian:1979gd}).  However, there is substantial evidence that the kinematics of cumulative hadrons is only weakly sensitive to the incident projectile and the nucleus \cite{PhysRevC.20.764}.  Data on cumulative hadron production in scattering of 400 GeV protons \cite{PhysRevC.20.764} (up to $p=1.52$~GeV [$T=0.85$~GeV] for the emitted hadron), in 14.5~GeV incident electronuclear reactions\cite{Degtyarenko:1994tt} (up to $T\approx 0.6$~GeV), and in 5~GeV incident electronuclear interactions \cite{StepanPrivateCommunication} (up to $p=2$~GeV [$T=1.25$~GeV])  all show an exponentially falling hadron energy distribution at angles above $60^{\circ}$, with a steeper fall-off at more backwards angles throughout the measured kinematic range.  

\begin{figure}
\includegraphics[height=0.4\textwidth]{sections/performance/figures/UMD_CLAS_vs_Geant4_electronuclear.png}
\includegraphics[height=0.4\textwidth]{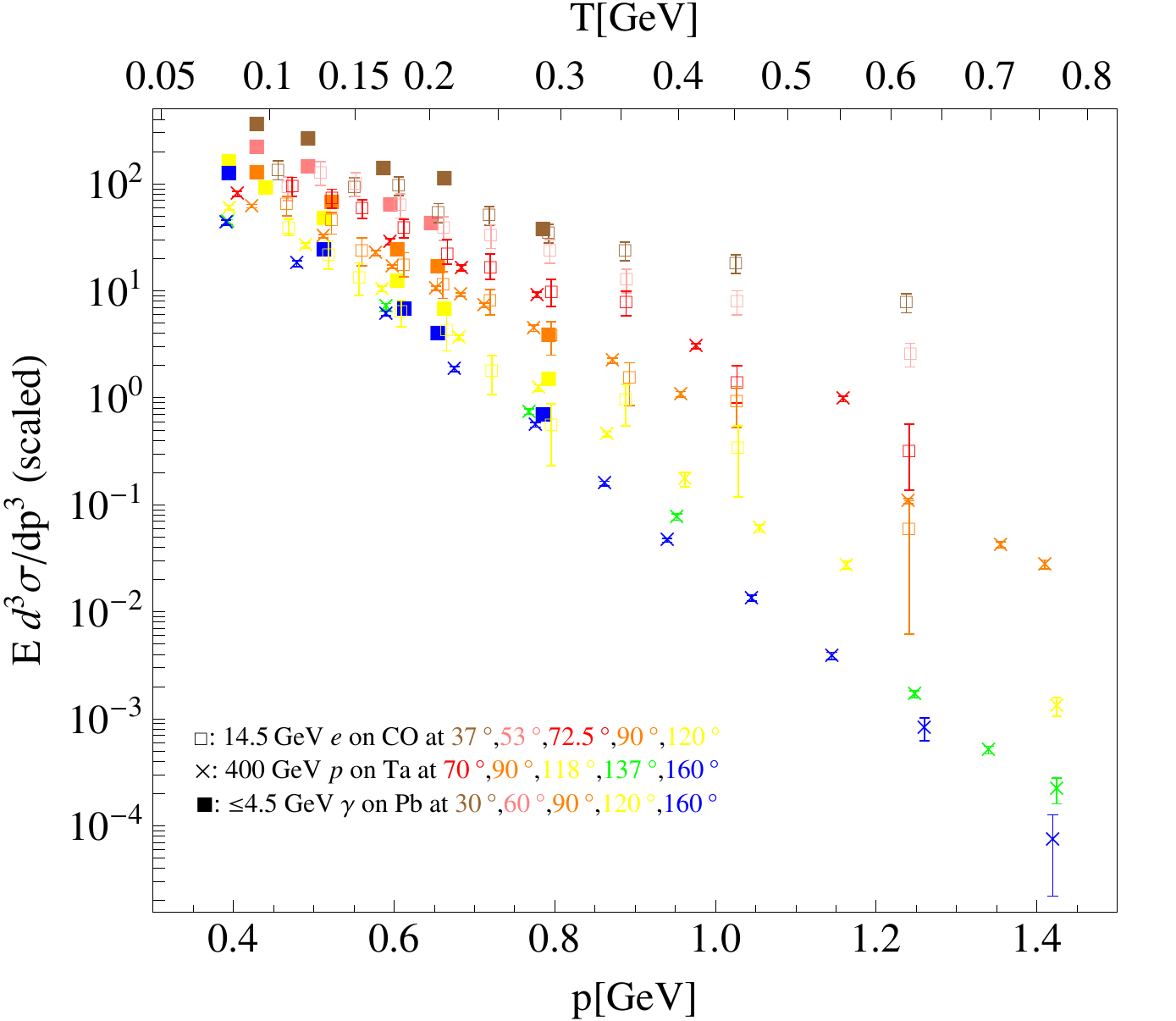}
\caption{\label{fig:dataVsGeant} Left: Distribution of proton momenta at $\theta_p > 100^\circ$ in 5 GeV electron scattering off $^{208}$Pb.  The black histogram shows the measured yield in a skim of CLAS eg2 data \cite{StepanPrivateCommunication}, for events with the electron in the CLAS acceptance.  The blue histogram shows the results of a \geant simulation before final-state down-weighting.  Right: Distribution of final-state proton momenta at angles from $30^\circ$ to $160^\circ$ for three different reactions (from \cite{Alanakian:1979gd} as plotted in \cite{Gudima:2006pn}, \cite{PhysRevC.20.764}, and \cite{Degtyarenko:1994tt}), illustrating the approximate universality of these distributions' shapes and their compatibility with an exponentially falling high-energy tail.}
\end{figure}

By contrast, \geant simulations of photonuclear and electro-nuclear reactions (with the physics list described above) show clear excesses over this exponential behavior for $p\gtrsim 1\text{-}1.3$~GeV.  This pattern of over-populating the wide-angle/backwards high energy tails is illustrated in Figure \ref{fig:dataVsGeant}.  The left panel compares the distribution of back-scattered protons ($>100^\circ$ polar angles) in electro-nuclear scattering from the CLAS eg2 experiment at Jefferson Lab \cite{StepanPrivateCommunication} to a \geant simulation with the same incident electron energy and target material (but not the same event selection).  The right panel shows the back-scattered and wide-angle proton yields in exclusive angle bins, for three different datasets from very different reactions,  illustrating in addition both the the angle-dependence of the physical distribution and the universality of these distributions for different reactions. This universality is further discussed in \cite{PhysRevC.20.764,Degtyarenko:1994tt}.

The artifact seen in Fig.~\ref{fig:dataVsGeant} (left) affects only a small fraction of photonuclear events (around $10^{-4}$), but events far out on this tail typically have no high-energy forward-going hadrons.  Thus, modeling these events more accurately informs the balance in designing the hadron veto between vetoing multi-GeV hadrons and those with much lower energies, as well as a potential floor to the experiments sensitivity.

By re-simulating these events with verbose debugging output turned on in the Bertini cascade, it was found that almost all of the events on the unphysical ``cumulative tail'' arose from a common (unphysical) mechanism: The \geant Bertini Cascade models the nucleus as a series of nested shells or ``zones'' with different densities, Fermi momenta, and potentials for nucleons in each region.  Accurately propagating low-energy hadrons in the nucleus requires a model for transport from one region to another, with reflections at the zone boundary if they do not have sufficient energy to penetrate the boundary.  In a thin-wall model, the condition for penetrating the boundary depends only on the radial component of the momentum.  Therefore, energetic hadrons hitting the boundary at grazing incidence, with large transverse momentum but negligible radial momentum, would reflect repeatedly until they interacted, transferring momentum to the remaining nucleus in the process.  This mechanism for reflection and momentum transfer is clearly unphysical --- in a realistic (continuously varying) nuclear potential, the angular momentum of a multi-GeV hadron is always more than sufficient to escape the nuclear potential.
  
\begin{figure}
    \includegraphics[height=0.3\textwidth]{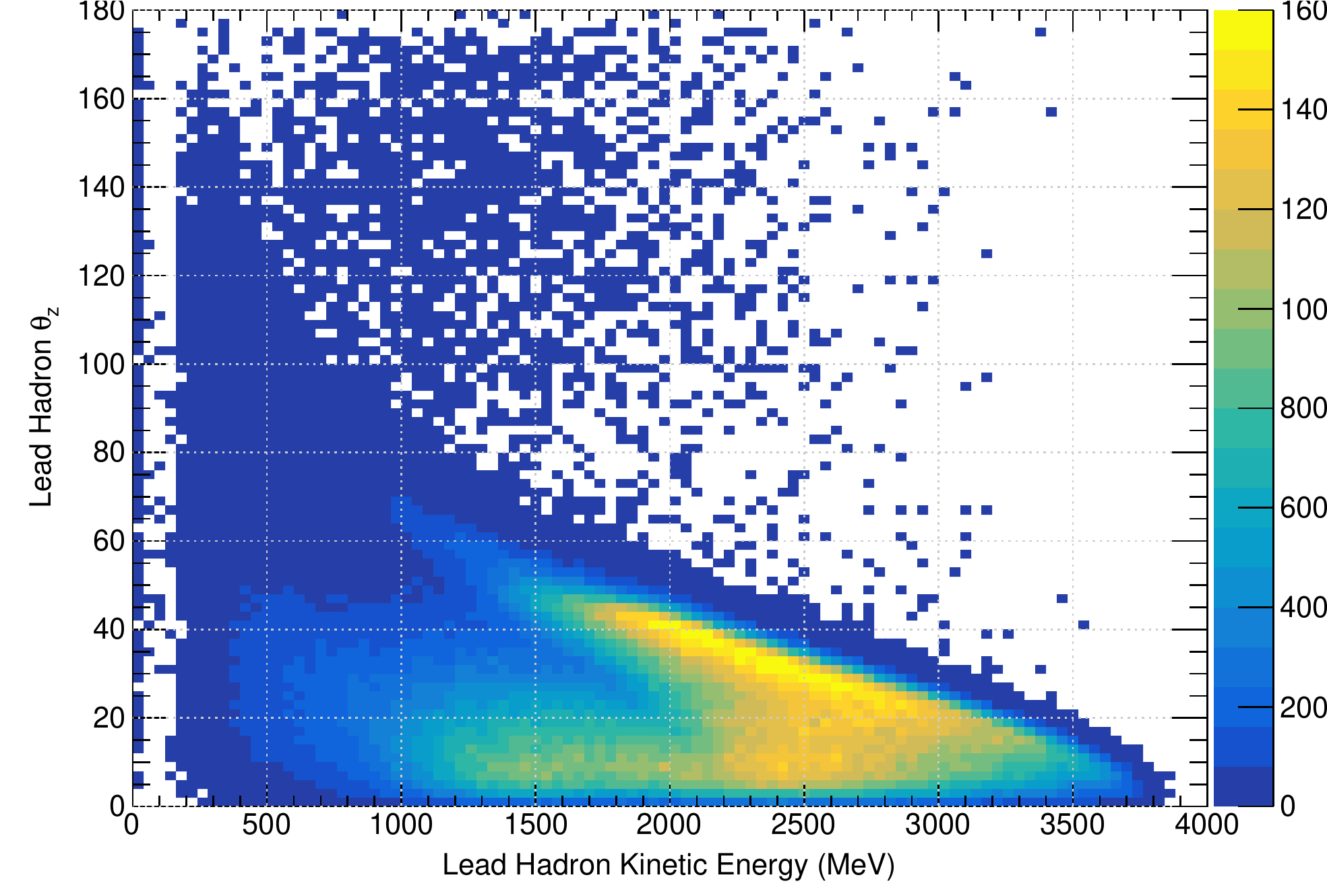}
    \includegraphics[height=0.3\textwidth]{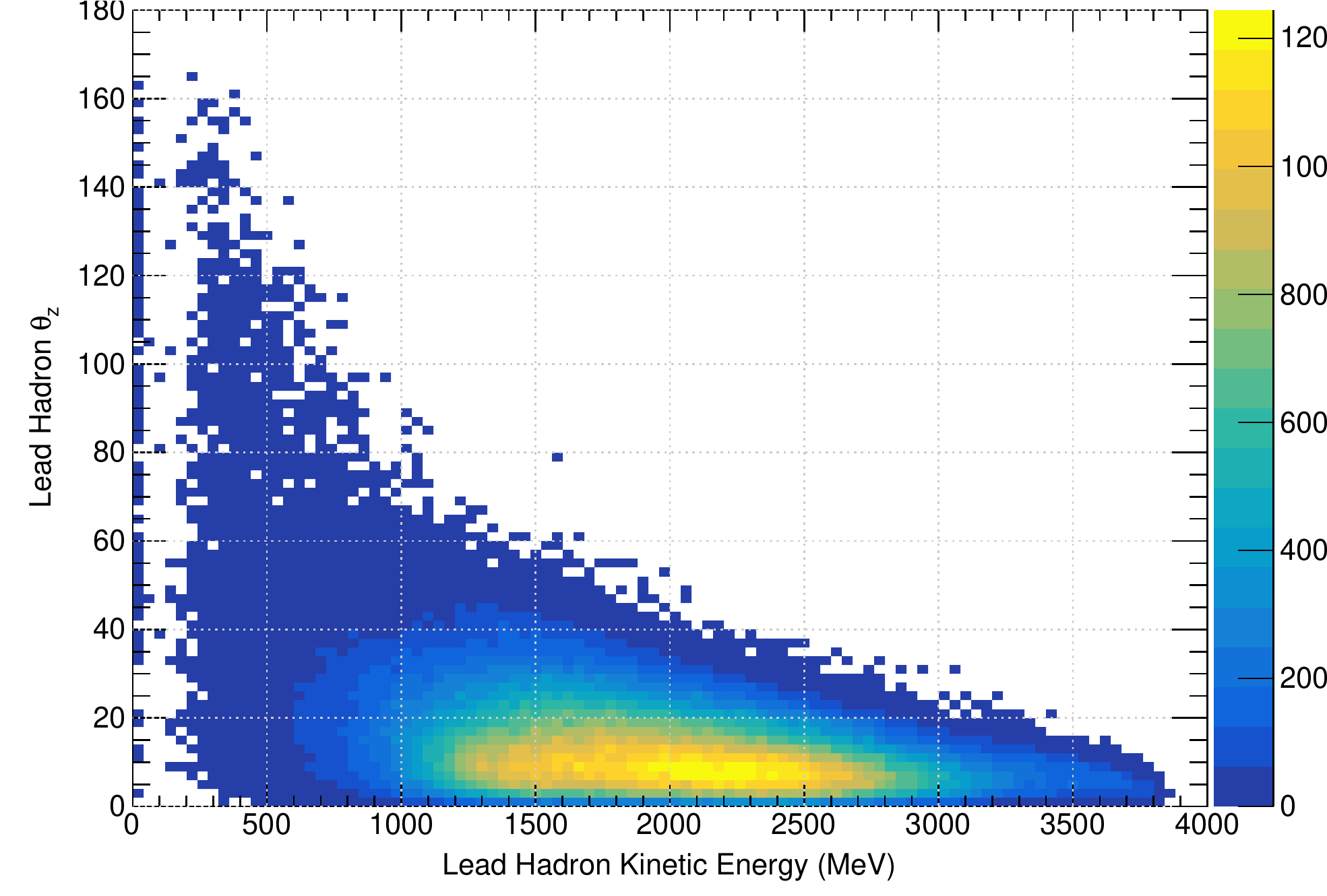}
    \caption{The energy-angle distribution of the leading hadron in a \pn 
             reaction prior to the implementation of a modified reflection
             condition (left) and after re-generation (right). After 
             re-generation, the cumulative tail is greatly reduced.}
    \label{fig:lead_hadron}
\end{figure}

We have implemented a modified reflection condition that accounts 
(conservatively) for the contribution of angular momentum to penetration of the
nuclear potential barrier, and are testing this in collaboration with the SLAC 
\geant group.  By re-generating a small numbers of events with this 
modification, we have verified that the tails shown in 
Fig.~\ref{fig:dataVsGeant} (left) is removed.  As shown in 
Fig.~\ref{fig:lead_hadron},  the impact of the change is also apparent in the 
energy-angle distribution of the leading hadron in the reaction.  The left 
plot in the figure shows the distribution prior to the implementation of the 
modified reflection condition, while right plot the distribution after 
re-generation. After the regeneration, the cumulative tail is greatly reduced.
We are still testing this code to see whether the more physical contributions 
to the cumulative tail, arising from the non-trivial momentum distribution of 
nucleons in the nucleus, is consistent with data, and plan to eventually 
incorporate this fix into the main LDMX photonuclear (and electro-nuclear) 
simulations.  

For the larger-scale simulations, the energy-angle distribution of the 
leading hadron after re-generation is parameterized and used to tag unphysical
events.  Specifically, if the angle of the leading hadron in an event with 
kinetic energy $T$ is greater than the expected value given by 
\begin{equation}
    \theta_{exp}(T) = \exp(-6\times10^{-4}\times T + 5.3)
\end{equation}
the event is tagged as unphysical.  Note that this parameterization is constructed
to be conservative. Unphysical events are disregarded in the 
background event counts given throughout the paper.

\end{appendices}

\bibliography{bibliography}

\end{document}